\documentclass[11pt,a4paper,oneside,openright]{book}
\usepackage[inner=4cm, outer=2cm, top=2.5cm, bottom=3cm]{geometry}

\usepackage{hyperref}
\usepackage{multimedia}
\usepackage{epsfig}
\usepackage{amsmath}
\usepackage{amssymb}
\usepackage{float}
\usepackage{relsize}
\usepackage{graphicx}
\usepackage{amsfonts}
\usepackage{amstext}
\usepackage{xcolor}
\usepackage{color}
\usepackage{fancybox}
\usepackage{tcolorbox}
\usepackage{url}
\usepackage{enumitem}
\usepackage{anyfontsize}
\usepackage{ctable}
\usepackage{bm}
\usepackage{array}
\usepackage{harpoon}
\usepackage{indentfirst}
\usepackage{calligra}
\usepackage[T1]{fontenc}
\usepackage{mathabx}
\usepackage[titletoc]{appendix}
\usepackage[italicdiff]{physics}
\usepackage{tgpagella}

\input{tcilatex}
\begin{document}
	
	\tolerance=1
	\emergencystretch=\maxdimen
	\hyphenpenalty=10000
	\hbadness=10000

\frontmatter

\begin{titlepage}
    \begin{center}
        
        \vspace*{0.5cm}
        
        
        \huge
        \textbf{Nonlinear Classical and Quantum Integrable Systems with $\mathcal{PT}$-symmetries}

        \vspace{1cm}
        
        \Large
        \textbf{Julia Cen}
        
        \vspace{0.5cm}
        
        \textbf{A Thesis Submitted for the Degree of Doctor of Philosophy} 
		
		\vspace{1.2cm}
		
	    \includegraphics[width=0.5\textwidth]{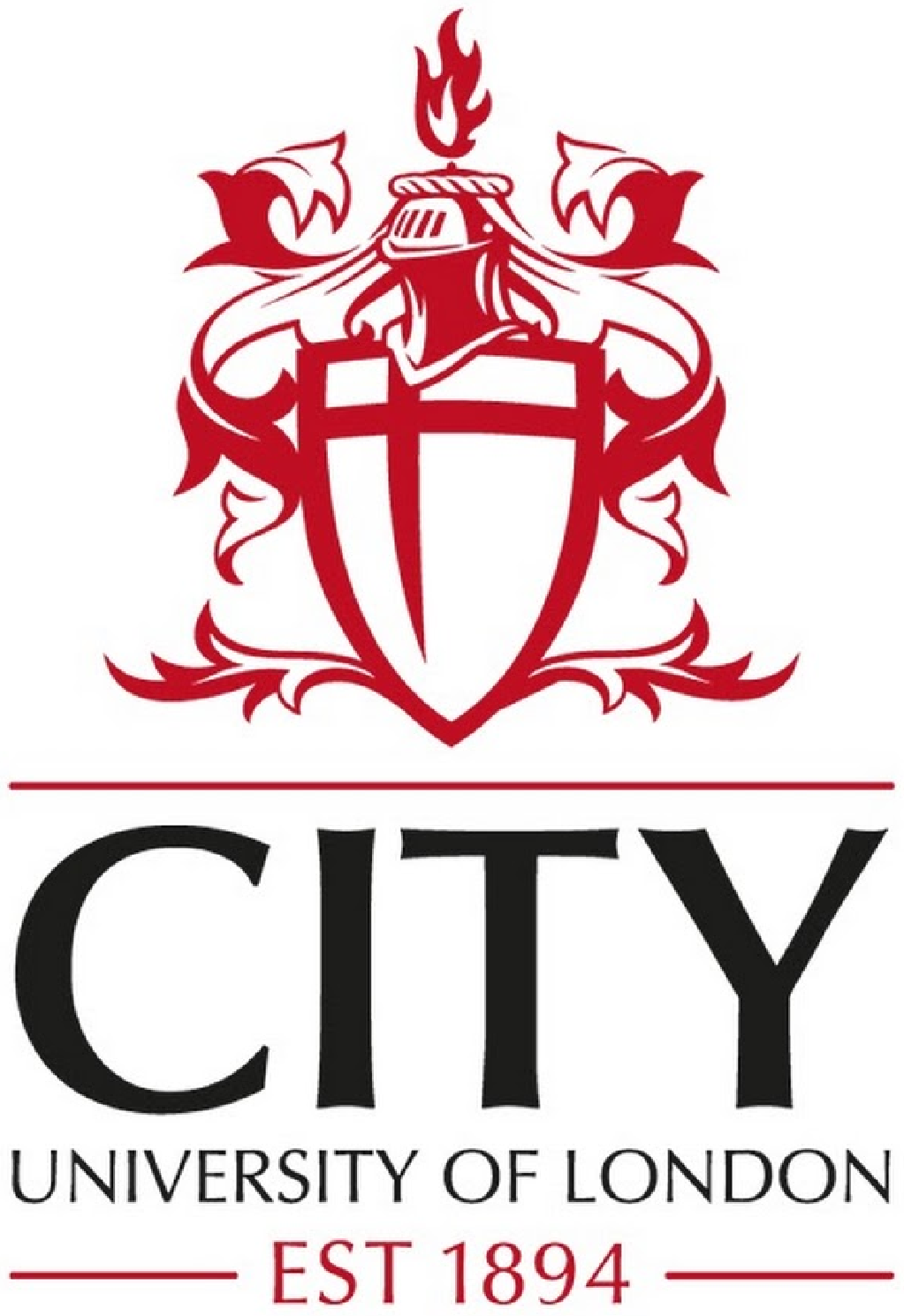}

		\vspace{1cm}
		
		\textbf{School of Mathematics, Computer Science and Engineering}\\
		
        \textbf{Department of Mathematics}\\
        
        \vspace{1cm}
        
        \textbf{Supervisor : Professor Andreas Fring}\\
        
        \vspace{1cm}
        
        \textbf{September 2019}
        
    \end{center}
\end{titlepage}

\chapter*{Thesis Commission}

\noindent Internal examiner\\
{\large\textbf{Dr Olalla Castro-Alvaredo}}\\
Department of Mathematics,\\ 
City, University of London, UK\\

\noindent External examiner\\
{\large\textbf{Professor Andrew Hone}}\\
School of Mathematics, Statistics and Actuarial Science\\
University of Kent, UK \\

\noindent Chair\\
{\large\textbf{Professor Joseph Chuang}}\\
Department of Mathematics,\\ 
City, University of London, UK


\newpage
\null\vspace{5cm}
\begin{center}
	{\Huge {\calligra{To my parents.}}}
\end{center}
\vspace{\fill}
\newpage


\tableofcontents

\chapter*{Acknowledgements}
\addcontentsline{toc}{chapter}{Acknowledgements}
First and foremost I would like to give my deepest heartfelt thanks to the person I am most indebted to, Professor Andreas Fring, my teacher, mentor and friend. For introducing me to the topics in this research and teaching me so much. For his immense support, advice, encouragement and countless hours devoted to me. For always making time for me in times of need, even during his busiest schedule. For his patience and enthusiasm to help me overcome the many hurdles on this journey and keep me on the right track. I am so blessed to have him as my supervisor with such insightful, invaluable guidance. I have greatly enjoyed our many hours of inspiring discussions. It has been a great pleasure and honour to work with him. \\

Second, I would like to extend my gratitude to my other collaborator and friends over the past few years, Dr Francisco Correa and Thomas Frith. For all the stimulating, fruitful, useful discussions and contributions, which without a doubt has greatly enhanced work in this thesis. \\

I wish to give many thanks also to my examiners, Dr Olalla Castro-Alvaredo and Professor Andrew Hone for kindly accepting the laborious task of assessing this thesis, their careful, thorough reading and insightful comments . \\

I am extremely grateful for all the opportunities, organisers and financial support from City, which has allowed me to participate and present at many great workshops, conferences and seminars. All of this has given me the chance to meet many incredible people and have invaluable discussions with. In particular, to Dr Clare Dunning and the funding from London Mathematical Society for the opportunity to organise and hold an ECR SEMPS workshop at City. Also to the Pint of Science organisation, for allowing me to be part of coordinating such a special international festival. \\

\vspace{1cm}

I owe special thanks to Dr Vincent Caudrelier, my tutor and supervisor during my undergraduate days, who introduced me to the fascinating world of solitons, challenging and teaching me to analyse with rigour like a mathematician. This extends to all my professors, lecturers and teachers. Without them, I would not be where I am today. \\

I thank everyone from the Mathematics Department, in particular my fellow officemates, for all the good times, memories shared and great friendships formed. \\

Lastly, I am most grateful to my friends and family outside of university. Most important of all, to my parents, for giving me my roots and wings, for their unconditional love and support.

\chapter*{Declaration}
\addcontentsline{toc}{chapter}{Declaration}

I declare that work presented in this thesis is my own except where stated otherwise and have given the appropriate reference or acknowledgement to the best of my knowledge to the work of others. This thesis has not been submitted elsewhere for a similar degree or qualification.

\vspace{-5cm}
\chapter*{Publications}
\addcontentsline{toc}{chapter}{Publications}

\begin{enumerate}
	\item J.Cen and A.Fring\\
	 \textbf{Complex solitons with real energies}\\
	 Journal of Physics A: Mathematical and Theoretical 49(36), 365202 (2016)\\
	 \textbf{Chapter 3}
	\item J. Cen, F. Correa, and A. Fring\\
	 \textbf{Time-delay and reality conditions for complex solitons}\\
	  Journal of Mathematical Physics 58(3), 032901 (2017) \\
	  \textbf{Chapters 3,5}
	\item J. Cen, F. Correa, and A. Fring\\
	 \textbf{Degenerate multi-solitons in the sine-Gordon equation}\\
	  Journal of Physics A: Mathematical and Theoretical 50(43), 435201 (2017) \\
	  \textbf{Chapter 5}
	  \item J.Cen and A.Fring\\
	  \textbf{Asymptotic and scattering behaviour for degenerate multi-solitons in the Hirota equation}\\
	  Physica D: Nonlinear Phenomena 397, 17-24 (2019)  \\
	  \textbf{Chapter 6}
	  \item J.Cen, F.Correa, and A.Fring\\
	  \textbf{Integrable nonlocal Hirota equations}\\
	  Journal of Mathematical Physics 60 (8), 081508 (2019) \\
	  \textbf{Chapter 7}
	  
	  \vspace{2cm}
	  
	\item J.Cen, A.Fring and T.Frith\\
	 \textbf{Time-dependent Darboux (supersymmetric) transformations for non-Hermitian quantum systems}\\
	  Journal of Physics A: Mathematical and Theoretical 52 (11), 115302 (2019) \\
	  \textbf{Chapter 9}
	\item J.Cen and A.Fring\\
	 \textbf{Multicomplex solitons}\\
	 Journal of Nonlinear Mathematical Physics 27 (1), 17-35 (2020)\\
	  \textbf{Chapter 4}
	\item J.Cen, F.Correa, and A.Fring\\
	\textbf{ Nonlocal gauge equivalence: Hirota versus extended
	continuous Heisenberg and Landau-Lifschitz equation}\\
	arXiv:1910.07272 (2019)\\
	\textbf{Chapter 8}
\end{enumerate}


\chapter*{List of abbreviations}
\addcontentsline{toc}{chapter}{List of abbreviations}

\begin{enumerate}
	\item [AKNS]   \quad Ablowitz, Kaup, Newell and Segur
	\item [BT]   \quad B\"{a}cklund transformation
	\item [DCT]   \quad Darboux-Crum transformation
	\item [DT]   \quad Darboux transformation
	\item [ECH]   \quad Extended continuous Heisenberg 
	\item [ELL]   \quad Extended Landau-Lifschitz 
	\item [HDM]   \quad Hirota's direct method
	\item [KdV]   \quad Korteweg-de Vries
	\item [mKdV]   \quad modified Korteweg-de Vries
	\item [NLS]   \quad nonlinear Schr\"{o}dinger
	\item [NPDE]   \quad nonlinear partial differential equation
	\item [PDE]   \quad partial differential equation
	\item [$\mathcal{PT}$]  \quad joint parity and time reversal
	\item [SG]   \quad sine-Gordon
	\item [TD]   \quad time-dependent
	\item [ZC]   \quad zero-curvature
\end{enumerate}

\chapter*{Abstract}
\addcontentsline{toc}{chapter}{Abstract}

A key feature of integrable systems is that they can be solved to obtain exact analytical solutions. In this thesis we show how new models can be found through generalisations of some well known nonlinear partial differential equations including the Korteweg-de Vries, modified Korteweg-de Vries, sine-Gordon, Hirota, Heisenberg and Landau-Lifschitz types with joint parity and time symmetries whilst preserving integrability properties. 

The first joint parity and time symmetric generalizations we take are extensions to the complex and multicomplex fields, such as bicomplex, quaternionic, coquaternionic and octonionic types. Subsequently, we develop new methods from well-known ones, such as Hirota's direct method, B\"{a}cklund transformations and Darboux-Crum transformations to solve for these new systems to obtain exact analytical solutions of soliton and multi-soliton types. Moreover, in agreement with the reality property present in joint parity and time symmetric non-Hermitian quantum systems, we find joint parity and time symmetries also play a key role for reality of conserved charges for the new systems, even though the soliton solutions are complex or multicomplex.

Our complex extensions have proved to be successful in helping one to obtain regularized degenerate multi-soliton solutions for the Korteweg-de Vries equation, which has not been realised before. We extend our investigations to explore degenerate multi-soliton solutions for the sine-Gordon equation and Hirota equation. In particular, we find the usual time-delays from degenerate soliton solution scattering are time-dependent, unlike the non-degenerate multi-soliton solutions, and provide a universal formula to compute the exact time-delay values for scattering of N-soliton solutions.

Other joint parity and time symmetric extensions of integrable systems we take are of nonlocal nature, with nonlocalities in space and/or in time, of time crystal type. Whilst developing new methods for the construction of soliton solutions for these systems, we find new types of solutions with different parameter dependence and qualitative behaviour even in the one-soliton solution cases. We exploit gauge equivalence between the Hirota system with continuous Heisenberg and Landau-Lifschitz systems to see how nonlocality is inherited from one system to another and vice versa.

In the final part of the thesis, we extend some of our investigations to the quantum regime. In particular we generalize the scheme of Darboux transformations for fully time-dependent non-Hermitian quantum systems, which allows us to create an infinite tower of solvable models.

\mainmatter

\chapter{Introduction}\label{ch_1}

The area of theoretical physics is the marriage of mathematics and physics with the purpose of formulating mathematical descriptions of the reality we live in. A key area of investigation is the modelling of natural phenomena and finding solutions to the models obtained. For realistic systems, due to the natural nonlinearity and many factors influencing the system, it is often difficult to find solutions and we usually have to resort to numerical or perturbation methods to solve them. This is one of the most important motivations for the study of integrable systems. The speciality and main aspect of integrable systems are that they can be solved to obtain exact analytical solutions, which are rare and powerful. They form the basis of finding numerical or approximate solutions in perturbation theory for new realistic models we find. Hence, the increasing interest to look not only at new solutions for some typical or representative models, but also to discover new integrable systems. 

Over the past two decades, an immense amount of investigations has been done for non-Hermitian systems with joint parity and time symmetries \cite{bender_pt_2019}. The interest comes from breaking the long-held belief from quantum mechanics that only Hermitian systems ensures unitary time evolution and possess real energies. Non-Hermitian systems with joint parity and time symmetries were also found to possess real energies. It is the interest of this thesis to explore how these symmetries can help us develop new classical nonlinear integrable models of complex, multicomplex, degenerate and nonlocal types, then extending to continuous spin models and the quantum regime for investigations of time-dependent non-Hermitian systems.

In \textbf{Chapter 2} we present a very brief selective history of the vast and rich area of classical integrable systems and list some of the key equations to be investigated in this thesis. Following on, we give a short introduction also to joint parity and time-symmetric non-Hermitian systems. In the second part of the chapter, we give a review for various types of well-established methods developed over the years in integrable systems for the construction of soliton solutions, that forms the foundation of a large part of our work.

Our story begins in \textbf{Chapter 3} \cite{cen_complex_2016,cen_time-delay_2017} on the investigation of complex extensions to some well-known real nonlinear integrable systems.  Integrability is a very delicate property; usually taking deformations destroys this property. However, we show under some particular complex joint parity and time-symmetric deformations, new models can be constructed from real integrable ones, which preserves integrability properties. Utilising well-established construction methods to construct soliton solutions, we derive new complex soliton solutions for these systems. In particular, despite the solutions being complex, the resulting energies are real, hence physically meaningful, and we present the argument for this.

After carrying out work in the complex regime, a good question to ask at this point is, what are the results and properties of higher order complex extensions, for example with bicomplex, quaternionic, coquaternionic and octonionic types? In \textbf{Chapter 4} \cite{cen_multicomplex_2018}, we take the extensions for those and develop further models of multicomplex types. In particular with the investigation of solution methods, new methods and solutions are developed using idempotent bases and a 'combined' imaginary unit for noncommutative extensions. Here, we find fascinating properties associated with spectral parameters and also reality of conserved quantities.

In \textbf{Chapter 5}, we provide an application for complex soliton solutions from Chapter 3. In particular, we see in this chapter how extensions to the complex domain can even help regularise singularities that come about when taking degeneracies of real multi-soliton solutions i.e. a single spectral parameter multi-soliton solution, through a review for the Korteweg-de Vries equation following \cite{correa_regularized_2016}. Inspired by the success of this application, we are led to investigate degeneracy for the sine-Gordon equation \cite{cen_degenerate_2017}. However, in the sine-Gordon case, we are lucky to find no singularities are developed when taking degeneracies of real multi-soliton solutions. Nevertheless, we develop an easier and a much more convenient method of producing degenerate solutions, which is a new recursive formula. Furthermore, we obtain new degenerate solutions with Jacobi elliptic and theta functions. It is one of the defining features of classical multi-soliton solutions to nonlinear integrable equations that individual one-soliton contributions maintain their overall shape before and after a scattering event. The only net effect is that they are delayed or advanced in time as a result of the scattering with other solitons when compared to the undisturbed motion of a single one-soliton solution. For non-degenerate multi-soliton solutions, time-delays tend to some constant in the asymptotes. Investigating this feature for degenerate multi-soliton solutions, time-delays are found to be time-dependent. Moreover, a universal formula for time-delays is shared between the Korteweg-de Vries and sine-Gordon cases \cite{cen_time-delay_2017,cen_degenerate_2017}.

Another interesting example of degeneracy to think about is in the Hirota equation, which can be seen as a joint parity and time-symmetric extension of the nonlinear Schr\"{o}dinger equation, and is already a complex system. We investigate in \textbf{Chapter 6} \cite{cen_asymptotic_2019} the degeneracy for this equation and compute new degenerate soliton solutions. Conducting detailed analysis of scattering and asymptotic properties, interesting scattering behaviours are found, and although the time-delays are time-dependent, they are no longer of the same universal form as for the Korteweg-de Vries and sine-Gordon cases.

 \textbf{Chapter 7} \cite{cen_integrable_2019} presents some new types of models from various transformations with a combination of parity and/or time-symmetries. As a result, new Hirota systems are found that are 'nonlocal' in space and/or in time, whilst retaining integrability properties. We also develop here new methods which implement nonlocality to find 'nonlocal' solutions. In particular, unlike the local case, we discover there two types of solutions for the nonlocal scenario.

Interestingly, the nonlinear Schr\"{o}dinger equation is related with the continuous Heisenberg spin model through a gauge equivalence. Knowing that the Hirota equation is an extension of the nonlinear Schr\"{o}dinger equation, it is a natural step to extend our investigations to look at the connection with an extended continuous Heisenberg spin model and an extended Landau-Lifschitz model forming the work in \textbf{Chapter 8} \cite{cen_gauge}. Independently from the Hirota case, we also develop new methods to construct solutions for 'nonlocal' extended continuous Heisenberg and Landau-Lifschitz equations. Then we extend the gauge equivalence for our new nonlocal Hirota systems and their solutions to find the corresponding gauge equivalence in the spin models and vice versa.

Up to now, we have only focused our investigations on the classical regime. In \textbf{Chapter 9} \cite{cen_time-dependent_2019}, we move into the quantum regime, looking at fully time-dependent non-Hermitian quantum systems. In particular, we develop a new scheme utilising Darboux transformations from our classical investigations and the Dyson equation to construct a hierarchy of solvable time-dependent joint parity and time-symmetric non-Hermitian potentials. Extensions of the scheme to Lewis-Riesenfeld invariants are presented, which are sometimes useful tools for solving time-dependent quantum systems. As a result, we obtain a powerful scheme, presenting us with various paths to build a solvable hierarchy of time-dependent non-Hermitian potentials, which we can choose from depending which path is easier.

\chapter{Integrability, $\mathcal{PT}$-symmetry and soliton solution methods}\label{ch_2}

\section{Classical Integrability}

The early notion of integrable systems dates back to the mid 19th century in the sense of Liouville integrability. Liouville's theorem states that given a Hamiltonian system on a 2n dimensional phase space (n being a finite number), if we are able to find n number of independent conserved quantities in involution, then the system can be solved analytically by quadratures \cite{liouville_note_1855,babelon_introduction_2003}. 

Towards the end of the 19th century, the experimental discovery of the solitary wave by Russell \cite{russell_report_1845}, a wave which behaves like a particle, preserving speed and shape as it travels, along with theoretical work of Boussinesq and Rayleigh \cite{boussinesq_essai_1877,rayleigh_xxxii._1876}, Korteweg and de Vries \cite{korteweg_change_1895} to describe such a phenomenon was the start of the development on the theory of integrability for Hamiltonian systems with infinitely many degrees of freedom, in particular integrable nonlinear partial differential equations (NPDEs). Since the 1960s, extensive work has been done in the area of integrable NPDEs. In 1965, Zabusky and Kruskal carried out numerical investigations to look at scattering of multi-soliton solutions to the Korteweg-de Vries (KdV) equation \cite{zabusky_interaction_1965}. A interesting property they discovered was that after scattering, each multi-soliton constituent preserved its speed and shape apart from a phase difference. Later, Gardner, Green, Kruskal and Miura developed what we now call the inverse scattering transform to solve the KdV equation for exact soliton solutions \cite{gardner_method_1967}. In 1972, Wadati and Toda computed the exact form of phase shifts from scattering, generally for a N-soliton solution (N arbitrary) \cite{wadati_exact_1972}. Meanwhile, many other new methods were being developed to find exact analytical solutions for the KdV and other NPDEs and there have been a lot of discussions around different types of definitions of integrability and there is in general no complete one definition. Some common, necessary characteristics of integrability for NPDEs includes possessing:

\begin{enumerate}
	\item An infinite number of conserved quantities \cite{miura_korteweg-vries_1968, shabat_exact_1972}.
	\item Lax pair representation \cite{gardner_method_1967, lax_integrals_1968, shabat_exact_1972}.
	\item Exact analytical soliton solutions \cite{gardner_method_1967, hirota_exact_1971, lamb_jr_analytical_1971, matveev_darboux_1991}.
	\item Infinitely many local commuting symmetries \cite{mikhailov_the_1991}.
\end{enumerate}

Before we move on to introducing joint parity and time ($\mathcal{PT}$) symmetries, we list first various well-known integrable NPDEs that will form the basis for development of new integrable models with $\mathcal{PT}$-symmetric deformations. Some of the systems we will explore are:

\vspace{0.5cm}

\noindent \large\textbf{The Korteweg-de Vries (KdV) equation}
\begin{equation}
u_{t}+6uu_{x}+u_{xxx}=0 \label{kdv}
\end{equation}

\noindent This equation was first theoretically developed by Boussinesq and Rayleigh and later Korteweg and de Vries \cite{boussinesq_essai_1877,rayleigh_xxxii._1876,korteweg_change_1895}. It is famous for describing shallow water waves.

\vspace{0.5cm}

\noindent\large\textbf{The modified Korteweg-de Vries (mKdV) equation}
\begin{equation}
v_{t}+24v^{2}v_{x}+v_{xxx}=0
\label{mkdv1}
\end{equation}

\noindent In 1968, the Miura transformation was found to relate the KdV equation with the mKdV equation \cite{miura_kdvandmkdv_1968}. In Chapter 3, we present the map between (\ref{kdv}) and (\ref{mkdv1}).

\vspace{0.5cm}

\noindent\large\textbf{The sine-Gordon (SG) equation}
\begin{equation}
\phi _{xt}=\sin \phi \label{sg}
\end{equation}

The SG equation was first discovered by Bour \cite{bour_th_1861} in mathematics in the context pseudospherical surfaces in differential geometry. Later, this equation proved to be of great physical significance in other areas such as particle physics \cite{perring_a_1961} and Josephson junctions \cite{josephson_supercurrents_1965}.

\vspace{0.5cm}

\noindent\large\textbf{The Hirota equation}
\begin{equation}
iq_{t}+\alpha\left(q_{xx}-2q^{2}r\right)+i\beta\left(q_{xxx}-6qrq_{x}\right)=0, \label{HE}
\end{equation}

\noindent with $r=\pm q^{*}$ is the Hirota equation \cite{hirota_exact_1973}, which reduces to the nonlinear Schr\"{o}dinger (NLS) equation for $\beta=0$ and complex mKdV equation for $\alpha=0$. The Hirota equation is a higher order extension of the NLS equation originally proposed by Kodama and Hasegawa \cite{kodama_nonlinear_1987} to model high-intensity and short pulse femtosecond wave pulses.

\vspace{0.5cm}

\noindent\large\textbf{The extended continuous Heisenberg (ECH) equation}

\begin{equation}
	S_{t}=i \frac{\alpha}{2}\left[S,S_{xx}\right]-\frac{\beta}{2}\left(3S_{x}^{3}+S\left[S,S_{xxx}\right]\right) \label{ech}
\end{equation}
The ECH equation, where $S$ as a $2 \times 2$ matrix of $SU(2)$ type, is the first member of the corresponding Heisenberg hierarchy \cite{wang_darboux_2005}. For $\beta=0$, it reduces to the well-known continuous limit of the Heisenberg spin chain \cite{nakamura_solitons_1974,lakshmanan_dynamics_1976,tjon_solitons_1977,takhtajan_integration_1977}.

\vspace{0.5cm}

\noindent\large\textbf{The extended Landau-Lifschitz (ELL) equation}

\begin{equation}
\mathbf{\overrightharp{s}}_{t}=-\alpha \mathbf{\overrightharp{s}\times \overrightharp{s}}_{xx}-\frac{3}{2}\beta \left( 
\mathbf{\overrightharp{s}}_{x}\cdot \mathbf{\overrightharp{s}}_{x}\right) \mathbf{\overrightharp{s}}_{x}+\beta \mathbf{%
	\overrightharp{s}\times }\left( \mathbf{\overrightharp{s}\times \overrightharp{s}}_{xxx}\right)  \label{ELLE}
\end{equation}

\noindent The ELL equation is an interesting vector variant of the ECH equation (\ref{ech}) with many physical applications that arises when decomposing $S$ in the standard
fashion as $S=\textbf{\overrightharp{s}}\cdot \textbf{\overrightharp{$\mathbf{\sigma}$}}$ with Pauli matrices vector $\textbf{\overrightharp{$\mathbf{\sigma}$}} = \left( \sigma _{1},\sigma _{2},\sigma _{3} \right) $. For $\beta \rightarrow 0$ this
equation reduces to the standard Landau-Lifschitz equation \cite{landau_zur_1935,baryakhtar_landau-lifshitz_2015}.

\section{Joint parity and time ($\mathcal{PT}$) symmetry}

In quantum mechanics, it is well-known that Hermitian systems possess real energy eigenvalues and have unitary time evolution/conservation of probability. These are closed systems, which are isolated and do not have any interaction with their environment. The other case are open systems, which do interact with their environment and probability is not conserved in the system. These systems are termed non-Hermitian and have been long known to describe dissipation with generically complex energy eigenvalues. 

In 1998, Bender and Boettcher discovered a wider class of quantum systems which can possess real energy eigenvalues, under the restriction of $\mathcal{PT}$-symmetry \cite{bender_real_1998}. In particular, they discovered a range of non-Hermitian Hamiltonians with $\mathcal{PT}$-symmetry of the form 
\begin{equation}
H = p^{2} + x^{2}(i x)^{\epsilon} ,   \quad \epsilon > -1
\end{equation}

\noindent possessing real energy eigenvalues. Here, $\mathcal{PT}$-symmetry implies that the Hamiltonian is invariant under the action of the $\mathcal{PT}$ operator defined as
\begin{equation}
\mathcal{P} : x \rightarrow -x , p \rightarrow -p ,
\end{equation}
\vspace{-0.4cm}
\begin{equation}
\mathcal{T} : i \rightarrow -i , p \rightarrow -p ,
\end{equation}
\vspace{-0.4cm}
\begin{equation}
\mathcal{PT} : i \rightarrow -i , x \rightarrow -x , p \rightarrow p \text{.}
\end{equation}

\noindent The conjugation of the $i$, for instance can be made plausible by requirement of the canonical commutation relation to be satisfied, i.e. 
\begin{equation}
\mathcal{PT} : \left[x,p\right]=i \hbar \rightarrow -\left[x,p\right]=-i \hbar.
\end{equation}
\noindent Note that when taking the time-dependent Schr\"{o}dinger equation, the $\mathcal{T}$-operator will also involve $t \rightarrow -t$ for operators involving $t$. For the full time-dependent case where we have an explicit time-dependence such as in Chapter 9, we must be careful in distinguishing whether $t$ is part of a quantum mechanical operator or just a classical parameter. In the latter case, we do not take $t \rightarrow -t$ \cite{figueira_de_morisson_faria_time_2006,bender_PT_2011,  cen_time-dependent_2019}.

Reality of energy eigenvalues is the first indication that there is a possibility for a non-Hermitian $\mathcal{PT}$-symmetric systems to be a consistent quantum mechanical system. The reasoning for reality of energy eigenvalues from $\mathcal{PT}$-symmetry could be explained by an argument already presented by Wigner in 1960 \cite{wigner_normal_1960}. The $\mathcal{PT}$ operator is actually a special case of an antilinear operator, which is some operator $A$, with the properties

\vspace{0.5cm}

{\bfseries{(1)}} \quad \quad $A(f+g)=A f + A g$ ,

{\bfseries{(2)}} \quad \quad $A(c f)= c^{*} Af $ ,

\vspace{0.5cm}

\noindent where $f$ and $g$ are any functions and $c$, an arbitrary complex constant with $c^{*}$ denoting its complex conjugate. If we take a Hamiltonian $H$, with eigenstates $\psi$, eigenvalues $E$ and it satisfies the conditions

\vspace{0.5cm}

{\bfseries{(1)}} \quad \quad $ \left[ H, A \right] =0$ ,

{\bfseries{(2)}} \quad \quad $A \psi = \psi$ ,

\vspace{0.5cm}

\noindent then
\begin{center}
	$E \psi = H \psi 
	= H A \psi
	= A H \psi 
	= A E \psi 
	= E^{*} A \psi
	= E^{*} \psi \text{.}$
\end{center}

\noindent Hence, we have proved when the Hamiltonian and its eigenfunctions are both invariant under the $\mathcal{PT}$ operator, real energy eigenvalues are obtained. We call this the unbroken $\mathcal{PT}$ regime. 

When the Hamiltonian is invariant under the $\mathcal{PT}$ operator, but the eigenfunctions are not $\mathcal{PT}$ invariant, we no longer have real eigenvalues. Instead, we have conjugate pairs of eigenvalues and we are in the broken $\mathcal{PT}$ regime. The critical points where there is a transition from the unbroken to the broken $\mathcal{PT}$ regime and real eigenvalues coalesce in the parameter space, are called exceptional points \cite{kato_perturbation_1966,heiss_physics_2012}.

For the unbroken $\mathcal{PT}$ regime, with real eigenvalues, we now have a new possibility of finding meaningful quantum mechanics from non-Hermitian systems. To show a $\mathcal{PT}$-symmetric non-Hermitian system is a consistent quantum mechanical theory, one needs also to have unitary time evolution/conservation of probability and a well-defined inner product with completeness for the Hilbert space of the system. 

In Hermitian systems, Hermiticity guarantees orthogonality. For unbroken $\mathcal{PT}$ regime, since $H^{\mathcal{PT}}=H$, the natural choice of an appropriate metric for the construction of a well-defined inner product would be the $\mathcal{PT}$ inner product

\begin{equation}
\bra{\psi_{m}}\ket{\psi_{n}}_{\mathcal{PT}}:=\int\psi_{m}^{*}(-x)\psi_{n}(x) dx.
\end{equation}

\noindent Taking $\psi_{m},\psi_{n}$ to be eigenfunctions with eigenvalues $\lambda_{m},\lambda_{n}$ respectively, the orthogonality condition is satisfied as
\begin{eqnarray}
\bra{\psi_{m}}\ket{H\psi_{n}}_{\mathcal{PT}}&=&\bra{H^{\mathcal{PT}}\psi_{m}}\ket{\psi_{n}}_{\mathcal{PT}},\\
\bra{\psi_{m}}\ket{\lambda_{n}\psi_{n}}_{\mathcal{PT}}&=&\bra{\lambda_{m}\psi_{m}}\ket{\psi_{n}}_{\mathcal{PT}}, \notag\\
(\lambda_{n}-\lambda_{m})\bra{\psi_{m}}\ket{\psi_{n}}_{\mathcal{PT}}&=&0, \notag
\end{eqnarray}

\noindent hence 
\begin{equation}
\bra{\psi_{m}}\ket{\psi_{n}}_{\mathcal{PT}}=0 \quad \text{for} \quad m \neq n.
\end{equation}

 For $m = n$, we need the inner product to be positive definite, which may not always be the case. To remedy this, Bender, Brody and Jones \cite{bender_complex_2002} introduced the operator $\mathcal{CPT}$ where $\mathcal{C}$, with properties similar to the charge-conjugation operator, will cancel the negativity from negative normed states by multiplying them with $-1$. This provides us with positive definiteness 
\begin{equation}
\bra{\psi_{m}}\ket{\psi_{n}}_{\mathcal{CPT}}
\end{equation}
\noindent and hence the completeness relation
\begin{equation}
\sum_{n}\ket{n}\bra{n}_{\mathcal{CPT}}=1.
\end{equation}

The calculation for the $\mathcal{C}$ operator is generally a difficult problem as it relies on knowledge of the complete set of eigenfunctions. One can use perturbation theory to find $\mathcal{C}$, utilising properties of the $\mathcal{C}$ operator such as being another symmetry of the Hamiltonian and commuting with the $\mathcal{PT}$ operator
\begin{equation}
\mathcal{C}^{2}=1, \quad \left[H,\mathcal{C}\right]=0, \quad \left[\mathcal{C},\mathcal{PT}\right]=0.
\end{equation}

Finally, we also have unitary time evolution with $U=e^{-iHt}$ under the $\mathcal{CPT}$ metric and hence probability is conserved
\begin{eqnarray}
\bra{\psi(t)}\ket{\psi(t)}_{\mathcal{CPT}}&=&\bra{e^{-iHt}\psi(0)}\ket{e^{-iHt}\psi(0)}_{\mathcal{CPT}}\\
&=&\left\langle \psi(0) \left\vert e^{iH^{\mathcal{CPT}}t}e^{-iHt}\right\vert \psi(0)\right\rangle_{\mathcal{CPT}} \notag\\
&=&\bra{\psi(0)}\ket{\psi(0)}_{\mathcal{CPT}} \notag
\end{eqnarray}
where
\begin{equation}
i\psi_{t}=H\psi \quad \text{with} \quad \psi(t)=e^{-iHt}\psi(0).
\end{equation}
Therefore, we have found a well defined, positive definite inner product for a $\mathcal{PT}$-symmetric system to be the $\mathcal{CPT}$ inner product.

An alternative approach to find the $\mathcal{CPT}$ metric is to use the more general concept of quasi/pseudo-Hermiticity \cite{pauli_on_1943,dieudonne_quasi-hermitian_1961,scholtz_quasi-hermitian_1992,mostafazadeh_pseudo-hermiticity_2002,mostafazadeh_pseudo-hermiticity_2002-2,mostafazadeh_pseudo-hermiticity_2002-3,mostafazadeh_exact_2003}. The key idea is to find an operator $\eta$ such that it becomes a similarity transformation operator for the non-Hermitian Hamiltonian $H$ with a Hermitian Hamiltonian $h$
\begin{equation}
h=\eta H\eta^{-1}=(\eta^{-1})^{\dagger} H^{\dagger}\eta^{\dagger}=h^{\dagger} \quad \Leftrightarrow \quad H^{\dagger}=\rho H \rho^{-1},
\end{equation}
where $\rho=\eta^{\dagger}\eta$ is Hermitian, invertible and positive definite. The eigenfunctions $\phi$ and $\psi$ of $h$ and $H$ respectively are related by
\begin{equation}
\phi=\eta \psi.
\end{equation}
In addition, as $H$ and $h$ are related by a similarity transformation, both Hamiltonians have the same eigenvalues.

We can prove when taking $\rho$ as the new metric, $H$ is Hermitian with respect to the metric since 
\begin{center}
	$\bra{\psi_{m}}\ket{H\psi_{n}}_{\rho}:=\bra{\psi_{m}}\ket{\eta^{\dagger}\eta H\psi_{n}}=\bra{\phi_{m}}\ket{h^{\dagger}\phi_{n}}=\bra{H\psi_{m}}\ket{\eta^{\dagger}\eta\psi_{n}}=\bra{H\psi_{m}}\ket{\psi_{n}}_{\rho}$.
\end{center}
Therefore, eigenvalues are also real, eigenfunctions are orthogonal and we have a well-defined inner product, hence a consistent quantum mechanical framework.

Taking a $\mathcal{PT}$-symmetric non-Hermitian Hamiltonian the metric is
\begin{equation}
\rho=\mathcal{PC},
\end{equation}
since we can show 
\begin{center}
	$\rho^{-1}H^{\dagger}\rho=H=(\mathcal{CPT})H(\mathcal{CPT})^{-1}=(PC)^{-1}H^{\dagger}(PC),$
\end{center}

and the $\mathcal{C}$ operator is
\begin{equation}
\mathcal{C}=\rho^{-1}\mathcal{P}.
\end{equation}

The study of $\mathcal{PT}$-symmetry has grown tremendously over the past two decades in many areas of physics, including the classical side, which we will make a contribution to in this thesis.\\

\noindent {\Large\textbf{Soliton solution methods}}

\vspace{0.4cm}

NPDEs are generally difficult to solve due to their nonlinearity. Over the years, different methods have been investigated to help us construct exact analytical soliton solutions for various NPDEs. In the following sections, we will introduce various well-known methods that will form the basis for development of new methods to solve new NPDEs.

\section{Hirota's direct method (HDM)}

 HDM was first developed by Hirota in 1971 \cite{hirota_exact_1971} to directly construct exact N-soliton solutions for the KdV equation. Later, this method was developed for many other NPDEs, some of which we will explore, such as the mKdV \cite{hirota_exact_1972-1}, SG \cite{hirota_exact_1972}, Hirota \cite{hirota_exact_1973} equations. Given a NPDE, the key idea is to convert the nonlinear problem to a bilinear one through a transformation of the dependent variable. Then using the Hirota D-operator definition along with the various properties and identities arising from it, we can transform ordinary derivatives to Hirota derivatives. The resulting equation(s) will be called Hirota bilinear equation(s).

Before we present some examples, let us first present the definition of Hirota D-operator and some of its properties.

\vspace{0.5cm}

\noindent{\large \bfseries{Hirota D-operator for one independent variable}}
	
\noindent The Hirota D-operator for functions $f$ and $g$ of one independent variable $x$ reads
	\begin{equation}
	D_{x}^{n}(f\cdot g)=\left. \frac{\partial ^{n}}{\partial y^{n}}%
		f(x+y)g(x-y)\right\vert _{y=0}\text{ .}  \label{1.7}
	\end{equation}

\vspace{0.5cm}

Recalling the Taylor expansion for functions $f(x)$ and $g(x)$ of one independent variable $x$ around points $y$ and $-y$ respectively and multiplying them together, we can rewrite the Taylor expansion of a product of two functions using the definition of the Hirota D-operator as a generating function, that is
\begin{equation}
f(x+y)g(x-y)=\sum\limits_{n=0}^{\infty }\frac{y^{n}}{n!}D_{x}^{n}(f\cdot g)=e^{yD_{x}}f\cdot g%
\text{ .}  \label{1.8}
\end{equation}

\noindent With this, we can explicitly write out the first few Hirota derivatives.
\begin{eqnarray}
D_{x}\left( f\cdot g\right) &=&f_{x}g-g_{x}f\text{ ,}  \label{1.9} \\
D_{x}^{2}\left( f\cdot g\right) &=&f_{xx}g-2f_{x}g_{x}+fg_{xx}\text{ ,}
\label{1.10} \\
D_{x}^{3}\left( f\cdot g\right)
&=&f_{xxx}g-3f_{xx}g_{x}+3f_{x}g_{xx}-fg_{xxx}\text{ ,}  \label{1.11} \\
D_{x}^{4}\left( f\cdot g\right)
&=&f_{xxxx}g-4f_{xxx}g_{x}+6f_{xx}g_{xx}-4f_{x}g_{xxx}+fg_{xxxx}\text{ .}
\label{1.12}
\end{eqnarray}

\noindent More generally, the derivatives can be written similarly to the Leibniz rule, but with alternating signs
\begin{equation}
D_{x}^{n}\left( f\cdot g\right) =\sum\limits_{k=0}^{n}\binom{n}{k}(-1)^{k}%
\frac{\partial ^{n-k}}{\partial _{x^{n-k}}}f(x)\frac{\partial ^{k}}{\partial
	_{x^{k}}}g(x)\text{ .}  \label{1.13}
\end{equation}

\noindent Let us now consider some properties and identities of the Hirota D-operator, which will help us to convert our NPDEs to Hirota bilinear equation(s).

\noindent \begin{itemize}[leftmargin=*]
	\item \textbf{Property one}
	
	Due to the alternating signs of the Leibniz rule (\ref{1.13}), switching the order of the functions $f(x)$ and $g(x)$, we find the property
	\begin{equation}
	\fbox{$D_{x}^{n}\left( f\cdot g\right) =(-1)^{n}D_{x}^{n}\left( g\cdot
		f\right) $}\text{ ,}  \label{1.14}
	\end{equation}
	
	\noindent obtaining also
	\begin{equation}
	\fbox{$D_{x}^{2n-1}\left( f\cdot f\right) =0$}\text{ .}  \label{1.15}
	\end{equation}
	
	\noindent \item \textbf{Property two}
	
	Take the Taylor expansions of function $f(x)$ around points $y$ and $-y$, then applying the logarithm on the expansions and adding them together yields
	\begin{eqnarray}
	\ln \left[ f(x+y)f(x-y)\right]&=&e^{y\frac{\partial }{\partial _{x}}}\ln f(x)+e^{-y\frac{\partial }{\partial _{x}}}\ln f(x)\text{ ,}\label{1.18}\\
	&=&2\cosh \left( y\frac{\partial }{\partial x}\right) \ln f(x)\text{ .}  \notag
	\end{eqnarray}
	
	\noindent In addition, the left hand side of (\ref{1.18}) can also be expressed as
	\begin{eqnarray}
	\hspace{-1cm}
	\ln \left[ f(x+y)f(x-y)\right] &=&  \ln \left[ \frac{1}{2}e^{yD_{x}}f \cdot f+\frac{1}{2}e^{-yD_{x}}f \cdot f \right] \text{ ,}  \label{1.19} \\
	&=&\ln \left[ \cosh \left( yD_{x}\right) \left( f\cdot f\right) \right]
	\text{ .}  \notag
	\end{eqnarray}
	
	\noindent As a result, we obtain the following property in terms of $\cosh$ functions
	\begin{equation}
	\fbox{$2\cosh (y\frac{\partial }{\partial x})\ln f=\ln \left[ \cosh
		(yD_{x})\left( f\cdot f\right) \right] $}\text{ .}  \label{1.20}
	\end{equation}
	
	\noindent Taylor expanding the $\cosh$ functions on the right- and left-hand sides of (\ref{1.20}) and comparing the coefficients of the $y$ terms, we can obtain the following identities:
	\begin{equation}
	\fbox{$2\partial _{x}^{2}\ln f=\frac{D_{x}^{2}\left( f\cdot f\right) }{f^{2}}
		$}\text{ ,}  \label{1.23}
	\end{equation}%
	\begin{equation}
	\fbox{$2\partial _{x}^{4}\ln f=\frac{D_{x}^{4}\left( f\cdot f\right) }{f^{2}}%
		-3\left( \frac{D_{x}^{2}\left( f\cdot f\right) }{f^{2}}\right) ^{2}$}\text{ .%
	}  \label{1.24}
	\end{equation}
	
	\noindent Identities for higher order derivatives may be derived similarly. 
\end{itemize}

	\vspace{0.3cm}
	\noindent {\large \bfseries{Hirota D-operator for two independent variables}}
	
	\noindent The Hirota D-operator for functions $f(x,t)$ and $g(x,t)$ of two independent variables reads
	\begin{equation}
		D_{x}^{n}D_{t}^{m} f \cdot g = \left. \partial_{y}^{n}  \partial_{s}^{m} f \left(x+y,t+s\right) g\left(x-y,t-s\right) \right\vert _{y=s=0}\text{ .} \label{1.27}
	\end{equation}
	
	 Using this definition and looking at the Taylor expansions of the functions $f(x,t)$ and $g(x,t)$ with two independent variables $x$ and $t$ around points $(y,s)$ and $(-y,-s)$ respectively, then multiplying together, the Taylor expansion can be rewritten in terms of Hirota derivatives as
	\begin{eqnarray}
	f(x+y,t+s)g(x-y,t-s) &=&\sum\limits_{k=0}^{\infty }\frac{1}{k!}\left(
	yD_{x}+sD_{t}\right) ^{k}\left( f\cdot g\right) \text{ ,}  \label{1.28}\\
	&=&e^{yD_{x}+sD_{t}}\left( f\cdot g\right) \text{ .}  \notag
	\end{eqnarray}
	
	\noindent From definition (\ref{1.27}) or Taylor expansion of (\ref{1.28}), we also have the following Hirota derivatives
	\begin{eqnarray}
	D_{x}D_{t}\left( f\cdot g\right) &=&f_{xt}g-f_{x}g_{t}-f_{t}g_{x}+fg_{xt}%
	\text{ ,}  \label{1.30} \\
	D_{t}\left( f\cdot g\right) &=&fg_{t}-f_{t}g\text{ .}  \label{1.31}
	\end{eqnarray}
 
\noindent Letting $g=f$, equation (\ref{1.30}) produces the following identity for a function of two variables
	\begin{equation}
	\fbox{$2\partial _{x}\partial _{t}\ln f=\frac{D_{x}D_{t}\left( f\cdot
			f\right) }{f^{2}}$}\text{ .}  \label{1.35}
	\end{equation}

\vspace{0.5cm}

 With these definitions and properties, we can convert ordinary derivatives into Hirota derivatives and vice versa. Subsequently, this allows us to convert NPDEs into Hirota bilinear equation(s), which we shall see in the following.

\subsection{Hirota bilinear equation for the KdV equation}

Applying the logarithmic transformation $u=2(\ln \tau )_{xx}$ to the KdV equation (\ref{kdv}), a bilinear form is obtained, as follows:
\begin{equation}
2\left(\ln \tau \right)_{txx}+24\left(\ln \tau \right)_{xx}\left(\ln \tau \right)_{xxx}+2\left(\ln \tau \right)_{xxxxx}= 0 \text{ .}  \label{3.2}
\end{equation}
Integrating and taking the integration constant to be zero for soliton solutions leads to
\begin{equation}
\left(\ln \tau \right)_{tx}+6\left[\left(\ln \tau \right)_{xx}\right]^{2}+\left(\ln \tau \right)_{xxxx} =0 \text{ .}
\end{equation}

\noindent Then using Hirota properties (\ref{1.23}) and (\ref{1.24}), (\ref{1.35})
transforms the bilinear form into the Hirota bilinear equation
\begin{eqnarray}
\hspace{-0.5cm}
\frac{D_{x}D_{t}(\tau \cdot \tau )}{\tau ^{2}}+3\left[ \frac{D_{x}^{2}(\tau
	\cdot \tau )}{\tau ^{2}}\right] ^{2}+\frac{D_{x}^{4}(\tau \cdot \tau )}{\tau
	^{2}}-3\left[ \frac{D_{x}^{2}(\tau \cdot \tau )}{\tau ^{2}}\right] ^{2} &=&0%
\text{ ,} \\
\hspace{-0.5cm}
\left( D_{x}^{4}+D_{x}D_{t}\right) \left( \tau \cdot \tau \right) &=&0\text{ .}  \label{3.3}
\end{eqnarray}

\subsection{Hirota bilinear equation for the mKdV equation}

Similarly, but using an arctangent transformation
\begin{eqnarray}
v &=&\partial _{x}\arctan \left( \frac{\tau }{\sigma }\right) \text{,}
\label{4.2} \\
&=&\frac{1}{1+\left( \frac{\tau }{\sigma }\right) ^{2}}\frac{\tau _{x}\sigma
	-\sigma _{x}\tau }{\sigma ^{2}}\text{ ,}  \notag \\
&=&\frac{D_{x}(\tau \cdot \sigma )}{\tau ^{2}+\sigma ^{2}}\text{ ,}  \notag
\end{eqnarray}

\noindent the mKdV equation (\ref{mkdv1}) becomes
\begin{equation}
\partial _{t}\left[ \frac{D_{x}(\tau \cdot \sigma )}{\tau ^{2}+\sigma ^{2}}%
\right] +24\left[ \frac{D_{x}(\tau \cdot \sigma )}{\tau ^{2}+\sigma ^{2}}%
\right] ^{2}\partial _{x}\left[ \frac{D_{x}(\tau \cdot \sigma )}{\tau
	^{2}+\sigma ^{2}}\right] +\partial _{x}^{3}\left[ \frac{D_{x}(\tau \cdot
	\sigma )}{\tau ^{2}+\sigma ^{2}}\right] =0\text{ ,}  \label{4.3}
\end{equation}

\noindent which can be simplified to,
\begin{equation}
(\sigma ^{2}+\tau ^{2})\left[ \left( D_{x}^{3}+D_{t}\right) (\tau \cdot
\sigma )\right] +3(D_{x}(\tau \cdot \sigma ))\left[ D_{x}^{2}(\tau \cdot
\tau +\sigma \cdot \sigma )\right] =0\text{ .}  \label{4.4}
\end{equation}

\noindent Taking the following two Hirota bilinear equations to solve
\begin{eqnarray}
\left( D_{x}^{3}+D_{t}\right) (\tau \cdot \sigma )&=&0 \text{ ,} \label{4.5} \\ 
D_{x}^{2}(\tau \cdot \tau +\sigma \cdot \sigma )&=&0 \text{ ,}
\end{eqnarray}
is a particular way to obtain soliton solutions.

\subsection{Hirota bilinear equation for the SG equation}

 Letting the variable transformation be $\phi =4\arctan \frac{\tau }{\sigma }$ and using Hirota properties with some trigonometric identities, the left hand side of SG equation (\ref{sg}) becomes
\begin{eqnarray}
	\phi _{xt} &=&4\frac{\partial }{\partial t}\frac{\partial }{\partial x}%
	\arctan \frac{\tau }{\sigma }\text{ ,} \\
	&=&4\frac{\left( \sigma ^{2}+\tau ^{2}\right) \left( \tau _{x}\sigma -\sigma
		_{x}\tau \right) _{t}-\left( \tau _{x}\sigma -\sigma _{x}\tau \right) \left(
		\sigma ^{2}+\tau ^{2}\right) _{t}}{\left( \sigma ^{2}+\tau ^{2}\right) ^{2}}%
	\text{ .} \notag
\end{eqnarray}

\noindent Then taking $\theta =\arctan \frac{\tau }{\sigma }$, which gives $\cos
\theta =\frac{\sigma }{\sqrt{\sigma ^{2}+\tau ^{2}}}$ and $\sin \theta =%
\frac{\tau }{\sqrt{\sigma ^{2}+\tau ^{2}}}$ , the right hand side of SG equation (\ref{sg}) becomes
\begin{eqnarray}
	\sin \phi &=&\sin \left( 4\arctan \frac{\tau }{\sigma }\right) \text{ ,} \\
	&=&2\left( 2\sin \theta \cos \theta \right) \left( \cos ^{2}\theta -\sin
	^{2}\theta \right) \text{ ,} \notag\\
	&=&\frac{4\left( \sigma ^{2}-\tau ^{2}\right) \tau \sigma }{\left( \tau
		^{2}+\sigma ^{2}\right) ^{2}}\text{ .} \notag
\end{eqnarray}

\noindent Equating the left hand side with the right hand side and using properties of the Hirota derivative, we obtain
\begin{equation}
\tau \sigma D_{x}D_{t}\left( \tau \cdot \tau -\sigma \cdot \sigma \right)
+\left( \sigma ^{2}-\tau ^{2}\right) D_{x}D_{t}\left( \tau \cdot \sigma
\right) =\left( \sigma ^{2}-\tau ^{2}\right) \tau \sigma \text{ .}
\label{5.2}
\end{equation}

\noindent Here, a natural splitting is the following Hirota bilinear equations
\begin{eqnarray}
\left( D_{x}D_{t}-1\right) \left( \tau \cdot \sigma \right) &=&0\text{ ,}
\label{5.3} \\
D_{x}D_{t}\left( \tau \cdot \tau -\sigma \cdot \sigma \right) &=&0\text{ .}
\label{5.4}
\end{eqnarray}

\subsection{Hirota bilinear equation for the Hirota equation}

Taking the Hirota's equation (\ref{HE}), as $q$ is a complex field, we apply the transformation $q=\frac{g}{f}$, with $g$, a complex and $f$, a real function, then we have the identity%
\begin{eqnarray}
&&f^{3}\left[ iq_{t}+\alpha q_{xx}-2\kappa \alpha \left\vert q\right\vert
^{2}q+i\!\beta \!\left( q_{xxx}-6\kappa \left\vert q\right\vert
^{2}q_{x}\right) \right] =  \label{BINLSE} \\
&&\resizebox{.8\hsize}{!}{$f\left[ iD_{t}g\cdot f+\alpha D_{x}^{2}g\cdot f+i\beta D_{x}^{3}g\cdot f%
\right] +\left[ 3i\beta \left( \frac{g}{f}f_{x}-g_{x}\right) -\alpha g\right]
\left[ D_{x}^{2}f\cdot f+2\kappa \left\vert g\right\vert ^{2}\right]$},
\notag
\end{eqnarray}%
then we can make the choice of solving the following Hirota bilinear equations
\begin{eqnarray}
iD_{t}g\cdot f+\alpha D_{x}^{2}g\cdot f+i\beta D_{x}^{3}g\cdot f &=&0,
\label{bi1} \\
D_{x}^{2}f\cdot f+2\kappa \left\vert g\right\vert ^{2} &=&0. \label{bi2}
\end{eqnarray}

\vspace{0.2in}

 With the Hirota bilinear forms of NPDEs, the solution construction process that follows is similar to perturbation theory. However, with a finite truncation of a perturbative series, we obtain an exact analytical solution, which is remarkable. After solving the Hirota bilinear problem, we carry out an inverse transformation back to the original dependent variable and obtain the solution for the NPDE.

\section{B\"{a}cklund transformations (BTs)}

The BT is a method that started very early on in the development of nonlinear integrable theory, but coming from a different origin, the area of differential geometry. It developed from the investigation of pseudospherical surfaces, to explore how one can find a new pseudospherical surface described by the SG equation from an old one \cite{bour_th_1861}.

For us, the key point is that a BT reduces the NPDE to a simpler lower order problem by relating two solutions from the same NPDE as a pair of first order PDEs. Then with Bianchi's permutability theorem, which serves as a 'nonlinear superposition principle' for solutions of the NPDEs, in analogy to the 'linear superposition principle' for solutions of linear equations, a fourth solution to a NPDE can be found from three known solutions of the NPDE \cite{lamb_backlund_1976}.

Let us demonstrate the process with two examples, the KdV equation \cite{wahlquist_backlund_1973} and the SG equation \cite{lamb_backlund_1976}.

\subsection{B\"{a}cklund transformation for the KdV equation}

When taking the transformation $u=w_{x}$ for the KdV equation (\ref{kdv}) and integrating with respect to $x$, then letting the integration constant be zero, the KdV equation transforms to
\begin{equation}
w_{t}+3w_{x}^{2}+w_{xxx}=0\text{ .}  \label{2.2}
\end{equation}

\noindent The BT is a pair of equations relating the two solutions $w$ and $\widetilde{w}$ to (\ref{2.2}), which reads
\begin{eqnarray}
w_{x}+\widetilde{w}_{x}&=&k-\frac{1}{2}\left( w-\widetilde{w}\right) ^{2}%
\text{ ,}  \label{2.26}\\
w_{t}+\widetilde{w}_{t}&=&(w-\widetilde{w})\left( w_{xx}-\widetilde{w}_{xx}\right)
	-2\left( w_{x}^{2}+\widetilde{w}_{x}^{2}+w_{x}\widetilde{w}_{x}\right) \text{ .}
\label{2.27}
\end{eqnarray}

\noindent where $k$ is a constant. For verification the BT is correct, we can see that repeatedly differentiating (\ref{2.26}) and using (\ref{2.2}) produces (\ref{2.27}).

\begin{figure}[h]
	\centering
	
	\includegraphics[width=0.4\linewidth]{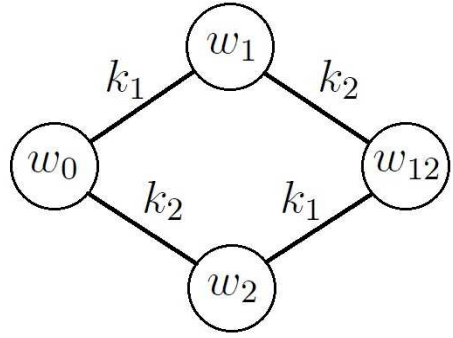}\\
	
	\caption{2x2 Bianchi-Lamb diagram of four arbitrary solutions $w_{0}$, $w_{1}$, $w_{2}$, $w_{12}$ of the KdV equation with each link representing a BT with a constant $k_{1}$ or $k_{2}$. }\label{fig2.1}
	\end{figure}

To create the 'nonlinear superposition principle' for solutions of the KdV equation, we need to make use of Bianchi's permutability theorem. This theorem can be nicely represented through the Bianchi-Lamb diagram in Figure \ref{fig2.1}. This diagram illustrates how to construct a new solution of the KdV equation $w_{12}$ through three known solutions $w_{0},w_{1}$ and $w_{2}$.

The four solutions are related as shown in the diagram through two constants $k_{1}$ and $k_{2}$ in the following way
\begin{equation}
\begin{array}{cc}
(1) &\left( w_{0}\right) _{x}+\left( w_{1}\right) _{x}=k_{1}-\frac{1}{2}\left(
w_{0}-w_{1}\right) ^{2} \text{ ,}\\ 
(2) &\left( w_{0}\right) _{x}+\left( w_{2}\right) _{x}=k_{2}-\frac{1}{2}\left(
w_{0}-w_{2}\right) ^{2} \text{ ,}\\ 
(3) &\left( w_{1}\right) _{x}+\left( w_{12}\right) _{x}=k_{2}-\frac{1}{2}\left(
w_{1}-w_{12}\right) ^{2} \text{ ,}\\ 
(4) &\left( w_{2}\right) _{x}+\left( w_{12}\right) _{x}=k_{1}-\frac{1}{2}\left(
w_{2}-w_{12}\right) ^{2} \text{ ,}
\end{array}%
  \label{2.29}
\end{equation}

\noindent together with (\ref{2.27}). Taking the differences (\ref{2.29} (1))-(\ref{2.29} (2)) and (\ref{2.29} (3))-(\ref{2.29} (4)), we can find the relation
\begin{equation}
w_{12}=w_{0}+2\frac{k_{1}-k_{2}}{w_{1}-w_{2}}\text{ .}  \label{2.33}
\end{equation}

\noindent This is the 'nonlinear superposition' relation, which we can use to construct a fourth solution $w_{12}$ to the KdV equation given three known solutions $w_{0},w_{1}$ and $w_{2}$.

\subsection{B\"{a}cklund transformation for the SG equation}

The 'nonlinear superposition principle' for the SG equation (\ref{sg}) works similarly. Suppose $\phi _{0}$ and $\phi _{1}$ are two solutions of the SG equations, then the pair of equations for the BT are
\begin{eqnarray}
\frac{1}{2}\partial _{x}\left( \phi _{1}+\phi _{0}\right) &=&\frac{1}{a}\sin
\left( \frac{\phi _{1}-\phi _{0}}{2}\right) \text{ ,}  \label{2.35} \\
\frac{1}{2}\partial _{t}\left( \phi _{1}-\phi _{0}\right) &=&a\sin \left( 
\frac{\phi _{1}+\phi _{0}}{2}\right) \text{ .}  \label{2.36}
\end{eqnarray}

We can verify the BT by cross differentiating the pair (\ref{2.35}),(\ref{2.36}) and we will see $\phi _{0}$ and $\phi _{1}$ are both solutions to the SG equation.

\begin{figure}[h]
	\centering
	
	\includegraphics[width=0.4\linewidth]{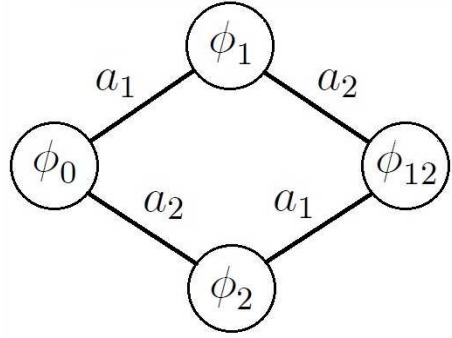}\\
	
	\caption{2x2 Bianchi-Lamb diagram of four arbitrary solutions $\phi_{0}$, $\phi_{1}$, $\phi_{2}$, $\phi_{12}$ of the SG equation with each link representing a BT with a constant $a_{1}$ or $a_{2}$. } \label{fig2.2}
	\end{figure}

Now using (\ref{2.35}) and (\ref{2.36}) together with Bianchi's
permutability theorem to relate four different solutions $\phi _{0},\phi
_{1},\phi _{2},\phi _{12}$ of the SG equation together as diagrammatically shown in Figure \ref{fig2.2},

\noindent we obtain the relations
\begin{equation}
\begin{array}{ccc}
(1) &\partial _{t}\left( \phi _{1}-\phi _{0}\right) =2a_{1}\sin \left( \frac{\phi
	_{1}+\phi _{0}}{2}\right), & \partial _{x}\left( \phi _{1}+\phi _{0}\right) =%
\frac{2}{a_{1}}\sin \left( \frac{\phi _{1}-\phi _{0}}{2}\right) \text{ ,}\\ 
(2) &\partial _{t}\left( \phi _{2}-\phi _{0}\right) =2a_{2}\sin \left( \frac{\phi
	_{2}+\phi _{0}}{2}\right), & \partial _{x}\left( \phi _{2}+\phi _{0}\right) =%
\frac{2}{a_{2}}\sin \left( \frac{\phi _{2}-\phi _{0}}{2}\right) \text{ ,}\\ 
(3) &\partial _{t}\left( \phi _{12}-\phi _{1}\right) =2a_{2}\sin \left( \frac{\phi
	_{12}+\phi _{1}}{2}\right), & \partial _{x}\left( \phi _{12}+\phi _{1}\right) =%
\frac{2}{a_{2}}\sin \left( \frac{\phi _{12}-\phi _{1}}{2}\right) \text{ ,}\\ 
(4) &\partial _{t}\left( \phi _{12}-\phi _{2}\right) =2a_{1}\sin \left( \frac{\phi
	_{12}+\phi _{2}}{2}\right), & \partial _{x}\left( \phi _{12}+\phi _{2}\right) =%
\frac{2}{a_{1}}\sin \left( \frac{\phi _{12}-\phi _{2}}{2}\right) \text{ .}
\end{array}%
  \label{2.39}
\end{equation}

\noindent Computing (\ref{2.39} (1))-(\ref{2.39} (2))+(\ref{2.39} (3))-(\ref{2.39} (4))=0 for $\partial _{t}$ equations, then (\ref{2.39} (1))-(\ref{2.39} (2))-(\ref{2.39} (3))+(\ref{2.39}
(4))=0 for $\partial _{x}$ equations and adding them together we obtain
\begin{equation}
a_{1}\sin \left( \frac{\phi _{12}-\phi _{0}}{4}+\frac{\phi _{2}-\phi _{1}}{4}%
\right) =a_{2}\sin \left( \frac{\phi _{12}-\phi _{0}}{4}-\frac{\phi _{2}-\phi
	_{1}}{4}\right) \text{ .}  \label{2.41a}
\end{equation}

\noindent Using the trigonometric identity $\sin \left( A\pm B\right) =\sin A\cos B\pm \cos A\sin B$ and dividing both sides of (\ref{2.41a}) by $\cos \left( \frac{\phi _{21}-\phi _{0}}{4}\right) \cos \left( \frac{\phi _{2}-\phi _{1}}{4}\right) $, results in
\begin{eqnarray}
\hspace{-1cm}
a_{1}\tan \left[ \resizebox{.075\hsize}{!}{$\frac{\phi _{12}-\phi _{0}}{4}$}\right] +a_{1}\tan \left[ 
\resizebox{.075\hsize}{!}{$\frac{\phi _{2}-\phi _{1}}{4}$}\right] \!&=&\! a_{2}\tan \left[ \resizebox{.075\hsize}{!}{$\frac{\phi
	_{12}-\phi _{0}}{4}$}\right] -a_{2}\tan \left[ \resizebox{.075\hsize}{!}{$\frac{\phi _{2}-\phi _{1}}{4}$}
\right] ,  \\
\hspace{-1cm}
\phi _{12}\!&=&\! \resizebox{.4\hsize}{!}{$\phi _{0}+4\arctan \frac{\left(
	a_{2}+a_{1}\right) }{\left( a_{2}-a_{1}\right) }\tan \left[ \frac{\phi
	_{2}-\phi _{1}}{4}\right]$} \text{ .}  \label{sg bianchi} 
\end{eqnarray}

\noindent This is the 'nonlinear superposition' for the SG equation, so knowing three SG solutions, we can easily construct a fourth one.

In the following chapters, we will use the above derived 'nonlinear superposition principles' to construct multi-soliton solutions from known single soliton solutions. 

\section{Darboux transformations (DTs) and Darboux-Crum transformations (DCTs)}

DTs are another powerful method to construct multi-soliton solutions. The initial investigation was started in 1882 by Darboux \cite{darboux_proposition_1882}. It was proposed that taking a solvable time independent Schr\"{o}dinger equation
\begin{equation}
	\frac{d^{2}y}{dx^{2}}=y \left[f(x)+m\right],
\end{equation}

\noindent one can construct infinitely many solvable Schr\"{o}dinger equations all with the same eigenvalue spectrum, $m$, possibly apart from a finite set of eigenvalues, but a new potential function, $f(x)$. Later, DTs were largely applied as a successful tool in constructing solutions of many types of linear and NPDEs \cite{matveev_darboux_1991,cooper_supersymmetry_1995, bagrov_supersymmetry_1996, bagrov_darboux_1997,song_generalization_2003,suzko_darboux_2009,correa_pt-symmetric_2015,correa_regularized_2016,correa_confluent_2017,cen_degenerate_2017,cen_integrable_2019,cen_asymptotic_2019,cen_time-dependent_2019}. Continually applying DTs, we can derive a recursive formula known as DCT, which helps us build multi-soliton solutions in a convenient way. In the following, we show the method of DT for the KdV, SG, Hirota and ECH equations following \cite{matveev_darboux_1991,wang_darboux_2005}.

\subsection{Darboux and Darboux-Crum transformation for the KdV equation}

To start, we introduce the Lax representation, an idea originating from Gardner, Greene, Kruskal and Miura \cite{gardner_method_1967}, later formally presented by Lax \cite{lax_integrals_1968}. This representation is often required as a necessary condition for integrability of a NPDE. The Lax representation reads
\begin{equation}
	L_{t}=[M,L],\label{lax}
\end{equation}

\noindent which is the compatibility condition of the following two linear equations
\begin{equation}
	L \psi = \lambda \psi, \quad
	M \psi = \psi_{t}.\label{lax1}
\end{equation}

For specific choices of the operators $L$ and $M$, also referred to as the Lax pair, (\ref{lax}) is satisfied up to different NPDEs. In particular for the KdV equation (\ref{kdv}), we take the choices
\begin{equation}
L = -\partial_{x}^{2}-u, \quad M = -4 \partial_{x}^{3}-6u \partial_{x}-3u_{x},
\end{equation}

\noindent where $u$ is the KdV solution field. Substituting these choices back into (\ref{lax}), this reproduces the KdV equation. The linear Schr\"{o}dinger form for the KdV can be seen from the $L$ operator equation of the Lax pair. Taking $\lambda=-\frac{\alpha^{2}}{4}$, this reads as 
\begin{equation}
-\psi_{xx}-u \psi =-\frac{\alpha^{2}}{4} \psi. \label{dar1}
\end{equation}

 To start the iteration we need an initial condition, so we take $u=0$, the trivial solution to the KdV equation, then $\psi=\cosh \frac{1}{2} \left( \alpha x - \alpha^{3} t\right)$ would be a resulting solution to the system (\ref{lax1}). To construct the Darboux operator, we need another independent solution that can be obtained with different eigenvalue parameter, so we take $\lambda=-\frac{\alpha_{1}^{2}}{4}$, then we have the solution $\psi_{1}=\cosh \frac{1}{2} \left( \alpha_{1} x - \alpha_{1}^{3} t\right)$ and the Darboux operator 
\begin{equation}
	L^{(1)}=\partial_{x}-\frac{\psi_{1x}}{\psi_{1}}.
\end{equation}

\noindent The new Schr\"{o}dinger equation after the first iteration of DT is
\begin{equation}
	-\psi^{(1)}_{xx}-u^{(1)}\psi^{(1)}=-\frac{\alpha^{2}}{4} \psi ^{(1)}, \label{iteratedkdv}
\end{equation} 

\noindent where $u^{(1)}$ is the new potential and one-soliton solution to the KdV equation
\begin{eqnarray}
 u^{(1)}&=&u+2\left(\ln \psi_{1} \right)_{xx}=2\partial_{x}^{2}\ln W_{1}\left(\psi_{1}\right),\\
 &=&\frac{\alpha_{1}^{2}}{2} \func{sech}^{2}\frac{1}{2}\left(\alpha_{1} x-\alpha_{1}^{3} t\right) \notag ,
\end{eqnarray}

\noindent with corresponding solution
\begin{equation}
  	\psi ^{(1)}=L^{(1)}\psi=\frac{W_{2}\left( \psi_{1} , \psi \right) }{W_{1}(\psi_{1}) }.
\end{equation}

\noindent $W_{1}$ and $W_{2}$ are Wronskians with respect to $x$. Denoting $N=1,2,\cdots$, the Wronskian $W_{N}$ is of the form
\begin{equation}
	W_{N}=\begin{vmatrix}
	\psi_{1}  & \cdots & \psi_{N-1} & \psi\\ 
	\left[\psi_{1}\right]^{(1)} & \cdots & \left[\psi_{N-1}\right]^{(1)} & \left[\psi\right]^{(1)}\\
	\vdots  & \cdots & \vdots & \vdots\\
	\left[\psi_{1}\right]^{(N-1)} & \cdots & \left[\psi_{N-1}\right]^{(N-1)} & \left[\psi \right]^{(N-1)}
	\end{vmatrix}
\end{equation}
\noindent where $\left[\psi_{j}\right]^{(k)}=\partial_{x}^{k}\psi_{j}$ and $\psi_{j}=\cosh \frac{1}{2}\left(\alpha_{j}x-\alpha_{j}^{3}t\right)$, $j=1, \dots, N-1$. As well as the covariance of the stationary Schr\"{o}dinger equation under DT, the equation (\ref{lax1}) governing the time dependence of $\psi$ is also covariant.

Note we can read the initial (\ref{dar1}) and iterated (\ref{iteratedkdv}) systems as equations involving Hamiltonians
\begin{equation}
H_{0}=-\partial_{x}^{2}-u, \quad \text{and} \quad
H_{1}=-\partial_{x}^{2}-u^{(1)},
\end{equation}
respectively. In this situation, the Darboux operator more generally can be viewed as an intertwining operator \cite{delsarte_transmutations_1958,deift_applications_1978,pursey_new_1986,cooper_supersymmetry_1995,bagrov_supersymmetry_1996, bagrov_darboux_1997,nieto_intertwining_2003} intertwining the two Hamiltonians as
\begin{equation}
	L^{(1)}H_{0}=H_{1}L^{(1)}.
\end{equation}
 
 To construct a multi-soliton solution to the KdV equation, one has to carry out DT multiple times, yielding what is known as DCT. So for a two-soliton solution, we carry out DT twice. The first step is to find another independent solution to the iterated Schr\"{o}dinger equation, (\ref{iteratedkdv}). For this, we take another solution to the original Schr\"{o}dinger equation, (\ref{dar1}), with eigenvalue parameter $\lambda=-\frac{\alpha_{2}^{2}}{4}$
\begin{equation}
\psi_{2}=\cosh \frac{1}{2}\left(\alpha_{2} x-\alpha_{2}^{3} t\right),
\end{equation}

\noindent then with $\psi_{1}$ we can obtain another independent solution to (\ref{iteratedkdv}) as
\begin{equation}
	\psi_{2}^{\left(1\right)}=L^{(1)}\psi_{2}=\frac{W_{2}\left( \psi_{1} , \psi_{2} \right) }{W_{1}\left(\psi_{1} \right)}.
\end{equation}

\noindent Now we can take the second DT, which involves taking the Darboux operator
\begin{equation}
L^{(2)}=\frac{\partial}{\partial x}-\frac{\left[\psi_{2}^{\left( 1\right) } \right] _{x}}{\psi_{2}^{\left(1\right)}}
\end{equation}

\noindent and the resulting second iterated Schr\"{o}dinger equation is
\begin{equation}
	-\psi^{(2)}_{xx}-u^{(2)}\psi^{(2)}=-\frac{\alpha^{2}}{4} \psi ^{(2)},
\end{equation}

\noindent with potential
\begin{eqnarray}
u^{\left(2\right)}&=& u^{\left(1\right)}+2\left[\ln \psi_{2}^{\left(1\right)}\right]_{xx},\\
&=& u+2\partial^{2}_{x}\ln \left[W_{2}\left(\psi_{1},\psi_{2}\right)\right], \notag
\end{eqnarray}

\noindent and solution
\begin{eqnarray}
	\psi^{(2)}&=&L^{(2)} L^{(1)} \psi,\\
	&=&\frac{W_{3}\left( \psi_{1} , \psi_{2}, \psi \right) }{W_{2}\left( \psi_{1}, \psi_{2} \right)}. \notag
\end{eqnarray}

\noindent The function $u^{\left(2\right)}$ is the two-soliton solution to the KdV equation with two parameters $\alpha_{1}$ and $\alpha_{2}$, which is usually called spectral or speed parameters, as the speed of the soliton solutions depend on these parameters. Repeatedly applying the Darboux iteration N times results in the Nth iterated Schr\"{o}dinger equation
\begin{equation}
	-\psi^{(N)}_{xx}-u^{(N)}\psi^{(N)}=-\frac{\alpha^{2}}{4} \psi ^{(N)},
	\end{equation}

\noindent with new potential and N-soliton solution
	\begin{equation}
			u^{(N)}=u+2\partial _{x}^{2}\ln \left[ W_{N}\left( \psi_{1} , \psi_{2} , \ldots , \psi _{N}\right) \right] ,
	\end{equation}

\noindent and eigenfunction
\begin{equation}
\psi^{(N)}=L^{(N)} L^{(N-1)} \cdots L^{(1)}\psi=\frac{W_{N+1}\left( \psi_{1} , \psi _{2} , \ldots  , \psi_{N} , \psi \right) }{W_{N}\left( \psi_{1} , \psi _{2} , \ldots ,  \psi_{N}\right) }.
\end{equation}

\subsection{Darboux and Darboux-Crum transformation for the SG equation}

For the SG equation (\ref{sg}), we show a more generalised method of DT, which involves carrying out the transformations on the zero curvature (ZC) representation of the SG equation. The ZC representation of NPDEs can be derived from taking the matrix formalism of the Lax representation \cite{ablowitz_the_1974}. In geometry, when the curvature of a connection on a vector bundle is flat, we say it has ZC. 

Take a pair of first order linear differential equations for an auxiliary function $\Phi$ and the pair of operators $U$ and $V$ such that
\begin{eqnarray}
	\Phi_{x}=U\Phi \quad , \quad \Phi_{t}=V\Phi \quad , \quad \Phi=\begin{pmatrix} \psi \\ \varphi \end{pmatrix}, \label{sgzc}
\end{eqnarray}
then the resulting ZC condition reads
\begin{equation}
U_{t}-V_{x}+\left[U,V\right]=0. \label{zc}
\end{equation}

\noindent To rewrite the SG equation as a ZC representation we can take the matrices $U$ and $V$ to be of the form
\begin{equation}
U= \begin{pmatrix} \frac{i}{2} \phi_{x} & \frac{\alpha}{2} \\ \frac{\alpha}{2} & -\frac{i}{2} \phi_{x} \end{pmatrix} ,
V=\begin{pmatrix} 0 & \frac{1}{2 \alpha}e^{i \phi} \\ \frac{1}{2 \alpha}e^{-i \phi} & 0 \end{pmatrix},
\end{equation} 

\noindent and the ZC condition holds if and only if $\phi$ satisfies the SG equation. This can be verified by substituting $U$ and $V$ into the ZC condition, which reproduces the SG equation. Taking the ZC representation (\ref{sgzc}), we obtain the following system of auxiliary equations
\begin{eqnarray}
\psi_{x}&=&\frac{i}{2}\phi_{x}\psi+\frac{\alpha}{2}\varphi, \qquad \varphi_{x}=-\frac{i}{2}\phi_{x}\varphi+\frac{\alpha}{2}\psi, \label{SGlinear}\\
\psi_{t}&=&\frac{1}{2 \alpha}e^{i \phi}\varphi,\hspace{1.8cm} \varphi_{t}=\frac{1}{2 \alpha}e^{-i \phi}\psi. \notag
\end{eqnarray}

\noindent Note, we can also decouple the equations (\ref{SGlinear}) into two Schr\"{o}dinger equations by computing $\psi_{xx}$ and $\varphi_{xx}$ 
\begin{equation}
	-\psi_{xx}+V_{+}\psi=-\frac{\alpha^{2}}{4}\psi \quad , \quad -\varphi_{xx}+V_{-}\varphi=-\frac{\alpha^{2}}{4}\varphi, \label{sglse}
\end{equation}

\noindent with potentials
\begin{equation}
	V_{\pm}=-\frac{1}{4}\left(\phi_{x}\right)^{2}\pm \frac{i}{2}\phi_{xx}. \label{sgpotentials}
\end{equation}

However, in this case, we will not carry out DT directly with the coupled Schr\"{o}dinger equations as the potential is not directly a solution to the SG equation. Instead, we apply DT to the ZC representation. To initiate DT we can again take the initial condition to be the trivial solution to the SG equation, $\phi=0$, then the above system of equations (\ref{SGlinear})  can be solved by
\begin{eqnarray}
\psi&=&e^{\frac{1}{2}\left(\alpha x+\frac{t}{\alpha}\right)}+i e^{-\frac{1}{2}\left(\alpha x+\frac{t}{\alpha}\right)} ,\\ \varphi&=&e^{\frac{1}{2}\left(\alpha x+\frac{t}{\alpha}\right)}-i e^{-\frac{1}{2}\left(\alpha x+\frac{t}{\alpha}\right)}.
\end{eqnarray}

\noindent For DT, we need another set of solutions with a different eigenvalue parameter $\alpha=\alpha_{1}$
\begin{eqnarray}
\psi_{1}&=&e^{\frac{1}{2}\left(\alpha_{1} x +\frac{t}{\alpha_{1}}\right)}+i e^{-\frac{1}{2}\left(\alpha_{1} x +\frac{t}{\alpha_{1}}\right)},\\
\varphi_{1}&=&e^{\frac{1}{2}\left(\alpha_{1} x +\frac{t}{\alpha_{1}}\right)}-i e^{-\frac{1}{2}\left(\alpha_{1} x +\frac{t}{\alpha_{1}}\right)},
\end{eqnarray}

\noindent then the first Darboux iteration can be performed to produce the new set
\begin{equation}
	\psi ^{(1)}=\frac{\alpha}{2} \varphi-\frac{\alpha_{1}}{2}\frac{\varphi_{1}}{\psi_{1}}\psi,\quad
	\varphi ^{(1)}=\frac{\alpha}{2} \psi-\frac{\alpha_{1}}{2}\frac{\psi_{1}}{\varphi_{1}}\varphi,
\end{equation}
\noindent accompanied by the new one-soliton solution for the SG equation
\begin{eqnarray}
	\phi ^{(1)}&=&-2i \ln \frac{\varphi_{1}}{\psi_{1}}, \label{dctsg1}\\
	&=& 4 \arctan e^{\alpha_{1} x +\frac{t}{\alpha_{1}}}. \notag 
\end{eqnarray}

Similarly to the KdV case, we can carry out multiple Darboux iterations, i.e. DCT. For the Nth iteration we have
\begin{equation}
	\psi^{(N)}=\frac{\Delta^{\psi}_ {N+1}}{\Delta^{\psi}_{N}},\quad
	\varphi^{(N)}=\frac{\Delta^{\varphi} _{N+1}}{\Delta^{\varphi}_{N}},
\end{equation}
\noindent together with the N-soliton solution to the SG equation, with $\varphi_{i}=\psi_{i}^{\ast}$
\begin{eqnarray}
\phi^{(N)}&=&-2 i \ln \frac{\Delta^{\varphi}_{N}}{\Delta^{\psi}_{N}}, \label{sgndct}\\
&=&-4\arctan \frac{Im \left[W_{N} \left(\psi_{1} \dots \psi_{N}\right)\right]}{Re \left[W_{N} \left(\psi_{1} \dots \psi_{N}\right)\right]}. \notag
\end{eqnarray}
\noindent where $\Delta^{\psi}_ {N}$ and $\Delta^{\varphi}_{N}$ are determinants, $det$, of the following matrices
\begin{equation}
\Delta^{\psi}_{N}=det \left[\left(\frac{1}{2}\alpha_{i}\right)^{j-1}\chi_{ij}\right],\quad
\Delta^{\varphi}_{N}=det \left[\left(\frac{1}{2}\alpha_{i}\right)^{j-1}\xi_{ij}\right],
\end{equation}
\begin{equation}
\chi_{ij}=\left\{ 
\begin{array}{cl}
\psi_{i}  \quad \left(j \,\, \text{odd}\right) \\ 
\noindent \varphi_{i}  \quad \left(j \,\, \text{even}\right)
\end{array}
\right. , \quad
\xi_{ij}=\left\{ 
\begin{array}{cl}
\varphi_{i}  \quad \left(j \,\, \text{odd}\right) \\ 
\noindent \psi_{i}  \quad \left(j \,\, \text{even}\right)
\end{array}
\right. ,
\end{equation}
\noindent with eigenfunctions of different eigenvalue parameters denoted by index $i$
\begin{eqnarray}
	\psi_{i}&=&e^{\frac{1}{2}\left(\alpha_{i} x+\frac{t}{\alpha_{i}}\right)}+i e^{-\frac{1}{2}\left(\alpha_{i} x+\frac{t}{\alpha_{i}}\right)} ,\\
	\varphi_{i}&=&e^{\frac{1}{2}\left(\alpha_{i} x+\frac{t}{\alpha_{i}}\right)}-i e^{-\frac{1}{2}\left(\alpha_{i} x+\frac{t}{\alpha_{i}}\right)},
\end{eqnarray}
\noindent for $ j=1,2, \dots N$ and $i=1,2, \dots N $.

\subsection{Darboux and Darboux-Crum transformation for the Hirota equation}
To show the construction of DT and DCT for the Hirota equation (\ref{HE}), we first present the Hamiltonian for carrying out the iterations. Let us take the ZC representation of the form
\begin{equation}
\psi_{x}=U \psi \quad , \quad \psi_{t}=V \psi \quad , \quad \psi=\begin{pmatrix}\varphi\\ \phi \end{pmatrix}, \label{hezc1}
\end{equation}
\noindent where
\begin{equation}
U=\begin{pmatrix} -i \lambda & q \\ r & i \lambda \end{pmatrix} ,\quad
V=\begin{pmatrix} A & B \\ C & -A \end{pmatrix}.\label{hezc12}
\end{equation}

\noindent It is then easy to see that the first differential equation in $x$ of the  ZC representation (\ref{hezc1}), can be rewritten as a one-dimensional stationary Dirac type Hamiltonian system \cite{nieto_intertwining_2003,correa_twisted_2014,correa_confluent_2017,cen_integrable_2019}
\begin{equation}
H \psi = -\lambda \psi \quad , \quad \psi=\begin{pmatrix}\varphi\\ \phi \end{pmatrix}, \label{dirac1}
\end{equation}
\noindent where
\begin{equation}
H= \begin{pmatrix} -i\partial_{x} & iq \\ -ir & i \partial_{x} \end{pmatrix}= -i \sigma_{3} \partial_{x}+Q_{H},
\end{equation}
\noindent $\sigma_{3}$, one of the Pauli matrices and $Q$, acting as a matrix potential given as
\begin{equation}
	\sigma_{3}=\begin{pmatrix} 1 & 0 \\ 0 & -1 \end{pmatrix},\quad
	Q_{H}=\begin{pmatrix} 0 & iq \\ -ir & 0 \end{pmatrix}.
\end{equation}
\noindent  In $Q_{H}$, the terms $q$ and $r$ satisfies the ZC representation (\ref{hezc1}) and will also be solutions of the NPDE.

Now suppose that we have a new Hamiltonian system of the same structure as the original Hamiltonian being
\begin{equation}
H^{(1)} \psi^{(1)} = -\lambda \psi^{(1)}
\end{equation}
\noindent where
\begin{equation}
H^{(1)}=-i \sigma_{3}+Q_{H}^{(1)},\quad
Q_{H}^{(1)}=\begin{pmatrix} 0 & iq^{(1)} \\ -ir^{(1)} & 0 \end{pmatrix}.
\end{equation}
\noindent We assume the original $H$ and iterated Hamiltonian $H^{(1)}$ are related by a Darboux or intertwining operator $L^{(1)}$ 
\begin{equation}
H^{(1)}L^{(1)}=L^{(1)}H,\label{intertwining}
\end{equation}
\noindent then the new solution $\psi^{(1)}$ to $H^{(1)}$ will be
\begin{equation}
	\psi^{(1)}=L^{(1)} \psi.
\end{equation}

\noindent Up to this point, we still have to find the form of $L^{(1)}$ that will give us such a relation, so we take the ansatz
\begin{equation}
	L^{(1)}=G_{L}^{(1)}+\partial_{x}
\end{equation}
\noindent with function $G_{L}^{(1)}$ to be found. Substituting the ansatz $L^{(1)}$ back into the intertwining relation (\ref{intertwining}), we obtain the equations
\begin{eqnarray}
Q_{H}^{(1)}-Q_{H}+ i\left[G_{L}^{(1)},\sigma_{3}\right]&=&0, \label{2.105}\\
-i \sigma_{3}\left(G_{L}^{(1)}\right)_{x}+Q_{H}^{(1)}G_{L}^{(1)}-G_{L}^{(1)}Q_{H}-\left(Q_{H}\right)_{x}&=&0,
\end{eqnarray}

\noindent which we can combine as
\begin{equation}
	\left(Q_{H}+i\sigma_{3}G_{L}^{(1)}\right)_{x}+\left[G_{L}^{(1)},Q_{H}\right]+i\left[G_{L}^{(1)},\sigma_{3}\right]G_{L}^{(1)}=0. \label{combine}
\end{equation}

\noindent If we take as suggested in \cite{nieto_intertwining_2003}
\begin{equation}
G_{L}^{(1)}=-u_{x}u^{-1},
\end{equation}
\noindent then this enables us to find the following result
\begin{eqnarray}
\hspace{-1.2cm}
\left(u^{-1}H u\right)_{x}&=&u^{-1}\left[\left(Q_{H}+i\sigma_{3}G_{L}^{(1)}\right)_{x}+\left[G_{L}^{(1)},Q_{H}\right]+i\left[G_{L}^{(1)},\sigma_{3}\right]G_{L}^{(1)}\right]u\\
\hspace{-1.2cm}
&=&0 \notag
\end{eqnarray}

\noindent and integrating this equation leads to 
\begin{equation}
	H u=u\Lambda,
\end{equation}
\noindent where $\Lambda=$ constant matrix, $u$, the solution matrix.  Taking two eigenfunctions of (\ref{dirac1}) with independent eigenvalues
\begin{equation}
	H \psi_{1}=-\lambda_{1} \psi_{1} \quad , \quad H \psi_{2}=-\lambda_{2} \psi_{2},
\end{equation}
\noindent we define
\begin{equation}
	u=\left(\psi_{1},\psi_{2}\right) \quad , \quad \Lambda=\begin{pmatrix}
	-\lambda_{1} & 0\\ 0 & -\lambda_{2}
	\end{pmatrix}.
\end{equation}

\noindent Consequently, $G_{L}^{(1)}$ will be
\begin{equation}
	G_{L}^{(1)}=-u_{x}u^{-1}=-\frac{1}{\det \Omega_{1}}\begin{pmatrix}
	\det D^{1}_{1} & -\det D^{q}_{1}\\ \det D^{r}_{1} & \det D^{2}_{1}
	\end{pmatrix} \label{2.113}
\end{equation}

\noindent where
\begin{equation}
\Omega_{1}= \begin{pmatrix}
	 \varphi_{1} & \phi_{1}\\ \varphi_{2} & \phi_{2}
	 \end{pmatrix}, \quad D^{q}_{1}= \begin{pmatrix}
	 \varphi^{'}_{1} & \varphi_{1}\\ \varphi^{'}_{2} & \varphi_{2}
	 \end{pmatrix}, \quad D^{r}_{1}= \begin{pmatrix}
	\phi^{'}_{1} & \phi_{1}\\ \phi^{'}_{2} & \phi_{2}
	 \end{pmatrix},
\end{equation}
\begin{equation}
D^{1}_{1}=  \begin{pmatrix}
\varphi^{'}_{1} & \phi_{1}\\ \varphi^{'}_{2} & \phi_{2}
\end{pmatrix}, \quad D^{2}_{1}= \begin{pmatrix}
\varphi_{1} & \phi^{'}_{1}\\ \varphi_{2} & \phi^{'}_{2}
\end{pmatrix} \notag
\end{equation}

\noindent with
\begin{equation}
\varphi_{j}=\varphi\left(\lambda_{j}\right) \quad , \quad  \phi_{j}=\phi\left(\lambda_{j}\right).
\end{equation}

\noindent So the first Darboux iteration results in the new potential matrix $Q_{H}^{(1)}$ and Hamiltonian system $H^{(1)}$
\begin{equation}
	H^{(1)}=-i \sigma_{3}+Q_{H}^{(1)},\quad
	Q_{H}^{(1)}=\begin{pmatrix} 0 & iq^{(1)} \\ -ir^{(1)} & 0 \end{pmatrix},
\end{equation}
\noindent where
\begin{equation}
	q^{(1)}=q+2\frac{det D_{1}^{q}}{det \Omega_{1}} \quad, \quad r^{(1)}=r-2\frac{det D_{1}^{r}}{det \Omega_{1}}. \label{hirotaq}
\end{equation}

\noindent From substituting (\ref{2.113}) into (\ref{2.105}). Taking $N$ iterations, we have in general
\begin{equation}
q^{(N)}=q+2\frac{det D_{N}^{q}}{det \Omega_{N}} \quad, \quad r^{(N)}=r-2\frac{det D_{N}^{r}}{det \Omega_{N}}, \label{hirotaqn}
\end{equation} 

\noindent where
\begin{eqnarray}
(W_{N})_{ij}&=& \left\{\begin{array}{c}
\left[\varphi_{i}\right]^{(N-j)} \quad (j\leq N)\\
\left[\phi_{i}\right]^{(2N-j)} \quad (j > N)
\end{array} \right. ,\\
(D_{N}^{q})_{ij}&=& \left\{\begin{array}{c}
\left[\phi_{i}\right]^{(N-j-1)} \quad (j< N)\\
\left[\varphi_{i}\right]^{(2N-j)} \quad (j \geq N)
\end{array} \right. ,\\
(D_{N}^{r})_{ij}&=& \left\{\begin{array}{c}
\left[\phi_{i}\right]^{(N-j+1)} \quad (j\leq N+1)\\
\left[\varphi_{i}\right]^{(2N-j)} \quad (j > N+1)
\end{array} \right. ,
\end{eqnarray}

\noindent with $\left[\varphi_{j}\right]^{(k)}=\partial_{x}^{(k)}\varphi_{j}$, $\left[\phi_{j}\right]^{(k)}=\partial_{x}^{(k)}\phi_{j}$ and $i,j=1, \dots, 2N$.

Now we can easily implement this Darboux procedure for the Hirota equation with the ZC representation being a special case of (\ref{hezc12}) where
\begin{eqnarray}
A&=& -i \alpha q r -2 i \alpha \lambda^{2}+\beta \left(r q_{x}-q r_{x}-4 i \lambda^{3}-2 i \lambda q r\right),\label{2.1251}\\
B&=&  i \alpha q_{x} +2 \alpha \lambda q+\beta \left(2 q^{2}r-q_{xx}+2 i \lambda q_{x}+4 \lambda^{2} q \right), \label{2.1252}\\
C&=&  -i \alpha r_{x} +2 \alpha \lambda r+\beta \left(2 qr^{2}-r_{xx}-2 i \lambda r_{x}+4 \lambda^{2} r \right).\label{2.1253}
\end{eqnarray}

\noindent Taking the ZC condition, we obtain the Ablowitz, Kaup, Newell and Segur (AKNS) equations \cite{ablowitz_nonlinear-evolution_1973}
\begin{eqnarray}
	iq_{t}+\alpha\left(q_{xx}-2rq^{2}\right)+i\beta\left(q_{xxx}-6rqq_{x}\right)&=&0, \label{2.126}\\
	ir_{t}-\alpha\left(r_{xx}-2qr^{2}\right)+i\beta\left(r_{xxx}-6qrr_{x}\right)&=&0.\label{2.127}
\end{eqnarray}

\noindent Letting $r=q^{\ast}$, where the asterisk denotes conjugation, the AKNS equations reduces to the Hirota equation. DT and DCT can be taken as before with a trivial solution to the Hirota equation, $q=0$ and solutions to the ZC representation (\ref{hezc1}-\ref{hezc12}) 

\begin{eqnarray}
	\varphi_{j}&=&e^{-i \lambda_{j} x -2 i \lambda_{j}^{2}\left(\alpha+2 \beta \lambda_{j}\right)t},\\
	\phi_{j}&=&e^{i \lambda_{j} x +2 i \lambda_{j}^{2}\left(\alpha+2 \beta \lambda_{j}\right)t}.
\end{eqnarray}

\subsection{Darboux and Darboux-Crum transformation for the ECH equation}
The final type of DT we present will be for a matrix form NPDE, being the ECH equation (\ref{ech}). In particular, we show here that DT is a specialisation of gauge transformations. 

Many integrable systems are related to each other by means of gauge
transformations, often in an unexpected way. Such type of correspondences
can be exploited to gain insight into either system from the other, for
instance by transforming solutions of one system to solutions of the other.
Often this process can only be carried out in one direction.

In general, we consider here two systems whose auxiliary functions $\psi$ and $\psi^{(1)}$ are related to each other by means of an gauge operator $D^{(1)}$ 
\begin{equation}
\psi^{(1)}=D^{(1)}\psi.
\end{equation}
\noindent Formally, the systems can be cast into two gauge equivalent ZC representations for
the two sets of operators, $U$, $V$ and $U^{(1)}$, $V^{(1)}$, involving the auxiliary function $\psi$ and $\psi^{(1)}$ by
 \begin{equation}
U\psi=\psi_{x} , \quad V\psi=\psi_{t} , \quad \psi=\begin{pmatrix}\varphi\\ \phi \end{pmatrix}, \label{ECHzc}
\end{equation}
and
\begin{equation}
U^{(1)}\psi^{(1)}=\psi^{(1)}_{x} , \quad V^{(1)}\psi^{(1)}=\psi^{(1)}_{t} , \quad \psi^{(1)}=\begin{pmatrix}\varphi^{(1)}\\ \phi^{(1)} \end{pmatrix}. \label{zc1}
\end{equation}

\noindent Given the transformation from $\psi$ to $\psi^{(1)}$, the operators $U$, $V$ and $U^{(1)}$, $V^{(1)}$ are related as
\begin{eqnarray}
U^{(1)}&=&D^{(1)}U\left[D^{(1)}\right]^{-1}+D^{(1)}_{x}\left[D^{(1)}\right]^{-1}, \label{2ndzcg1}\\
V^{(1)}&=&D^{(1)}V\left[D^{(1)}\right]^{-1}+D^{(1)}_{t}\left[D^{(1)}\right]^{-1}.
\end{eqnarray} 
\noindent This is the gauge transformation and is entirely generic, providing a connection between two types
of integrable systems, assuming the invertible gauge transformation map G exists. Specific
systems are obtained by concrete choices of the two sets of operators $U$, $V$ and $U^{(1)}$, $V^{(1)}$.

For special choices of  $D^{(1)}$, we can tune the relation (\ref{2ndzcg1}) into an intertwining relation between two ZC representations and this is DT. We proceed to see this through constructing DT for the ECH equation.

Let us take $U$, $V$ to be the ZC representation for the ECH equation, hence $U$, $V$ will be matrices of the form
\begin{equation}
U=\lambda U_{1} \quad , \quad
V=\lambda V_{1}+ \lambda^{2} V_{2} +\lambda^{3} V_{3},
\end{equation}
\noindent where
\begin{equation}
U_{1}=-iS,
\end{equation}
$S$ is the solution to the ECH equation, of the form
\begin{equation}
S= \begin{pmatrix} -w & u \\ v & w \end{pmatrix} \quad \text{with} \quad w^{2}+uv=1 \label{sform}
\end{equation}
and
\begin{eqnarray}
V_{1}&=&\left(\frac{3}{2}i \beta -\alpha \right)U_{1}U_{1x}-\beta U_{1xx} ,\\
V_{2}&=&2\alpha U_{1}-2\beta U_{1} U_{1x}, \\
V_{3}&=&4 \beta U_{1}.
\end{eqnarray}

The relation (\ref{2ndzcg1}) becomes an intertwining relation
\begin{equation}
U^{(1)}Q^{(1)}=Q^{(1)}U
\end{equation}
when taking the special choice of gauge operator $D^{(1)}$ to be
\begin{equation}
	D^{(1)}=\lambda Q^{(1)}-I,
\end{equation}
\noindent where
\begin{equation}
	Q^{(1)}=H\Lambda^{-1}H^{-1}, \quad
	H=\resizebox{.2\hsize}{!}{$\begin{pmatrix} \varphi\left(\lambda_{1}\right) & \varphi\left(\lambda_{2}\right) \\ \phi\left(\lambda_{1}\right) & \phi\left(\lambda_{2}\right) \end{pmatrix}$}, \quad
\Lambda=\resizebox{.15\hsize}{!}{$\begin{pmatrix} \lambda_{1} & 0 \\ 0 & \lambda_{2} \end{pmatrix}$}.  \label{H}
\end{equation}
\noindent Then it follows that we can iterate the solution $S$, to the ECH equation as
\begin{eqnarray}
	S^{(1)}&=&Q^{(1)}S\left[Q^{(1)}\right]^{-1}
\end{eqnarray}
\noindent and the $N^{th}$ iteration reads
\begin{equation}
	S^{(N)}=\widetilde{\mathcal{Q}}^{(N)}S\left[\widetilde{\mathcal{Q}}^{(N)}\right]^{-1}, \quad \text{where} \quad
	\widetilde{\mathcal{Q}}^{(N)}=Q^{(N)}\dots Q^{(1)}, \label{sng}
\end{equation}
where $\widetilde{\mathcal{Q}}^{(N)}$ is a $2 \times 2$ matrix with terms
\begin{equation}
\widetilde{\mathcal{Q}}^{(N)}=\frac{1}{\det \Omega^{(N)}}\begin{pmatrix}
\det \widetilde{\mathcal{Q}}^{(N)}_{11}& \det \widetilde{\mathcal{Q}}^{(N)}_{12} \\
\det \widetilde{\mathcal{Q}}^{(N)}_{21}& \det \widetilde{\mathcal{Q}}^{(N)}_{22}
\end{pmatrix},
\end{equation}
\begin{equation}
	(\Omega^{(N)})_{ij}= \left\{\begin{array}{c}
	\lambda_{i}^{N+1-j}\varphi_{i} \quad (j\leq N)\\
	\lambda_{i}^{2N+1-j}\phi_{i} \quad (j > N)
	\end{array} \right.,
\end{equation}
\begin{eqnarray}
\hspace{-1.2cm}
(\widetilde{\mathcal{Q}}^{(N)}_{11})_{ij}\!\!&=&\!\! \resizebox{.25\hsize}{!}{$\left\{\begin{array}{c}
\lambda_{i}^{j-1}\varphi_{i} \quad (j \leq N)\\
\lambda_{i}^{j-N}\phi_{i} \quad (j > N)
\end{array} \right.$},\quad
(\widetilde{\mathcal{Q}}^{(N)}_{22})_{ij}= \resizebox{.25\hsize}{!}{$\left\{\begin{array}{c}
\lambda_{i}^{j}\varphi_{i} \quad (j\leq N)\\
\lambda_{i}^{j-N-1}\phi_{i} \quad (j > N)
\end{array} \right.$},\\
\hspace{-1.2cm}
(\widetilde{\mathcal{Q}}^{(N)}_{12})_{ij}\!\!&=&\!\! \resizebox{.25\hsize}{!}{$\left\{\begin{array}{c}
\lambda_{i}^{N-j}\phi_{i} \quad (j < N)\\
\lambda_{i}^{2N-j}\varphi_{i} \quad (j \geq N)
\end{array} \right.$},\quad
(\widetilde{\mathcal{Q}}^{(N)}_{21})_{ij}= \resizebox{.3\hsize}{!}{$\left\{\begin{array}{c}
\lambda_{i}^{j-1}\phi_{i} \quad (j\leq N+1)\\
\lambda_{i}^{j-N-1}\varphi_{i} \quad (j > N+1)
\end{array} \right.$},
\end{eqnarray}

\noindent and $\phi_{k}=\phi \left(\lambda_{k}\right)$, $\varphi_{k}=\varphi \left(\lambda_{k}\right)$ and $i,j=1, \dots, 2N$. As a result, $S^{(N)}$ becomes

\begin{equation}
	S^{(N)}= \begin{pmatrix} -w_{N} & u_{N} \\ v_{N} & w_{N} \end{pmatrix} \label{suvwn} 
\end{equation}
\noindent where 
\begin{eqnarray}
u_{N} &=&\left( A_{N}^{2}u_{0}-B_{N}^{2}v_{0}+2A_{N}B_{N}w_{0}\right)
/\chi _{N},  \label{uvw1} \\
v_{N} &=&\left( D_{N}^{2}v_{0}-C_{N}^{2}u_{0}-2C_{N}D_{N}w_{0} \right)
/\chi _{N},  \label{uvw2} \\
w _{N} &=&\left[
A_{N}C_{N}u_{0}-B_{N}D_{N}v_{0}+(A_{N}D_{N}+B_{N}C_{N})w_{0}\right]
/\chi _{N},  \label{uvw3}
\end{eqnarray}
\noindent and
\begin{eqnarray}
A_{N}&=&\frac{\det \widetilde{\mathcal{Q}}_{11}^{(N)}}{\det \Omega^{(N)}} , \quad B_{N}=\frac{\det \widetilde{\mathcal{Q}}_{12}^{(N)}}{\det \Omega^{(N)}}, \quad C_{N}=\frac{\det \widetilde{\mathcal{Q}}_{21}^{(N)}}{\det \Omega^{(N)}}, \\
 D_{N}&=&\frac{\det \widetilde{\mathcal{Q}}_{22}^{(N)}}{\det \Omega^{(N)}},
	\quad \chi_{N}=\det \widetilde{\mathcal{Q}}^{(N)}.
\end{eqnarray}
\noindent We also observe the identity 
\begin{equation}
\left( A_{N}\right) _{x}D_{N}-B_{N}\left( C_{N}\right) _{x}=A_{N}\left(
D_{N}\right) _{x}-\left( B_{N}\right) _{x}C_{N}=0, \label{AB}
\end{equation}
which becomes useful in Chapter 8, where we use them to simplify gauge relations between the Hirota and ECH equations.

Now taking Darboux iterations works analogously to previous examples. For  example, taking an initial seed solution to the ECH equation as
\begin{equation}
S=\begin{pmatrix}
-1 & 0\\0 & 1
\end{pmatrix},
\end{equation}
\noindent $U$, $V$ ZC matrices become
\begin{equation}
U=\begin{pmatrix}
i \lambda & 0\\0 & -i \lambda
\end{pmatrix} \quad , \quad
V=\begin{pmatrix}
2 i \alpha \lambda^{2}+4i \beta \lambda^{3} & 0\\0 & -2 i \alpha \lambda^{2}-4i \beta \lambda^{3}
\end{pmatrix}.
\end{equation}
As a result, we can solve the ZC representation equations (\ref{ECHzc}) by
\begin{eqnarray}
\varphi&=&e^{i \lambda x +2 i \lambda^{2}\left(\alpha+2 \beta \lambda\right)t},\\
\phi&=&e^{-(i \lambda x +2 i \lambda^{2}\left(\alpha+2 \beta \lambda\right)t)}
\end{eqnarray}
\noindent and utilise these for Darboux iterations.

\chapter{ Complex soliton solutions and reality conditions for conserved charges}\label{ch_3}

With investigations showing how $\mathcal{PT}$-symmetry can help us with the generalisation of quantum systems, it is worth investigating whether we can also generalise classical systems with $\mathcal{PT}$-symmetries, in particular integrable NPDEs. 

Taking the concept of $\mathcal{PT}$-symmetry from quantum mechanics, $\mathcal{PT}$-symmetry for the classical side will also be of space and time reversal with conjugation, $\mathcal{PT}$: $x \rightarrow -x, t \rightarrow -t, i \rightarrow -i$. It is natural to question whether the properties seen for $\mathcal{PT}$-symmetric quantum models can be extended for classical models. In particular reality of energy, when both the Hamiltonian and solutions are $\mathcal{PT}$-symmetric. 

Some $\mathcal{PT}$-symmetric deformations have been explored for various classical integrable models, including for the KdV equation \cite{bender_$calpt$-symmetric_2007,fring_$calpt$-symmetric_2007,bagchi_pt-symmetric_2008,cavaglia_$calpt$-symmetry_2011} and Burgers equation \cite{assis_integrable_2009}. Integrable models are very delicate systems, deformations (including $\mathcal{PT}$-symmetric ones) usually destroy the integrability properties, although some rare cases pass the Painlev\'{e} test \cite{assis_integrable_2009} indicating that they are likely to remain integrable.

More recently, Khare and Saxena found some interesting novel $\mathcal{PT}$-symmetric soliton solutions to various types of NPDEs that appear to have been overlooked this far \cite{khare_novel_2015}.  Their
approach is to start off from some well-known real solutions to these equations and then by adding a term build around that solution a suitable complex ansatz including various constants. In many cases they succeeded to determine those constants in such a way that their expressions constitute solutions to the different types of complex nonlinear wave equations considered.

In this chapter, we will introduce some new $\mathcal{PT}$-symmetric deformations of well-known integrable NPDEs by extending real solution fields to the complex plane \cite{cen_complex_2016}. Applying some methods for soliton solution construction from the previous chapter, we will also show how we can generalise these to help us in the construction of complex soliton solutions for these new systems. In particular, we find that some special cases of our complex soliton solutions are exactly the ones obtained in \cite{khare_novel_2015}, moreover they can be viewed as physical as they possess real conserved charges \cite{cen_complex_2016,cen_time-delay_2017}.

\vspace{0.5cm}

\section{The complex KdV equation}

The complex KdV equation is obtained from taking the solution field of the KdV equation $u(x,t)$ to be the complex field as $u(x,t)=p(x,t)+i q(x,t)$, to obtain
\begin{equation}
u_{t}+6uu_{x}+u_{xxx}=0\quad \Leftrightarrow \quad \left\{ 
\begin{array}{r}
p_{t}+6pp_{x}+p_{xxx}-6qq_{x}=0 \text{ ,}\\ 
\hspace{0.5cm} q_{t}+6\left( pq\right) _{x}+q_{xxx}=0 \hspace{0.1cm} \text{.}
\end{array}%
\right.  \label{KdVcomplex}
\end{equation}
The coupled equations reduce to the Hirota-Satsuma \cite{hirota_soliton_1981} or Ito system \cite{ito_symmetries_1982} when setting $\left( pq\right) _{x}\rightarrow pq_{x}$ or $q_{xxx}\rightarrow 0$ in the second equation, respectively. Other complex $\mathcal{PT}$-symmetric deformations of the KdV equation that have been previously explored includes deformations on the nonlinear term by taking $uu_{x} \rightarrow -iu(iu_{x})^{\epsilon}$ \cite{bender_$calpt$-symmetric_2007}, deformations for derivatives $u_{x} \rightarrow -i(iu_{x})^{\epsilon}$ \cite{fring_$calpt$-symmetric_2007} and deformations on the solution field $u \rightarrow -i(iu)^{\epsilon}$ \cite{cavaglia_$calpt$-symmetry_2011}, where $\epsilon \in \mathbb{R}$ in all cases.

In our investigation, we consider the space, time and speed parameters to be real. However, one may still wonder if the complex deformation on the solution field could still have physical meaning. Indeed it does, because from the deformation, we obtain a pair of real coupled equations and finding a solution to the complex system is then equal to solving the coupled real system and vice versa.  

These equations (\ref{KdVcomplex}) remain $\mathcal{PT}$ invariant under $\mathcal{PT}$: $x\rightarrow -x$,\ $t\rightarrow -t$, $i\rightarrow -i$, $u\rightarrow u$, $p\rightarrow p$, $q\rightarrow -q$. For invariance, it is important to note that not all solutions to (\ref{KdVcomplex}) need to be $\mathcal{PT}$-symmetric since the symmetry could map one solution, say $u_{1}(x,t)$, into a new one $u_{1}(-x,-t)=u_{2}(x,t)\neq u_{1}(x,t)$.

\subsection{Complex one-soliton solution from Hirota's direct method}

Following the explanation of HDM from Chapter 2, we show the explicit construction of soliton solutions. We take the bilinear Hirota form derived for the KdV equation (\ref{3.3}) and expanding the Hirota variable $\tau $ as a power series
\begin{equation}
\begin{array}{c}
\tau =1+\lambda \tau _{1}+\lambda ^{2}\tau _{2}+\lambda ^{3}\tau _{3}+\ldots%
\end{array}%
\text{ , }%
\begin{array}{c}
\lambda \in 
\mathbb{R}
\end{array}%
\text{ .}  \label{3.4}
\end{equation}

\noindent Extracting different orders of $\lambda $ from expansion of the Hirota  bilinear form, the following equations are obtained
\begin{eqnarray}
\hspace{-1cm}
\lambda ^{0}: \hspace{5.6cm} (D_{x}^{4}+D_{x}D_{t})(1\cdot 1)&=&0 \text{ ,} \label{3.5a} \\ 
\hspace{-1.2cm}
\lambda ^{1}:\hspace{4.0cm} (D_{x}^{4}+D_{x}D_{t})(1\cdot \tau _{1}+\tau _{1}\cdot
1)&=&0\text{ ,} \notag\\ 
\hspace{-1.4cm}
\lambda ^{2}:\hspace{2.4cm} (D_{x}^{4}+D_{x}D_{t})(1\cdot \tau _{2}+\tau _{1}\cdot \tau
_{1}+\tau _{2}\cdot 1)&=&0 \text{ ,} \notag\\ 
\hspace{-1.6cm}
\lambda ^{3}: \hspace{0.8cm} (D_{x}^{4}+D_{x}D_{t})(1\cdot \tau _{3}+\tau _{1}\cdot \tau
_{2}+\tau _{2}\cdot \tau _{1}+\tau _{3}\cdot 1)&=&0 \text{ ,} \notag\\ 
\hspace{-1.4cm}
\vdots \hspace{9.4 cm} &\vdots& \notag\\ 
\hspace{-1.4cm}
\lambda^{n}: \hspace{7.2cm} n^{th} \hspace{0.1cm} \text{ order} \,\, \lambda \!\!\!\!\!\!&&\!\!\!\!\!\! \text{equation} \text{ .} \notag
\end{eqnarray}

\noindent Equation of order $\lambda ^{0}$ is trivially satisfied. Taking  equation of order $\lambda ^{1}$ and integrating this equation with respect to $x$, then from asymptotically vanishing boundary conditions for soliton solutions, we also let the integration constant to be zero to give
\begin{equation}
(\tau _{1})_{xxx}+(\tau _{1})_{t}=0\text{ .}  \label{3.7}
\end{equation}

\noindent We take $\tau _{1}$ to be a complex function solution by choosing the integration constant to be complex, that is
\begin{equation}
\begin{array}{c}
\tau _{1}=e^{\beta x-\beta ^{3}t+\mu _{0}}%
\end{array}%
\text{ , }%
\begin{array}{c}
\mu _{0}\in 
\mathbb{C}%
\end{array}%
\text{ .}  \label{3.8}
\end{equation}

\noindent Letting $z_{\alpha}=\alpha x-\alpha^{3}t+\nu_{0}$ and $z_{\beta}=\beta x-\beta^{3}t+\mu_{0}$ and looking at general Hirota derivatives of the form
\begin{eqnarray}
D_{x}^{m}D_{t}^{n}(e^{z_{\alpha }}\cdot e^{z_{\beta }})
&=&D_{x}^{m}D_{t}^{n}(e^{\alpha x-\alpha ^{3}t+\nu _{0}}\cdot e^{\beta
	x-\beta ^{3}t+\mu _{0}})\text{ ,}  \label{3.9} \\
&=&\partial _{y}^{m}\partial _{s}^{n}\left( e^{\alpha (x+y)-\alpha
	^{3}(t+s)+\nu _{0}}e^{\beta (x-y)-\beta ^{3}(t-s)+\mu _{0}}\right) 
\Bigr|_{y=s=0}%
\text{ ,}  \notag \\
&=&(\alpha -\beta )^{m}(\beta^{3} -\alpha^{3} )^{n}e^{z_{\alpha }}e^{z_{\beta }}%
\text{ ,}  \notag
\end{eqnarray}
\noindent if we choose $\beta =\alpha $, then we have the following property
\begin{equation}
D_{x}^{m}D_{t}^{n}(e^{z_{\alpha}}\cdot e^{z_{\alpha}})=0\text{ ,}  \label{3.10}
\end{equation}

\noindent so that
\begin{equation}
(D_{x}^{4}+D_{x}D_{t})(\tau _{1}\cdot \tau _{1})=0\text{ .}  \label{3.11}
\end{equation}

\noindent This allows the $\lambda ^{2}$ equation to become
\begin{eqnarray}
(D_{x}^{4}+D_{x}D_{t})(1\cdot \tau _{2}+\tau _{2}\cdot 1) &=&0\text{ ,}
\label{3.12} \\
(\tau _{2})_{xxxx}+(\tau _{2})_{xt} &=&0\text{ .}  \label{3.13}
\end{eqnarray}

\noindent Letting $\tau _{n}=0,\forall n\geq 2$, we are truncating the power series and this is the key step to obtain an exact analytical soliton solution. The remaining higher order $\lambda $ equations can be shown to be satisfied. Taking $\mu =\mu _{0}+\ln \lambda $, we have
\begin{eqnarray}
\tau &=&1+\lambda \tau _{1}\text{ ,}  \label{3.14} \\
&=&1+e^{\beta x-\beta ^{3}t+\mu }\text{ .}  \notag
\end{eqnarray}

\noindent Applying the inverse bilinear transformation gives a complex one-soliton solution to the KdV equation
\begin{eqnarray}
u_{\beta }^{H}(x,t) &=&2(\ln \tau )_{xx}\text{ ,}  \label{3.15} \\
&=& \frac{\beta^{2} }{2}\limfunc{sech}\nolimits^{2}\frac{1}{2}(\beta x-\beta ^{3}t+\mu )%
\text{ .} \notag
\end{eqnarray}

\subsection{Properties of the KdV complex one-soliton solution}

The complex one-soliton solution (\ref{3.15}) is a generalisation of well known real hyperbolic KdV soliton and cusp solutions
\begin{equation}
\resizebox{.38\hsize}{!}{$u(x,t)=\frac{\beta ^{2}}{2}\func{sech}^{2}\left[ \frac{1}{2}(\beta
x-\beta ^{3}t)\right]$}\quad \text{and\quad } \resizebox{.38\hsize}{!}{$u(x,t)=\frac{%
	\beta ^{2}}{2}\func{csch}^{2}\left[ \frac{1}{2}(\beta x-\beta ^{3}t)\right]$}%
\text{,}
\end{equation}

\noindent which can be obtained with the special choices of $\mu=0$ or $\mu=i \pi$ respectively. In general, with $\mu$ being purely imaginary, the solution is $\mathcal{PT}$-symmetric with $x\rightarrow -x,t\rightarrow -t,i\rightarrow -i$ , so $u\rightarrow u$. For the special case $\mu =\pm i \frac{\pi }{2}$ yields
\begin{eqnarray}
\hspace{-0.7cm}
u_{\beta ,\pm \frac{\pi }{2}}^{H}(x,t) &=&\frac{\beta ^{2}}{2}\limfunc{sech}%
\nolimits^{2}\frac{1}{2}(\beta x-\beta ^{3}t\pm i\frac{\pi }{2})\text{ ,}
\label{3.18} \\
\hspace{-0.7cm}
&=&\beta ^{2}\limfunc{sech}\nolimits^{2}(\beta x-\beta ^{3}t)\mp i\beta ^{2}%
\limfunc{sech}(\beta x-\beta ^{3}t)\tanh (\beta x-\beta ^{3}t)\text{ .} 
\notag
\end{eqnarray}

\noindent This is the complex $\mathcal{PT}$-symmetric soliton solution found by Khare and Saxena \cite{khare_novel_2015} up to a minus sign difference in the KdV equation.

By letting $\mu =\eta +i\theta $ , $\eta ,\theta \in \mathbb{R}$ , we find infinitely many new complex soliton solutions that have been overlooked so far. $\mathcal{PT}$-symmetry, excluding the special cases is broken for these solutions, however can be mended with either a space shift
\begin{equation}
	x\rightarrow  x+\frac{\eta }{\beta } ,
\end{equation}

\noindent which leaves the solution invariant under $\mathcal{PT}$ : $x+\frac{\eta }{\beta }\rightarrow -\left( x+\frac{\eta }{\beta }\right) ,t\rightarrow -t,i\rightarrow -i$ , so $u\rightarrow u$ . Or we can apply a time shift
\begin{equation}
t\rightarrow  t-\frac{\eta }{\beta^{3} } ,
\end{equation} 
 
\noindent which leaves the solution invariant under $\mathcal{PT}$ : $t-\frac{\eta }{\beta^{3} }\rightarrow -\left( t-\frac{\eta }{\beta^{3} }\right) ,x\rightarrow -x,i\rightarrow -i$ , so $u\rightarrow u$ .
 
Expressing the general complex soliton solution in terms of real and imaginary parts gives
\begin{eqnarray}
u_{\beta }^{H}(x,t)
 &=& \begin{array}{c} 
 \frac{\beta ^{2}+\beta ^{2}\cos \theta \cosh (\beta x-\beta ^{3}t+\eta )}{%
	\left[ \cos \theta +\cosh (\beta x-\beta ^{3}t+\eta )\right] ^{2}}-i\frac{%
	\beta ^{2}\sin \theta \sinh (\beta x-\beta ^{3}t+\eta )}{\left[ \cos \theta
	+\cosh (\beta x-\beta ^{3}t+\eta )\right] ^{2}}
\end{array}
 \text{ .}  \label{3.16}
\end{eqnarray}

\vspace{0.5cm}

\begin{figure}[h]
	\centering
	
	\includegraphics[width=0.48\linewidth]{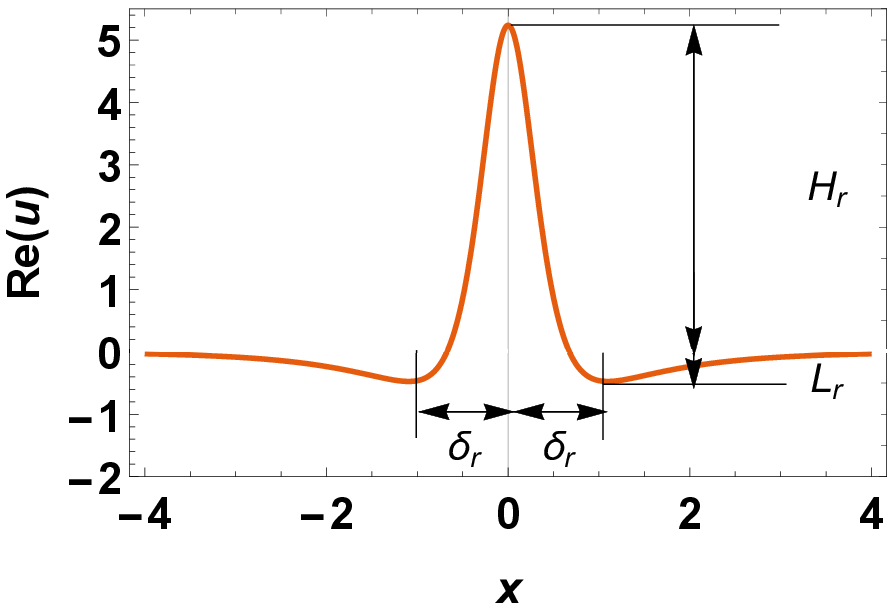} 
	\hspace{0.2cm}
	\includegraphics[width=0.48\linewidth]{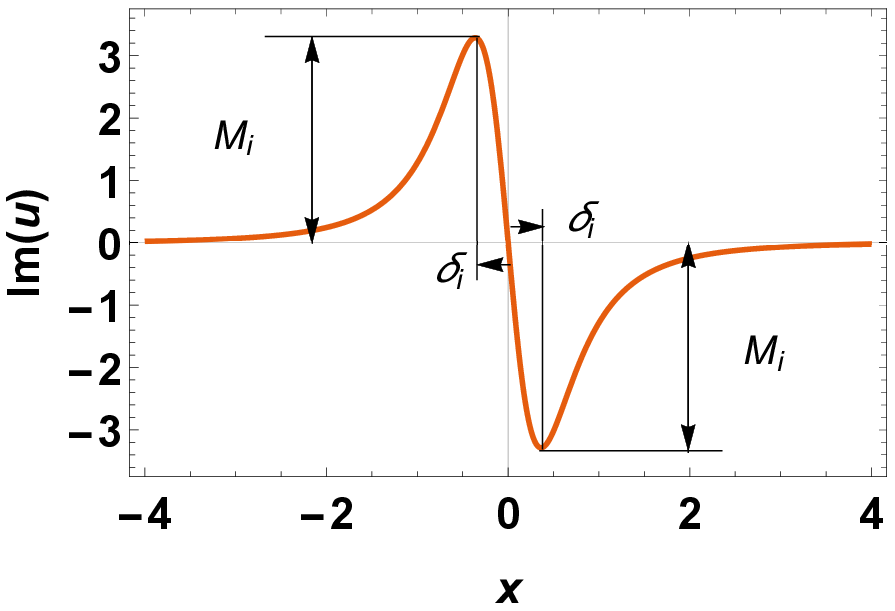}\\
	
	\caption{KdV complex one-soliton solution's real (left) and imaginary (right) parts for $\beta=1, \, \mu=2+i\frac{8 \pi}{10} \, \text{and} \, t=2$} \label{fig3.1}
\end{figure}

\noindent We can plot the real and imaginary parts separately as in Figure \ref{fig3.1}. For the real part, the solution has one maximum and two equal minima given by
\begin{equation}
	\begin{array}{c}
		H_{r}=\frac{\beta ^{2}}{2}\sec ^{2}\frac{\theta }{2},%
	\end{array}
	\\
	\begin{array}{c}
		L_{r}=\frac{\beta ^{2}}{4}\cot ^{2}\theta %
	\end{array}
	\\
	\begin{array}{c}
	\end{array}%
\end{equation}

\noindent respectively, with $\delta _{r}=\frac{1}{\beta }\func{arccosh}(\cos \theta -2\sec \theta )$. For the imaginary part, the magnitude of the maximum and minimum is
 \begin{equation}
 	\begin{array}{c}
 		M_{i}=\frac{8\beta ^{2}\sin \theta \sqrt{5+\cos 2\theta +\sqrt{2}\cos
 				\theta \sqrt{17+\cos 2\theta }}}{\left[ 6\cos \theta +\sqrt{2}\sqrt{17+\cos
 				2\theta }\right] ^{2}}%
 	\end{array}
 	\\
 \end{equation}

\noindent with $\delta _{i}=\frac{1}{\beta }\func{arccosh}\left[ \frac{1}{2}\cos \theta +%
\frac{\sqrt{2}}{4}\sqrt{17+\cos 2\theta }\right]$ . These values shall help us when computing displacements and time-delays for multi-soliton solutions as a result of scattering.

\subsection{KdV complex two-soliton solution from Hirota's direct method}

For the construction of a complex two-soliton solution, we just take a different $\tau_{1}$ to solve the equation at order $\lambda^{1}$ (\ref{3.7}). In particular, as this is a linear differential equation, a linear superposition of two solutions is also a solution. As a general two-soliton solution has two speeds, we take two speed parameters $\alpha$ and $\beta$, hence
\begin{eqnarray}
\tau _{1} &=&e^{\alpha x-\alpha ^{3}t+\upsilon _{0}}+e^{\beta x-\beta
	^{3}t+\mu _{0}}\text{ ,}  \label{3.33} \\
&=&e^{z_{\alpha}}+e^{z_{\beta}}\text{ ,}  \notag
\end{eqnarray}

\noindent where $\upsilon _{0}\in \mathbb{C}$ and $\mu _{0}\in \mathbb{C}$. Along with (\ref{3.9}), the equation resulting at order $\lambda ^{2}$ is

\begin{equation}
\left( \tau _{2}\right) _{xt}+\left( \tau _{2}\right) _{xxxx}=-\left[
(\alpha -\beta )^{4}+(\alpha -\beta )(\alpha ^{3}-\beta ^{3})\right]
e^{z_{\alpha}+z_{\beta}}\text{ .}  \label{3.37}
\end{equation}

\noindent Comparing both sides of the equation, we assume a $\tau _{2}$ solution to be of the form
\begin{equation}
\tau _{2}=\gamma e^{z_{\alpha}+z_{\beta}}\text{ \, with \, } \gamma=\gamma (\alpha ,\beta )\text{
	,}  \label{3.38}
\end{equation}

\noindent which when substituting into (\ref{3.37}), rearranging, then simplifying we find to be
\begin{equation}
\gamma =\frac{(\alpha -\beta )^{2}}{(\alpha +\beta )^{2}}\text{ .}
\label{3.40a}
\end{equation}

\noindent Next, we solve the equation at order $\lambda ^{3}$
\begin{equation}
(D_{x}^{4}+D_{x}D_{t})(1\cdot \tau _{3}+\tau _{1}\cdot \tau _{2}+\tau
_{2}\cdot \tau _{1}+\tau _{3}\cdot 1)=0\text{ .}  \label{3.43}
\end{equation}

\noindent Note that the second and third terms vanish, which is easily seen using the identity (\ref{3.9}), so the equation becomes
\begin{equation}
\left( \tau _{3}\right) _{xt}+\left( \tau _{3}\right) _{xxxx}=0,
\label{3.46}
\end{equation}

\noindent for which we can truncate the power series from here, with the choice $\tau _{n}=0,\forall n\geq 3$, satisfying the remaining higher order $\lambda$ equations and to obtain an exact analytical solution
\begin{eqnarray}
\tau &=&1+\lambda (e^{z_{\alpha}}+e^{z_{\beta}})+\lambda ^{2}\frac{(\alpha -\beta
	)^{2}}{(\alpha +\beta )^{2}}e^{z_{\alpha}+z_{\beta}}\text{ ,}  \label{3.47} \\
&=&1+e^{\alpha x-\alpha ^{3}t+\upsilon }+e^{\beta x-\beta ^{3}t+\mu }+\frac{%
	(\alpha -\beta )^{2}}{(\alpha +\beta )^{2}}e^{(\alpha x-\alpha
	^{3}t+\upsilon )+(\beta x-\beta ^{3}t+\mu )}\text{ ,}  \notag
\end{eqnarray}

\noindent after taking $\lambda=1$.

Using the inverse bilinear transformation, the KdV complex two-soliton solution from HDM is
\begin{eqnarray}
\hspace{-0.5cm}
u_{\alpha \beta }^{H} &=&2\partial _{x}^{2}(\ln \tau )\text{ ,}  \label{3.48}\\
\hspace{-1cm}
&=&\frac{2\left( \beta ^{2}e^{2\alpha ^{3}t+\upsilon +\beta x+\beta
		^{3}t}+\alpha ^{2}e^{\alpha x+\alpha ^{3}t+2\beta ^{3}t+\mu }\right) }{%
	(e^{\alpha ^{3}t+\beta ^{3}t}+e^{\alpha ^{3}t+\upsilon +\beta x}+e^{\alpha
		^{3}t+\beta x+\mu }+\gamma e^{\alpha x+\upsilon +\beta x+\mu })^{2}}\notag \\
\hspace{-1cm}
&&\resizebox{.83\hsize}{!}{$+\frac{2\gamma e^{\nu +\mu }\left[ 2(\alpha +\beta )^{2}e^{\alpha x+\alpha
		^{3}t+\beta x+\beta ^{3}t}+\alpha ^{2}e^{\alpha x+\alpha ^{3}t+\upsilon
		+2\beta x}+\beta ^{2}e^{2\alpha x+\beta x+\beta ^{3}t+\mu }\right] }{%
	(e^{\alpha ^{3}t+\beta ^{3}t}+e^{\alpha ^{3}t+\upsilon +\beta x}+e^{\alpha
		^{3}t+\beta x+\mu }+\gamma e^{\alpha x+\upsilon +\beta x+\mu })^{2}}\text{ .}$} \notag
\end{eqnarray}

\noindent This solution is generally not $\mathcal{PT}$-symmetric. However, using the arbitrariness of $\mu$ and $\nu$, if we take $\mu \rightarrow \mu +\ln \left(-\frac{\alpha+\beta}{\alpha -\beta}\right)$ and $\nu \rightarrow \nu +\ln \left(\frac{\alpha+\beta}{\alpha -\beta}\right)$, the solution will become the same solution as the complex two-soliton solution obtained from the BT and $\mathcal{PT}$-symmetry can be restored with space-time shifts, as we shall discuss in the next section. In addition, we will also see later that their corresponding energies are real.  

\subsection{KdV complex two-soliton solution from B\"{a}cklund transformation}

Recalling from the methods chapter, construction of multi-soliton solutions by BT uses a 'nonlinear superposition' of soliton solutions by (\ref{2.33}). For a complex two-soliton solution, we take a trivial solution and two complex one-soliton solutions of the KdV equation
\begin{eqnarray}
u_{\alpha } &=&\frac{\alpha ^{2}}{2}\limfunc{sech}\nolimits^{2}\frac{1}{2}%
(\alpha x-\alpha ^{3}t+\nu )\text{ ,}  \label{3.55} \\
u_{\beta } &=&\frac{\beta ^{2}}{2}\limfunc{sech}\nolimits^{2}\frac{1}{2}%
(\beta x-\beta ^{3}t+\mu )  \label{3.56}
\end{eqnarray}

\noindent and integrate the soliton solutions with respect to $x$\, setting the integration constant to zero to obtain
\begin{eqnarray}
w_{\alpha } &=&\alpha \tanh \frac{1}{2}(\alpha x-\alpha ^{3}t+\nu )\text{ ,}
\label{3.57} \\
w_{\beta } &=&\beta \tanh \frac{1}{2}(\beta x-\beta ^{3}t+\mu )\text{ .}
\label{3.58}
\end{eqnarray}

\noindent Using (\ref{2.29}), we can compute the relation constants
\begin{equation}
k_{\alpha } = \frac{\alpha ^{2}}{2}\text{ ,} \quad k_{\beta }=\frac{\beta ^{2}}{2}\text{ .} \label{3.59}
\end{equation}

Now we have all the ingredients to construct a KdV two-soliton solution using Bianchi's permutability theorem (\ref{2.33})
\begin{eqnarray}
w_{\alpha \beta }^{B} &=&2\frac{k_{\alpha }-k_{\beta }}{w_{\alpha }-w_{\beta
}}\text{ ,}  \label{3.61} \\
&=&\frac{\alpha ^{2}-\beta ^{2}}{\alpha \tanh \frac{1}{2}(\alpha x-\alpha
	^{3}t+\nu )-\beta \tanh \frac{1}{2}(\beta x-\beta ^{3}t+\mu )} \text{ ,} \notag
\end{eqnarray}

\noindent and taking the derivative with respect to $x$, the KdV two-soliton solution from the BT is
\begin{equation}
u_{\alpha \beta }^{B}=\frac{\left( \alpha ^{2}-\beta ^{2}\right) \left(
	\beta ^{2}\limfunc{sech}\nolimits^{2}\frac{1}{2}(\beta x-\beta ^{3}t+\mu
	)-\alpha ^{2}\limfunc{sech}\nolimits^{2}\frac{1}{2}(\alpha x-\alpha
	^{3}t+\nu )\right) }{2\left[ \alpha \tanh \frac{1}{2}(\alpha x-\alpha
	^{3}t+\nu )-\beta \tanh \frac{1}{2}(\beta x-\beta ^{3}t+\mu )\right] ^{2}}\text{ ,}
\label{3.62}
\end{equation}

\noindent where
\begin{equation}
\nu =\eta _{\alpha }+i\theta _{\alpha }\text{ ,} \quad \mu =\eta _{\beta }+i\theta _{\beta }\text{ .}%
\label{3.63}
\end{equation}

\noindent We notice that when $\mu$ and $\nu$ are chosen to be real, singularities will appear for certain $x$ and $t$ values, thus complex values for $\mu$ and $\nu$ can be used to regularize this expression. 

Furthermore, like the complex one-soliton solution from HDM, this solution is $\mathcal{PT}$-invariant for purely imaginary choices of $\mu$ and $\nu$. However, when the real parts of $\mu$ and $\nu$ are nonzero, $\mathcal{PT}$-symmetry is broken, but with a real space and time shift, we can restore $\mathcal{PT}$-symmetry.

Take $a$ and $b$ to be the space and time shifts respectively
\begin{equation}
a =\frac{\beta ^{3}\eta _{\alpha }-\alpha ^{3}\eta _{\beta }}{\alpha
	^{3}\beta -\alpha \beta ^{3}}\text{ ,} 
\quad
b=\frac{\beta \eta _{\alpha }-\alpha \eta _{\beta }}{\alpha ^{3}\beta
	-\alpha \beta ^{3}}\text{ ,}   \label{3.66}
\end{equation}

\noindent the solution (\ref{3.62}) can be rewritten as
\begin{equation}
\begin{array}{c}
u_{\alpha \beta }^{B}=\frac{\left( \alpha ^{2}-\beta ^{2}\right) \left(
	\beta ^{2}\limfunc{sech}\nolimits^{2}\frac{1}{2}\left[ \beta (x+a)-\beta
	^{3}(t+b)+i\theta _{\beta }\right] -\alpha ^{2}\limfunc{sech}\nolimits^{2}%
	\frac{1}{2}\left[ \alpha (x+a)-\alpha ^{3}(t+b)+i\theta _{\alpha }\right]
	\right) }{2\left( \alpha \tanh \frac{1}{2}\left[ \alpha (x+a)-\alpha
	^{3}(t+b)+i\theta _{\alpha }\right] -\beta \tanh \frac{1}{2}\left[ \beta
	(x+a)-\beta ^{3}(t+b)+i\theta _{\beta }\right] \right) ^{2}}%
\end{array}%
\text{ .}  \label{3.68}
\end{equation}

\noindent In this form, it is evident $\mathcal{PT}$-symmetry is present with $x+a\rightarrow -(x+a),t+b\rightarrow -(t+b),i\rightarrow -i$ so $u\rightarrow u$.

Following the same principles of the construction of two-soliton solutions, higher order multi-solitons from the HDM or BT can be constructed. However, $\mathcal{PT}$-symmetry is now generally broken and cannot be repaired for N-soliton solution, with N greater than 2. The reason being is that now we generally have N integration constants and with only two real variables $x$ and $t$, so we do not have enough variables to absorb all the real parts coming from the integration constants to restore $\mathcal{PT}$-symmetry of the N-soliton solution.

\section{The complex mKdV equation}

The mKdV equation possesses two types of  $\mathcal{PT}$-symmetries.  $\mathcal{PT}$ $x\rightarrow -x,t\rightarrow -t,i\rightarrow -i$, $v\rightarrow \pm v$. This equation is closely related to the KdV equation and through a Miura transformation, we can relate a solution to the KdV equation with a solution of the mKdV equation. We can complexify the equation with $v(x,t)= n(x,t)+i m(x,t)$ to obtain the coupled real equations
\begin{equation}
v_{t}+24v^{2}v_{x}+v_{xxx}=0 \Leftrightarrow \left\{ 
\begin{array}{r}
\hspace{-0.2cm}
n_{t}-48nmm_{x}+24(n^{2}-m^{2})n_{x}+n_{xxx}=0  \text{ ,}\\ 
\hspace{-0.2cm}
m_{t}+48nmn_{x}+24(n^{2}-m^{2})m_{x}+m_{xxx}=0 \hspace{0.1cm} \text{.} %
\end{array}%
\right.  \label{mkdv}
\end{equation}

\subsection{Complex one-solition solution from Hirota's direct method}

Taking the bilinear form of the mKdV equation (\ref{4.5}) and $\sigma =1$, the bilinear form simplifies to
\begin{eqnarray}
\tau _{t}+\tau _{xxx}=0 \text{ ,}   \label{4.7}\\ 
\left( \tau _{x}\right) ^{2}-\tau _{xx}\tau =0 \text{ .} \notag
\end{eqnarray}

\noindent Without much effort, we can easily find a particular solution $\tau =e^{\beta x-\beta ^{3}t+\mu }$ to the coupled equations (\ref{4.7}) and carrying out an inverse bilinear transformation, the complex mKdV solution from the HDM is obtained as
\begin{eqnarray}
v &=&\partial _{x}\arctan e^{\beta x-\beta ^{3}t+\mu }\text{ ,}  \label{4.8}
\\
&=&\frac{\beta }{2}\limfunc{sech}(\beta x-\beta ^{3}t+\mu )  \notag
\end{eqnarray}

\noindent with $\mu =\eta +i\theta \text{ , }\eta ,\theta \in \mathbb{R}$ . Now we can re-write the solution in terms of real and imaginary parts for general $\mu =\eta +i\theta $ as
\begin{equation}
v(x,t)=\frac{\beta \cos \theta \cosh \left( \beta x-\beta ^{3}t+\eta \right)
	-i\beta \sin \theta \sinh \left( \beta x-\beta ^{3}t+\eta \right) }{2\cos
	^{2}\theta \cosh ^{2}\left( \beta x-\beta ^{3}t+\eta \right) +2\sin
	^{2}\theta \sinh ^{2}\left( \beta x-\beta ^{3}t+\eta \right) }\text{ .} 
\end{equation}

\subsection{Complex Miura transformation}

Using an ansatz of the Miura transformation of the form 
\begin{equation}
u=\alpha v^{2}+\beta v_{x}
\text{ ,}  \label{4.12}
\end{equation}

\noindent where $\alpha$ and $\beta$ are arbitrary constants, we find the KdV (\ref{KdVcomplex}) and mKdV (\ref{mkdv}) equations are related through complex Miura transformation
\begin{equation}
u=4v^{2}\pm 2iv_{x} \text{ .}  \label{4.16}
\end{equation}

\noindent This provides us a new way to obtain complex KdV soliton solutions from real or complex mKdV soliton solutions. Taking the mKdV solution
\begin{equation}
v=\frac{\beta }{2}\limfunc{sech}(\beta x-\beta ^{3}t+\mu ) \text{ , } \label{4.17}
\end{equation}

\noindent with $\mu =\eta +i\theta$ and $\eta ,\theta \in \mathbb{R}$ and the Miura transformations, the corresponding KdV solutions are
\begin{equation}
u^{M} =\frac{\beta ^{2} \mp \beta ^{2}\sin \theta \cosh (\beta x-\beta ^{3}t+\eta )}{%
	\left[ \sin \theta -\cosh (\beta x-\beta ^{3}t+\eta )\right] ^{2}} \mp i\frac{%
	\beta ^{2}\cos \theta \sinh (\beta x-\beta ^{3}t+\eta )}{\left[ \sin \theta
	-\cosh (\beta x-\beta ^{3}t+\eta )\right] ^{2}}\text{ .}  
\end{equation}

\noindent We may notice that this solution is precisely the KdV complex one-soliton solution obtained from the HDM after a phase shift of $\theta \rightarrow \theta \mp \frac{\pi }{2}$, hence we have a new way of constructing the same set of KdV complex soliton solutions. Next, we look at some different kinds of solutions in terms of Jacobi elliptic functions.

\subsection{Complex Jacobi elliptic soliton solutions}

There are many ways to understand Jacobi elliptic functions, one of them is from the equation of motion for a pendulum \cite{lawden_elliptic_1989,whittaker_course_1996}. Let us start with the rewritten form of the equation of a pendulum as

\begin{equation}
	\left(\frac{d \theta}{d u}\right)^{2}-1+m \sin^{2} \theta = 0,
\end{equation}
where $0<m<1$, then we can show
\begin{equation}
	u\left(\phi,m\right)=\int^{\phi}_{0}\frac{d \theta}{\sqrt{1-m \sin^{2} \theta}},
\end{equation}
which is called the elliptic integral of the first kind. If we solve $\phi$ as a function of $u$, this gives the amplitude function
\begin{equation}
	\phi = am \left(u,m\right)
\end{equation}
and we can define the Jacobi elliptic function
\begin{equation}
	sn\left(u,m\right)=\sin \phi.
\end{equation} 

A lot of formulae for Jacobi elliptic functions are similar to trigonometric formulae. One key features of Jacobi elliptic functions is they are doubly periodic on the complex plane. Jacobi elliptic functions are written in the form
\begin{equation}
pq \left[u,m\right]
\end{equation}

\noindent where $p$ or $q$ can be the letters $c,d,n$ or $s$. Three basic Jacobi elliptic functions are
\begin{eqnarray}
\limfunc{cn%
}(u,m)& &\text{with periods  } 4K,2iK' \text{ ,}\\
\limfunc{sn}(u,m)& &\text{with periods  } 4K,2(K+iK') \text{ ,}\\
\limfunc{dn}(u,m)& &\text{with periods  } 2K,4iK' \text{ ,}
\end{eqnarray}

\noindent where $K$ and $K'$ are defined as 
\begin{eqnarray}
K(m)&=&\int_{0}^{\frac{\pi}{2}} \frac{d \theta}{\sqrt{1-m \sin^{2} \theta}} \text{ .}\\
K'(m)&=&K(1-m) \text{.}
\end{eqnarray}

\noindent $K$ is known as the complete elliptic integral of the first kind. The elliptic integral of the second kind is defined as
\begin{equation}
E(\phi,m)=\int_{0}^{\phi} \sqrt{1-m \sin^{2} \theta}. \label{ellipticsecond}
\end{equation}
Similarly, the complete elliptic integral of the second kind is defined by taking $\phi=\frac{\pi}{2}$.

The other Jacobi elliptic functions are related with trigonometric functions as 
\begin{eqnarray}
\limfunc{cn}(u,m)&= & \cos [am(u,m)]\text{ ,}\\
\limfunc{dn}(u,m)&= & \sqrt{ 1-m \sin^{2}(am(u,m)) } \text{ .}	
\end{eqnarray}

\vspace{1cm}

After this brief introduction to Jacobi elliptic functions, we now look at some solutions formulated from these functions. With the shifted Jacobi elliptic solution to the mKdV equation \cite{khare_novel_2015}
\begin{equation}
v^{dn}(x,t)=\frac{\beta }{2}\func{dn}\left[ \beta x-\beta
^{3}t(2-m)+\mu ,m\right] ,  \label{el1}
\end{equation}

\noindent from the Miura transformation we obtain the corresponding solution to the KdV equation to be
\begin{equation}
u^{dn}_{\pm}(x,t)=\beta ^{2}\func{dn}\left[ \widehat{z},m\right]
^{2}\pm im\beta ^{2}\func{cn}\left[ \widehat{z},m\right] \func{sn}\left[ \widehat{z}%
,m\right] ,
\end{equation}

\noindent where we abbreviated the argument $\widehat{z}=\beta x-\beta ^{3}t(2-m)+\mu $.
The elliptic parameter is denoted by $m$ as usual. Likewise from the shifted
known solution to the mKdV equation \cite{khare_novel_2015}%
\begin{equation}
v^{cn}(x,t)=\frac{\beta }{2}\sqrt{m}\func{cn}\left[ \beta
x-\beta ^{3}t(2m-1)+\mu ,m\right]  \text{ ,} \label{el2}
\end{equation}

\noindent we construct
\begin{equation}
u^{cn}_{\pm }(x,t)=m\beta ^{2}\func{cn}\left[ \widetilde{z},m%
\right] ^{2}\pm i\sqrt{m}\beta ^{2}\func{dn}\left[ \widetilde{z},m\right] \func{%
	sn}\left[ \widetilde{z},m\right] 
\end{equation}

\noindent with $\widetilde{z}:=\beta x-\beta ^{3}t(2m-1)+\mu $. In particular, taking $v^{dn}(x,t)$ and $v^{cn}(x,t)$ as solution to the mKdV equation, with the choice of $\mu=0$, leads to the complex $\mathcal{PT}$-symmetric solutions for the KdV equation reported in \cite{khare_novel_2015} up to a minus sign in the equation. 

\section{The complex SG equation}

The quantum field theory version of the complex sine-Gordon model has been studied for some time \cite{lund_unified_1976,lund_example_1977,de_vega_semiclassical_1983,dorey_quantum_1995,aratyn_complex_2000,okamura_perspective_2007}. Here we demonstrate that its classical version also admits interesting $\mathcal{PT}$-symmetric solutions. By taking the solution field of real SG equation to be complex, $\phi (x,t)=\varphi (x,t)+i\psi (x,t)$, we obtain the coupled real equations
\begin{equation}
\phi _{xt}=\sin \phi \quad \Leftrightarrow \quad \left\{ 
\begin{array}{r}
\varphi _{xt}=\sin \varphi \cosh \psi \text{ ,} \\ 
\psi _{xt}=\cos \varphi \sinh \psi \text{ .}
\end{array}%
\right.  \label{SG}
\end{equation}
\noindent We observe that the equations admit an infinite number of $\mathcal{PT}$-symmetries, $\mathcal{PT}_{\pm }^{(n)}$ :
$x\rightarrow -x$,\ $t\rightarrow -t$, $%
i\rightarrow -i$, $\phi \rightarrow \pm\phi+2\pi n$, $\varphi \rightarrow \pm\varphi+2\pi n$, $\psi \rightarrow \mp\psi$ with $n\in \mathbb{Z}$ .

\subsection{Complex one-soliton solution from Hirota's direct method}

Taking the Hirota bilinear form for the SG equation and setting $\sigma =1,$ results in
\vspace{-0.5cm}
 \begin{eqnarray}
 \tau _{xt} &=&\tau \text{ ,}  \label{5.5} \\
 \tau \tau _{xt} &=&\tau _{x}\tau _{t}\text{ .}  \label{5.6}
 \end{eqnarray}
 
\noindent Solving (\ref{5.5}) and (\ref{5.6}), we obtain $\tau =e^{\beta x+\frac{t}{\beta }+\mu }$with $\mu =\eta +i\theta$ and with the inverse bilinear transform, we obtain
 \begin{equation}
 \phi _{\beta }^{H}=4\arctan e^{\beta x+\frac{t}{\beta }+\eta +i\theta }\text{
 	.}  \label{5.7}
 \end{equation}
 
 \noindent This is the complex one-soliton solution for the SG equation from the HDM. 
 
 \subsection{Properties of the SG complex one-soliton solution}
 
This solution can be separated into real and imaginary
 parts with $z=\beta x+\frac{t}{\beta }+\eta +i\theta$ and the well-known relation $\arctan z=-i/2\ln \left[ (i-z)/(i+z)\right]$
 \begin{eqnarray}
 \phi _{\beta }^{H} &=&\frac{2}{i}\ln \frac{i-e^{z}}{i+e^{z}}\text{ ,}  \label{5.9} \\
 &=&\frac{2}{i}\left[ \ln \left\vert \frac{i-e^{z}}{i+e^{z}}\right\vert
 +i\arg \left( \frac{i-e^{z}}{i+e^{z}}\right) \right] \text{ .}  \notag
 \end{eqnarray}
 
 \noindent $\frac{i-e^{z}}{i+e^{z}}$ can be rewritten with real and imaginary parts using trigonometric identities yielding
 \begin{eqnarray}
 \phi _{\beta }^{H} &=&%
 \begin{array}{c}
 2\arg \left[ \frac{-\sinh \left( \beta x+\frac{t}{\beta }+\eta \right)
 	+i\cos \theta }{\cosh \left( \beta x+\frac{t}{\beta }+\eta \right) +\sin
 	\theta }\right] -i\ln \left[ \frac{\sinh ^{2}\left( \beta x+\frac{t}{\beta }%
 	+\eta \right) +\cos ^{2}\theta }{\left( \cosh \left( \beta x+\frac{t}{\beta }%
 	+\eta \right) +\sin \theta \right) ^{2}}\right]%
 \end{array}%
 \text{ .}  \label{5.12}
 \end{eqnarray}
 
 \noindent Then the argument can be expressed as an arctangent function, separating the cases for $\theta =\pm \frac{\pi }{2}$ for the function to be defined for all $\theta $ as
 \begin{equation}
\phi _{\beta }^{H}=\left\{ 
 \begin{array}{rc}
  \!\!\!\!\begin{array}{r}
 4\arctan \left[ \frac{\sqrt{\sinh ^{2}\left( \beta x+\frac{t}{\beta }+\eta
 		\right) +\cos ^{2}\theta }+\sinh \left( \beta x+\frac{t}{\beta }+\eta
 	\right) }{\cos \theta }\right] \\
 
 \vspace{-0.5cm}\\
 
 -i\ln \left[ \frac{\sinh ^{2}\left( \beta x+\frac{t}{\beta }+\eta \right)
 	+\cos ^{2}\theta }{\left( \cosh \left( \beta x+\frac{t}{\beta }+\eta \right)
 	+\sin \theta \right) ^{2}}\right]%
 \end{array}
 , & \theta \neq \pm \frac{\pi }{2} \text{ ,}\\ 
 &  \\ 
-i\ln \left[ \frac{\sinh ^{2}\left( \beta x+\frac{t}{\beta }+\eta \right) }{\left( \cosh \left( \beta x+\frac{t}{\beta }+\eta \right)
 	\pm 1 \right) ^{2}}\right] , & \theta =\pm \frac{\pi }{2} \text{ .}
 \end{array}%
 \right.  \label{5.13}
 \end{equation}
 
For the case where $\eta=0$ and $\theta \ne 0$ , we note that the solution is $\mathcal{PT}$ invariant under $\mathcal{PT}$ :  $x\rightarrow -x ,t\rightarrow
-t,i\rightarrow -i$ and $\phi _{\beta
}^{H}\rightarrow \pm\phi _{\beta }^{H}+$ $2n\pi $ , $n \in \mathbb{Z}$ . Similarly as for the KdV case, when in general for  $\eta \ne 0$ and $\theta \ne 0$, we have $\mathcal{PT}$-symmetry broken, however, this can be mended with a space or time shift.

Note that for the choice of $\theta=\frac{\pi}{2}$ the imaginary part of the SG complex soliton solution will have singularities for certain values of $x$ and $t$. In fact the imaginary part becomes a cusp solution.

\begin{figure}[h]
 	\centering
 	
 	\includegraphics[width=0.48\linewidth]{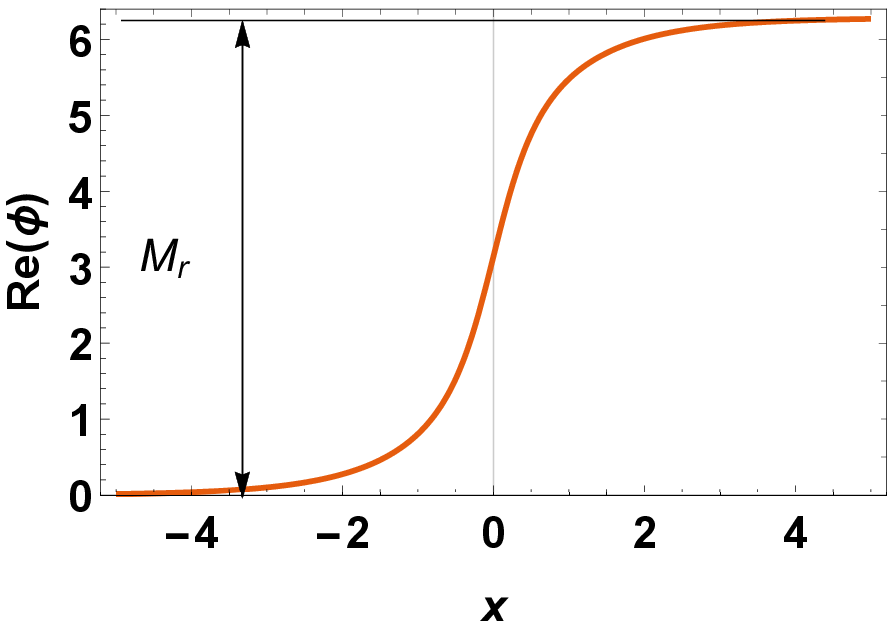} 
 	\hspace{0.2cm} 
 	\includegraphics[width=0.48\linewidth]{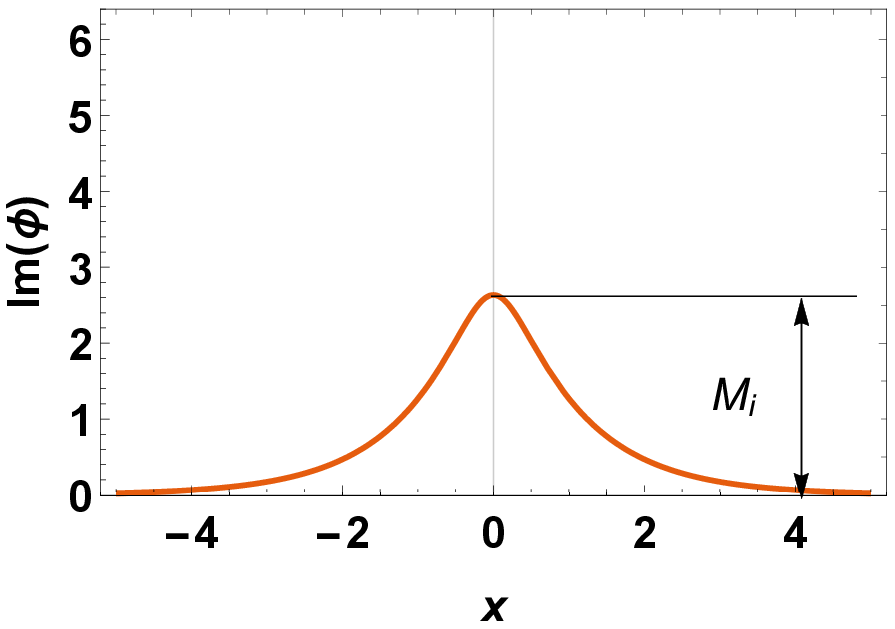}\\
 	
 	\caption{SG complex one-soliton solution's real (left) and imaginary (right) parts for $\beta=1, \, \mu=2+i\frac{\pi}{3} \, \text{and} \, t=-2$}. \label{fig3.2}
 \end{figure}

We can again plot the real and imaginary parts of the SG complex one-soliton solution, Figure \ref{fig3.2}, \, and compute their extrema 
 
 \begin{equation*}
 M_{r}=2\pi \hspace{0.5in} \text{and} \hspace{0.5in} M_{i}=\ln \left( \frac{1-\sin \theta }{1+\sin \theta }\right) .
 \end{equation*}
 
 \subsection{SG complex two-soliton solution from Hirota's direct method}
 
To obtain a two-soliton solution, we take the solutions $\tau $ and $\sigma $ of the Hirota bilinear form as
 \begin{equation}
 \begin{array}{l}
 \tau =e^{\alpha x+\frac{t}{\alpha }+\nu }+e^{\beta x+\frac{t}{\beta }+\mu }\text{ ,}
 \\ 
 \sigma =1+Ae^{\alpha x+\frac{t}{\alpha }+\nu }e^{\beta x+\frac{t}{\beta }%
 	+\mu }\text{ ,}%
 \end{array}%
   \label{5.38}
 \end{equation}
 
 \noindent with
 \begin{equation}
 A=-\frac{(\alpha -\beta )^{2}}{(\alpha +\beta )^{2}}\text{ ,}  \label{5.39}
 \end{equation}
 
 \noindent and $\upsilon ,\mu \in \mathbb{C}.$ So carrying out the inverse bilinear transformation, the complex two-soliton solution for SG is
 \begin{eqnarray}
 \phi _{\alpha \beta }^{H} &=&4\arctan \frac{\tau }{\sigma }\text{ ,}
 \label{5.41} \\
 &=&4\arctan \left( \frac{e^{\alpha x+\frac{t}{\alpha }+\nu }+e^{\beta x+%
 		\frac{t}{\beta }+\mu }}{1-\frac{(\alpha -\beta )^{2}}{(\alpha +\beta )^{2}}%
 	e^{\alpha x+\frac{t}{\alpha }+\nu }e^{\beta x+\frac{t}{\beta }+\mu }}\right) 
 \text{ .}  \notag
 \end{eqnarray}
 
\noindent This solution is also generally not $\mathcal{PT}$-symmetric like for the KdV case, but becomes the solution obtained by BT, as we shall see in the next section, after taking $\mu \rightarrow \mu +\ln \left(\frac{\alpha+\beta}{\alpha -\beta}\right)$ and $\nu \rightarrow \nu +\ln \left(-\frac{\alpha+\beta}{\alpha -\beta}\right)$. Consequently, the solution can be made $\mathcal{PT}$-symmetric under some space-time shifts, which we will also discuss in the following.
 
 \subsection{SG complex two-soliton solution from B\"{a}cklund transformation}
 
Using two known soliton solutions and taking one trivial solution, then we can derive a new solution from the BT for the SG
equation. For a complex two-soliton solution, we take two one-soliton solutions with different speeds
 \begin{eqnarray}
 \phi _{\alpha } &=&4\arctan e^{\alpha x+\frac{t}{\alpha }+\eta _{\alpha
 	}+i\theta _{\alpha }}\text{ ,}  \label{5.46} \\
 \phi _{\beta } &=&4\arctan e^{\beta x+\frac{t}{\beta }+\eta _{\beta
 	}+i\theta _{\beta }}\text{.}  \label{5.47}
 \end{eqnarray}
 
\noindent With the nonlinear superposition principle (\ref{sg bianchi}), the two-soliton solution can be found as 
 \begin{eqnarray}
 \phi _{\alpha \beta }^{B} &=&-4\arctan \left( \frac{\left( \alpha +\beta \right) }{\left( \alpha
 	-\beta \right) }\frac{\left( e^{z_{\alpha }}-e^{z_{\beta }}\right) }{\left(
 	1+e^{z_{\alpha }+z_{\beta }}\right) }\right) \text{ ,} \label{5.48}
 \end{eqnarray}
 
\noindent where $z_{\alpha }=\alpha x+\frac{t}{\alpha }+\eta _{\alpha }+i\theta
 _{\alpha }$ and $z_{\beta }=\beta x+\frac{t}{\beta }+\eta _{\beta }+i\theta_{\beta }.$
 
 Similarly, through some real space and time shift, $a$ and $b$ respectively, this solution is $\mathcal{PT}$-symmetric with $\mathcal{PT}$ : $x+a\rightarrow -\left( x+a\right) ,t+b\rightarrow -\left( t+b\right) ,i\rightarrow -i$ giving $\phi _{\alpha \beta }^{B}\rightarrow \pm\phi _{\alpha \beta }^{B}$$+2n\pi$, $n \in \mathbb{Z}$, with no phase shift. The space and time shift values are
 \begin{equation}
 \left. a=\frac{\alpha \eta _{\alpha }-\beta \eta _{\beta }}{\alpha
 	^{2}-\beta ^{2}}\right. ,\left. b=\frac{\alpha \beta (\alpha \eta _{\beta
 	}-\beta \eta _{\alpha })}{\alpha ^{2}-\beta ^{2}}\right. \text{ .}
 \label{5.53}
 \end{equation}

$\mathcal{PT}$-symmetry for general higher order SG N-soliton solutions is also generally broken for the same reason as for KdV N-soliton solutions.

\section{$\mathcal{PT}$-symmetry and reality of conserved charges}

In the previous sections, we have seen how to construct various complex soliton solutions to the KdV, mKdV and SG equations. In the following, we find it surprising at first, that all these complex soliton solutions, with some of them not $\mathcal{PT}$-symmetric, possess real energies. We will provide here a detailed analysis of scattering and asymptotic properties of complex soliton solutions in order to find the explanation for reality of energies. In the latter part, we show in fact, that for KdV complex soliton solutions, all conserved charges are real.

First, we look at the energy from the different equations for complex one-soliton solutions.

\subsection{Energy of the KdV complex one-soliton solution}

The Hamiltonian ${\mathcal{H}}$ leading to the KdV equation (\ref{KdVcomplex}) reads
\begin{equation}
{\mathcal{H}}=-u^{3}+\frac{1}{2}u_{x}^{2}\text{ .}  \label{3.24}
\end{equation}

\noindent We can verify this yields the equation (\ref{KdVcomplex}) using the Hamiltonian form \cite{gardner_korteweg-vries_1971} with Hamiltonian operator $\partial_{x}$ 
\begin{eqnarray}
u_{t} &=&\partial_{x}\left(\fdv{H}{u}\right), \label{3.25}\\
&=&-6uu_{x}-u_{xxx}  \notag
\end{eqnarray}
where $\fdv{H}{u}$ denotes the standard functional derivatives
\begin{equation}
\fdv{H}{u}=\sum\limits_{n=0}^{\infty
}\left( -1\right) ^{n}\frac{d^{n}}{dx^{n}}\frac{\partial {\mathcal{H}}}{%
	\partial u_{nx}}.
\end{equation}
\noindent With (\ref{3.24}) as the Hamiltonian density function for the KdV equation, we can calculate the energy of a soliton solution as
\begin{equation}
E=\int\nolimits_{-\infty }^{\infty }{\mathcal{H}}[u(x,t),u_{x}(x,t)]dx\text{ .}
\label{3.23}
\end{equation}

\noindent Taking the derivative of the KdV complex one-soliton solution (\ref{3.15}), we find
\begin{eqnarray}
\left[ u_{\beta}^{H}\right] _{x}
&=&\mp \left[ u_{\beta }^{H}\right] \sqrt{\beta ^{2}-2\left[
	u_{\beta }^{H}\right] }\text{ ,}  \label{3.28}
\end{eqnarray}

\noindent the energy is computed to be
\begin{eqnarray}
E &=&\int\nolimits_{-\infty }^{\infty }\left(-\left[ u_{\beta  }^{H}\right]
^{3}+\frac{1}{2}\left[ u_{\beta  }^{H}\right] _{x}^{2}\right)dx\text{ ,} 
\notag \\
&=&\left[ \frac{\beta ^{4}}{10}\left( \ln \left[ u_{\beta  }^{H}%
\right] \right) _{x}+\frac{\beta ^{3}}{5}\left[ u_{\beta  }^{H}%
\right]_{x} +\frac{2}{5}\left[ u_{\beta  }^{H}\right] \left[ u_{\beta
	}^{H}\right] _{x}\right] _{-\infty }^{\infty }\text{ ,}  \notag \\
&=&-\frac{\beta ^{5}}{5}\text{ ,}  \notag
\end{eqnarray}

\noindent where we have $z=\frac{1}{2}(\beta x- \beta^{3}t+ \mu)$ and $\left( \ln \left[ u_{\beta  }^{H}\right] \right)
_{x}\rightarrow \mp \beta ,\left[ u_{\beta  }^{H}\right]
_{x}\rightarrow 0,\left[ u_{\beta  }^{H}\right] \rightarrow 0$ as $x\rightarrow \pm \infty$.

As $\beta$, the speed parameter is real, this shows the complex KdV one-soliton solutions has real energies for any choice of complex $\mu$ constant. The reason behind this is the fact that the complex one-soliton solution is either $\mathcal{PT}$-symmetric in the case $Re[\mu]=0$, or in the case $Re[\mu] \neq 0$, the solution is $\mathcal{PT}$-symmetric up to a shift in space or time, as explained in Section 3.1.2. Both types of shift are permitted, as the shift in $x$ can be absorbed in the limits of the integral and the shift in $t$ is allowed since $\mathcal{H}$ is a conserved quantity in time. With a $\mathcal{PT}$-symmetric integrand, despite the Hamiltonian density function being complex, reality of energy is ensured on symmetric intervals \cite{fring_$calpt$-symmetric_2007} as one can check
\begin{equation}
E=\int_{-a}^{a}\mathcal{H}dx=\int_{-a}^{a}\mathcal{H}^{\dagger}dx=E^{\dagger}.
\end{equation}

\subsection{Energy of the mKdV complex one-soliton solution}

For the mKdV equation (\ref{mkdv}), the Hamiltonian density function is given by
\begin{equation}
{\mathcal{H}}=-2v^{4}+\frac{1}{2}v_{x}^{2}\text{ ,}  \label{4.20}
\end{equation}

\noindent which can be verified to yield the mKdV equation by using Hamiltonian form, similarly as for KdV case
\begin{eqnarray}
v_{t} &=& \partial_{x}\left(\fdv{H}{u}\right),  \label{4.21}\\
&=&-24v^{2}v_{x}-v_{xxx}\text{ .}  \notag
\end{eqnarray}

\noindent As a result, energy evaluated for the hyperbolic complex one-soliton solution (\ref{4.8}) is
\begin{eqnarray}
E &=&\int\nolimits_{-\infty }^{\infty }{\mathcal{H}}[v^{H}(x,t),v^{H}_{x}(x,t)]dx\text{ ,}
\label{4.23} \\
&=&-\frac{\beta ^{3}}{12}\text{ .}  \notag
\end{eqnarray}

For Jacobi elliptic solutions, they have the two periods $4K(m)/\beta $ and $i4K(1-m)/\beta $ in $x$. Thus we have to restrict the domain of integration for $E$ in order to obtain finite energies. For the solution $v^{dn}$ in (\ref{el1}) we have
\begin{eqnarray}
E &=&\int\nolimits_{-2K(m)/\beta }^{2K(m)/\beta }{%
	\mathcal{H}}\left[ v^{dn}(x,t),\left( 
v^{dn}(x,t)\right) _{x}\right] dx , \\
&=&\frac{\beta ^{3}}{24}\left[ (m-2)E\left[ \limfunc{am}\left(
4K(m)|m\right) ,m\right] +4K(m)(m-1)\right] ,  \notag
\end{eqnarray}

\noindent where $\limfunc{am}\left( u|m\right) $ denotes the amplitude of the Jacobi
elliptic function and $E\left[ \phi ,m\right] $ the elliptic integral of the
second kind. Similarly for the solution $v^{cn}$ in (\ref%
{el2}) we find 
\begin{eqnarray}
E &=&\int\nolimits_{-2K(m)/\beta }^{2K(m)/\beta }{%
	\mathcal{H}}\left[ v^{cn}(x,t),\left( 
v^{cn}(x,t)\right) _{x}\right] dx ,\\
&=&\frac{\beta ^{3}}{24}\left[ (1-2m)E\left[ \limfunc{am}\left(
2K(m)|m\right) ,m\right] -4K(m)(3m^{2}-4m+1)\right] .  \notag
\end{eqnarray}

\noindent We observe that in the limit $m \rightarrow 1$ for the energies computed from the Jacobi elliptic solutions, we obtain twice the energy of the hyperbolic solution. Again all energies are real with the same reasoning as the KdV case in the above section.

\subsection{Energy of the SG complex one-soliton solution}

For the SG equation (\ref{SG}), the Hamiltonian density reads
\begin{equation}
{\mathcal{H}}=1-\cos \phi 
\text{ .}  \label{5.26}
\end{equation}

\noindent Again it can be verified that this is the Hamiltonian density using the Hamiltonian form with Hamiltonian operator $\partial_{x}^{-1}$
\begin{eqnarray}
\phi_{t}&=&\partial_{x}^{-1}\left(\fdv{H}{u}\right) \text{ ,}  \notag\\
\phi_{xt}&=&\sin \phi.
\end{eqnarray}

Using the Hamiltonian density function (\ref{5.26}), the energy of the
complex one-soliton solution (\ref{5.7}) to the SG equation can be computed and is again real
\begin{eqnarray}
E &=&\int\nolimits_{-\infty }^{\infty }{\mathcal{H}}[\phi _{\beta
}^{H}(x,t)]dx\text{ ,}  \label{5.29} \\
&=&\frac{4}{\beta}\text{ .}  \notag
\end{eqnarray}

\subsection{Energy of the complex multi-soliton solutions}

From numerical calculations, we can also confirm that the energy values for all our complex two-soliton and three-soliton solutions are real, in particular the energy values are found to be the sum of the energies from the corresponding one-soliton solutions. This result remains true whether or not the solution is $\mathcal{PT}$-symmetric. 

Recalling properties of complex soliton solutions in the sections above, we can explain the reasoning for this result for complex two-soliton solutions, because we know that any $\mathcal{PT}$-broken symmetry solutions can be made $\mathcal{PT}$-symmetric again through space and time shifts.

However, for higher order complex multi-soliton solutions, they are generally not $\mathcal{PT}$-symmetric, as with three or more complex constants, we have not enough variables for us to absorb the real parts of the constants to mend $\mathcal{PT}$ broken symmetry.

To see the reality of energies for general complex N-soliton solutions, we resort to looking at lateral displacements or time-delays, which is a result of the scattering of an $N$-soliton solution compound, simultaneously with the asymptotic properties and structures of the energy densities. We will conduct such analysis for solutions of the KdV equation.

\vspace{0.3cm}

\noindent \large{\bfseries{Lateral displacements or time-delays for KdV complex two-soliton solutions}}

One of the prominent features of multi-soliton solutions is that the single soliton constituents within the compound preserve their shape after they scatter with the only net effect being a lateral displacement or time-delay when compared with the corresponding one-soliton solutions for each constituent.

Reviewing the classical scattering picture \cite{fring_vertex_1994}, the lateral displacement for a single particle or soliton constituent is defined to be the difference, $\Delta_{x}$, of the asymptotic trajectories before, $x_{b}=vt+x^{(i)}$ and after, $x_{a}=vt+x^{(f)}$ collision
\begin{equation}
	\Delta_{x} := x^{(f)}-x^{(i)},
\end{equation}

\noindent consequently the time-delay is defined as
\begin{equation}
	\Delta_{t} := t^{(f)}-t^{(i)} = -\frac{\Delta_{x}}{v},
\end{equation}

\noindent where $v$ is the speed of the particle or constituent.

Negative and positive time-delays are interpreted as attractive and
repulsive forces, respectively. In a multi-particle scattering process of particles, or soliton constituents, of type $k$, the corresponding lateral displacements and time-delays $(\Delta _{x})_{k}$ and $(\Delta_{t})_{k}$ respectively, have to satisfy certain consistency conditions \cite{fring_vertex_1994}. Demanding for instance that the total centre of mass coordinate
\begin{equation}
X=\frac{\sum\nolimits_{k}m_{k}x_{k}}{\sum\nolimits_{k}m_{k}}
\end{equation}

\noindent remains the same before and after the collision, i.e. $X^{(i)}=X$ $^{(f)}$, immediately implies that
\begin{equation}
\sum\nolimits_{k}m_{k}(\Delta _{x})_{k}=0,  \label{SM}
\end{equation}

\noindent with $m_{k}$ being the mass of the $k$ type particle or constituent.

\noindent Furthermore, given that $m\Delta _{x}=-mv\Delta _{t}=-p\Delta _{t}$ yields
\begin{equation}
\sum\nolimits_{k}p_{k}(\Delta _{t})_{k}=0,  \label{SP}
\end{equation}

\noindent where $p_{k}$ is the momentum of a particle of type $k$.

\vspace{0.3cm}

\noindent {\large{\bfseries{KdV complex two-soliton solutions}}}

\begin{figure}[h]
	\centering
	
	\includegraphics[width=0.48\linewidth]{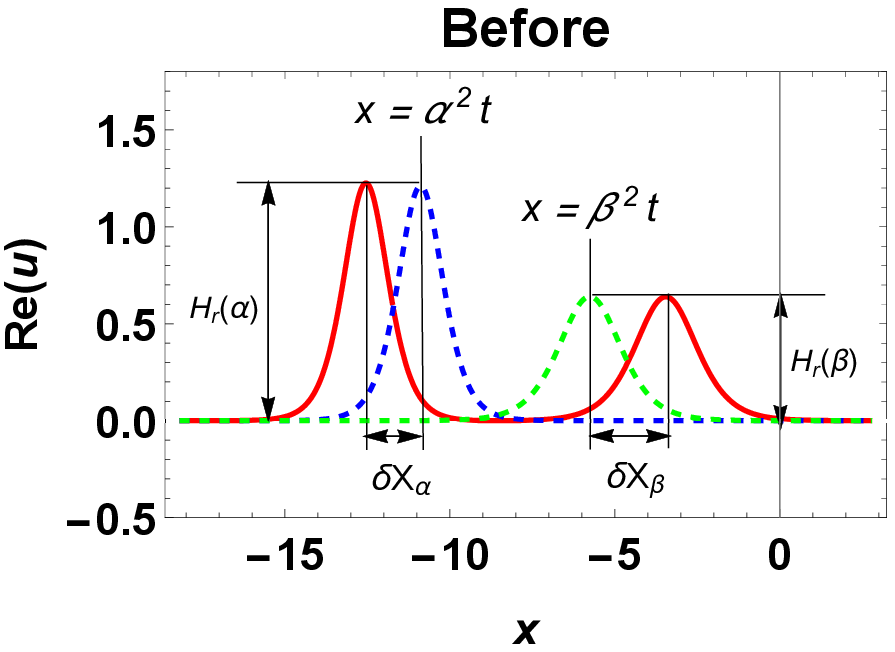}
	\hspace{0.2cm}
	\includegraphics[width=0.48\linewidth]{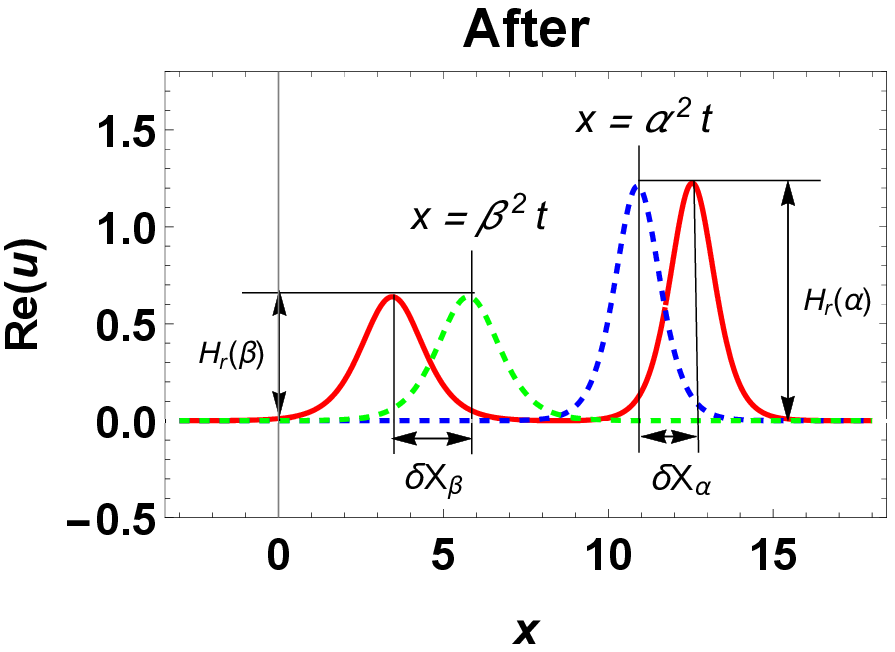}\\
	
	\caption{Snapshots of before $t=-9$ and after $t=9$ scattering of the real part of KdV complex two-soliton solution from BT (red), with corresponding real parts of complex one-soliton solutions (blue/green), where $\alpha=1.1$, $\beta=0.8$ and $\mu=\nu=i\frac{\pi}{2}$.} \label{fig3.3}
\end{figure}

Let us consider the KdV complex two-soliton solution from BT with speed parameters $\alpha$ and $\beta$ along with its corresponding two one-soliton solutions, one with speed parameter $\alpha$ and the other $\beta$, matching the multi-compound constituents. If we plot the three solutions at large times before and after scattering as in Figure \ref{fig3.3}, we see that the shapes of each multi-compound constituent matches its corresponding one-soliton solutions, but there is a distance between them, $\delta X_{\alpha}$ for the faster peak or $\delta X_{\beta}$ for the slower one. This is a result of the lateral displacement or time-delay. Note that although Figure \ref{fig3.3} shows only the real part; the imaginary part has the same properties.

To calculate these distances, we need to carry an asymptotic analysis of the two-soliton solution constituents one at a time and make use of the properties we found in the previous section for the KdV complex one-soliton solution.

Let us calculate, for example, the distance $\delta X_{\alpha}$ before scattering, for the two-soliton constituent with speed $\alpha^{2}$.  We need to first decide on the reference frame to track the soliton solutions; this will be decided by choosing a point on the one-soliton solution we want to track. For simplicity of expressions, let us take the maximum point and consequently we take $x=\alpha^{2}t$. Now we want to match the constituent of speed $\alpha^{2}$ with the speed $\alpha^{2}$ one-soliton solution asymptotically, hence we want to solve the asymptotic relation
\begin{equation}
	Re [u^{B}_{\alpha \beta}(\alpha^{2}t+\delta X_{\alpha},t) ] \sim Re [ u^{H}_{\alpha}(\alpha^{2}t,t) ] = \alpha^{2}
\end{equation} 

\noindent as $t \rightarrow - \infty$ for $\delta X_{\alpha}$, where $Re$ denotes the real part. As a result, we find 
\begin{equation}
\delta X_{\alpha}=\frac{2}{\alpha} \ln \left( \frac{\alpha+\beta}{\alpha-\beta}\right)
\end{equation}

\noindent and similarly, 
\begin{equation}
\delta X_{\beta}=\frac{2}{\beta} \ln \left( \frac{\alpha+\beta}{\alpha-\beta}\right) ,
\end{equation}

\noindent for the two-soliton constituent with speed $\beta^{2}$.

Utilising the snapshot of the soliton solutions for large time before and after scattering, we can compare the distances between the two-soliton constituents with the corresponding one-soliton solutions and find the lateral displacements and time-delays as
\begin{eqnarray}
	\Delta_{x} &=& 2 \delta X_{\alpha} ,\\
	\Delta_{t} &=& - \frac{2}{\alpha^{2}} \delta X_{\alpha} ,
\end{eqnarray}  

\noindent for the constituent with speed $\alpha^{2}$ and
\begin{eqnarray}
\Delta_{x} &=& -2 \delta X_{\beta} ,\\
\Delta_{t} &=& \frac{2}{\beta^{2}} \delta X_{\beta}
\end{eqnarray}  

\noindent for the constituent with speed $\beta^{2}$. It is easily checked that the consistency relations for masses (\ref{SM}) and momenta (\ref{SP}) are also satisfied.

\begin{figure}[h]
	\centering
	
	\includegraphics[width=0.48\linewidth]{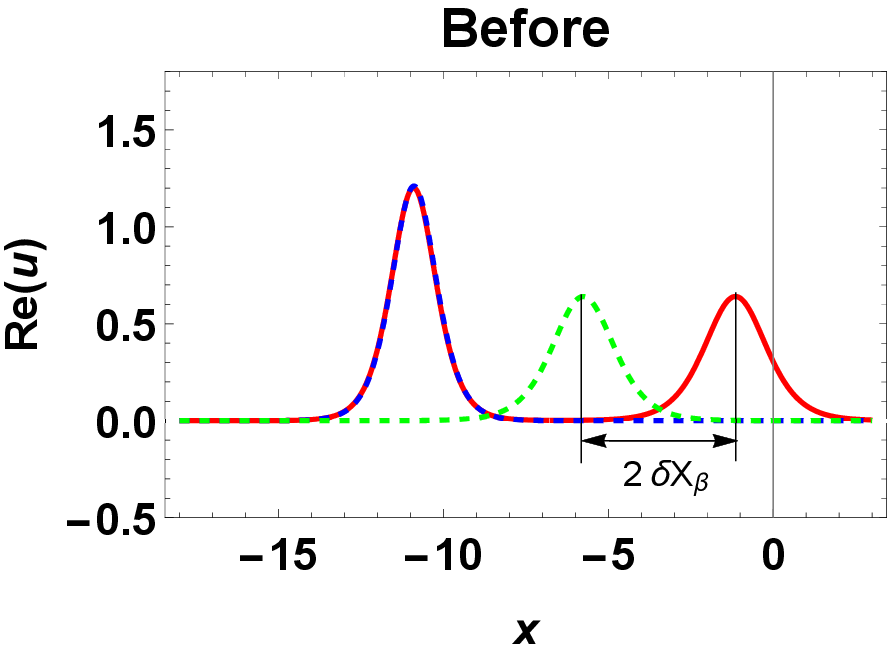}
	\hspace{0.2cm}
	\includegraphics[width=0.48\linewidth]{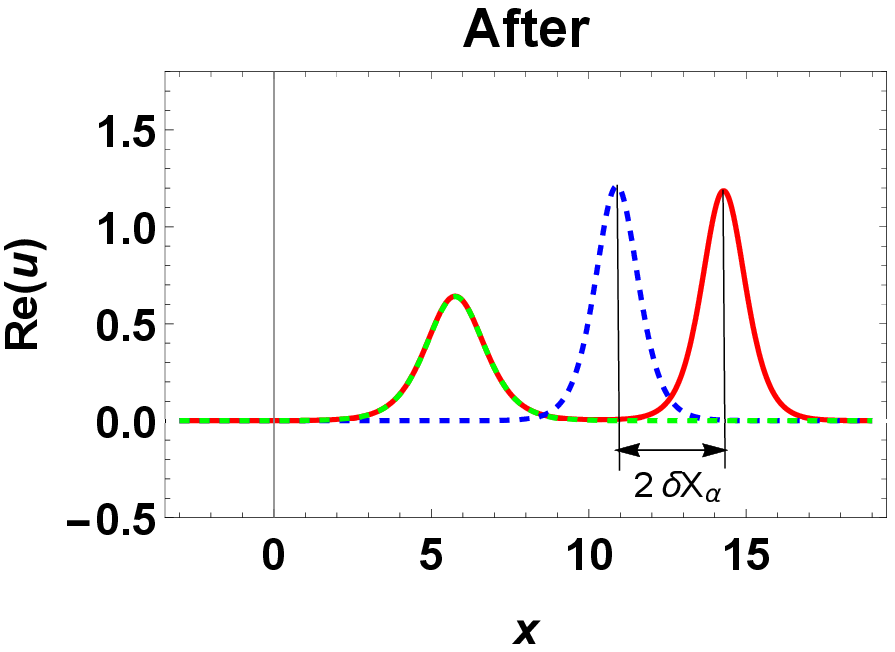}\\
	
	\caption{Snapshots of before $t=-9$ and after $t=9$ scattering of real part of the KdV complex two-soliton solution from HDM (red), with corresponding real parts of complex one-soliton solutions (blue/green), where $\alpha=1.1$, $\beta=0.8$ and $\mu=\nu=i \frac{\pi}{2}$.} \label{fig3.4}
\end{figure}

With the same asymptotic analysis, we can take the complex two-soliton solution from HDM and also compute for large time, before and after scattering, the distances between the two-soliton compound constituents and their corresponding one-soliton solutions, as shown in Figure \ref{fig3.4}. Although for the HDM case, the distances before and after scattering are different compared with the BT case, the lateral displacements and time-delays are found to be the same in both cases.

\vspace{0.3cm}

\noindent \large{\bfseries{Reality of conserved charges for the KdV equation}}

\label{realitycond}

 The snapshots of the complex soliton solution are in fact, as we shall see shortly, mass densities of these solutions. The value of mass is then computed from taking the integral of the mass density on the whole real line in space. 

For each of the complex one-soliton solutions in their moving reference frames, we see the real and imaginary parts are always an even and odd function respectively, as they are $\mathcal{PT}$-symmetric or can be made $\mathcal{PT}$-symmetric with a shift in space or time. With this fact, along with the fact that asymptotically, the complex two-soliton solution can be seen as sum of one-soliton solutions up to some displacements, which is also a property of integrability, we can conclude the imaginary part's contribution to mass from the complex two-soliton compound from any method is always zero. Furthermore, the value of mass will be sum of corresponding real parts of the complex one-soliton solutions, explaining reality of mass for complex two-soliton solutions. This reality of mass explanation can be extended for general complex N-soliton solutions from HDM or BT.

We now proceed to provide an argument that $\mathcal{PT}$-symmetry together with integrability will guarantee that we have reality for all conserved charges of the KdV equation through looking at the structure of charge densities.

First we provide a brief review of the construction of conserved charges from the Gardner transformation \cite{miura_korteweg-vries_1968,miura_korteweg-vries_1976,kupershmidt_nature_1981,cen_time-delay_2017}. The central idea is to expand the KdV-field $u(x,t)$ in terms of a new field $w(x,t)$
\begin{equation}
u(x,t)=w(x,t)+\varepsilon w_{x}(x,t)-\varepsilon ^{2}w^{2}(x,t),  \label{uw}
\end{equation}

\noindent for some deformation parameter $\varepsilon \in \mathbb{R}$. The
substitution of $u(x,t)$ into the KdV equation (\ref{KdVcomplex}) yields
\begin{equation}
\left( 1+\varepsilon \partial _{x}-2\varepsilon ^{2}w\right) \left[
w_{t}+\left( w_{xx}+3w^{2}-2\varepsilon ^{2}w^{3}\right) _{x}\right] =0.
\end{equation}

\noindent Since the last bracket is in form of a conservation law and needs to vanish by itself, one concludes that $\int\nolimits_{-\infty }^{\infty }w(x,t)dx=$ constant (independent of $t$). Expanding the new field as
\begin{equation}
w(x,t)=\sum\limits_{n=0}^{\infty }\varepsilon ^{n}w_{n}(x,t)
\end{equation}

\noindent implies that also the quantities $I_{n}:=\int\nolimits_{-\infty
}^{\infty }w_{2n-2}(x,t)dx$ are conserved. We may then use the relation (\ref{uw}) to construct the charge densities in a recursive manner
\begin{equation}
w_{n}=u\delta _{n,0}-\left( w_{n-1}\right)
_{x}+\sum\limits_{k=0}^{n-2}w_{k}w_{n-k-2}.  \label{rec1}
\end{equation}

\noindent Solving (\ref{rec1}) recursively, by taking $w_{n}=0$ for $n<0$, we obtain easily the well known expressions for the first few charge densities, namely
\begin{eqnarray}
w_{0} &=&u, \\
w_{1} &=&-\left( w_{0}\right) _{x}=-u_{x}, \\
w_{2} &=&-\left( w_{1}\right) _{x}+w_{0}^{2}=u_{xx}+u^{2}, \\
w_{3} &=&-\left( w_{2}\right) _{x}+2w_{0}w_{1}=-u_{xxx}-2(u^{2})_{x}, \\
w_{4} &=&-\left( w_{3}\right)
_{x}+2w_{0}w_{2}+w_{1}^{2}=u_{xxxx}+6(uu_{x})_{x}+2u^{3}-u_{x}^{2}.
\end{eqnarray}

\noindent The expressions simplify substantially when we drop surface terms and we
recover the first three charges of the KdV equation, given by $I_{0}$, $I_{1}$, $I_{2}$. 

For the charges constructed from the KdV complex one-soliton solution  we obtain real expressions
\begin{equation}
I_{n}=\int\nolimits_{-\infty }^{\infty }w_{2n-2}(x,t)dx=\frac{2}{2n-1}%
\alpha ^{2n-1}\qquad \text{and\qquad }I_{n/2}=0.  \label{III}
\end{equation}

The reality of all charges built on one-soliton solutions is guaranteed by $\mathcal{PT}$-symmetry alone: When realizing the $\mathcal{PT}$-symmetry as $\mathcal{PT}$: $u\rightarrow u$, $x\rightarrow -x$, $t\rightarrow -t$, $i\rightarrow -i$ it is easily seen from (\ref{rec1}) that the charge
densities transform as $w_{n}\rightarrow (-1)^{n}w_{n}$. This mean when $u(x,t)$ is $\mathcal{PT}$-symmetric so are the even graded charge densities $w_{2n}(x,t)$. Changing the argument of the functional dependence to the travelling wave coordinate $\zeta _{\alpha }=x-\alpha ^{2}t$ this means we can separate $w_{2n}(\zeta _{\alpha })$ into a $\mathcal{PT}$-even and $\mathcal{PT}$-odd part $w_{2n}^{e}(\zeta _{\alpha })\in \mathbb{R}$ and $w_{2n}^{o}(\zeta _{\alpha })\in \mathbb{R}$, respectively, as $w_{2n}(\zeta_{\alpha })=w_{2n}^{e}(\zeta _{\alpha })+iw_{2n}^{o}(\zeta _{\alpha })$, which allows us to conclude
\begin{eqnarray}
I_{n}(\alpha )&=&\int\nolimits_{-\infty }^{\infty
}w_{2n-2}(x,t)dx\\
&=&\int\nolimits_{-\infty }^{\infty }\left[
w_{2n-2}^{e}(\zeta _{\alpha })+iw_{2n-2}^{o}(\zeta _{\alpha })\right] d\zeta
_{\alpha } \nonumber \\
&=&\int\nolimits_{-\infty }^{\infty }w_{2n-2}^{e}(\zeta _{\alpha
})d\zeta _{\alpha }\in \mathbb{R}. \nonumber
\end{eqnarray}

It is easily seen that the previous argument applies directly to the charges built from the KdV complex one-soliton solution, i.e.
the real part and imaginary part are even and odd in $\zeta _{\alpha }$,
respectively. When the parameter $\mu $ has a nonvanishing real part the $\mathcal{PT}$-symmetry is broken, but it can be restored by absorbing the real part by a shift either in $t$ or $x$.

In order to ensure the same for the multi-soliton solutions we use the fact that the multi-soliton solutions separate asymptotically into single solitons with distinct support. As the charges are conserved in time, we may compute $I_{n}$ at any time. In the asymptotic regime, any charges built from an $N$-soliton solution $u_{i\theta _{1},\ldots, i\theta_{N};\alpha _{1},\ldots,\alpha _{N}}^{(N)}$, decomposes into the sum of charges built on the one-soliton solutions, that is
\begin{eqnarray}
I_{n}(\alpha _{1},\ldots ,\alpha _{N}) &=&\int\nolimits_{-\infty }^{\infty
}\left( w_{i\theta _{1},\ldots ,i\theta _{N};\alpha _{1},\ldots ,\alpha
	_{N}}^{(N)}\right) _{2n-2}(x,t)dx,  \label{I1} \\
&=&\int\nolimits_{-\infty }^{\infty }\sum\nolimits_{k=1}^{N}\left[ \left(
w_{i\theta _{k};\alpha _{k}}^{(1)}\right) _{2n-2}(\zeta _{\alpha _{k}})%
\right] d\zeta _{\alpha _{k}},  \label{I2} \\
&\sim&\sum\nolimits_{k=1}^{N}I_{n}(\alpha _{k}),  \label{I3} \\
&=&\frac{2}{2n-1}\sum\nolimits_{k=1}^{N}\alpha _{k}^{2n-1}.  \label{I4}
\end{eqnarray}

\noindent We used here the decomposition of the N-soliton solution into a sum of one-solitons in the asymptotic regime $u_{i\theta _{1},\ldots ,i\theta _{N};\alpha	_{1},\ldots ,\alpha_{N}}^{(N)}\sim \dsum\nolimits_{k=1}^{N}\left( u_{i\theta_{k};\alpha _{k}}^{(1)}\right) $, which we have seen in detail above. Since each of the one-solitons is well localized we always have $u_{i\theta_{k};\alpha _{k}}^{(1)}\cdot u_{i\theta_{l};\alpha_{l}}^{(1)}\sim 0$ for $N\geq 2$ when $k\neq l$, which implies that 
\begin{equation}
\left[ u_{i\theta _{1},\ldots ,i\theta _{N};\alpha _{1},\ldots ,\alpha
	_{N}}^{(N)}\right] ^{m}\sim\left[ \sum\nolimits_{k=1}^{N}\left( u_{i\theta
	_{k};\alpha _{k}}^{(1)}\right) \right] ^{m}\sim\sum\nolimits_{k=1}^{N}\left(
u_{i\theta _{k};\alpha _{k}}^{(1)}\right) ^{m}.
\end{equation}

\noindent As all the derivatives are finite and the support is the same as for the $u$s, this also implies
\begin{equation}
\left[ \left( u_{i\theta _{1},\ldots ,i\theta _{N};\alpha _{1},\ldots
	,\alpha _{N}}^{(N)}\right) _{nx}\right] ^{m}\sim\left[ \sum\nolimits_{k=1}^{N}%
\left( u_{i\theta _{k};\alpha _{k}}^{(1)}\right) _{nx}\right]
^{m}\sim\sum\nolimits_{k=1}^{N}\left( u_{i\theta _{k};\alpha
	_{k}}^{(1)}\right) _{nx}^{m},
\end{equation}

\noindent and similarly for mixed terms involving different types of derivatives. As all charge densities are made up from $u$ and its derivatives we obtain
\begin{equation}
\left( w_{i\theta _{1},\ldots ,i\theta _{N};\alpha _{1},\ldots ,\alpha
	_{N}}^{(N)}\right) _{2n-2}\sim\sum\nolimits_{k=1}^{N}\left( w_{i\theta
	_{k};\alpha _{k}}^{(1)}\right) _{2n-2}
\end{equation}

\noindent in the asymptotic regime, which is used in the step from (\ref{I1}) to (\ref{I2}). In the remaining two steps (\ref{I3}) and (\ref{I4}) we use (\ref{III}).

\noindent Thus, \emph {\large \bfseries $\mathbb{\mathcal{PT}}$-symmetry} and \emph {\large\bfseries integrability} guarantee the \emph {\large\bfseries reality of all charges}.

\section{Conclusions}

In this chapter, we have shown how one can generalise some well-known NPDEs including the KdV, mKdV and SG equations to the complex field whilst preserving $\mathcal{PT}$-symmetries and in particular, integrability in the sense of possessing soliton solutions and in the KdV case, also an infinite number of conserved charges. 

We are able to derive new complex soliton and multi-soliton solutions for these models through making adjustments with HDM and BT. For all the complex soliton solutions derived, we found they possess real energies. In the one-soliton case, this reasoning is due to $\mathcal{PT}$-symmetry of the Hamiltonian density and solution. However, for complex multi-soliton solutions, whether  $\mathcal{PT}$-symmetric or not, we found they all possessed real energies due to the additional property of integrability; how each complex multi-soliton solution asymptotically separates into complex one-soliton solutions which are $\mathcal{PT}$-symmetrizable up to some lateral displacements or time-delays. In particular, for the KdV equation, we proved all charges are real.

\chapter{Multicomplex soliton solutions of the KdV equation}\label{ch_4}

Similar to the previous chapter, we will investigate here further extensions of the real KdV equation not in the complex \cite{cen_complex_2016,cen_time-delay_2017}, but the multicomplex regime \cite{cen_multicomplex_2018}. These are higher order complex extensions, in particular they will be of bicomplex, quaternionic, coquaternionic and octonionic types. 

Extending quantum systems to the multicomplex regime has been proved useful in different ways. The application of bicomplex extension to extend the inner product space over which the Hilbert space is defined was found to help unravel the structure of the neighbourhood of higher order exceptional points \cite{dast_eigenvalue_2013,dizdarevic_cusp_2015,gutohrlein_bifurcations_2016}, where we have more than two eigenvalues coalescing. Quaternions and coquaternions have been long studied in the quantum regime, as it was found they are related to many important algebras and groups in physics \cite{finkelstein_foundations_1962,girard_quaternion_1984,adler_quaternionic_1995} and have recently been suggested to offer a unifying framework for complexified classical and quantum mechanics \cite{brody_complexified_2011}. Octonionic Hilbert spaces have been utilised in the study of quark structures \cite{gunaydin_quark_1973}. 

We first review some properties of multicomplex numbers. For more detailed introductions, we refer the reader to \cite{cockle_on_1849,kantor_hypercomplex_1989,price_introduction_1991}.

\subsection{Bicomplex and hyperbolic numbers}

\begin{table}[h]
	\centering
	\begin{tabular}{|c|c|c|c|c|}
		\hline
		* & $1$ & $\imath$ & $\jmath$ & $k$  \\ \hline
		$1$ & $1$ & $\imath$ & $\jmath$ & $k$  \\ \hline
		$\imath$ & $\imath$ & $-1$ & $k$ & $-\jmath$ \\ \hline
		$\jmath$ & $\jmath$ & $k$ & $-1$ & $-\imath$ \\ \hline
		$k$ & $k$ & $-\jmath$ & $-\imath$ & $1$  \\ \hline
	\end{tabular}
\caption{Bicomplex Cayley table} \label{bicomplexcayley}
\end{table}

\noindent Denoting the field of complex numbers with imaginary unit $\imath $ as
\begin{equation}
\mathbb{C}(\imath )=\left\{ x+\imath y|x,y\in \mathbb{R}\right\} ,
\end{equation}

\noindent the bicomplex numbers $\mathbb{B}$ form an algebra over the complex numbers admitting various equivalent types of representations
\begin{eqnarray}
\mathbb{B} &\mathbb{=}&\left\{ z_{1}+\jmath z_{2}|z_{1},z_{2}\in \mathbb{C}(\imath )\right\} ,  \label{B1} \\
&=&\left\{ w_{1}+\imath w_{2}|w_{1},w_{2}\in \mathbb{C}(\jmath)\right\} , \label{B2} \\
&=&\left\{ a_{0} +a_{1}\imath +a_{2}\jmath +a_{3}k|a_{0},a_{1},a_{2},a_{3}\in \mathbb{R}\right\} ,  \label{B3} \\
&=&\left\{ v_{1}e_{1}+v_{2}e_{2}|v_{1}\in \mathbb{C}(\imath ),v_{2}\in \mathbb{C}(\jmath )\right\} .  \label{B4}
\end{eqnarray}

The canonical basis is spanned by the units $1$, $\imath $, $\jmath $, $k$, involving the two imaginary units $\imath $ and $\jmath $ squaring to $-1$, so that the representations in equations (\ref{B1}) and (\ref{B2}) naturally prompt the notion to view these numbers as a doubling of the complex numbers. The real unit $1$ and the hyperbolic unit $k=\imath \jmath $ square to $1$. The multiplication of these units is commutative and we can represent the
products in the Cayley multiplication table \ref{bicomplexcayley}. The idempotent representation (\ref{B4}) is an orthogonal
decomposition obtained by using the orthogonal idempotents
\begin{equation}
e_{1}:=\frac{1+k}{2},\qquad \text{and\qquad }e_{2}:=\frac{1-k}{2},
\end{equation}

\noindent with properties $e_{1}^{2}=e_{1}$, $e_{2}^{2}=e_{2}$, $e_{1}e_{2}=0$ and $e_{1}+e_{2}=1$. All four representations (\ref{B1}) - (\ref{B4}) are uniquely related to each other. For instance, given a bicomplex number in the canonical representation (\ref{B3}) in the form
\begin{equation}
n_{a}=a_{0} +a_{1}\imath +a_{2}\jmath +a_{3}k,  \label{ar}
\end{equation}

\noindent the equivalent representations (\ref{B1}), (\ref{B3}) and (\ref{B4}) are obtained with the identifications
\begin{equation}
\begin{array}{ll}
z_{1}=a_{0}+\imath a_{1}, & z_{2}=a_{2}+\imath a_{3}, \\ 
w_{1}=a_{0}+\jmath a_{2}, & w_{2}=a_{1}+\jmath a_{3}, \\ 
v_{1}^{a}=(a_{0}+a_{3}) +(a_{1}-a_{2})\imath ,  & 
v_{2}^{a}=(a_{0}-a_{3}) +(a_{1}+a_{2})\jmath .
\end{array}
\label{rel}
\end{equation}

Arithmetic operations are most elegantly and efficiently carried out in the idempotent representation (\ref{B4}). For the composition of two arbitrary numbers $n_{a}$ and $n_{b}$ we have 
\begin{equation}
n_{a}\circ n_{b}=v_{1}^{a}\circ v_{1}^{b}e_{1}+v_{2}^{a}\circ
v_{2}^{b}e_{2}\quad \text{with }\circ \equiv \pm ,\cdot , \divisionsymbol 
\label{comp}
\end{equation}

\begin{figure}[h]
 	\centering
 	
 	\includegraphics[width=0.48\linewidth]{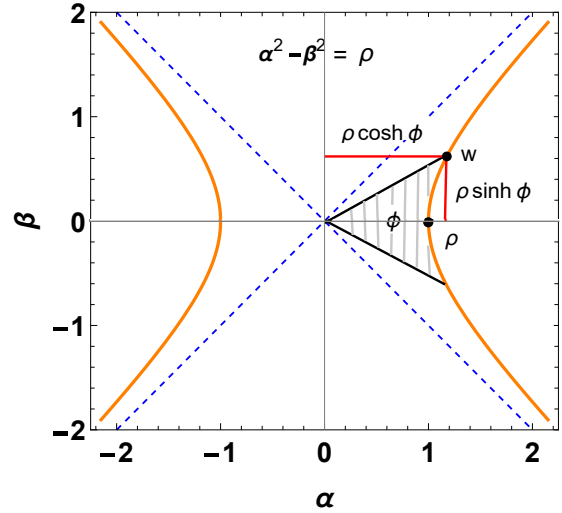} 
 	
 	\caption{Geometrical representation of Hyperbolic numbers}
 	
 	\label{hyperbola}
 \end{figure}

The hyperbolic numbers (or split-complex numbers) 
\begin{equation}
	\mathbb{D=} \left\{ a_{0} +a_{3}k|a_{0},a_{3}\in \mathbb{R} \right\}
\end{equation}

 \noindent are an important special case of $\mathbb{B}$ obtained in the absence of the imaginary units $\imath $ and $\jmath $, or when taking $a_{1}=a_{2}=0$. Similar to how we can represent complex numbers in polar form, we have the same for hyperbolic numbers \cite{sobczyk_hyperbolic_1995}, as show in Figure \ref{hyperbola}. $W$ represents a hyperbolic number with several representations as
 \begin{eqnarray}
 	w &=& \alpha + k \beta , \\
 	&=& \rho e^{k \phi} , \\
 	&=& \rho (\cosh \phi + k \sinh \phi) , 
 \end{eqnarray} 
 
 \noindent where $\rho = \sqrt{\alpha^{2}-\beta^{2}}$ and $\phi =\func { arctanh } \frac{\beta}{\alpha}$.
 
\vspace{0.3cm}

\noindent {\large{\bfseries{Bicomplex functions}}}

For bicomplex functions, we have the same arithmetical rules as for numbers. In what follows we are most interested in functions depending on two real variables $x$ and $t$ of the form $f(x,t)= p(x,t)+\imath q(x,t)+\jmath r(x,t)+ks(x,t)\in \mathbb{B}$ involving four real fields $p(x,t)$, $q(x,t)$, $r(x,t)$, $s(x,t)\in \mathbb{R}$. Having kept the functional variables real, we also keep our derivatives real, so that we can differentiate $f(x,t)$ component-wise as $\partial _{x}f(x,t)= \partial _{x}p(x,t)+\imath \partial _{x}q(x,t)+\jmath \partial_{x}r(x,t)+k\partial _{x}s(x,t)$ and similarly for $\partial _{t}f(x,t)$. 

\vspace{0.2cm}

\noindent {\large \bfseries{{Bicomplex extended $\mathcal{PT}$-symmetries}}}

As there are two different imaginary units, there are three different types of conjugations for bicomplex numbers, corresponding to conjugating only $\imath $, only $\jmath $ or conjugating both $\imath $ and $\jmath $ simultaneously. This is reflected in different symmetries that leave the Cayley multiplication table invariant. As a consequence we also have three different types of bicomplex $\mathcal{PT}$-symmetries, acting as
\begin{eqnarray}
\mathcal{PT}_{\imath \jmath } &:&\imath \rightarrow -\imath ,\jmath \rightarrow -\jmath ,k\rightarrow k,x\rightarrow -x,t\rightarrow -t,\ \   \label{PT1} \\
\mathcal{PT}_{\imath k} &:&\imath \rightarrow -\imath ,\jmath \rightarrow \jmath ,k\rightarrow -k,x\rightarrow -x,t\rightarrow -t, \label{PT2} \\
\mathcal{PT}_{\jmath k} &:&\imath \rightarrow \imath,\jmath \rightarrow -\jmath ,k\rightarrow -k,x\rightarrow -x,t\rightarrow -t,\   \label{PT3}
\end{eqnarray}

\noindent see also \cite{bagchi_bicomplex_2015,cen_multicomplex_2018}.

\subsection{Quaternionic numbers and functions}

\begin{table}[h]
	\centering
	\begin{tabular}{|c|c|c|c|c|}
		\hline
		* & $1$ & $\imath$  & $\jmath$ & $k$  \\ \hline
		$1$ & $1$ & $\imath$  & $\jmath$  & $k$  \\ \hline
		$\imath$ & $\imath$ & $-1$ & $k$  & $-\jmath$ \\ \hline
		$\jmath$ & $\jmath$ & $-k$ & $-1$ & $\imath$  \\ \hline
		$k$ & $k$ & $\jmath$  & $-\imath$ & $-1$ \\ \hline
	\end{tabular}
\caption{Quaternion Cayley table} \label{quaternioncayley}
\end{table}

\noindent The quaternions in the canonical basis are defined as the set of elements
\begin{equation}
\mathbb{H}=\left\{ a_{0} +a_{1}\imath +a_{2}\jmath
+a_{3}k|a_{0},a_{1},a_{2},a_{3}\in \mathbb{R}\right\} .
\end{equation}

\noindent The multiplication of the basis $\{1 ,\imath ,\jmath ,k\}$ is
noncommutative, with $\imath ,\jmath ,k$ denoting the three imaginary units with $\imath ^{2}=\jmath^{2}=k^{2}=-1$. The remaining multiplication rules are shown in table \ref{quaternioncayley}. The multiplication table remains invariant under the symmetries $\mathcal{PT}_{\imath \jmath }$, $\mathcal{PT}_{\imath k}$ and $\mathcal{PT}_{\jmath k}$. Using these rules for the basis, two quaternions in the canonical basis $n_{a}=a_{0} +a_{1}\imath +a_{2}\jmath +a_{3}k\in \mathbb{H}$ and $n_{b}=b_{0} +b_{1}\imath +b_{2}\jmath +b_{3}k\in \mathbb{H}$ are multiplied as
\begin{eqnarray}
n_{a}n_{b} \!\!&=&\!\!\left( a_{0}b_{0}-a_{1}b_{1}-a_{2}b_{2}-a_{3}b_{3}\right)
+\left( a_{0}b_{1}+a_{1}b_{0}+a_{2}b_{3}-a_{3}b_{2}\right) \imath 
\label{m2} \\
&&+\left( a_{0}b_{2}-a_{1}b_{3}+a_{2}b_{0}+a_{3}b_{1}\right) \jmath +\left(
a_{0}b_{3}+a_{1}b_{2}-a_{2}b_{1}+a_{3}b_{0}\right) k.  \notag
\end{eqnarray}

There are various representations for quaternions, see e.g. \cite
{sangwine_fundamental_2011}, of which the complex form will be especially useful for what follows. With the help of (\ref{m2}) one easily verifies that
\begin{equation}
\xi :=\frac{1}{\mathcal{N}}\left( a_{1}\imath +a_{2}\jmath +a_{3}k\right) \quad \text{with} \quad \mathcal{N=}\sqrt{a_{1}^{2}+a_{2}^{2}+a_{3}^{2}}
\label{ima}
\end{equation}

\noindent constitutes a new imaginary unit with\ $\xi ^{2}=-1$. This means that in this representation we can formally view a quaternion, $n_{a}\in \mathbb{H}$, as an element in the complex numbers
\begin{equation}
n_{a}=a_{0} +\xi \mathcal{N\in \mathbb{C}(\xi ),}  \label{cf}
\end{equation}

\noindent with real part $a_{0}$ and imaginary part $\mathcal{N}$. Notice that a $\mathcal{PT}_{\xi }$-symmetry can only be achieved with a $\mathcal{PT}_{\imath \jmath k}$-symmetry acting on the unit vectors in the canonical representation. Unlike the bicomplex numbers or the coquaternions, the quaternionic algebra does not contain any idempotents.

\subsection{Coquaternionic numbers and functions}

\begin{table}[h]
	\centering
	\begin{tabular}{|c|c|c|c|c|}
		\hline
		* & $1$ & $\imath$  & $\jmath$ & $k$  \\ \hline
		$1$ & $1$ & $\imath$  & $\jmath$ & $k$  \\ \hline
		$\imath$ & $\imath$ & $-1$ & $k$ & $-\jmath$ \\ \hline
		$\jmath$ & $\jmath$ & $-k$ & $1$ & $-\imath$ \\ \hline
		$k$ & $k$ & $\jmath$  & $\imath$ & $1$  \\ \hline
	\end{tabular}
\caption{Coquaternion Cayley table} \label{coquaternioncayley}
\end{table}

\noindent The coquaternions, often also referred to as split-quaternions in the
canonical basis, are defined as the set of elements
\begin{equation}
\mathbb{P}=\left\{ a_{0} +a_{1}\imath +a_{2}\jmath
+a_{3}k|a_{0},a_{1},a_{2},a_{3}\in \mathbb{R}\right\} .  \label{co1}
\end{equation}

The multiplication of the basis $\{1 ,\imath ,\jmath ,k\}$ is
noncommutative with two
hyperbolic unit elements $\jmath ,k$, $\jmath ^{2}=k^{2}=1$, and one
imaginary unit $\imath ^{2}=-1$. The remaining multiplication rules are shown in the Cayley table \ref{coquaternioncayley}. The multiplication table remains invariant under the symmetries $\mathcal{PT}_{\imath \jmath }$, $\mathcal{PT}_{\imath k}$ and $\mathcal{PT}_{\jmath k}$. Using these rules for the basis, two
coquaternions in the canonical basis $n_{a}=a_{0}+a_{1}\imath
+a_{2}\jmath +a_{3}k\in \mathbb{P}$ and $n_{b}=b_{0}+b_{1}\imath
+b_{2}\jmath +b_{3}k\in \mathbb{P}$ are multiplied as
\begin{eqnarray}
n_{a}n_{b}\!\!&=&\!\!\left( a_{0}b_{0}-a_{1}b_{1}+a_{2}b_{2}+a_{3}b_{3}\right)
+\left( a_{0}b_{1}+a_{1}b_{0}-a_{2}b_{3}+a_{3}b_{2}\right) \imath  \label{m3} \\
&&+\left( a_{0}b_{2}-a_{1}b_{3}+a_{2}b_{0}+a_{3}b_{1}\right) \jmath +k\left(a_{0}b_{3}+a_{1}b_{2}-a_{2}b_{1}+a_{3}b_{0}\right) k.  \notag
\end{eqnarray}

There are various coquaternionic representations for numbers and functions. Similar as a quaternion one can formally view a coquaternion, $n_{a}\in \mathbb{P}$, as an element in the complex numbers
\begin{equation}
n_{a}=a_{0}+\zeta \mathcal{M\in \mathbb{C}(\zeta )},  \label{cocompl}
\end{equation}

\noindent with real part $a_{0}$ and imaginary part $\mathcal{M}$. The new imaginary unit, $\zeta ^{2}=-1$, 
\begin{equation}
\zeta :=\frac{1}{\mathcal{M}}\left( a_{1}\imath +a_{2}\jmath +a_{3}k\right)
\quad \text{with }\mathcal{M}=\sqrt{a_{1}^{2}-a_{2}^{2}-a_{3}^{2}}
\end{equation}

\noindent is, however, only defined for $\mathcal{M}\neq 0$. Similarly as for the $\mathcal{PT}_{\xi }$-symmetry also the $\mathcal{PT}_{\zeta }$-symmetry requires a $\mathcal{PT}_{\imath \jmath k}$-symmetry. Unlike the quaternions, the coquaternions possess a number of idempotents $e_{1}=\frac{1+k}{2}$, $e_{2}=\frac{1-k}{2}$ with $e_{1}^{2}=e_{1}$, $e_{2}^{2}=e_{2}$, $e_{1}e_{2}=0$ or $e_{3}=\frac{1+\jmath}{2}$, $e_{4}=\frac{1-\jmath}{2}$ with $e_{3}^{2}=e_{3}$, $e_{4}^{2}=e_{4}$, $e_{3}e_{4}=0$. So for
instance, $n_{a}$ is an element in 
\begin{equation}
\mathbb{P}=\left\{ e_{1}v_{1}+e_{2}v_{2}|v_{1}\in \mathbb{D}(\jmath
),v_{2}\in \mathbb{D}(\jmath )\right\} ,  \label{idco}
\end{equation}

\noindent where the hyperbolic numbers in (\ref{idco}) are related to the coefficient in the canonical basis as $v_{1}=(a_{0}+a_{3}) +(a_{1}+a_{2})\jmath $ and $v_{2}=(a_{0}-a_{3}) +(a_{2}-a_{1})\jmath $.

\subsection{Octonionic numbers and functions}

\begin{table}[h]
	\centering
	\begin{tabular}{|c|c|c|c|c|c|c|c|c|}
		\hline
		*  & $e_{0}$ & $e_{1}$  & $e_{2}$  & $e_{3}$  & $e_{4}$  & $e_{5}$  & $e_{6}$  & $e_{7}$  \\ \hline
		$e_{0}$ & $e_{0}$ & $e_{1}$ & $e_{2}$  & $e_{3}$  & $e_{4}$  & $e_{5}$  & $e_{6}$  & $e_{7}$  \\ \hline
		$e_{1}$ & $e_{1}$ & $-e_{0}$ & $e_{3}$  & $-e_{2}$ & $e_{5}$  & $-e_{4}$ & $-e_{7}$ & $e_{6}$  \\ \hline
		$e_{2}$ & $e_{2}$ & $-e_{3}$ & $-e_{0}$ & $e_{1}$  & $e_{6}$  & $e_{7}$  & $-e_{4}$ & $-e_{5}$ \\ \hline
		$e_{3}$ & $e_{3}$ & $e_{2}$  & $-e_{1}$ & $-e_{0}$ & $e_{7}$  & $-e_{6}$ & $e_{5}$  & $-e_{4}$ \\ \hline
		$e_{4}$ & $e_{4}$ & $-e_{5}$ & $-e_{6}$ & $-e_{7}$ & $-e_{0}$ & $e_{1}$  & $e_{2}$  & $e_{3}$  \\ \hline
		$e_{5}$ & $e_{5}$ & $e_{4}$  & $-e_{7}$ & $e_{6}$  & $-e_{1}$ & $-e_{0}$ & $-e_{3}$ & $e_{2}$  \\ \hline
		$e_{6}$ & $e_{6}$ & $e_{7}$  & $e_{4}$  & $-e_{5}$ & $-e_{2}$ & $e_{3}$  & $-e_{0}$ & $-e_{1}$ \\ \hline
		$e_{7}$ & $e_{7}$ & $-e_{6}$ & $e_{5}$  & $e_{4}$  & $-e_{3}$ & $-e_{2}$ & $e_{1}$  & $-e_{0}$ \\ \hline
	\end{tabular}
\end{table}

Octonions or Cayley numbers have a double the dimensions of quaternions and not only non-commutative, but also non-associative. In the canonical basis they can be represented as
\begin{equation}
\mathbb{O}=\left\{
a_{0}e_{0}+a_{1}e_{1}+a_{2}e_{2}+a_{3}e_{3}+a_{4}e_{4}+a_{5}e_{5}+a_{6}e_{6}+a_{7}e_{7}|a_{i}\in 
\mathbb{R}\right\} .
\end{equation}

The multiplication of the units is defined by noting that each of the seven quadruplets $(e_{0},e_{1},e_{2},e_{3})$, $(e_{0},e_{1},e_{4},e_{5})$, $(e_{0},e_{1},e_{7},e_{6})$, $(e_{0},e_{2},e_{4},e_{6})$, $(e_{0},e_{2},e_{5},e_{7})$, $(e_{0},e_{3},e_{4},e_{7})$ and $(e_{0},e_{3},e_{6},e_{5})$, constitutes a canonical basis for the quaternions in one-to-one correspondence with $(1 ,\imath ,\jmath ,k)$. Hence, the octonions have one real unit, $7$ imaginary units and the
multiplication of two octonions is noncommutative. Similarly as for
quaternions and coquaternions we can view an octonion $n_{a}\in \mathbb{O}$ as a complex number 
\begin{equation}
n_{a}=a_{0} +o\mathcal{O\in \mathbb{C}}(o)  \label{pctn}
\end{equation}

\noindent with real part $a_{0}$, imaginary part $\mathcal{O}$ and newly defined
imaginary unit, $o^{2}=-1$, 
\begin{equation}
o:=\frac{1}{\mathcal{O}}\dsum\nolimits_{i=1}^{7}a_{i}e_{i}\qquad \text{where} \qquad \mathcal{O=}\sqrt{\dsum\nolimits_{i=1}^{7}a_{i}^{2}}.
\end{equation}

In order to obtain a $\mathcal{PT}_{o}$-symmetry we require a $\mathcal{PT}_{e_{1}e_{2}e_{3}e_{4}e_{5}e_{6}e_{7}}$-symmetry in the canonical basis.

\section{The bicomplex KdV equation}

Using the multiplication law for bicomplex functions, the KdV equation for a bicomplex field in the canonical form 
\begin{equation}
u(x,t)= p(x,t)+\imath q(x,t)+\jmath r(x,t)+k s(x,t)\in \mathbb{B},
\label{ub}
\end{equation}

\noindent can either be viewed as a set of coupled equations for the four real fields $p(x,t)$, $q(x,t)$, $r(x,t)$, $s(x,t)\in \mathbb{R}$ 
\begin{equation}
u_{t}+6uu_{x}+u_{xxx}=0 \Leftrightarrow \left\{ 
\begin{array}{r}
p_{t}+6pp_{x}-6qq_{x}-6rr_{x}+6ss_{x}+p_{xxx}=0 ,  \\ 
q_{t}+6qp_{x}+6pq_{x}-6sr_{x}-6rs_{x}+q_{xxx}=0 ,  \\ 
r_{t}+6rp_{x}+6pr_{x}-6qs_{x}-6sq_{x}+r_{xxx}=0 , \\ 
s_{t}+6sp_{x}+6ps_{x}+6qr_{x}+6rq_{x}+s_{xxx}=0 ,
\end{array} \right.
\label{KdVbi}
\end{equation}

\noindent or when using the representation (\ref{B4}) as a couple of complex KdV
equations
\begin{equation}
v_{t}+6vv_{x}+v_{xxx}=0,\qquad \text{and\qquad }w_{t}+6ww_{x}+w_{xxx}=0,
\label{SKdV}
\end{equation}

\noindent related to the canonical representation as
\begin{eqnarray}
v(x,t) &=&\left[ p(x,t)+s(x,t)\right] +\imath \left[ q(x,t)-r(x,t)\right]
\in \mathbb{C}(\imath ), \\
w(x,t) &=&\left[ p(x,t)-s(x,t)\right] +\jmath \left[ q(x,t)+r(x,t)\right]
\in \mathbb{C}(\jmath ).
\end{eqnarray}

\noindent We recall that we keep here our space and time variables, $x$ and $t$, to be both real so that also the corresponding derivatives $\partial _{x}$ and $\partial _{t}$ are not bicomplexified.

When acting on the component functions the $\mathcal{PT}$-symmetries (\ref{PT1})-(\ref{PT3}) are implemented in (\ref{KdVbi}) as 
\begin{eqnarray}
\mathcal{PT}_{\imath \jmath } &:&x\rightarrow -x,\ t\rightarrow
-t,p\rightarrow p,q\rightarrow -q,r\rightarrow -r,s\rightarrow
s,u\rightarrow u,  \\
\mathcal{PT}_{\imath k} &:&x\rightarrow -x,\ t\rightarrow -t,p\rightarrow
p,q\rightarrow -q,r\rightarrow r,s\rightarrow -s,u\rightarrow u, \\
\mathcal{PT}_{\jmath k} &:&x\rightarrow -x,\ t\rightarrow -t,p\rightarrow
p,q\rightarrow q,r\rightarrow -r,s\rightarrow -s,u\rightarrow u,
\end{eqnarray}

\noindent ensuring that the KdV-equation remains invariant for all of the
transformations. Notice that the representation in (\ref{SKdV}) remains only invariant under $\mathcal{PT}_{\imath \jmath }$, but does not respect the symmetries $\mathcal{PT}_{\imath k}$ and $\mathcal{PT}_{\jmath k}$.

\subsection{Hyperbolic scaled KdV equation}

 We observe that (\ref{KdVbi}) allows for a scaling of space by the hyperbolic unit $k$ as $x\rightarrow kx$, leading to a new type of KdV-equation with $u\rightarrow h$
\begin{equation}
kh_{t}+6hh_{x}+h_{xxx}=0 \Leftrightarrow \left\{ 
\begin{array}{r}
s_{t}+6pp_{x}-6qq_{x}-6rr_{x}+6ss_{x}+p_{xxx}=0 , \\ 
r_{t}-6qp_{x}-6pq_{x}+6sr_{x}+6rs_{x}-q_{xxx}=0 , \\ 
q_{t}-6rp_{x}-6pr_{x}+6qs_{x}+6sq_{x}-r_{xxx}=0 , \\ 
p_{t}+6sp_{x}+6ps_{x}+6qr_{x}+6rq_{x}+s_{xxx}=0 ,
\end{array} \right.
 \label{KdVH}
\end{equation}

\noindent that also respects the $\mathcal{PT}_{\imath \jmath }$-symmetry. The
interesting consequence of this modification is that travelling wave
solutions $u(\xi )$ of (\ref{KdVbi}) depending on real combination of $x$ and $t$ as $\xi =x+ct\in \mathbb{R}$, with $c$ denoting the speed, become solutions $h(\zeta )$ dependent on the hyperbolic number $\zeta =kx+ct\in \mathbb{D}$ instead. 

Interestingly a hyperbolic rotation the number $\zeta $ with $\phi =\func{ arctanh } (v/c)$ constitutes a Lorentz transformation \cite{sobczyk_hyperbolic_1995,ulrych_relativistic_2005}. Taking $\zeta$ and performing a hyperbolic rotation, we have
\begin{eqnarray}
\zeta ^{\prime } &=& \zeta e^{-\phi k} , \\
&=& k (x \cosh \phi - ct \sinh \phi)+c(t \cosh \phi -\frac{x}{c} \sinh \phi),
\end{eqnarray}

\noindent then with 
\begin{equation}
	\cosh \left( \func{arctanh}\frac{v}{c}\right)=\frac{1}{\sqrt{1-\frac{v^{2}}{c^{2}}}} \quad \text{,} \quad 	\sinh \left(\func{arctanh}\frac{v}{c}\right)=\frac{v}{c\sqrt{1-\frac{v^{2}}{c^{2}}}}
\end{equation}

\noindent we have a Lorentz transformed $\zeta$ given by
\begin{equation}
\zeta ^{\prime } =  k x^{\prime } + c t^{\prime },
\end{equation}

\noindent with $t^{\prime }=\gamma (t-\frac{v}{c^{2}}x)$, $x^{\prime }=\gamma (x-vt)$ and $\gamma =1/\sqrt{1-v^{2}/c^{2}}$.

Next we consider various solutions to these different versions of the
bicomplex KdV-equation, discuss how they may be constructed and their key properties.

\subsection{Bicomplex soliton solutions}

\noindent {\large\bfseries{Bicomplex one-soliton solution with $\mathcal{PT}$-symmetry broken}}

We start from the well known one-soliton solution of the real KdV equation 
\begin{equation}
u_{\mu ;\alpha }(x,t)=\frac{\alpha ^{2}}{2}\func{sech}^{2}\left[ \frac{1}{2}
(\alpha x-\alpha ^{3}t+\mu )\right] ,  \label{us}
\end{equation}

\noindent when $\alpha ,\mu \mathbb{\in R}$. Since our differentials have not been bicomplexified we may take $\mu $ to be a bicomplex number $\mu^{\mathbb{B}} =\eta_{0} +\theta_{0} \imath +\theta_{1} \jmath +\eta_{1} k\mathbb{\in B}$ with $\eta_{0} ,\theta_{0} ,\theta_{1},\eta_{1} \in \mathbb{R}$, so that (\ref{us}) becomes a solution of the bicomplex equation (\ref{KdVbi}). Expanding the hyperbolic function, we can separate the bicomplex function $u_{\mu^{\mathbb{B}} ;\alpha }(x,t)$ after some lengthy computation into its different canonical components
\begin{equation}
\! u_{\mu^{\mathbb{B}} ;\alpha } =\resizebox{.8\hsize}{!}{$\frac{1 }{2}\left( p_{+;\alpha}+p_{-;\alpha}\right) +\frac{ \imath }{2}\left( q_{+;\alpha}+q_{-;\alpha}\right) +\frac{\jmath }{2}\left( q_{+;\alpha}-q_{-;\alpha}\right) +\frac{k}{2}\left( p_{+;\alpha}-p_{-;\alpha}\right)$} ,  \label{unpt}
\end{equation}

\noindent when using the two functions 
\begin{eqnarray}
p_{\pm;\alpha}(x,t) &=&\frac{\alpha ^{2}+\alpha ^{2}\cos \theta_{\mp} \cosh (\alpha x-\alpha ^{3}t+\eta_{\pm})}{\left[ \cos \theta_{\mp}+\cosh (\alpha x-\alpha^{3}t+\eta_{\pm})\right] ^{2}},  \label{newS} \\
q_{\pm;\alpha}(x,t) &=&\frac{\alpha ^{2}\sin \theta_{\pm}\sinh (\alpha x-\alpha
^{3}t+\eta_{\mp})}{\left[ \cos \theta_{\pm}+\cosh (\alpha x-\alpha ^{3}t+\eta_{\mp})\right] ^{2}},
\end{eqnarray}

\noindent where $\eta_{\pm}\!=\eta_{0}\pm \eta_{1}$ \, and \, $\theta_{\pm}\!=\theta_{0}\pm \theta_{1}$.

 Noting that if we let $\eta_{1}\!\!=\!\theta_{1}\!\!=\!0$,\, $\eta_{0}\!\!=\!\eta$,\, $\theta_{0}\!\!=\!\theta$, the complex solution, i.e. $\mu=\eta+i \theta$ as studied in Chapter 3, (\ref{3.16}), can be expressed as 
 \begin{equation}
 	u_{i \theta;\alpha}(x,t)=\widehat{p}_{\eta,\theta;\alpha}(x-\eta/\alpha ,t)-i\widehat{q}_{\eta,\theta;\alpha}(x-\eta/\alpha ,t),
 \end{equation}
 where
 \begin{eqnarray}
\widehat{p}_{\eta,\theta;\alpha}(x ,t)&=&\{p_{\pm;\alpha}(x,t)| \eta_{1}\!\!=\!\theta_{1}\!\!=\!0,\, \eta_{0}=\eta,\, \theta_{0}=\theta\},\\
\widehat{q}_{\eta,\theta;\alpha}(x ,t)&=&\{q_{\pm;\alpha}(x,t)| \eta_{1}\!\!=\!\theta_{1}\!\!=\!0,\, \eta_{0}=\eta,\, \theta_{0}=\theta\}.
 \end{eqnarray}
We can also expand the bicomplex solution (\ref{unpt}) in terms of the complex solution as
\begin{eqnarray}
u_{\mu^{\mathbb{B}} ;\alpha } \!\!\!&=&\!\!\! \frac{1}{2}\left[u_{i\theta_{-} ;\alpha }\left( x+\frac{\eta_{+}}{\alpha},t\right) +u_{i\theta_{+} ;\alpha}\left(x+\frac{\eta_{-} }{\alpha },t\right) \right] \label{uco} \\
&&+\frac{k }{2}\left[ u_{i\theta_{-};\alpha}\left(x+\frac{\eta_{+} }{\alpha },t\right) -u_{i\theta_{+};\alpha}\left(x+\frac{\eta_{-} }{\alpha },t\right) \right] .  \notag
\end{eqnarray}

\begin{figure}[h]
	\centering
	
	\includegraphics[width=0.48\linewidth]{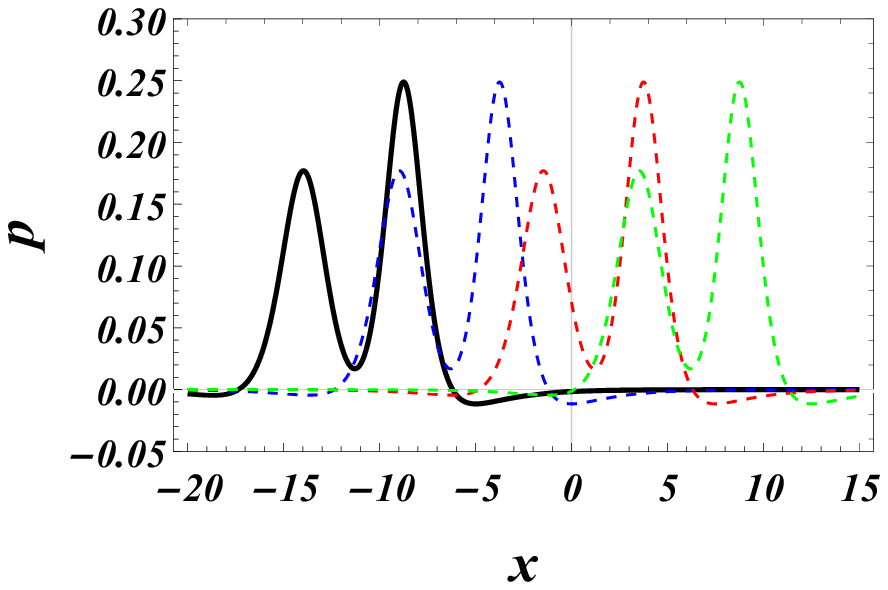} 
	\hspace{0.2cm}
	\includegraphics[width=0.48\linewidth]{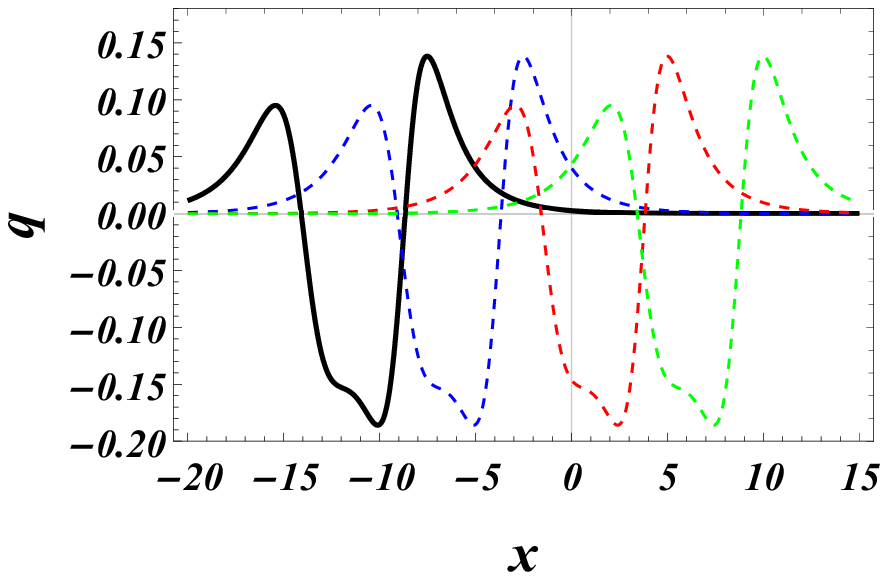}\\
	\hspace{0.2cm}\\
	
	\includegraphics[width=0.48\linewidth,height=0.31\linewidth]{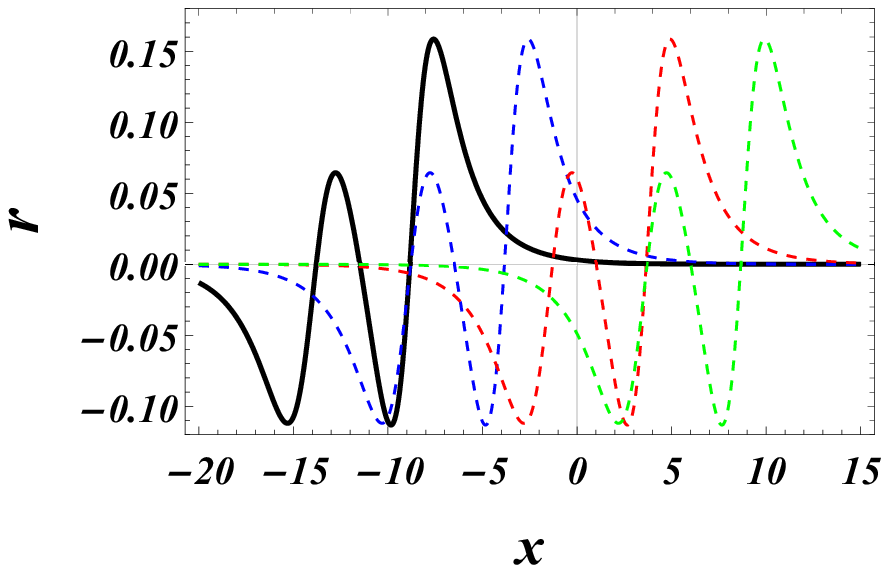}
	\hspace{0.2cm}
	\includegraphics[width=0.48\linewidth]{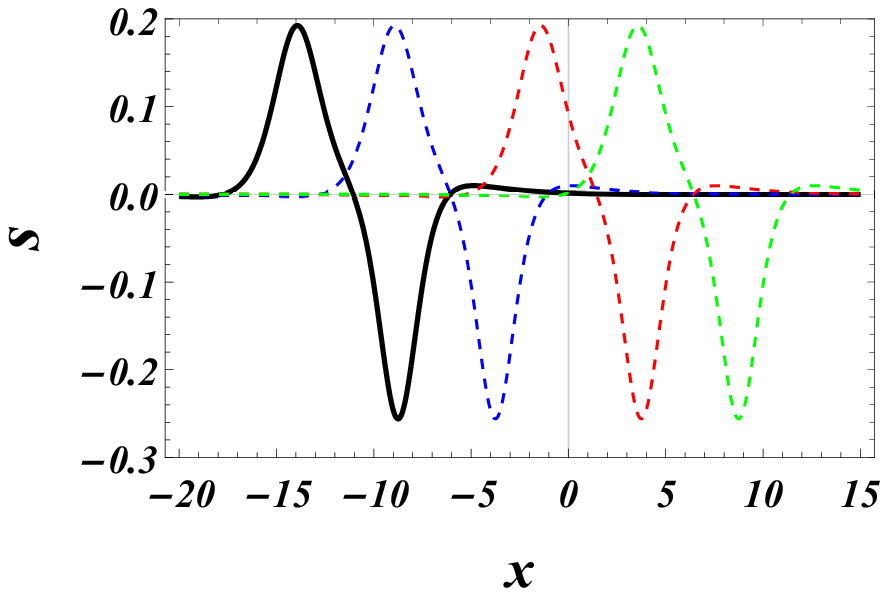}\\
	\vspace{0.2cm}
	\hspace{0.2cm}
	\includegraphics[width=0.7\linewidth]{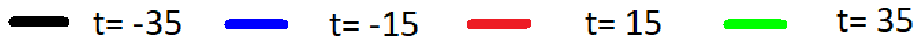}\\
	
	\caption{Canonical component functions $p$, $q$, $r$ and $s$ (clockwise starting in the top left corner) of the decomposed one-soliton solution $u_{\mu^{\mathbb{B}} ;\alpha }$ to the bicomplex KdV equation (\ref{KdVbi}) with broken $\mathcal{PT}$-symmetry at different times for $\alpha =0.5 $, $\eta_{0}=1.3 $, $\theta_{0}=0.4 $, $\theta_{1}=2.0 $ and $\eta_{1}=1.3 $.}

\label{fig4.2}
\end{figure}

In Figure \ref{fig4.2} we depict the canonical components of this solution at different times. We observe in all of them that the one-soliton solution is split into two separate one-soliton-like components moving parallel to each other with the same speed. The real $p$-component can be viewed as the sum of two bright solitons and the hyperbolic $s$-component is the sum of a bright and a dark soliton. This effect is the results of the decomposition of each of the components into a sum of the functions $p_{\pm;\alpha }$ or $q_{\pm;\alpha }$, as defined in (\ref{newS}) with $\theta_{\pm}$ and $\eta_{\pm}$ controlling the amplitude and distance, whereas $\alpha $ regulates the speed, so the constituents travel at the same speed. This is a novel type of phenomenon for soliton solutions previously not observed.

In general, the solution (\ref{us}) is not $\mathcal{PT}$-symmetric with regard to any of the possibilities defined above. It becomes $\mathcal{PT}_{\imath \jmath }$-symmetric when $\eta_{0} =\eta_{1} =0$, $\mathcal{PT}_{\imath k}$-symmetric when $\eta_{0} =\theta_{1} =0$ and $\mathcal{PT}_{jk}$-symmetric when $\eta_{0} =\theta_{0} =0$.

A solution to the hyperbolic scaled KdV equation (\ref{KdVH}) is constructed as 
\begin{equation}
h_{\mu;\alpha}(x,t)=\frac{\alpha^{2}}{2}\func{sech}^{2}\left[\frac{1}{2} (\alpha xk-\alpha ^{3}t+\mu )\right] ,
\end{equation}

\noindent which in component form reads
\begin{equation}
\! h_{\mu^{\mathbb{B}} ;\alpha } =\resizebox{.8\hsize}{!}{$\frac{1}{2}\left(\bar{p}_{-;\alpha}+p_{+;\alpha}\right)+\frac{\imath}{2}\left( \bar{q}_{+ ;\alpha }+q_{-;\alpha}\right) +\frac{\jmath }{2}\left( \bar{q}_{+ ;\alpha
}-q_{- ;\alpha }\right) +\frac{k}{2}\left( p_{+ ;\alpha }-\bar{p}_{- ;\alpha }\right)$} ,
\end{equation}

\noindent \sloppy where we introduced the notations $\bar{p}_{-;\alpha}(x,t)=p_{-;\alpha}(-x,t)$ and $\bar{q}_{+;\alpha }(x,t) = q_{+;\alpha }(-x,t)$. 

\begin{figure}[h]
	\centering
	
	\includegraphics[width=0.492\linewidth]{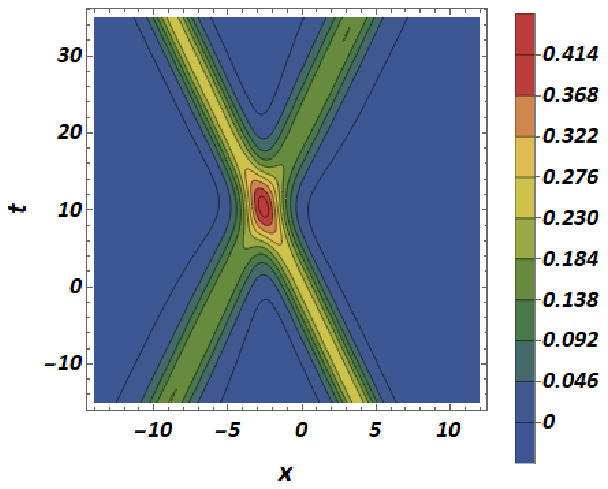} 
	\hspace{0cm}
	\includegraphics[width=0.492\linewidth]{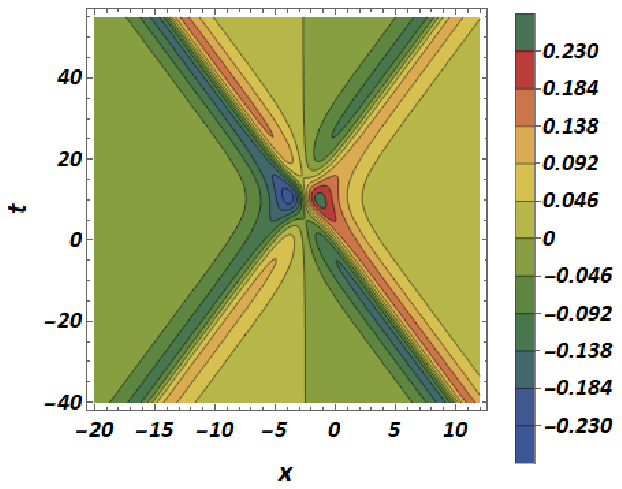}\\
	\hspace{0cm}\\
	
	\includegraphics[width=0.492\linewidth]{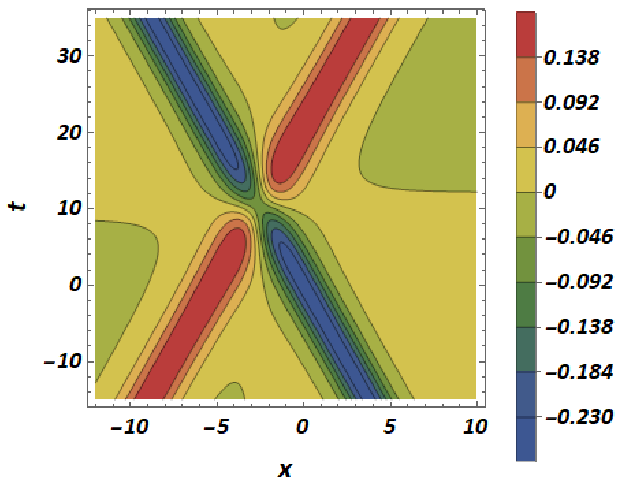}
	\hspace{0cm}
	\includegraphics[width=0.492\linewidth]{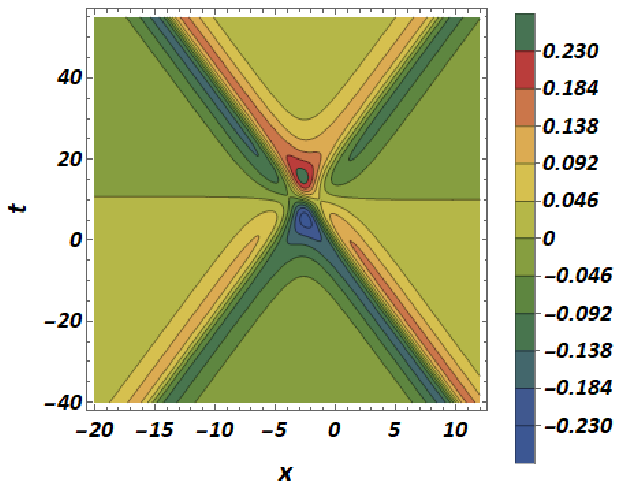}\\
	
	\caption{Head-on collision of a bright soliton with a dark soliton in the canonical components $p$, $q$, $r$, $s$ (clockwise starting in the top left corner) for the one-soliton solution $h_{\rho ,\theta ,\phi ,\chi ;\alpha }$ 
		to the bicomplex KdV equation (\ref{KdVbi}) with broken $\mathcal{PT}$-symmetry for $\alpha =0.5
		$, $\eta_{0} =1.3$, $\theta_{0} =0.1$, $\theta_{1} =2.0$ and $\eta_{1} =1.3$. Time is running vertically, space horizontally and contours of the amplitudes are colour-coded indicated as in the legends.} 
	\label{fig4.3}
\end{figure}

In Figure \ref{fig4.3} we depict the canonical component functions of this solution. We observe that the one-soliton solution is split into two one-soliton-like structures that scatter head-on with each other. The real $p$-component consists of a head-on scattering of two bright solitons and the hyperbolic $s$-component is a head-on collision of a bright and a dark soliton. Given that\ $u_{\mu^{\mathbb{B}} ;\alpha }$ and $h_{\mu^{\mathbb{B}} ;\alpha }(x,t)$ differ in the way that one of its constituent functions is space-reversed this is to be expected.

\noindent {\large\bfseries{Bicomplex one-soliton solution with $\mathcal{PT}_{ij}$-symmetry}}

\label{subptij}

An interesting solution can be constructed when we start with a complex $\mathcal{PT}_{\imath k}$ and a complex $\mathcal{PT}_{\jmath k}$ symmetric solution to assemble the linear decomposition of an overall $\mathcal{PT}_{\imath \jmath }$-symmetric solution with different velocities. Taking in the decomposition (\ref{SKdV}) $v(x,t)=u_{\imath \theta_{1} ,\alpha_{1} }(x,t)$ and $w(x,t)=u_{\jmath \theta_{2} ,\alpha_{2} }(x,t)$, we can build the bicomplex KdV-solution in the idempotent representation 
\begin{equation}
\widehat{u}_{\imath \theta_{1},\jmath \theta_{2} ;\alpha_{1} ,\alpha_{2} }(x,t)=u_{\imath \theta_{1} ,\alpha_{1}
}(x,t)\,e_{1}+u_{\jmath \theta_{2} ,\alpha_{2} }(x,t)\,e_{2}. \label{biidem}
\end{equation}

\begin{figure}[h]
	\centering
	
	\includegraphics[width=0.492\linewidth]{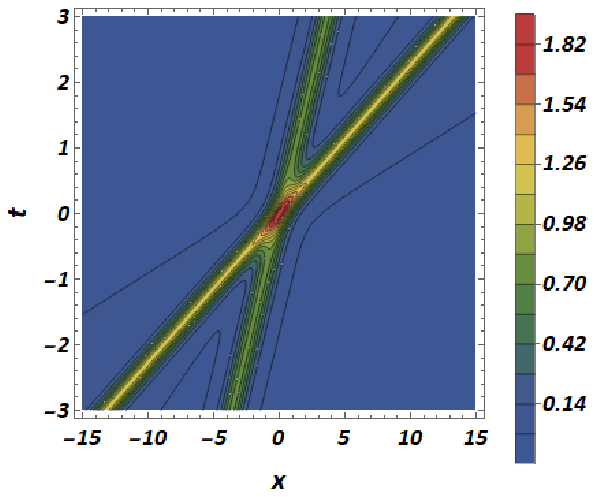} 
	\hspace{0cm}
	\includegraphics[width=0.492\linewidth]{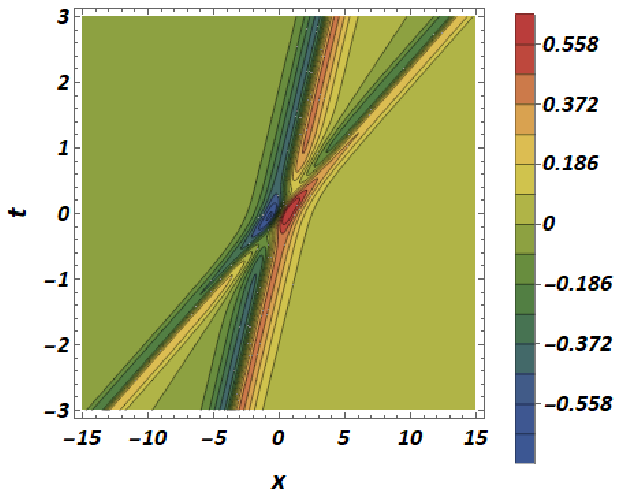}\\
	\hspace{0cm}\\

	\includegraphics[width=0.492\linewidth]{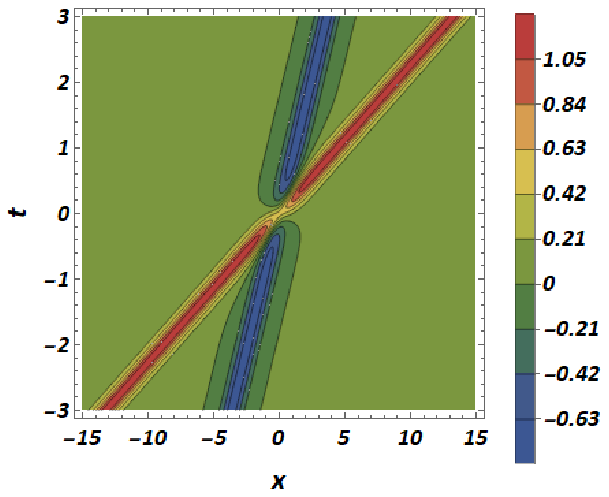}
	\hspace{0cm}
	\includegraphics[width=0.492\linewidth]{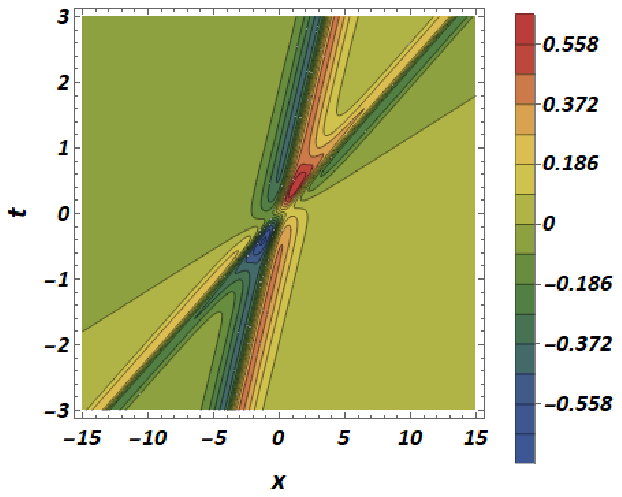}\\

\caption{A fast bright soliton overtaking a slower bright soliton in the canonical component functions $p$, $q$, $r$ and $s$ (clockwise starting in the top left corner) for the one-soliton solution $\widehat{u}_{\imath \theta_{1},\jmath \theta_{2} ;\alpha_{1} ,\alpha_{2} }$  
			to the bicomplex KdV equation (\ref{KdVbi}) with $\mathcal{PT}_{ij}$-symmetry for $\alpha_{1} =2.1$, $\alpha_{2} =1.1$, $\theta_{1} =0.6$ and $\theta_{2} =1.75$.}
		\label{fig4.4}
\end{figure}

The expanded version in the canonical representation becomes in this case
\begin{eqnarray}
\! \widehat{u}_{\imath \theta_{1},\jmath \theta_{2} ;\alpha_{1} ,\alpha_{2} }= & &\frac{1 }{2}\left[ p_{0,\theta_{1};\alpha_{1} }+p_{0,\theta_{2} ;\alpha_{2} }\right] +\frac{\imath }{2}\left[ q_{0,\theta_{1};\alpha_{1} }+q_{0,\theta_{2} ;\alpha_{2} }\right] \label{upt}\\
& &+\frac{\jmath }{2}\left[ q_{0,\theta_{2};\alpha_{2} }-q_{0,\theta_{1} ;\alpha_{1}}\right] +\frac{k}{2}\left[ p_{0,\theta_{1} ;\alpha_{1}}-p_{0,\theta_{2} ;\alpha_{2} }\right] ,  \notag
\end{eqnarray}

\noindent which is evidently $\mathcal{PT}_{\imath \jmath }$-symmetric. This solution contains an arbitrary multicomplex shift, however in each component, we now have two solitonic contributions with different amplitude and speed parameter. As we can see in Figure \ref{fig4.4}, in the real $p$-component a faster bright soliton is overtaking a slower bright solitons and in hyperbolic $s$-component a faster bright soliton is overtaking and a slower dark soliton.

\noindent {\large\bfseries{Bicomplex N-soliton solution}}

The most compact way to express the $N$-soliton solution for the real KdV equation in the form (\ref{kdv}) is seen from DC transformations
\begin{equation}
u_{\mu _{1},\mu _{2},\ldots ,\mu _{N};\alpha _{1},\alpha _{2},\ldots ,\alpha_{N}}^{(N)}(x,t)=2\partial_{x}^{2}\left[ \ln W_{N}(\psi_{\mu_{1},\alpha_{1}},\psi _{\mu_{2},\alpha _{2}},\ldots ,\psi _{\mu _{N},\alpha _{N}})\right] ,
\label{nsol}
\end{equation}
\noindent where $W_{N}$ denotes the Wronskian and the functions $\psi _{i}$ are solutions to the time-independent Schr\"{o}dinger equation for the free particle. Taking for instance $\psi _{\mu ,\alpha }(x,t)=\cosh \left[ (\alpha x-\alpha^{3}t+\mu)/2\right] $ for $N=1$ leads to the one-soliton solution (\ref{us}).

We could now take the shifts $\mu _{1},\mu _{2},\ldots ,\mu _{N}\in $ $\mathbb{B}$ and expand (\ref{nsol}) into its canonical components to obtain the $N$-soliton solution for the bicomplex equation. Alternatively we may also construct $N$-soliton solutions in the idempotent basis in analogy to (\ref{upt}). We demonstrate here the latter approach for the two-soliton solution. From (\ref{nsol}) we observe that the second derivative will not alter the linear bicomplex decomposition and it is therefore useful to introduce the quantity $w(x,t)$ as $u=w_{x}$. Thus a complex one-soliton solution can be obtained from 
\begin{equation}
w_{\eta,\theta;\alpha }=w_{\eta,\theta;\alpha }^{r}+\imath w_{\eta,\theta;\alpha }^{i}
\end{equation}

\noindent with
\begin{equation}
w_{\eta,\theta;\alpha }^{r}=\resizebox{.3\hsize}{!}{$\frac{\alpha \sinh (\alpha x-\alpha^{3}t+\eta)}{\cos \theta+\cosh (\alpha x-\alpha ^{3}t+\eta)}$}, \qquad w_{\eta,\theta;\alpha }^{i}=\resizebox{.3\hsize}{!}{$\frac{\alpha \sin \theta}{\cos \theta+\cosh (\alpha x-\alpha ^{3}t+\eta)}$}.  \label{w}
\end{equation}

Recalling now the expression
\begin{equation}
w_{a,b,c,d;\alpha,\beta}=\frac{\alpha^{2}-\beta^{2}}{w_{a,b;\alpha}-w_{c,d;\beta }},  \label{Baeck}
\end{equation}

\noindent from the BT of the complex two-soliton solution from Chapter 3, we can express this in terms of the functions in (\ref{w})
\begin{eqnarray}
\hspace{-1cm}
w_{\eta,\theta,\xi,\delta;\alpha ,\beta }&=&\frac{\left( \alpha ^{2}-\beta ^{2}\right) \left[\left( w_{\eta,\theta;\alpha }^{r}-w_{\xi,\delta;\beta }^{r}\right) -\imath \left( w_{\eta,\theta;\alpha }^{i}-w_{\xi,\delta;\beta }^{i}\right) \right] }{\left( w_{\eta,\theta;\alpha}^{r}-w_{\xi,\delta;\beta }^{r}\right) ^{2}+\left( w_{\eta,\theta;\alpha }^{i}-w_{\xi,\delta;\beta}^{i}\right)^{2}}\\
\hspace{-1cm}
&=&w_{\eta,\theta,\xi,\delta;\alpha,\beta}^{r}+\imath w_{\eta,\theta,\xi,\delta;\alpha,\beta }^{i}.  \label{ww}
\end{eqnarray}

Using (\ref{ww}) to define the two complex quantities $w_{1}=w_{\eta_{1},\theta_{1},\xi_{1},\delta_{1};\alpha_{1} ,\beta_{1}}\in \mathbb{C}(\imath )$ and $w_{2}=w_{\eta_{2},\theta_{2},\xi_{2},\delta_{2};\alpha_{2} ,\beta_{2}}\in \mathbb{C}(\jmath )$ we introduce the bicomplex function 
\begin{eqnarray}
w^{B}_{12} &=& w_{1}e_{1}+w_{2}e_{2}\\
&=&(w_{1}^{r}+\imath w_{1}^{i})e_{1}+(\widetilde{w}_{2}^{r}+\jmath \widetilde{w}_{2}^{i})e_{2} \\
&=&\frac{1 }{2}\left( w_{1}^{r}+w_{2}^{r}\right) +\frac{\imath }{2}\left( w_{1}^{i}+w_{2}^{i}\right) +\frac{\jmath }{2}\left( w_{2}^{i}-w_{1}^{i}\right) +\frac{k}{2}\left( w_{1}^{r}-w_{2}^{r}\right) .
\end{eqnarray}

\noindent Then by construction $u^{B}_{\eta_{1},\theta_{1},\xi_{1},\delta_{1},\eta_{2},\theta_{2},\xi_{2},\delta_{2};\alpha_{1} ,\beta_{1},\alpha_{2} ,\beta_{2}}\!\!=\!\partial_{x}\!\left(w_{12}^{B}\right)$ is a bicomplex two-soliton solution with four speed parameters. In a similar fashion we can proceed to construct higher order $N$-soliton solutions for $N>2$, i.e. independently constructing two complex $N$-soliton solutions, then using idempotent basis to form a bicomplex soliton solution.

\subsection{$\mathcal{PT}$-symmetry and reality of conserved charges for bicomplex soliton solutions}

When decomposing the bicomplex energy eigenvalue of a bicomplex Hamiltonian $H$ in the time-independent Schr\"{o}dinger equation, $H\psi =E\psi $, as $E=E_{0} +E_{1}\imath +E_{2}\jmath+E_{3}k$, Bagchi and Banerjee argued in \cite{bagchi_bicomplex_2015} that a $\mathcal{PT}_{\imath k}$-symmetry ensures that $E_{1}=E_{3}=0$, a $\mathcal{PT}_{\jmath k}$-symmetry forces $E_{2}=E_{3}=0$ and a $\mathcal{PT}_{\imath \jmath }$-symmetry sets $E_{1}=E_{2}=0$. In Chapter 3, we argued that for complex soliton solutions the $\mathcal{PT}$-symmetries together with the integrability of the model guarantees the reality of all physical conserved quantities. One of the main concerns in this section is to investigate the roles played by
the symmetries\ (\ref{PT1})-(\ref{PT3}) for the bicomplex soliton solutions and to clarify whether the implications are similar as observed in the quantum case.

Decomposing a density function for any conserved quantity as 
\begin{equation}
\rho (x,t)=\rho _{0}(x,t)+\imath \rho _{1}(x,t)+\jmath \rho
_{2}(x,t)+k\rho _{3}(x,t)\in \mathbb{B},
\end{equation}

\noindent and demanding it to be $\mathcal{PT}$-invariant, it is easily verified that a $\mathcal{PT}_{\imath k}$-symmetry implies that $\rho _{0}$, $\rho _{2}$ and $\rho _{1}$, $\rho _{3}$ are even and odd functions of $x$, respectively. A $\mathcal{PT}_{\jmath k}$-symmetry forces $\rho_{0}$, $\rho_{1}$ and $\rho _{2}$, $\rho _{3}$ to even and odd in $x$, respectively and a $\mathcal{PT}_{\imath \jmath }$-symmetry makes $\rho _{0}$, $\rho _{3}$ and $\rho _{1}$, $\rho _{2}$ even and odd in $x$, respectively. The corresponding conserved quantities must therefore be of the form
\begin{equation}
Q=\dint\nolimits_{-\infty }^{\infty }\rho (x,t)dx=\left\{ 
\begin{array}{ll}
Q_{0} +Q_{2}\jmath & \text{for }\mathcal{PT}_{\imath k}\text{-symmetric }\rho \\ 
Q_{0} +Q_{1}\imath \qquad & \text{for }\mathcal{PT}_{\jmath k}\text{-symmetric }\rho \\ 
Q_{0} +Q_{3}k & \text{for }\mathcal{PT}_{\imath \jmath }\text{-symmetric }\rho
\end{array}
\right. ,
\end{equation}

\noindent where we denote $Q_{i}:=\dint\nolimits_{-\infty}^{\infty}\rho_{i}(x,t)dx$ with $i=0,1,2,3$. Thus we expect the same property that forces certain quantum mechanical energies to vanish to hold similarly for all classical conserved quantities. We only regard $Q_{0}$ and $Q_{3}$ as physical, so that only a $\mathcal{PT}_{\imath \jmath }$-symmetric system is guaranteed to be physical.

\vspace{0.3cm}

\noindent {\large \bfseries{Real and hyperbolic conserved quantities}}

We compute the first conserved quantities, namely the mass $m$, the momentum $p$ and the energy $E$ of the KdV equation
\begin{eqnarray}
m(u) &=&\dint\nolimits_{-\infty }^{\infty }udx=m_{0} +m_{1}\imath
+m_{2}\jmath +m_{3}k, \\
p(u) &=&\dint\nolimits_{-\infty }^{\infty }u^{2}dx=p_{0} +p_{1}\imath +p_{2}\jmath +p_{3}k, \\
E(u) &=&\dint\nolimits_{-\infty }^{\infty }\left( \frac{1}{2}
u_{x}^{2}-u^{3}\right) dx=E_{0} +E_{1}\imath +E_{2}\jmath +E_{3}k
\end{eqnarray}

Decomposing the relevant densities into the canonical basis, $u$ as in (\ref{ub}), $u^{2}$ as 
\begin{equation}
u^{2}=\left( p^{2}-q^{2}-r^{2}+s^{2}\right)  +2(pq-rs)\imath
+2(pr-qs)\jmath +2(qr+ps)k
\end{equation}

\noindent and the Hamiltonian density $\mathcal{H}(u,u_{x})=u_{x}^{2}/2-u^{3}$ as 
\begin{eqnarray}
\mathcal{H} &=&\left[ 3p\left( q^{2}+r^{2}-s^{2}\right) +\frac{p_{x}^{2}-q_{x}^{2}-r_{x}^{2}+s_{x}^{2}}{2}-6qrs-p^{3}\right]   \\
&&+\left[ q^{3}-3p^{2}q+p_{x}q_{x}+6prs+3q\left( r^{2}-s^{2}\right)
-r_{x}s_{x}\right] \imath   \notag \\
&&+\left[ r^{3}+6pqs+3r\left( q^{2}-s^{2}-p^{2}\right) +p_{x}r_{x}-q_{x}s_{x} \right] \jmath   \notag \\
&&+\left[ 3s\left( r^{2}-p^{2}+q^{2}\right) -6pqr+p_{x}s_{x}+q_{x}r_{x}-s^{3} \right] k,  \notag
\end{eqnarray}

\noindent we integrate component-wise. For the solutions $u_{\mu^{\mathbb{B}} ;\alpha }$ and $h_{\mu^{\mathbb{B}} ;\alpha }$ with broken $\mathcal{PT}$-symmetry we obtain the real conserved quantities 
\begin{eqnarray}
m(u_{\mu^{\mathbb{B}} ;\alpha }) &=&m(h_{\mu^{\mathbb{B}};\alpha })=2\alpha  ,  \label{mr1} \\
p(u_{\mu^{\mathbb{B}} ;\alpha }) &=&p(h_{\mu^{\mathbb{B}};\alpha })=\frac{2}{3}\alpha ^{3} ,  \label{mr2} \\
E(u_{\mu^{\mathbb{B}} ;\alpha }) &=&E(h_{\mu^{\mathbb{B}};\alpha })=-\frac{1}{5}\alpha ^{5} .  \label{mr3}
\end{eqnarray}

\noindent These values are the same as presented in Chapter 3 for the complex
soliton solutions. Given that the $\mathcal{PT}$-symmetries are all broken, this is surprising at first sight. However, considering the representation (\ref{uco}) this is easily understood, as $m(u_{\mu^{\mathbb{B}} ;\alpha })$ is simply $ \frac{1}{2}(2\alpha+2\alpha )+\frac{\jmath}{2}(2\alpha -2\alpha )=2\alpha  $. We can argue similarly for the other conserved quantities.

For the $\mathcal{PT}_{ij}$-symmetric solution $\widehat{u}_{\imath \theta_{0}+\jmath \theta_{1}
;\alpha ,\beta }$ we obtain the following hyperbolic values for the
conserved quantities 
\begin{eqnarray}
m(\widehat{u}_{\imath \theta_{0}+\jmath \theta_{1} ;\alpha ,\beta }) &=&(\alpha +\beta ) +(\alpha-\beta )k, \\
p(\widehat{u}_{\imath \theta_{0}+\jmath \theta_{1} ;\alpha ,\beta }) &=&\frac{1}{3}\left( \alpha
^{3}+\beta ^{3}\right)+\frac{1}{3}\left(\alpha^{3}-\beta^{3}\right)
k, \\
E(\widehat{u}_{\imath \theta_{0}+\jmath \theta_{1} ;\alpha ,\beta }) &=&-\left( \frac{\alpha ^{5}}{10}+\frac{\beta ^{5}}{10}\right)  +\left( \frac{\beta ^{5}}{10}-\frac{\alpha^{5}}{10}\right) k.
\end{eqnarray}

\noindent The values become real and coincide with the expressions (\ref{mr1})-(\ref{mr3}) when we sum up the contributions from the real and hyperbolic component or when we take degeneracy, i.e. the limit $\beta \rightarrow \alpha $.

\section{The quaternionic KdV equation}

Applying now the multiplication law (\ref{m2}) to quaternionic functions, the KdV equation for a quaternionic field of the form $u(x,t)= p(x,t)+\imath q(x,t)+\jmath r(x,t)+ks(x,t)\in \mathbb{H}$ can also be viewed as a set of coupled equations for the four real fields $p(x,t)$, $q(x,t)$, $r(x,t)$, $s(x,t)\in \mathbb{R}$ 
\begin{equation}
\resizebox{.32\hsize}{!}{$u_{t}+3\left(uu_{x}+u_{x}u\right)+u_{xxx}=0$} \Leftrightarrow \left\{ 
\begin{array}{r}
\resizebox{.47\hsize}{!}{$p_{t}+6pp_{x}-6qq_{x}-6rr_{x}-6ss_{x}+p_{xxx}=0$} \\ 
\resizebox{.32\hsize}{!}{$q_{t}+6qp_{x}+6pq_{x}+q_{xxx}=0$} \\ 
\resizebox{.32\hsize}{!}{$r_{t}+6rp_{x}+6pr_{x}+r_{xxx}=0$} \\ 
\resizebox{.32\hsize}{!}{$s_{t}+6sp_{x}+6ps_{x}+s_{xxx}=0$}
\end{array}
\right. .  \label{qKdV}
\end{equation}

Notice that when comparing the above system with the bicomplex KdV equation (\ref{KdVbi}), the nonlinear term $6uu_{x}$ has been replaced with $3\left(uu_{x}+u_{x}u\right)$, which is a very natural modification when keeping in mind that the product of quaternionic functions is noncommutative \cite{olver_integrable_1998}. In the paper, it is shown that under some symmetry reductions, this equation and similar extensions to various equations, including mKdV and NLS equations, leads to Painlev\'{e} type equations.

The remaining set of
equations is in addition, the aforementioned $\mathcal{PT}_{\imath \jmath k}$-symmetric
\vspace{-0.2cm}
\begin{eqnarray}
\mathcal{PT}_{\imath \jmath k}:&& x\rightarrow -x,\,\, t\rightarrow -t,\,\,\imath
\rightarrow -\imath ,\,\,\jmath \rightarrow -\jmath ,\,\,k\rightarrow
-k,\\
&&p\rightarrow p,\,\,q\rightarrow -q,\,\,r\rightarrow -r,\,\,s\rightarrow
-s,\,\,u\rightarrow u. \notag
\end{eqnarray}

\subsection{Quaternionic N-soliton solution with $\mathcal{PT}_{\imath \jmath k}$-symmetry \label{comsol}}

Due to the noncommutative nature of the quaternions it appears difficult at first sight to find solutions to the quaternionic KdV equation. However, using the complex representation (\ref{cf}), and imposing the $\mathcal{PT}_{\imath \jmath k}$-symmetry, we may resort to our previous analysis on complex soliton solutions. Considering the shifted solution (\ref{us}) in the complex space $\mathbb{C}(\xi )$ yields the solution
\begin{eqnarray}
u^{Q}_{a_{0},\mathcal{N},\alpha } &=& \widehat{p}_{a_{0},\mathcal{N};\alpha}-\xi_{a} \,\widehat{q}_{a_{0},\mathcal{N};\alpha }  \label{sq1} \\
&=&\widehat{p}_{a_{0},\mathcal{N};\alpha } -\frac{a_{1}\imath +a_{2}\jmath +a_{3}k}{\mathcal{N}_{a}}\widehat{q}_{a_{0},\mathcal{N};\alpha } .
\label{sq2}
\end{eqnarray}

This solution becomes $\mathcal{PT}_{\imath \jmath k}$-symmetric when we carry out a shift in $x$ or $t$ to eliminate the real part of the shift. The real component is a one-solitonic structure similar to the real part of a complex soliton solution and the remaining component consists of the imaginary parts of a complex soliton solution with overall different amplitudes. It is clear that the conserved quantities constructed from this solution must be real, which follows by using the same argument as for the imaginary part in the complex case, as in Chapter 3, separately for each of the $\imath $,$\jmath $,$k$-components. By considering all functions to be
in $\mathbb{C}(\xi )$, it is also clear that multi-soliton solutions can be constructed in analogy to the complex case $\mathbb{C}(\imath)$ treated in Chapter 3, with a subsequent expansion into canonical components.

Since the quaternionic algebra does not contain any idempotents, a
construction similar to the one carried out for the bicomplex one-soliton solution with $\mathcal{PT}_{ij}$-symmetry (\ref{biidem}) does not seem to be possible for quaternions. However, we can use (\ref{Baeck}) for two complex solutions $w^{Q}_{a_{0},\mathcal{N};\alpha}= w_{a_{0},\mathcal{N};\alpha}^{r}+\xi_{a}w_{a_{0},\mathcal{N};\alpha }^{i}$ \,,\, $w^{Q}_{b_{0},\mathcal{N};\beta}= w_{b_{0},\mathcal{N};\beta}^{r}+\xi_{b}w_{b_{0},\mathcal{N};\beta }^{i}$\,, where the imaginary units are defined as in (\ref{ima}) with $\xi _{a}(a_{1},a_{2},a_{3})$ and $\xi_{b}(b_{1},b_{2},b_{3})$. Expanding that expression in the canonical basis we obtain 
\begin{equation}
w^{Q}_{a_{0},b_{0};\alpha,\beta}=\frac{\alpha ^{2}-\beta ^{2}}{\omega _{0}^{2}+\omega _{1}^{2}+\omega
	_{2}^{2}+\omega _{3}^{2}}\left( \omega _{0}-\imath \omega _{1}-\jmath
\omega _{2}-k\omega _{3}\right)  \label{wco}
\end{equation}

\noindent with
\begin{equation}
\omega _{0}=w_{a_{0},\mathcal{N};\alpha }^{r}-w_{b_{0},\mathcal{N};\beta }^{r},\quad \omega_{m}=\frac{a_{m}}{\mathcal{N}_{a}}w_{a_{0},\mathcal{N};\alpha}^{i}-\frac{b_{m}}{\mathcal{N}_{b}}w_{b_{0},\mathcal{N};\beta }^{i}\,, \quad m=1,2,3.
\end{equation}

\noindent A quaternionic two-soliton solution to (\ref{qKdV}) is then obtained from (\ref{wco}) as $u^{Q}_{a_{0},b_{0};\alpha,\beta}=\left(w^{Q}_{a_{0},b_{0};\alpha,\beta}\right)_{x}$.

\section{The coquaternionic KdV equation}

Applying now the multiplication law (\ref{m3}) to coquaternionic functions, the KdV equation for a coquaternionic field of the form $u(x,t)= p(x,t)+\imath q(x,t)+\jmath r(x,t)+ks(x,t)\in \mathbb{P}$ can also be viewed as a set of coupled equations for the four real fields $p(x,t)$, $q(x,t)$, $r(x,t)$, $s(x,t)\in \mathbb{R}$. The coquaternionic KdV equation then becomes 
\begin{equation}
\resizebox{.32\hsize}{!}{$u_{t}+3(uu_{x}+u_{x}u)+u_{xxx}=0$} \Leftrightarrow \left\{ 
\begin{array}{r}
\resizebox{.47\hsize}{!}{$p_{t}+6pp_{x}-6qq_{x}+6ss_{x}+6rr_{x}+p_{xxx}=0$} \\ 
\resizebox{.32\hsize}{!}{$q_{t}+6qp_{x}+6pq_{x}+q_{xxx}=0$} \\ 
\resizebox{.32\hsize}{!}{$r_{t}+6rp_{x}+6pr_{x}+r_{xxx}=0$} \\ 
\resizebox{.32\hsize}{!}{$s_{t}+6sp_{x}+6ps_{x}+s_{xxx}=0$}
\end{array}
\right. .  \label{KdVco}
\end{equation}

\noindent Notice that the last three equations of the coupled equation in (\ref{KdVco}) are identical to the quaternionic KdV equation (\ref{qKdV}).

\subsection{Coquaternionic N-soliton solution with $\mathcal{PT}_{\imath \jmath k}$-symmetry}

Using the representation (\ref{cocompl}) we proceed as in Subsection \ref{comsol} and consider the shifted solution (\ref{us}) in the complex space $\mathbb{C}(\zeta )$
\begin{eqnarray}
u^{CQ}_{a_{0},\mathcal{M};\alpha } &=&\widehat{p}_{a_{0},\mathcal{M};\alpha}-\zeta_{a}\, \widehat{q}_{a_{0},\mathcal{M};\alpha }  \label{soco2} \\
&=&\widehat{p}_{a_{0},\mathcal{M};\alpha}-\frac{a_{1}\imath +a_{2}\jmath +a_{3}k}{\mathcal{M}_{a}}\widehat{q}_{a_{0},\mathcal{M};\alpha }
\end{eqnarray}

\noindent that solves the\ coquaternionic KdV equation (\ref{KdVco}). There are two cases, for $\mathcal{M} \neq 0$, we obtain the solution 
 \begin{equation}
\hspace{-0.1cm}
 u_{a_{0},\mathcal{M};\alpha }^{CQ}  \!\!=\!\! \resizebox{.8\hsize}{!}{$\frac{\alpha ^{2}+\alpha ^{2}\cos \mathcal{M} \cosh (\alpha x-\alpha ^{3}t+a_{0})}{\left[ \cos \mathcal{M}+\cosh (\alpha x-\alpha^{3}t+a_{0})\right] ^{2}}-\zeta \frac{\alpha ^{2}\sin \mathcal{M}\sinh (\alpha x-\alpha
	^{3}t+a_{0})}{\left[ \cos \mathcal{M}+\cosh (\alpha x-\alpha ^{3}t+a_{0})\right] ^{2}}$}
\end{equation}
\noindent and for $\mathcal{M} \rightarrow 0$,
\begin{equation}
\hspace{-0.14cm}
	 u_{a_{0},\mathcal{M};\alpha }^{CQ}  \!\!=\!\! \resizebox{.8\hsize}{!}{$\frac{\alpha ^{2}}{ 1+\cosh (\alpha x-\alpha^{3}t+a_{0})}-\resizebox{.15\hsize}{!}{$\left(a_{1} \imath+a_{2} \jmath+a_{3}k\right)$} \frac{\alpha ^{2}\sinh (\alpha x-\alpha
			^{3}t+a_{0})}{\left[ 1+\cosh (\alpha x-\alpha ^{3}t+a_{0})\right] ^{2}}$} .\!
\end{equation}
\noindent Both solutions are $\mathcal{PT}_{\imath \jmath k}$-symmetric when we take a shift in $x$ or $t$ to absorb the real part. Multi-soliton solutions can be constructed in analogy to the complex case $\mathbb{C} (\imath )$ treated in Chapter 3.

\section{The octonionic KdV equation}

Taking now an octonionic field to be of the form $u(x,t)=p(x,t)e_{0}+q(x,t)e_{1}+r(x,t)e_{2}+s(x,t)e_{3}+t(x,t)e_{4}+v(x,t)e_{5}+w(x,t)e_{6}+z(x,t)e_{7}\in \mathbb{O}$ the octonionic KdV equation, in this form of (\ref{KdVco}) becomes a set of eight coupled equations 

\begin{equation}
\!\!\! \begin{array}{r}
 p_{t}+6pp_{x}-6qq_{x}-6rr_{x}-6ss_{x}-6tt_{x}-6vv_{x}-6ww_{x}-6zz_{x}+p_{xxx}=0, \\ 
\chi _{t}+6\chi p_{x}+6p\chi _{x}+\chi _{xxx}=0,
\end{array}
\label{KdVoct}
\end{equation}

\noindent with $\chi =q,r,s,t,v,w,z$. Setting any of four variables for $\chi $ to zero reduces (\ref{KdVoct}) to the coupled set of equations corresponding to the quaternionic KdV equation (\ref{qKdV}).

\subsection{Octonionic N-soliton solution with $\mathcal{PT}_{e_{1}e_{2}e_{3}e_{4}e_{5}e_{6}e_{7}}$-symmetry}

Using the representation (\ref{pctn}) we proceed as in Subsection \ref{comsol} and consider the shifted solution (\ref{us}) in the complex space $\mathbb{C}(o)$ 
\begin{eqnarray}
u_{a_{0}, \mathcal{O};\alpha }^{O} &=&\widehat{p}_{a_{0},\mathcal{O};\alpha
}-o\widehat{q}_{a_{0},\mathcal{O};\alpha }  \label{sooct} \\
&=&\widehat{p}_{a_{0},\mathcal{O};\alpha } -\frac{\dsum\nolimits_{i=1}^{7}a_{i}e_{i}}{\mathcal{O}}\widehat{q}_{a_{0},\mathcal{O};\alpha }
\end{eqnarray}

\noindent that solves the\ octonionic KdV equation (\ref{KdVoct}). The solution in (\ref{sooct}) is $\mathcal{PT}_{e_{1}e_{2}e_{3}e_{4}e_{5}e_{6}e_{7}}$-symmetric. Once more, multi-soliton solutions can be constructed in analogy
to the complex case $\mathbb{C}(\imath )$ treated in Chapter 3.

\section{Conclusions}

In this chapter, we have shown that the bicomplex, quaternionic, coquaternionic and octonionic extensions of the real KdV equation display properties that are typical of integrable systems, such as having multi-soliton solutions with novel qualitative behaviours. A particularly interesting case is the N-soliton solution from the idempotent basis decomposes into 2N one-soliton solutions, with each of the 2N constituents involving an independent speed parameter. Unlike for the real and complex soliton solutions, where degeneracy poses a non-trivial technical problem \cite{correa_regularized_2016,cen_degenerate_2017}, here these parameters can be trivially set to be equal. For all noncommutative versions of the KdV equation, i.e. quaternionic, coquaternionic and octonionic types, we found multi-soliton solutions based on complex representation in which the imaginary unit is built from specific combinations of the imaginary and hyperbolic units. In the bicomplex case, the first few conserved charges are also presented and higher order charges should also be possible. It would be interesting to conduct a more thorough investigation not only of higher order conserved charges for the bicomplex case, but also the other multicomplex cases where properties of noncommutativity and nonassociativity comes into play.

\chapter{Degenerate multi-soliton solutions for KdV and SG equations}\label{ch_5}

Up to now, the multi-soliton solutions we have been constructing are nondegenerate. This means the compound soliton solutions are made of one-soliton constituents that are all independent in terms of speed and amplitude. In this chapter, we look at the degenerate case, in which multi-soliton compounds have one-soliton constituents of the same speed and amplitude. In particular, we find an interesting property for degenerate multi-soliton solutions, namely that they have different properties at different timescales. 

At a small timescale the one-soliton constituents travel simultaneously at the same speed and with the same amplitude. Due to this property the collection of them could be regarded as an almost stable compound. In this regime, the solutions behave similarly to the famous tidal bore phenomenon, which consists of multiple wave amplitudes of heights up to several meters travelling jointly upstream a river and covering distances of up to several hundred kilometres, see e.g. \cite{chanson_tidal_2012}. At very large time the individual one-soliton constituents separate from each other with a time-dependent displacement, which can be computed exactly in closed analytical form for any number of one-solitons contained in the
solution. 
 
In this chapter, we explore degeneracy in the KdV \cite{correa_regularized_2016} and SG \cite{cen_degenerate_2017} equations with HDM, BT and DCT. The natural way to obtain a degenerate multi-soliton solution would be to take the limit of all speed parameters in a nondegenerate multi-soliton solution to one particular speed parameter, however, we shall see that this does not work in general. We investigate the construction of degenerate multi-soliton solutions with various methods including HDM, BT and DCT. Furthermore, when comparing the time-dependent displacements, we find a universal pattern for KdV and SG degenerate multi-soliton solutions \cite{cen_time-delay_2017,cen_degenerate_2017}.

\section{KdV degenerate multi-soliton solutions}

In the KdV case, the direct limiting process of the usual real multi-soliton solutions to one velocity leads to cusp type solutions, hence singularities. However, we will demonstrate that taking the degeneracy of complex multi-soliton solutions from the previous chapter produces finite degenerate multi-soliton solutions \cite{correa_regularized_2016}. The complex extensions help to regularize the singularities that can arise from imposing degeneracy.

\subsection{Degeneracy with Darboux-Crum transformation}

In Chapter 2, we demonstrated how  a nondegenerate N-soliton solution can be constructed via DCT with N independent solutions to the Schr\"{o}dinger equation with different eigenvalue parameters. For a degenerate N-soliton solution, the criteria of N independent solutions is still required for a non-trivial solution, but with the same eigenvalue parameter for degeneracy. We now show how these solutions will be constructed; they are the so-called Jordan states \cite{correa_pt-symmetric_2015}. To start, we take the Schr\"{o}dinger equation with Hamiltonian, $\widehat{H}$, solution $\psi$ and eigenvalue/energy $\lambda=E=\frac{\alpha^{2}}{4}$, to be the same as for the non-degenerate case
\begin{equation}
\widehat{H}\psi=\left(-\partial_{x}^{2}-u+\frac{\alpha^{2}}{4}\right)\psi=0. \label{kdvseed}
\end{equation}

For the next independent solution, rather than taking another $\psi$ with an independent eigenvalue parameter, we take a Jordan state, $\frac{\partial \psi}{\partial E}$, which is a null vector for $\widehat{H}^{2}$. In particular, $\frac{\partial \psi}{\partial \alpha}$ is also a solution
\begin{equation}
\widehat{H}\widehat{H}\left(\frac{\partial}{\partial\alpha }\psi\right)=-\frac{\alpha}{2}\widehat{H}\psi=0.
\end{equation}

\noindent Carrying on this procedure, we can construct N Jordan states for a N-soliton solution, which will be given by $\left\{ \psi, \frac{\partial}{\partial\alpha }\psi, \dots, \frac{\partial^{N }}{\partial\alpha^{N}}\psi \right\}$.

Hence, the $N^{th}$ DT iteration will give the iterated Schr\"{o}dinger equation
\begin{equation}
	-(\psi_{\alpha^{(N)}})_{xx}+u_{\alpha^{(N)}}\psi_{\alpha^{(N)}}=-\frac{\alpha^{2}}{4}\psi_{\alpha^{(N)}},
\end{equation}
\noindent with potential, $u_{\alpha^{(N)}}$, hence the degenerate N-soliton solution to the KdV equation is
\begin{equation}
	u_{\alpha^{(N)}}=2\partial_{x}^{2} \ln W_{N}\left(\psi, \frac{\partial}{\partial\alpha }\psi, \dots, \frac{\partial^{N-1 }}{\partial\alpha^{N-1}}\psi\right)
\end{equation}
\noindent and the corresponding wave function is
\begin{equation}
	\psi_{\alpha^{(N)}}
	=\frac{W_{N+1}\left(\psi, \frac{\partial}{\partial\alpha }\psi, \dots, \frac{\partial^{N-1 }}{\partial\alpha^{N-1}}\psi,\phi\right)}{W_{N}\left(\psi, \frac{\partial}{\partial\alpha }\psi, \dots, \frac{\partial^{N-1 }}{\partial\alpha^{N-1}}\psi\right)},
\end{equation}
\noindent where $\phi$ is the second independent solution to the Schr\"{o}dinger equation (\ref{kdvseed}), so we can take for example
\begin{eqnarray}
\psi&=&\cosh \frac{1}{2}(\alpha x-\alpha^{3}t+\mu),\\
	\phi&=&\frac{\alpha}{2} \psi \int \frac{dx}{\psi^{2}}=\sinh\frac{1}{2}\left(\alpha x-\alpha^{3}t+\mu\right).
\end{eqnarray}
\noindent It is important to point out that $\mu$ needs to have a non-zero imaginary part to regularise singularities from our degeneracy procedure.   

\subsection{Degeneracy with Hirota's direct method and B\"{a}cklund transformation}
Another two methods we can use to construct multi-soliton solutions are HDM and BT, as seen in Chapters 2 and 3. Taking the two types of two-soliton solutions from each case, if we try to carry out degeneracy by taking the direct limit of one speed parameter to the other in the generality, we obtain a one-soliton and trivial zero solution, respectively. To obtain a true degenerate two-soliton solution from HDM or BT, we need to implement some shifts at the initial stage of the methods before taking the equal speed limit.

For the case of using HDM, a two-soliton solution is known to be constructed with an initial $\tau_{1}$ function
\begin{equation}
\tau_{1}=c_{1}e^{\alpha x-\alpha^{3}t+\gamma_{1}}+c_{2}e^{\beta x-\beta^{3}t+\gamma_{2}}
\end{equation}
\noindent which is a solution to the order $\lambda^{1}$ equation from expansion of Hirota's bilinear form (\ref{3.5a}), where $\gamma_{1}$ , $\gamma_{2}$ are arbitrary constants and $c_{1}$ , $c_{2}$ are to be determined. In general, the degenerate limit will result in a one soliton solution. However, for the choices
\begin{equation}
	 \gamma_{1}=\mu+\frac{\alpha-\beta}{\alpha+\beta}\nu \quad , \quad \gamma_{2}=\mu-\frac{\alpha-\beta}{\alpha+\beta}\nu,
\end{equation}

\noindent carrying out HDM and then taking the limit $\beta\rightarrow\alpha$ the resulting expression is the degenerate two-soliton solution
\begin{equation}
u_{\alpha^{(2)}}^{H}=\resizebox{.8\hsize}{!}{$\frac{2\alpha^{2}\left[\left(\alpha x-3\alpha^{3}t+i \theta_{\nu}\right)\sinh\left(\alpha x-\alpha^{3}t+i \theta_{\mu}\right)-2\cosh\left(\alpha x-\alpha^{3}t+i \theta_{\mu}\right)-2 \right] }{\left[\alpha x-3\alpha^{3}t+i \theta_{\nu}+\sinh \left(\alpha x-\alpha^{3}t+i \theta_{\mu}\right)\right]^{2}}$} \label{degkdv2}
\end{equation}

\noindent where we have taken complex parameters $\mu=i \theta_{\mu}$, $\nu=i \theta_{\nu}$ with $\frac{\theta_{\nu}}{\sin \theta_{\mu}} >-1$ to obtain a $\mathcal{PT}$-symmetric solution that is without singularities and asymptotically finite.

For the case of BT, if we choose the two one-soliton solutions to have the same shifts, $\gamma_{1}$, $\gamma_{2}$, and conditions as taken for the HDM for the soliton solutions with speed parameter $\alpha$ and $\beta$ respectively:
\begin{eqnarray}
	u_{\alpha}&=&\frac{\alpha^{2}}{2} \func{sech}^{2}\frac{1}{2}\left(\alpha x-\alpha^{3}t+\mu+\frac{\alpha-\beta}{\alpha+\beta}\nu\right),\\
		u_{\beta}&=&\frac{\beta^{2}}{2} \func{sech}^{2}\frac{1}{2}\left(\beta x-\beta^{3}t+\mu-\frac{\alpha-\beta}{\alpha+\beta}\nu\right),
\end{eqnarray}
then the 'nonlinear superposition principle' and degenerate limiting results in the same degenerate two-soliton solution (\ref{degkdv2}). 

For higher order degeneracies in both the HDM and BT, we need to implement the right shifts to obtain a degenerate multi-soliton solution. Up to now, there are no known systematic methods to do this for general degenerate N-soliton solutions.

\subsection{Properties of degenerate multi-soliton solutions}

Similar to nondegenerate multi-solitons solutions as seen in Chapter 3, degenerate KdV multi-soliton solutions also admit lateral displacements and time-delays as a result of scattering. These can be computed as in the previous cases through tracking a particular point on the soliton solution, usually the maxima or minima for simplicity.

\begin{figure}[h]
	\centering
	
	\includegraphics[width=0.48\linewidth]{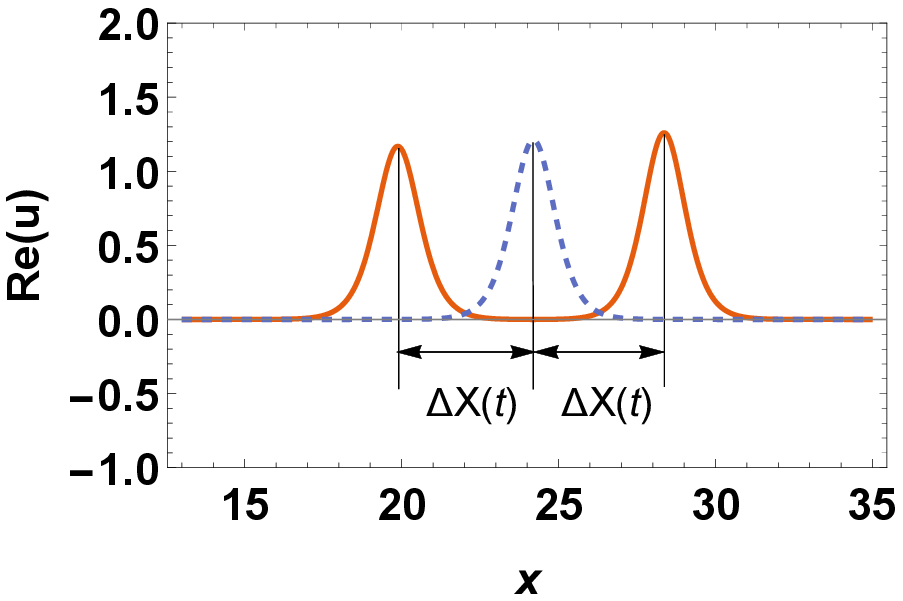}
	\hspace{0.2cm}
		\includegraphics[width=0.48\linewidth]{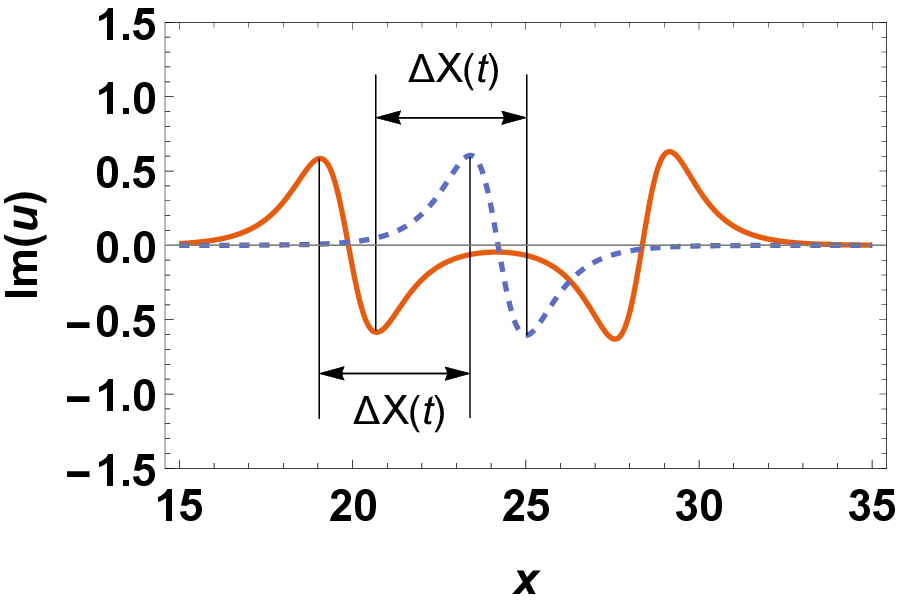}\\
	
	\caption{Real (left) and imaginary (right) parts of KdV degenerate two-soliton solution with $\alpha=1.1$, $\mu=\nu=i\frac{\pi}{2}$ and $t=20$.} \label{fig5.1}
\end{figure}

Taking the degenerate two-soliton solution (\ref{degkdv2}) in the asymptotic limit when $t \rightarrow \infty$, we can see the soliton constituents regain the same shape and amplitude as the corresponding one-soliton solution up to some displacements by plotting the solution with the corresponding one-soliton solution at large times in Figure \ref{fig5.1}. 

\noindent In particular, we find the lateral displacements will tend to an explicit time-dependent logarithmic function $\Delta X (t)$ in asymptotic time
\begin{equation}
\Delta X (t)=\frac{1}{\alpha }\ln \left( 4\alpha ^{3}\left\vert t\right\vert \right) .  \label{delt}
\end{equation}

\noindent For the right constituent in the imaginary part, the same displacement expression also holds with a shift of $i \pi$, in $\mu$, of the one-soliton solution.

For degenerate N-soliton solution, we conjecture the generalised displacement expression 
\begin{equation}
\Delta X_{mlk}\left(t\right)=\frac{1}{\alpha}\ln \left[\frac{(m-l)!}{(m+l-k)!}|4 \alpha^{3}t|^{2l-k}\right],
\end{equation}
\noindent with
\begin{eqnarray}
k&=&\left\{ \begin{array}{c}
1 \quad \text{, for N even}\\
0 \quad \text{, for N odd}
\end{array} \right. ,\\
m&=&\frac{N-1+k}{2},\\
l&=& l^{th} \text{constituent from soliton compound centre}.
\end{eqnarray}

\noindent As these shifts are logarithmic in time, the change is very slow and when confined to some finite regions they may be viewed as a $N$-soliton compound, hence similar to the tidal bore phenomenon. We verified these properties up to degenerate $10$-soliton solutions.

We can also compute conserved charges as in Section 3.4.4, for our complex degenerate N-soliton solution. The total charge will be N times the corresponding single soliton solution and real, due to the asymptotic limit of the compound solution being the sum of N one-solitons up to some lateral displacements or time-delays and the solution possessing $\mathcal{PT}$-symmetry.

\section{SG degenerate multi-soliton solutions}

Contrary to the KdV case, for the SG equation, the degeneracy limit could even be taken with real SG multi-soliton solutions and we will look in this section at various types of degeneracies including multi-kinks, multi-breathers and multi-Jacobi-elliptic solutions. We abbreviate solutions
with a $m$-fold degeneracy in $\alpha $ as \allowbreak $\phi _{\alpha^{(m)} \alpha_{m+1}\cdots \alpha_{N}}(x,t)$. This denotes an $N$-soliton solution
with $\alpha _{1}=$ $\alpha _{2}=\ldots =$ $\alpha _{m}=\alpha $.

\subsection{Degenerate multi-soliton solutions from B\"{a}cklund transformation}

From the nonlinear superposition principle for the SG equation (\ref{sg bianchi}), a new solution $\phi _{12}$ can be constructed from three known solutions $\phi_{0} $, $\phi _{1}$, $\phi _{2}$. However, one can easily see that degenerate solutions cannot be obtained directly in general from (\ref{sg bianchi}), as that would give the trivial zero solution. We will now demonstrate how the limits may be taken appropriately, thus leading to degenerate multi-soliton
solutions. Subsequently, we study the properties of this
solution.

\begin{figure}[h]
	\centering
	
	\includegraphics[width=0.4\linewidth]{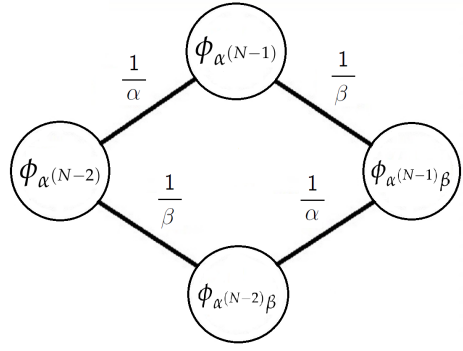}\\
	
	\caption{Bianchi-Lamb diagram for four arbitrary solutions $\phi_{\alpha^{(N-2)}} $, $\phi _{\alpha^{(N-2)}\beta }$, $\phi _{\alpha^{(N-1)} }$, $\phi _{\alpha^{(N-1)}\beta }$ of the SG equation, with each link representing a BT with constants $\alpha$ and $\beta$ chosen as indicated.} \label{fig5.2}
\end{figure}

At first we construct an $N$-soliton solution with an $(N-1)$-fold
degeneracy. For this purpose we start by relating four solutions to the
SG equation as depicted in the Bianchi-Lamb diagram in Figure \ref{fig5.2} with the choice $\phi_{0} =\phi _{\alpha^{(N-2)} }$, $\phi _{1 }=\phi _{\alpha^{(N-1)} }$, $\phi _{2 }=\phi _{\alpha^{(N-2)} \beta }$ and constants $a_{1 }=\frac{1}{\alpha} $ and $a _{2 }=\frac{1}{\beta} $, such that by (\ref{sg bianchi}) we obtain 
\begin{equation}
\phi _{\alpha^{(N-1)} \beta }=\phi _{\alpha^{(N-2)} }+4\arctan \left[ \frac{\alpha +\beta }{\alpha
-\beta }\tan \left( \frac{\phi _{\alpha^{(N-2)} \beta }-\phi_{\alpha^{(N-1)} }}{4}\right) \right] .  \label{super}
\end{equation}

\noindent Using the identity (see Appendix A for a derivation)
\begin{equation}
\lim_{\beta \rightarrow \alpha }\frac{\alpha +\beta }{\alpha -\beta }\tan \left( \frac{\phi _{\alpha^{(N-2)} \beta }-\phi _{\alpha^{(N-1)} }}{4}\right) =-\frac{\alpha }{2(N-1)}\frac{d }{d\alpha }\phi _{\alpha^{(N-1)}},  \label{Id1}
\end{equation}

\noindent we can perform the non-trivial limit $\beta \rightarrow \alpha $ in (\ref{super}), obtaining in this way the recursive equation 
\begin{equation}
\phi _{\alpha^{(N)} }=\lim_{\beta \rightarrow \alpha }\phi _{\alpha^{(N-1)} \beta }=\phi _{\alpha^{(N-2)} }-4\,\resizebox{.3\hsize}{!}{$\arctan \left[ \frac{\alpha }{2(N-1)}\frac{d}{d\alpha }\phi _{\alpha^{(N-1)} }\right]$} ,\text{\quad for }N\geq 2,
\label{rec}
\end{equation}

\noindent with $\phi _{\alpha^{(N)} }=0$ for $N\leq 0$. In principle equation (\ref{rec}) is sufficient to compute the degenerate solutions $\phi _{\alpha^{(N)} }$ recursively. However, it still involves a derivative term which evidently becomes more and more complicated for higher order. We eliminate this term next and replace it with combinations of just degenerate solutions. By iterating the BT (\ref{2.35}-\ref{2.36}) we compute the derivatives with respect to $x$ and $t$ to be
\begin{eqnarray}
\left( \phi _{\alpha^{(N)} }\right) _{x} &=& 2\alpha\sum\nolimits_{k=1}^{N}(-1)^{N+k}\sin \left( \frac{\phi_{\alpha^{(k)} }-\phi_{\alpha^{(k-1)} }}{2}\right) , \\
\left( \phi _{\alpha^{(N)} }\right) _{t} &=&\frac{2}{\alpha }\sum
\nolimits_{k=1}^{N}\sin \left( \frac{\phi _{\alpha^{(k)} }+\phi_{\alpha^{(k-1)} }}{2}\right) .
\end{eqnarray}

\noindent Taking the relation (see Appendix B for a derivation) 
\begin{equation}
\alpha \left( \phi _{\alpha^{(N)} }\right) _{\alpha }=x\left( \phi _{\alpha^{(N)}
}\right) _{x}-t\left( \phi _{\alpha^{(N)} }\right) _{t},  \label{axt}
\end{equation}

\noindent we convert this into the derivative with respect to $\alpha $ required in the recursive relation (\ref{rec}), that is
\begin{equation}
\! \resizebox{.9\hsize}{!}{$\frac{\alpha }{2}\frac{d}{d\alpha }\phi _{\alpha^{(N-1)} }=\sum\limits_{k=1}^{N}\left[ (-1)^{N+k}x\alpha \sin \left( \frac{\phi _{\alpha^{(k)} }-\phi_{\alpha^{(k-1)} }}{2}\right) -\frac{t}{\alpha }\sin \left( \frac{\phi_{\alpha^{(k)} }+\phi _{\alpha^{(k-1)} }}{2}\right) \right] .$}
\end{equation}

\noindent Therefore for $N\geq 2$ equation (\ref{rec}) becomes
\begin{equation}
\hspace{-0.20cm}
\resizebox{.91\hsize}{!}{$\phi _{\alpha^{(N)} }\!=\!\phi _{\alpha^{(N-2)} }-4\arctan \left[ \resizebox{.08\hsize}{!}{$\frac{1}{1-N}
\sum\limits_{k=1}^{N-1}$}\left[ \resizebox{.08\hsize}{!}{$(-1)^{N+k}$}x\alpha \sin \left( \frac{\phi
_{\alpha^{(k)} }-\phi _{\alpha^{(k-1)} }}{2}\!\right) +\frac{t}{\alpha }\sin \left( \frac{\phi _{\alpha^{(k)} }+\phi _{\alpha^{(k-1)} }}{2}\!\right) \right] \right] $} .
\label{tx}
\end{equation}

This equation can be solved iteratively with an appropriate choice for the initial condition $\phi _{\alpha }$. Taking this to be the well-known kink solution
\begin{equation}
\phi _{\alpha }=4\arctan \left( e^{\xi _{+}}\right) \text{,}\qquad \text{with }\xi _{\pm }:=\frac{t}{\alpha} \pm x\alpha ,
\end{equation}

\noindent we compute from (\ref{rec}) the degenerate multi-soliton solutions
\begin{eqnarray}
\hspace{-1cm}
\phi _{\alpha \alpha }\!\!&=&\!\! 4\arctan \left( \frac{\xi _{-}}{\cosh \xi _{+}}\right) ,  \label{deg2} \\
\hspace{-1cm}
\phi _{\alpha \alpha \alpha }\!\!&=&\!\! 4\arctan \left( \frac{\xi _{+}\cosh \xi_{+}-\xi _{-}^{2}\sinh \xi_{+}}{\xi_{-}^{2}+\cosh^{2}\xi_{+}}\right)
+\phi _{\alpha },  \label{deg3} \\
\hspace{-1cm}
\phi _{\alpha \alpha \alpha \alpha } \!\!&=&\!\! \resizebox{.75\hsize}{!}{$4\arctan \left[ \frac{-\xi _{-}}{3\cosh \xi _{+}}\ \frac{\xi _{-}^{4}+3\xi _{+}^{2}-(3+2\xi _{-}^{2})\cosh^{2}\xi _{+}+3\xi _{+}\sinh 2\xi _{+}}{\xi _{-}^{4}+\xi _{+}^{2}+2\xi_{-}^{2}+\cosh ^{2}\xi _{+}-2\xi _{-}^{2}\xi _{+}\tanh \xi _{+}}\right]+\phi _{\alpha \alpha }$}. \label{deg4}
\end{eqnarray}

\begin{figure}[h]
	\centering
	
	\includegraphics[width=0.48\linewidth]{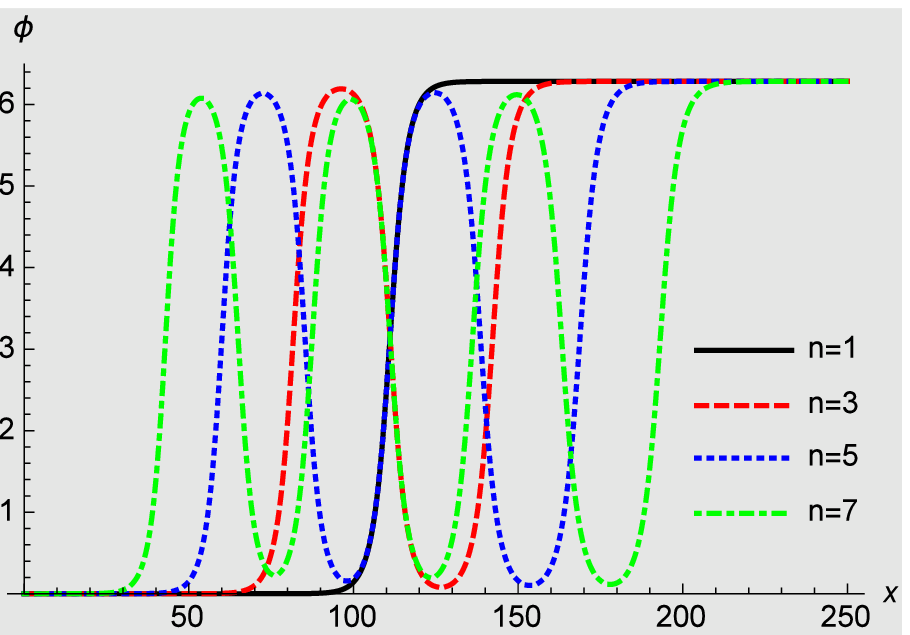}
	\hspace{0cm}
	\includegraphics[width=0.48\linewidth]{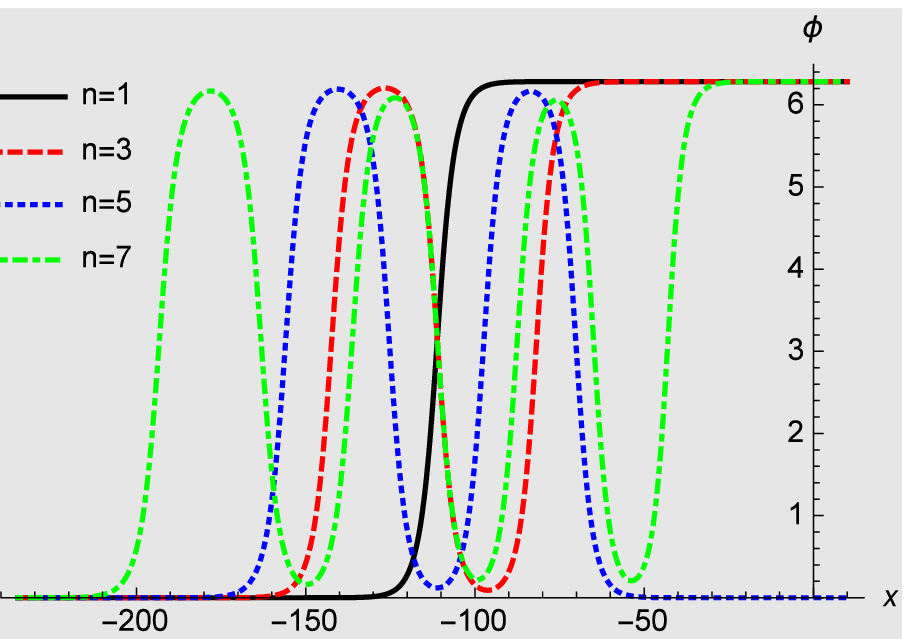}\\
	
	\vspace{0.2cm}
	
	\includegraphics[width=0.48\linewidth,height=0.325\linewidth]{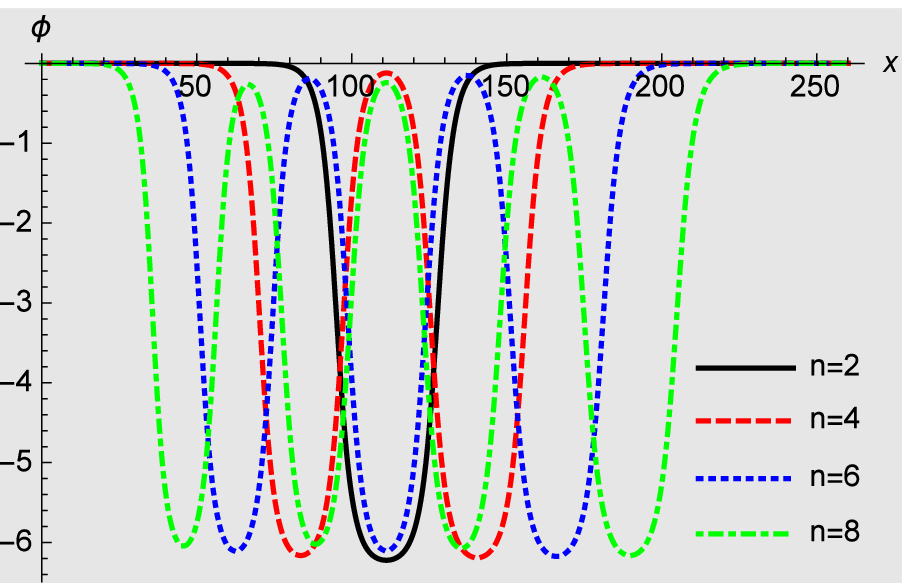}
	\hspace{0cm}
	\includegraphics[width=0.48\linewidth]{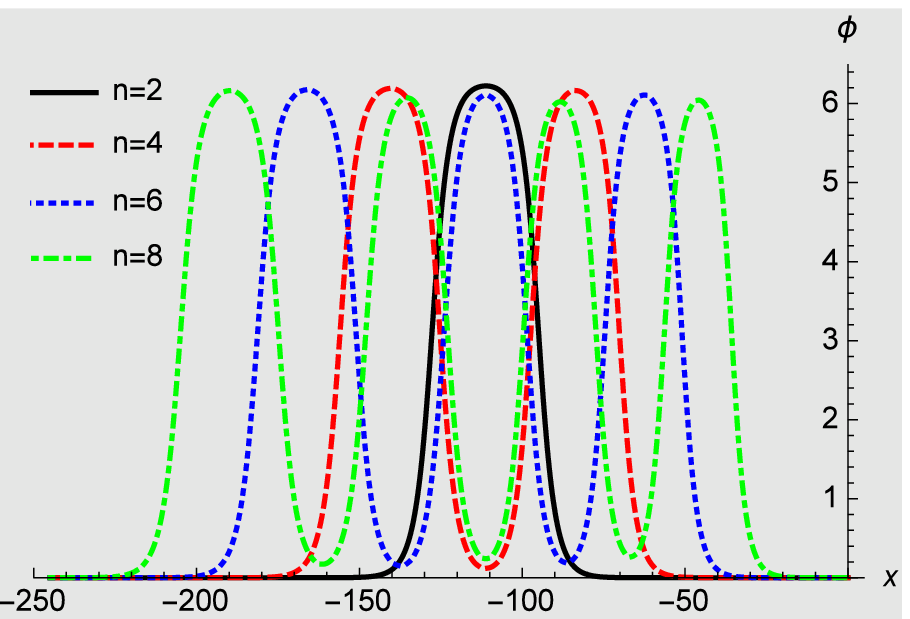}
	
	\caption{SG degenerate N-kink solutions at different times $t=-10$ (left panels) and $t=10$ (right panels) for spectral parameter $\alpha=0.3$.}
	\label{deg16}
\end{figure}

\noindent Snapshots of these solutions at two specific values in time are depicted in Figure \ref{deg16} for some concrete values of $\alpha $. The $N$-kink solution with $N=2n$ exhibits $n$ almost identical solitons travelling at nearly the same speed at small time scales. When $N=2n+1$ the solutions do not vanish asymptotically and have an additional kink at large values of $x$.

It is clear that solutions constructed in this manner, i.e. by iterating (\ref{tx}), will be of a form involving sums over $\arctan$-functions, which does not immediately allow to study properties such as the asymptotic behaviour we are interested in here. Of course one may combine these functions into one using standard identities, although these become increasingly nested for larger $N$ in $\phi_{\alpha^{(N)}}$. This can be avoided by deriving a recursive relation directly for the argument of just one $\arctan $-function. Defining for this purpose the functions $\tau _{N}$ via the relation
\begin{equation}
\phi _{\alpha^{(N)} }=4\arctan \tau _{N},  \label{ft}
\end{equation}

\noindent we convert the recursive relation (\ref{rec}) in $\phi $ into a recursive relation in $\tau $ as
\begin{equation}
\tau _{N}=\frac{\tau _{N-2}(1+\tau _{N-1}^{2})-\frac{2\alpha }{N-1}\frac{d\tau _{N-1}}{d\alpha }}{1+\tau _{N-1}^{2}+\frac{2\alpha }{N-1}\tau _{N-2}\frac{d\tau _{N-1}}{d\alpha }}.  \label{taurec}
\end{equation}

\noindent Similarly as to the derivatives for $\phi $, we may also compute them for the functions $\tau $ using (\ref{ft}). Computing first 
\begin{eqnarray}
\left(\tau_{N}\right)_{x}&=&\alpha(1+\tau_{N}^{2})\sum\nolimits_{k=1}^{N}(-1)^{N+k}\frac{(\tau _{k}-\tau_{k-1})(1+\tau _{k-1}\tau _{k})}{(1+\tau _{k-1}^{2})(1+\tau _{k}^{2})}, \\
\left( \tau _{N}\right) _{t} &=&\frac{1}{\alpha }(1+\tau
_{N}^{2})\sum\nolimits_{k=1}^{N}\frac{(\tau _{k}+\tau _{k-1})(1-\tau
_{k-1}\tau _{k})}{(1+\tau _{k-1}^{2})(1+\tau _{k}^{2})},
\end{eqnarray}

\noindent and using $\alpha \left( \tau _{N}\right) _{\alpha }=x\left( \tau_{N}\right) _{x}-t\left( \tau _{N}\right) _{t}$ we obtain the derivative of $\tau $ with respect to $\alpha $
\begin{equation}
\!
 \resizebox{.9\hsize}{!}{$\frac{\alpha }{(1+\tau _{N}^{2})}\frac{d\tau _{N}}{d\alpha }
=\sum\limits_{k=1}^{N}\left[\resizebox{.1\hsize}{!}{$ (-1)^{N+k}x\alpha$} \frac{(\tau _{k}-\tau
_{k-1})(1+\tau _{k-1}\tau _{k})}{(1+\tau _{k-1}^{2})(1+\tau _{k}^{2})}-\frac{t}{\alpha }\frac{(\tau _{k}+\tau _{k-1})(1-\tau _{k-1}\tau _{k})}{(1+\tau_{k-1}^{2})(1+\tau _{k}^{2})}\right]$} ,
\end{equation}

\noindent which we use to convert (\ref{taurec}) into
\begin{equation}
\tau _{N}=\frac{(N-1)\tau _{N-2}-2 S_{\tau}}{(N-1)+2\tau _{N-2}S_{\tau}},
\end{equation}
\noindent where
\begin{equation}
\hspace{-0.09cm} S_{\tau}=	\resizebox{.82\hsize}{!}{$\sum\limits_{k=1}^{N-1}\left[
	 \frac{\alpha x (-1)^{N+k-1}(\tau _{k}-\tau _{k-1})(1+\tau _{k}\tau _{k-1})-\frac{t}{\alpha }(\tau _{k}+\tau_{k-1})(1-\tau _{k}\tau _{k-1})}{(1+\tau _{k}^{2})(1+\tau _{k-1}^{2})}\right]$}.
\end{equation}

\noindent Using the variables $\xi _{\pm }$ instead of $x$, $t$ we obtain 
\begin{equation}
S_{\tau}=\sum\limits_{k=1}^{N-1}\left[\frac{ \xi _{(-1)^{N+k-1}}\tau _{k-1}(\tau _{k}^{2}-1)+\xi_{(-1)^{N+k}}\tau _{k}(\tau
_{k-1}^{2}-1)}{(1+\tau_{k-1}^{2})(1+\tau_{k}^{2})}\right] .
\label{tauc}
\end{equation}

These relations lead to simpler compact expressions allowing us to study the asymptotic properties of these functions more easily. Iterating (\ref{taurec}) we obtain the first solutions as
\begin{eqnarray}
\tau _{1} &=&e^{\xi _{+}},  \label{tau1} \\
\tau _{2} &=&\frac{2\xi _{-}\tau _{1}}{1+\tau _{1}^{2}}=\frac{\xi _{-}}{\cosh \xi _{+}}, \\
\tau _{3} &=&\frac{(1+2\xi _{+}+2\xi _{-}^{2})\tau _{1}+\tau _{1}^{3}}{1+(1-2\xi _{+}+2\xi _{-}^{2})\tau _{1}^{2}}, \\
\tau _{4} &=&\frac{4\xi _{-}(3+3\xi _{+}+\xi _{-}^{2})\tau _{1}+4\xi
_{-}(3-3\xi _{+}+\xi _{-}^{2})\tau _{1}^{3}}{3+(6+12\xi _{+}^{2}+4\xi
_{-}^{4})\tau _{1}^{2}+3\tau _{1}^{4}}, \\
\tau _{5} &=&\frac{3c_{+}\tau _{1}+2d_{+}\tau _{1}^{3}+9\tau _{1}^{5}}{9+2\tau _{1}^{2}d_{-}+3c_{-}\tau _{1}^{4}},  \label{tau5}
\end{eqnarray}

\noindent with
\begin{eqnarray*}
	\hspace{-0.7cm} 
	c_{\pm } \!\!&=&\!\! 6\xi _{+}^{2}\pm 12\xi _{+}\xi _{-}^{2}\pm 12\xi _{+}+2\xi_{-}^{4}+18\xi _{-}^{2}+3, \\
	\hspace{-0.7cm} 
	d_{\pm } \!\!&=&\!\! \resizebox{.85\hsize}{!}{$\pm 18\xi _{+}^{3}+18\xi _{+}^{2}\xi _{-}^{2}-9\xi _{+}^{2}\mp6\xi _{+}\xi _{-}^{4}\mp 36\xi _{+}\xi _{-}^{2}\pm 18\xi _{+}+2\xi_{-}^{6}+3\xi _{-}^{4}+27\xi _{-}^{2}+9.$}
\end{eqnarray*}

\noindent It is now straightforward to compute the $\tau _{N}$ for any larger value of $N$ in this manner.

It is clear that by setting up the nonlinear superposition equation (\ref{super}) for different types of solutions will produce recurrence relations for new types of degenerate multi-soliton solutions. We will not pursue this here, but instead compare the results obtained in this section with those obtained from different methods.

\subsection{Degenerate multi-soliton solutions from Darboux-Crum transformation}

 We can carry out the procedure to produce the degeneracy of various multi-soliton solutions to the SG equation by replacing eigenstates in the Wronskians of SG DCT (\ref{sgndct}) with Jordan states similar to what was done for the KdV case in Section 5.1.1 to obtain degenerate KdV multi-soliton solutions. 

\vspace{0.5cm}

\noindent\large\textbf{Degenerate kinks, antikinks, breathers and imaginary cusps from vanishing potentials}

We start by solving the four linear first order differential equations (\ref{SGlinear}) to the lowest level in the DC iteration procedure for some specific choices of $\phi ^{(0)}$. Considering the simplest case of vanishing potentials $V_{\pm }=0$, by taking $\phi^{(0)}=0 $, the equations in (\ref{SGlinear}) are easily solved by
\begin{equation}
\psi _{\alpha }(x,t)=c_{1}e^{\xi _{+}/2}+c_{2}e^{-\xi _{+}/2}\quad \text{and\quad }\varphi _{\alpha}(x,t)=c_{1}e^{\xi_{+}/2}-c_{2}e^{-\xi _{+}/2}.
\label{psiphi}
\end{equation}

\noindent Evidently, the constants $c_{1}$, $c_{2}\in \mathbb{C}$ determine the boundary conditions. Imposing the $\mathcal{PT}_{\mu}$-symmetry 
\begin{equation}
\mathcal{PT}_{\mu }:x\rightarrow -x\text{, }t\rightarrow -t\text{, }%
i\rightarrow -i\text{, }\phi \rightarrow -\phi \text{, }\chi \rightarrow
e^{i\mu }\chi \text{, }\text{for }\chi =\psi ,\varphi ,  \label{PTsg}
\end{equation}
 on each of these solutions selects some specific choices for the constants. For instance, for $c_{1}=c_{2}$ the fields in (\ref{psiphi}) obey the symmetries $\mathcal{PT}_{0}:\psi _{\alpha }\rightarrow \psi _{\alpha }$, $\mathcal{PT}_{\pi}:\varphi _{\alpha }\rightarrow -\varphi _{\alpha }$ and we obtain from the DT (\ref{dctsg1}) the purely imaginary cusp solution 
\begin{equation}
\phi _{\alpha }^{c}=-2i\ln \left( \tanh \frac{\xi _{+}}{2}\right) .  \label{s1}
\end{equation}

\noindent Imposing instead the symmetries $\mathcal{PT}_{-\pi /2}:\psi _{\alpha
}\rightarrow i\psi _{\alpha }$, $\mathcal{PT}_{-\pi /2}:\varphi_{\alpha
}\rightarrow -i\varphi _{\alpha }$ on the fields in (\ref{psiphi}) the kink and anti-kink solutions
\begin{equation}
\phi _{\alpha }=4\arctan \left( e^{\xi _{+}}\right) \qquad \text{and\qquad }\phi _{\bar{\alpha}}=4\func{arccot}\left( e^{\xi _{+}}\right) ,  \label{s2}
\end{equation}

\noindent are obtained from $c_{2}=ic_{1}\in \mathbb{R}$ and $c_{1}=ic_{2}\in \mathbb{R}$, respectively. We notice that the remaining constant $c_{1}$ or $c_{2}$ cancel out in all solutions in (\ref{s1}) and (\ref{s2}). Iterating these results leads for instance to the following:

\vspace{0.5cm}

\noindent\large{\textbf{Degenerate kink solutions}}

For the choice $c_{2}=ic_{1}$ we obtain the degenerate solutions 
\begin{equation}
\phi _{a^{(N)}}=-2i\ln \frac{W\left[ \varphi _{\alpha },\partial _{\alpha
}\varphi _{\alpha },\partial _{\alpha }^{2}\varphi _{\alpha },\ldots
,\partial _{\alpha }^{N-1}\varphi _{\alpha }\right] }{W\left[ \psi _{\alpha},\partial _{\alpha }\psi _{\alpha },\partial _{\alpha }^{2}\psi _{\alpha},\ldots ,\partial _{\alpha }^{N-1}\psi _{\alpha }\right] },
\end{equation}

\noindent which when evaluated explicitly coincide precisely with the expressions
previously obtained in (\ref{deg2}-\ref{deg4}) in a recursive manner.

\vspace{0.5cm}

\noindent\large{\textbf{Degenerate complex cusp solutions}}

In a similar way we can construct degenerate purely complex cusp solutions. Such type of solutions are also well-known in the literature, see for instance \cite{kawamoto_cusp_1984} for an early occurrence. These solutions appear to be non-physical at first sight, but they find applications for instance as an explanation for the entrainment of air \cite{eggers_air_2001}. For the choice $c_{1}=c_{2}$ we obtain
\begin{eqnarray}
\hspace{-0.6cm} \phi _{a^{(2)}}^{c} \!\!&=&\!\! -2i\ln \frac{W\left[ \varphi _{\alpha },\partial _{\alpha}\varphi _{\alpha }\right] }{W\left[ \psi _{\alpha },\partial _{\alpha }\psi_{\alpha }\right] }=-2i\ln \left( \frac{\sinh \xi _{+}+\xi _{-}}{\sinh \xi_{+}-\xi _{-}}\right) , \\
\hspace{-0.6cm} \phi _{a^{(3)}}^{c} \!\!&=&\!\! -2i\ln \frac{W\left[ \varphi _{\alpha },\partial _{\alpha}\varphi _{\alpha },\partial _{\alpha }^{2}\varphi _{\alpha }\right] }{W\left[ \psi _{\alpha },\partial _{\alpha }\psi _{\alpha },\partial _{\alpha}^{2}\psi _{\alpha }\right] } \\
\hspace{-0.6cm} \!\!&=&\!\! -2i\ln \resizebox{.7\hsize}{!}{$\left( \frac{\cosh \left( 3\xi _{+}/2\right) +2\xi _{+}\sinh
\left( \xi _{+}/2\right) -(1+2\xi _{-}^{2})\cosh \left( \xi_{+}/2\right) }{\sinh \left( 3\xi _{+}/2\right) -2\xi _{+}\cosh \left( \xi _{+}/2\right)+(1+2\xi _{-}^{2})\sinh \left( \xi _{+}/2\right) }\right)$} .  \notag
\end{eqnarray}

\noindent Similarly we can proceed to obtain the solutions $\phi _{a^{(N)}}^{c}$ for $N>3$.

\vspace{0.5cm}

\noindent\large{\textbf{Degenerate breather solutions}}

Breather solutions may be obtained in various ways. An elegant real solution can be constructed as follows: Taking as the starting point the two-kink solution with two distinct spectral parameters $\alpha $ and $\beta $, that is
\begin{equation}
\phi _{\alpha \beta }=-2i\ln \frac{W\left[ \varphi _{\alpha },\varphi
_{\beta }\right] }{W\left[ \psi _{\alpha },\psi _{\beta }\right] }=4\arctan \left[ \frac{\alpha +\beta }{\alpha -\beta }\frac{\sinh \left[ (\frac{1}{2\beta }-\frac{1}{2\alpha })(t-x\alpha \beta )\right] }{\cosh \left[ (\frac{1}{2\beta }+\frac{1}{2\alpha })(t+x\alpha \beta )\right] }\right] ,
\end{equation}

\noindent we obtain a breather by converting one of the functions in the argument into a trigonometric function. Taking first $\beta \rightarrow 1/\alpha $ we obtain a kink-antikink solution
\begin{equation}
\phi _{\alpha ,1/\alpha }=4\arctan \left[ \frac{\alpha ^{2}+1}{\alpha ^{2}-1}\frac{\sinh \left[ \frac{1}{2}(\alpha -\frac{1}{\alpha })(t-x)\right] }{\cosh \left[ \frac{1}{2}(\alpha +\frac{1}{\alpha })(t+x)\right] }\right] .
\end{equation}

\noindent Thus by demanding that $(\alpha ^{2}+1)/(\alpha ^{2}-1)=i\theta$ and $(\alpha -\frac{1}{\alpha })/2=-i\widetilde{\theta}$ for some constants for $\theta ,\widetilde{\theta}\in \mathbb{R}$ we obtain an oscillatory function in the argument of the $\arctan $. Solving for instance the first relation gives $\alpha=(\theta-i)/\sqrt{1+\theta^{2}}$ so that $\widetilde{\theta}=1/\sqrt{1+\theta ^{2}}$. The corresponding breather solution then results in the form
\begin{equation}
\phi _{\alpha,1/\alpha}=4\arctan \left[ \theta \frac{\sin \left[ (t-x)/\sqrt{1+\theta ^{2}}\right] }{\cosh \left[ \theta (t+x)/\sqrt{1+\theta ^{2}}\right] }\right] .  \label{b1}
\end{equation}

\noindent This solution evolves with a constant speed $-1$ modulated by some overall oscillation resulting from the sine function. Similarly we can construct a two-breather solution from two degenerated kink-solutions, given by
\begin{equation}
\phi _{\alpha \alpha \beta \beta }=-2i\ln \frac{W\left[ \varphi _{\alpha},\partial _{\alpha }\varphi _{\alpha },\varphi _{\beta },\partial _{\beta}\varphi _{\beta }\right] }{W\left[ \psi _{\alpha },\partial _{\alpha }\psi_{\alpha },\psi _{\beta },\partial _{\beta }\psi _{\beta }\right] },
\end{equation}

\noindent by using the same parametrisation $\phi_{\widetilde{\alpha},\widetilde{\alpha},1/\widetilde{\alpha},1/\widetilde{\alpha}}$.

\subsection{Cnoidal kink solutions from shifted Lam\'{e} potentials}

The sine-Gordon equation also admits a solution in terms of the Jacobi
amplitude $\limfunc{am}(x,m)$ depending on the parameter $0\leq m\leq 1$ in the form
\begin{equation}
\phi _{cn}^{(0)}=2\limfunc{am}\left( \frac{x-t}{\sqrt{\mu }},\mu \right)
\label{fcn}
\end{equation}

\noindent for any $0<\mu (m)<1$. 

 The potentials (\ref{sgpotentials}) following from the solution (\ref{fcn}) are 
\begin{eqnarray}
V_{\pm }^{cn} &=&-\frac{1}{\mu }\limfunc{dn}{}^{2}\left( \frac{x-t}{\sqrt{\mu }},\mu \right) \mp i\limfunc{cn}{}\left( \frac{x-t}{\sqrt{\mu }},\mu\right) \limfunc{sn}\left( \frac{x-t}{\sqrt{\mu }},\mu \right) \\
&=&\frac{\sqrt{m}}{2}\limfunc{sn}{}^{2}\left( \frac{x-t}{2m^{1/4}}\mp \frac{i}{2}K^{\prime },m\right) -\frac{1}{4}(m^{1/2}+m^{-1/2}),
\end{eqnarray}

\noindent where we used the parametrisation $\mu =4\sqrt{m}/(1+\sqrt{m})^{2}$ with $K(m)$ denoting the complete elliptic integral of the first kind and $K^{\prime }(m)=K(1-m)$. Notice that this is a complex shifted and scaled Lam\'{e} potential \cite{ince_viifurther_1940,whittaker_course_1996} invariant under any $\mathcal{PT}$-symmetry as defined in (\ref{PTsg}). Such type of potentials emerge in various contexts, e.g. they give rise to elliptic string solutions \cite{bakas_elliptic_2016} or the study of the origin of spectral singularities in periodic $\mathcal{PT}$-symmetric systems \cite{correa_spectral_2012}.

Next, we require the solutions $\psi$ and $\varphi$ to the
auxiliary equations from the SG ZC representation (\ref{SGlinear}) corresponding to the sine-Gordon solution $\phi_{cn}^{(0)}$. We have to to distinguish the two cases $0< \alpha < 1$ and $\alpha >1$. In the first case we parametrise $\alpha =m^{1/4}$ finding the solutions
\begin{equation}
\hspace{-0.1cm} \widecheck{\psi}_{m}(x,t)=c\limfunc{cn}{}\left[\resizebox{.07\hsize}{!}{$\frac{x-t}{2m^{1/4}}$}-\frac{i}{2}
K^{\prime },m\right] ,\quad \widecheck{\varphi}_{m}(x,t)=-ic\limfunc{cn}{}\left[\resizebox{.07\hsize}{!}{$\frac{x-t}{2m^{1/4}}$}+\frac{i}{2}K^{\prime },m\right],  \label{scon1}
\end{equation}

\noindent and for the second case we parametrise $\alpha =m^{-1/4}$ obtaining the
solutions
\begin{equation}
\hspace{-0.1cm}
\widehat{\psi}_{m}(x,t)=ic\limfunc{dn}{}\left[\frac{x-t}{2m^{1/4}}-\frac{i}{2}K^{\prime },m\right] ,\quad \widehat{\varphi}_{m}(x,t)=c\limfunc{dn}{}\left[ 
\frac{x-t}{2m^{1/4}}+\frac{i}{2}K^{\prime },m\right] ,  \label{scon2}
\end{equation}

\noindent with integration constant $c$. For real values of $c$ we observe the $\mathcal{PT}$-symmetries
\begin{equation}
	\mathcal{PT}_{0}\widecheck{\psi}_{m}=\widecheck{\psi}_{m}, \quad \mathcal{PT}_{\pi }\widecheck{\varphi} _{m}=\widecheck{\varphi} _{m}, \quad \mathcal{PT}_{\pi}\widehat{\psi} _{m}=\widehat{\psi} _{m}, \quad \mathcal{PT}_{0}\widehat{\varphi} _{m}=\widehat{\varphi}_{m}.
\end{equation}

The DT (\ref{dctsg1}), then yields the real solutions for the sine-Gordon equation
\begin{eqnarray}
\hspace{-1cm}
\widecheck{\phi} _{m}^{(1)}(x,t) \!\!&=&\!\! \resizebox{.73\hsize}{!}{$2\limfunc{am}\left( \frac{x-t}{\sqrt{\mu }},\mu\right) -4\arctan \left[ \frac{\limfunc{dn}{}\left( \frac{x-t}{2m^{1/4}},m\right) \limfunc{sn}\left( \frac{x-t}{2m^{1/4}},m\right) }{\limfunc{cn}{}\left( \frac{x-t}{2m^{1/4}},m\right) }\right] -\pi ,$}  \label{phs1} \\
\hspace{-1cm}
\widehat{\phi} _{m}^{(1)}(x,t) \!\!&=&\!\! \resizebox{.73\hsize}{!}{$2\limfunc{am}\left( \frac{x-t}{\sqrt{\mu }},\mu\right) -4\arctan \left[ \sqrt{m}\frac{\limfunc{cn}{}\left( \frac{x-t}{2m^{1/4}},m\right) \limfunc{sn}\left( \frac{x-t}{2m^{1/4}},m\right) }{\limfunc{dn}{}\left( \frac{x-t}{2m^{1/4}},m\right) }\right] -\pi ,$}  
\label{phs2}
\end{eqnarray}

\noindent after using the addition theorem for the Jacobi elliptic functions, the properties $\limfunc{cn}{}\left(iK^{\prime}/2,m\right)\!\!=\!\!\sqrt{1+\sqrt{m}}/m^{1/4}$,\,\,\,\,\,$\limfunc{sn}{}\left( iK^{\prime }/2,m\right) \!\!=\!\!i/m^{1/4}$,\,\,\,\,\, $\limfunc{dn}{}\left( iK^{\prime }/2,m\right) =\sqrt{1+\sqrt{m}}$, and the well known relation between the $\ln $ and the $\arctan$-functions. Notice that the $\limfunc{cn}$-function can be vanishing for real arguments, such that $\widecheck{\phi}^{(1)}$ is a discontinuous function. Furthermore, we observe that this solution has a fixed speed and does not involve any variable spectral parameter. For this reason we construct a different type of solution also related to $\phi_{cn}^{(0)}$ that involves an additional parameter, utilising Theta functions following \cite{whittaker_course_1996}.

These type of solutions can be obtained from 
\begin{equation}
\Psi _{m,\beta }^{\pm }(y)=\frac{H\left( y\pm \beta \right) }{\Theta (y)}e^{\mp y Z (\beta )},\qquad \Phi _{m,\beta }^{\pm }(y)=\frac{\Theta \left(y\pm \beta \right) }{\Theta (y)}e^{\mp y X \left(\beta \right)},  \label{lame}
\end{equation}

\noindent which are solutions of the Schr\"{o}dinger equation involving the Lam\'{e} potential $V_{L}$, that is
\begin{equation}
-\Psi _{yy}+V_{L}\Psi =E_{\beta }\Psi , \quad \text{ with} \quad V_{L}=2m\limfunc{sn}\left( y,m\right) ^{2}-(1+m)\text{,}
\end{equation}

\noindent with $E_{\beta }=-m\limfunc{sn}\left( \beta ,m\right) ^{2}$ and $E_{\beta}=-1/\limfunc{sn}\left( \beta ,m\right) ^{2}$, respectively. The functions $H $, $\Theta $, $Z$ and $X$ are defined in terms of Jacobi's theta functions $\vartheta _{i}(z,q)$ with $i=1,2,3,4$,\, $\kappa =\frac{\pi}{2K}$ and nome $q=\exp(-\pi K/K^{\prime })$ as
\begin{equation}
\resizebox{.9\hsize}{!}{$H\left( z\right) :=\vartheta _{1}\left( z\kappa ,q\right) ,\quad \Theta
\left( z\right) :=\vartheta _{4}\left( z\kappa ,q\right) ,\quad X(z):=\kappa \frac{H^{\prime }\left( z \right) }{H\left( z \right) },\quad
Z(z):=\kappa \frac{\Theta ^{\prime }\left( z \right) }{\Theta \left(z \right) }.$}
\end{equation}

\noindent With a suitable normalization factor and the introduction of a
time-dependence the function $\Psi ^{\pm }(y)$ can be tuned to solve the equations (\ref{SGlinear}). We find
\begin{eqnarray}
\hspace{-0.6cm} \widecheck{\Psi }_{\pm ,m,\beta }(x,t) \!\!&=&\!\! \Psi _{m,\beta }^{\pm }\left( \frac{x-t}{2m^{1/4}}-\frac{i}{2}K^{\prime }\right) e^{\pm \frac{t}{2m^{1/4}}\frac{\limfunc{cn}\left( \beta ,m\right) \limfunc{dn}\left( \beta ,m\right) }{\limfunc{sn}\left( \beta ,m\right) }},  \label{sp1} \\
\hspace{-0.6cm} \widecheck{\Phi} _{\pm ,m,\beta }(x,t) \!\!&=&\!\! \mp e^{\pm iK^{\prime }Z(\beta )\pm i\beta\kappa }\Psi _{m,\beta }^{\pm }\left( \frac{x-t}{2m^{1/4}}+\frac{i}{2}K^{\prime }\right) e^{\pm \frac{t}{2m^{1/4}}\frac{\limfunc{cn}\left( \beta,m\right) \limfunc{dn}\left( \beta ,m\right) }{\limfunc{sn}\left( \beta
,m\right) }},  \label{sp2}
\end{eqnarray}

\noindent for $\alpha =m^{1/4}\limfunc{sn}\left( \beta ,m\right) $ and
\begin{eqnarray}
\hspace{-0.6cm} \widehat{\Psi} _{\pm ,m,\beta }(x,t) \!\!&=&\!\! \Phi _{m,\beta }^{\pm }\left( \frac{x-t}{2m^{1/4}}-\frac{i}{2}K^{\prime }\right) e^{\mp \frac{t}{2m^{1/4}}\frac{\limfunc{cn}\left( \beta ,m\right) \limfunc{dn}\left( \beta ,m\right) }{\limfunc{sn}\left( \beta ,m\right) }},  \label{sp3} \\
\hspace{-0.6cm} 
\widehat{\Phi} _{\pm ,m,\beta }(x,t) \!\!&=&\!\! \mp e^{\pm iK^{\prime }X(\beta )\pm i\beta\kappa }\Phi _{m,\beta }^{\pm }\left( \frac{x-t}{2m^{1/4}}+\frac{i}{2} K^{\prime }\right) e^{\mp \frac{t}{2m^{1/4}}\frac{\limfunc{cn}\left( \beta,m\right) \limfunc{dn}\left( \beta ,m\right) }{\limfunc{sn}\left( \beta,m\right) }},  \label{sp4}
\end{eqnarray}

\noindent for $\alpha =1/(m^{1/4}\limfunc{sn}\left( \beta ,m\right) )$. The
corresponding solutions for the sine-Gordon equation resulting from the
DT (\ref{dctsg1}) are
\begin{equation}
\phi _{\pm ,m,\beta }^{\ell (1)}(x,t)=\phi _{cn}^{(0)}\pm 2\beta \kappa
-4\arctan\left[i\frac{M_{\pm}^{\ell}-(M_{\pm}^{\ell})^{\ast}}{M_{\pm}^{\ell }+(M_{\pm }^{\ell })^{\ast }}\right] , \quad \ell =\,\widecheck{ } \,\, , \,\,\widehat{ }
\end{equation}

\noindent with the abbreviations
\begin{eqnarray}
\widecheck{M}_{\pm } &=&H\left( \frac{x-t}{2m^{1/4}}\pm \beta +\frac{i}{2}K^{\prime}\right) \Theta \left( \frac{x-t}{2m^{1/4}}-\frac{i}{2}K^{\prime }\right) ,\\
\widehat{M}_{\pm } &=&\Theta \left( \frac{x-t}{2m^{1/4}}\pm \beta +\frac{i}{2} K^{\prime }\right) \Theta \left( \frac{x-t}{2m^{1/4}}-\frac{i}{2}K^{\prime}\right) .
\end{eqnarray}

\begin{figure}[h]
	\centering
	\includegraphics[width=0.48\linewidth]{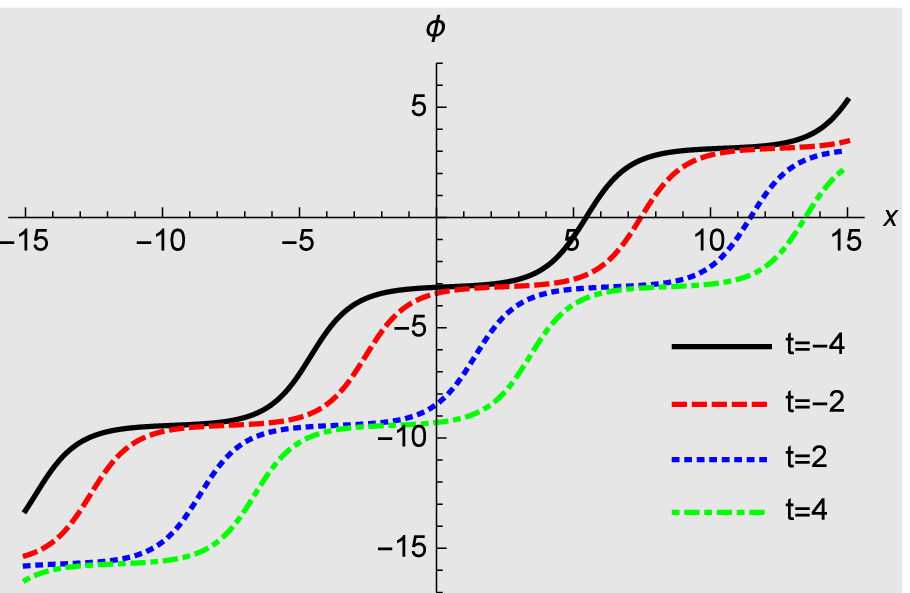} 
	\hspace{0cm}
	\includegraphics[width=0.48\linewidth]{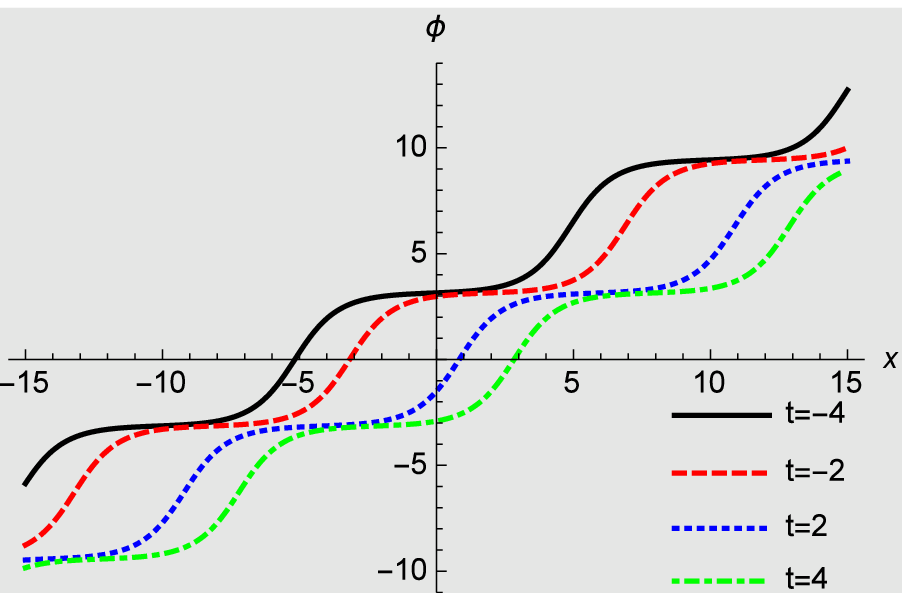}\\	
	\caption{ SG cnoidal kink solution $\widehat{\phi} _{+ ,m,\beta }^{ (1)}\left(x,t\right)$, left panel, and degenerate cnoidal kink solution $\widehat{\phi} _{+ ,m,\beta \beta }^{c}\left(x,t\right)$, right panel, for spectral parameter $\beta = 0.9$ and $m = 0.3$ at different times.}
	\label{degcniodal}
\end{figure}

\noindent We depict this solution in Figure \ref{degcniodal}. We notice that the two solutions depicted are qualitatively very similar and appear to be just translated in amplitude and $x$. However, these translations are not exact and even the approximations depend nontrivially on $\beta $ and $m$.

Taking the normalization constants in (\ref{scon1}) and (\ref{scon2})
respectively as $c=\pm m^{1/4}/(1-m)^{1/4}$ and $c=i/(1-m)^{1/4}$ we recover the simpler solution with constant speed parameter from the limits
\begin{eqnarray}
\lim_{\beta \rightarrow K}\Psi _{\pm ,m,\beta }^{\ell }(x,t)&=&\psi _{m}^{\ell}(x,t),\\
\lim_{\beta \rightarrow K}\Phi _{\pm ,m,\beta }^{\ell }(x,t)&=&\varphi_{m}^{\ell }(x,t),\\
\lim_{\beta \rightarrow K}\phi _{\pm ,m,\beta }^{\ell}(x,t)&=&\phi _{m}^{\ell }(x,t),
\end{eqnarray}

\noindent such that (\ref{sp1}) and (\ref{sp4}) can be viewed as generalizations of those solutions.

It is interesting to compare these type of solutions and investigate whether they can be used to obtain BT. It is clear that since $\phi _{cn}^{(0)}$ does not contain any spectral parameter it cannot be employed in the nonlinear superposition (\ref{sg bianchi}). However, taking $\phi_{0}=\phi_{\pm,m,\alpha}^{\ell(1)}(x,t)$, $\phi_{1}=\phi_{\pm,m,\beta }^{\ell (1)}(x,t)$ and $\phi_{2}=\phi_{\pm,m,\gamma }^{\ell(1)}(x,t)$ we identify from (\ref{2.39}) the constants $a_{1}=\pm m^{1/4}\limfunc{sn}\left(\alpha -\beta ,m \right) $ and $a_{2}=\pm m^{1/4}\limfunc{sn}\left( \alpha -\gamma ,m\right) $, such that by (\ref{sg bianchi}) we obtain the new three-parameter solution 
\begin{equation}
\resizebox{.88\hsize}{!}{$\phi _{\pm ,\alpha \beta \gamma ,m}^{\ell (3)}=\phi _{\pm ,m,\alpha }^{\ell(1)}+4\arctan \left[ \frac{\limfunc{sn}\left( \alpha -\beta ,m\right) +\limfunc{sn}\left( \alpha -\gamma ,m\right) }{\limfunc{sn}\left( \alpha-\beta ,m\right) -\limfunc{sn}\left( \alpha -\gamma ,m\right) }\tan \left( \frac{\phi _{\pm ,m,\beta }^{\ell (1)}-\phi _{\pm ,m,\gamma }^{\ell (1)}}{4}\right) \right]$} .
\end{equation}

\noindent As is most easily seen in the simpler solutions (\ref{phs1}) and (\ref{phs2}) the solutions for $\ell =\,\widehat{ }$\,\,\, are also regular in the cases with spectral parameter.

\vspace{0.5cm}

\noindent\large{\textbf{Degenerate cnoidal kink solutions}}

Using the solution $\phi _{cn}^{(0)}$ as initial solutions and the solutions (\ref{sp1}) and (\ref{sp4}) to the linear equations from SG ZC representation (\ref{SGlinear}), we are now in a position to compute the degenerate cnoidal kink solutions using the DT involving Jordan states from
\begin{equation}
\phi _{\pm ,m,\beta \beta }^{c\ell }=\phi _{cn}^{(0)}-2i\ln \frac{W\left[\Phi_{\pm,m,\beta}^{\ell},\partial_{\beta}\Phi_{\pm,m,\beta }^{\ell }\right] }{W\left[ \Psi _{\pm ,m,\beta }^{\ell },\partial _{\beta }\Psi _{\pm,m,\beta }^{\ell }\right] }.
\end{equation}

\noindent A lengthy calculation yields
\begin{equation}
\phi _{\pm ,m,\beta \beta }^{c\ell }=\phi _{cn}^{(0)}\pm 4\beta \kappa
-4\arctan \left[ i\frac{N_{\pm }^{\ell }-(N_{\pm }^{\ell })^{\ast }}{N_{\pm}^{\ell }+(N_{\pm }^{\ell })^{\ast }}\right] .
\end{equation}

\noindent where we defined the quantities
\begin{eqnarray}
\hspace{-1cm} 
\widecheck{N}_{\pm } \!\!&=&\!\! \Theta^{2}\left(\frac{x-t}{2m^{1/4}}-\frac{i}{2}K^{\prime
}\right) \left\{ H^{2}\left( \frac{x-t}{2m^{1/4}}\pm \beta +\frac{i}{2}
K^{\prime }\right) W_{\beta }\left[ \partial _{\beta }\Theta \left( \beta\right) ,\Theta \left( \beta \right) \right] \right. \\
\hspace{-1cm} 
\!\!&&\!\! +\left. \Theta ^{2}\left( \beta \right) W_{\beta }\left[ H\left( \frac{x-t}{2m^{1/4}}\pm \beta +\frac{i}{2}K^{\prime }\right) ,\partial _{\beta}H\left( \frac{x-t}{2m^{1/4}}\pm \beta +\frac{i}{2}K^{\prime }\right) \right] \right\} ,   \notag
\end{eqnarray}
\begin{eqnarray}
\hspace{-1cm} 
\widehat{N}_{\pm } \!\!&=&\!\! \Theta^{2}\left(\frac{x-t}{2m^{1/4}}-\frac{i}{2}K^{\prime
}\right) \left\{ \Theta ^{2}\left( \frac{x-t}{2m^{1/4}}\pm \beta +\frac{i}{2} K^{\prime }\right) W_{\beta }\left[ \partial _{\beta }H\left( \beta \right),H\left( \beta \right) \right] \right. \\
\hspace{-1cm} 
\!\!&&\!\! +\left. H^{2}\left( \beta \right) W_{\beta }\left[ \Theta \left( \frac{x-t}{2m^{1/4}}\pm \beta +\frac{i}{2}K^{\prime }\right) ,\partial _{\beta}\Theta \left( \frac{x-t}{2m^{1/4}}\pm \beta +\frac{i}{2}K^{\prime }\right) \right] \right\} .   \notag
\end{eqnarray}

\noindent Notice that the argument of the $\arctan $ is always real. These functions are regular for real values of $\beta $. Furthermore we observe that the additional speed spectral parameter is now separated from $x$ and $t$, so that the degenerate solution has only one speed, i.e. the degenerate solution is not displaced at any time. We depict this solution in Figure \ref{degcniodal}.

\subsection{Degenerate multi-soliton solutions from Hirota's direct method}

Finally we explore how the degenerate solutions may be obtained within the context of HDM for the SG equation. In Chapter 2, we saw how the SG equation could be converted into bilinear form with an $\arctan$ transformation. In this section, we introduce another transformation to help us convert the SG equation into bilinear form; this is the logarithmic parametrisation $\phi (x,t)=2i\ln [g(x,t)/f(x,t)]$ found in \cite{hirota_direct_2004} and hence converting the SG equation into the two equations
\begin{equation}
D_{x}D_{t}f\cdot f+\frac{1}{2}(g^{2}-f^{2})=\lambda f^{2},\quad \text{and\quad }D_{x}D_{t}g\cdot g+\frac{1}{2}(f^{2}-g^{2})=\lambda g^{2}, \label{HirotaSG}
\end{equation}

\noindent with $D_{x}$, $D_{t}$ denoting the Hirota derivatives. Explicitly we have $D_{x}D_{t}f\cdot f=2f^{2}(\ln f)_{xt}$. Taking $g=f^{\ast }$ the equations (\ref{HirotaSG}) become each other's conjugate and with $\lambda =0$ can be solved by the Wronskian 
\begin{equation}
f=W[\psi _{\alpha _{1}},\psi _{\alpha _{2}},\ldots ,\psi_{\alpha_{N}}],
\label{fW}
\end{equation}

\noindent where
\begin{equation}
\psi _{\alpha }=e^{\xi _{+}/2}+ic_{\alpha }e^{-\xi _{+}/2}.  \label{psi}
\end{equation}

\noindent For simplicity we ignore here an overall constant that may be cancelled out without loss of generality and also do not treat the possibility $\xi_{+}\rightarrow -\xi _{+}$ separately. This gives rise to the real valued $N$-soliton solutions
\begin{equation}
\phi =2i\ln \frac{f^{\ast }}{f}=4\arctan \left( i\frac{f^{\ast }-f}{f^{\ast}+f}\right) =4\arctan \frac{f_{i}}{f_{r}},  \label{dfg}
\end{equation}

\noindent where $f=f_{r}+if_{i}$ with $f_{r},f_{i}\in \mathbb{R}$. For instance the one, two and three-soliton solution obtained in this way are
\begin{eqnarray}
\hspace{-1cm} 
\phi _{\alpha } \!\!&=&\!\! 4\arctan (c_{\alpha }e^{-\xi _{+}^{\alpha }}), \\
\hspace{-1cm} 
\phi _{\alpha \beta } \!\!&=&\!\! 4\arctan \left[ \Gamma _{\alpha \beta }\frac{c_{\beta }e^{\xi _{+}^{\beta }}-c_{\alpha }e^{\xi _{+}^{\alpha }}}{1+c_{\alpha }c_{\beta }e^{\xi _{+}^{\alpha }+\xi _{+}^{\beta }}}\right] , \\
\hspace{-1cm} 
\phi _{\alpha \beta \gamma } \!\!&=&\!\! \resizebox{.75\hsize}{!}{$4\arctan \left[ \frac{c_{\alpha }c_{\beta}c_{\gamma }+c_{\alpha }\Gamma _{\alpha \beta }\Gamma_{\alpha \gamma}e^{\xi _{+}^{\beta }+\xi _{+}^{\gamma }}+c_{\beta}\Gamma_{\beta \alpha}\Gamma_{\beta\gamma}e^{\xi_{+}^{\alpha}+\xi_{+}^{\gamma}}+c_{\gamma}\Gamma_{\gamma\alpha}\Gamma_{\gamma\beta}e^{\xi_{+}^{\alpha}+\xi_{+}^{\beta }}}{c_{\beta }c_{\gamma }\Gamma _{\alpha \beta }\Gamma_{\alpha
		\gamma }e^{\xi _{+}^{\alpha }}+c_{\alpha }c_{\gamma }\Gamma _{\beta \alpha}\Gamma _{\beta \gamma }e^{\xi _{+}^{\beta }}+c_{\alpha}c_{\beta }\Gamma_{\gamma \alpha }\Gamma _{\gamma \beta }e^{\xi _{+}^{\gamma }}+e^{\xi_{+}^{\alpha }+\xi _{+}^{\beta }+\xi _{+}^{\gamma }}}\right]$} ,
\end{eqnarray}

\noindent where $\Gamma _{xy}:=(x+y)/(x-y)$. We kept here the constants $c_{\alpha},c_{\beta },c_{\gamma }$ generic as it was previously found in \cite{correa_regularized_2016} and discussed in Section 5.1.2, that they have to be chosen in a specific way to render the limits to the degenerate case finite.

Following the procedure outlined in \cite{correa_regularized_2016} and discussed in Section 5.1.1, we replace the standard solutions to the Schr\"{o}dinger equation in the non-degenerate solution by Jordan states in the computation of $f$ in (\ref{fW}) as
\begin{equation}
f=W[\psi _{\alpha },\partial _{\alpha }\psi _{\alpha },\partial _{\alpha}^{2}\psi _{\alpha },\ldots ,\partial _{\alpha }^{N}\psi _{\alpha }].
\end{equation}

\noindent We then recover from (\ref{dfg}) the degenerate kink solution $\phi_{\alpha\alpha }$ and $\phi _{\alpha \alpha \alpha }$ in (\ref{deg2}) and (\ref{deg3}), respectively, with $c_{\alpha }=-1$ in (\ref{psi}). Unlike as in the treatment of the KdV equation in \cite{correa_regularized_2016} or in Section 5.1.2, the equations are already in a format that allows to carry out the limits $\lim_{\beta\rightarrow \alpha }\phi _{\alpha \beta }=\phi _{\alpha \alpha }$ and $\lim_{\beta ,\gamma \rightarrow \alpha }\phi _{\alpha \beta \gamma }=\phi_{\alpha \alpha \alpha }$ with the simple choices $c_{\alpha }=c_{\beta }=1$
and $c_{\alpha }=c_{\beta }=c_{\gamma }=-1$, respectively.

\subsection{Asymptotic properties of degenerate multi-soliton solutions}

Let us now compute the time-dependent displacements by tracking the
one-soliton solution within a degenerate multi-soliton solution.

\vspace{0.5cm}

\noindent\large{\textbf{Time-dependent displacements for degenerate multi-kink solutions}}

Unlike standard multi-soliton solutions, one cannot track the maxima or minima for the kink solutions as they might have maximal or minimal amplitudes extending up to infinity. However, they have many intermediate points in between the extrema that are uniquely identifiable. For instance, for the solutions constructed in Sections 5.2.1 and 5.2.2, a suitable choice is the point of inflection at half the maximal value, that is at $\phi_{\alpha^{(N)} }=\pi $ corresponding to $\tau _{N}=1$. For an $N$-soliton solution $\phi_{\alpha^{(N)}}$ with $N$ parametrised as $N=2n+1-\kappa $ we
find the pattern that these $N$ points are reached asymptotically

\begin{equation}
\lim\limits_{t\rightarrow \infty }\tau _{2n+1-\kappa }\left( -\frac{t}{\alpha ^{2}}\pm \Delta _{m,\ell ,\kappa },t\right) =1, \label{limit}
\end{equation}

\noindent for the time-dependent displacements 
\begin{equation}
\Delta X_{m \ell \kappa}\left(t\right)=\frac{1}{\alpha}\ln \left[\frac{(m-\ell)!}{(m+\ell-\kappa)!}|4\alpha v t|^{2\ell-\kappa}\right] \label{SGDelta}
\end{equation}

\noindent with
\begin{eqnarray}
v&=&-\frac{1}{\alpha^{2}},\\
k&=&\left\{ \begin{array}{c}
1 \quad \text{ for N even}\\
0 \quad \text{ for N odd}
\end{array} \right. ,\\
m&=&\frac{N-1+k}{2},\\
l&=& l^{th} \text{constituent from soliton compound centre}.
\end{eqnarray}

\begin{figure}[h]
	\centering

	\includegraphics[width=0.7\linewidth]{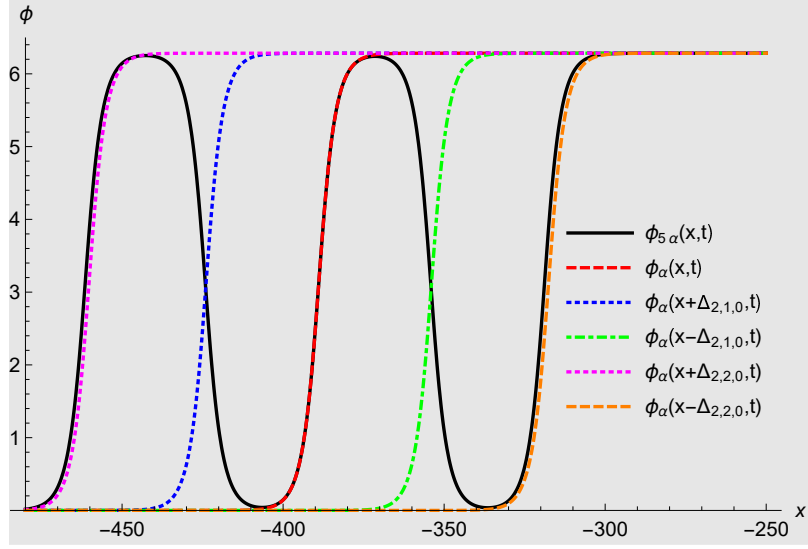}\\
	
	\caption{Degenerate $5$-soliton solution compared with five time-dependently laterally displaced one-soliton solutions for $\alpha=0.3$ at time $t=35$.}
	\label{displ}
\end{figure}

\noindent For example, given a degenerate $5$-soliton solution $\phi _{\alpha^{(5)}}$ we have $m=2$, $\kappa =0$ and $\ell =0,1,2$, so that we can compare it with five laterally displaced one-soliton solutions as depicted in Figure \ref{displ}.

Let us now derive this expression for the first five examples. Introducing the notation $x_{\pm }\!:=\!-\,\frac{t}{\alpha ^{2}}\pm \frac{1}{\alpha} \ln \delta $, assuming that $\delta \sim t^{\mu }$, $\mu \geq 1$ and taking $y$ to be a polynomial in $t$, we obtain the useful auxiliary limits 
\begin{eqnarray}
\lim\limits_{t\rightarrow \infty }\lim\limits_{x\rightarrow x_{\pm }}\left(y+\xi _{+}\right) &=&\lim\limits_{t\rightarrow \infty }\left( y\pm \ln\delta \right) \approx \lim\limits_{t\rightarrow \infty }y, \\
\lim\limits_{t\rightarrow \infty }\lim\limits_{x\rightarrow x_{\pm }}\left(y+\xi _{-}\right) &=&\lim\limits_{t\rightarrow \infty }\left( y+\frac{2t}{\alpha}\mp\ln\delta\right)\approx\lim\limits_{t\rightarrow \infty}\left(y+\frac{2t}{\alpha}\right).
\end{eqnarray}

\noindent Using these expressions in (\ref{tau1}-\ref{tau5}) and the notation $\delta _{m,\ell,\kappa}=\alpha\exp(\Delta X_{m,\ell,\kappa})$, $T=\left| \frac{2t}{\alpha}\right| $ we derive the asymptotic expressions for the $N$-soliton
solution for the lowest values of $N$ 
\begin{eqnarray}
\lim\limits_{\substack{ t\rightarrow \infty  \\ x\rightarrow x_{\pm }}}\tau_{2} \!\!&\approx &\!\! \lim\limits_{t\rightarrow \infty }\frac{2T\delta ^{\pm }}{1+(\delta ^{\pm })^{2}}= \lim\limits_{t\rightarrow \infty }\frac{2T}{\delta
^{\pm }}=1 \\
\lim\limits_{\substack{ t\rightarrow \infty  \\ x\rightarrow x_{\pm }}}\tau_{3} \!\!&\approx &\!\! \lim\limits_{t\rightarrow \infty }\frac{2T^{2}+(\delta ^{\pm})^{2}}{2T^{2}\delta ^{\pm }}=1 \\
\lim\limits_{\substack{ t\rightarrow \infty  \\ x\rightarrow x_{\pm }}}\tau_{4} \!\!&\approx &\!\! \lim\limits_{t\rightarrow \infty }\frac{4T^{3}\left[1+(\delta ^{\pm })^{2}\right] }{4T^{4}\delta ^{\pm }+3(\delta ^{\pm })^{3}}=1\\
\lim\limits_{\substack{ t\rightarrow \infty  \\ x\rightarrow x_{\pm }}}\tau_{5} \!\!&\approx &\!\! \lim\limits_{t\rightarrow \infty }\frac{4T^{6}\delta ^{\pm}+9(\delta ^{\pm })^{3}}{4T^{6}+6T^{4}(\delta ^{\pm })^{2}}=1
\end{eqnarray}
\noindent where
\begin{eqnarray}
 \tau_{2}: \quad	 \delta ^{\pm }\!\!&=&\!\!2T =\delta _{1,1,1},\\
 \tau_{3}:	\quad	\delta ^{\pm }\!\!&=&\!\!1
		\quad \text{ or} \quad \delta^{\pm}=2T^{2}=\delta _{1,1,0},\\
 \tau_{4}:\quad	\delta ^{\pm }\!\!&=&\!\! T=\delta _{2,1,1}
		\quad \text{ or} \quad \delta ^{\pm }=\frac{4}{3}T^{3}=\delta _{2,2,1}, \\
 \tau_{5}:	\quad	\delta ^{\pm }\!\!&=&\!\!1
		\quad \text{ or} \quad \delta ^{\pm }=\frac{2}{3}T^{2}=\delta _{2,1,0}
		\quad \text{ or} \quad \delta ^{\pm }=\frac{2}{3}T^{4}=\delta _{2,2,0},
\end{eqnarray}

\noindent for constituents counting outwards from the centre of the multi-solution compound respectively. The limits need to be carried out in consecutive order, i.e. first take $x\rightarrow x_{\pm }$ and then compute the limit $t\rightarrow\infty$. These are the first explicit examples for the asymptotic values all confirming the general expression for the time-dependent displacements (\ref{SGDelta}). Similarly we have computed examples for higher values, up to $N=10$ that may also be cast into the form of (\ref{SGDelta}). So far we have not obtained a general proof valid for any $N$.

Having computed the lateral displacement $\Delta_{x}$ the time-displacement is obtained as usual from $\Delta_{t}=-\Delta _{x}/v$, where $v=1/\alpha ^{2}$ in our case. 

\noindent Note also that the displacement (\ref{SGDelta}) is of the same form as for KdV degenerate multi-soliton solutions, where the velocity parameter is $v=\alpha^{2}$. So the expression of time-dependent displacements is universal for KdV and SG degenerate multi-soliton solutions and possibly other nonlinear systems.

\vspace{0.5cm}

\noindent\large{\textbf{Time-dependent displacements for breather solutions}

For the breathers it is even less evident what point in the solution is
suitable for tracking due to the overall oscillation. However, since we are only interested in the net movement we can neglect the internal oscillation and determine the displacement for an enveloping function that surrounds the breather and moves with the same overall speed. For the one-breather solution an enveloping function is obtained by setting the $\sin $-function in (\ref{b1}) to $1$, obtaining
\begin{equation}
\phi _{\alpha,1/\alpha}^{\text{env}}=4\arctan \left[ \frac{\theta }{\cosh \left[ \theta (t+x)/\sqrt{1+\theta ^{2}}\right] }\right] .
\end{equation}

\noindent This function is depicted together with the breather solution in Figure \ref{breatherenv} having a clearly identifiable maximum value $4\arctan \theta $ which we can track.
\begin{figure}[h]
	\centering
	
	\includegraphics[width=0.4\linewidth]{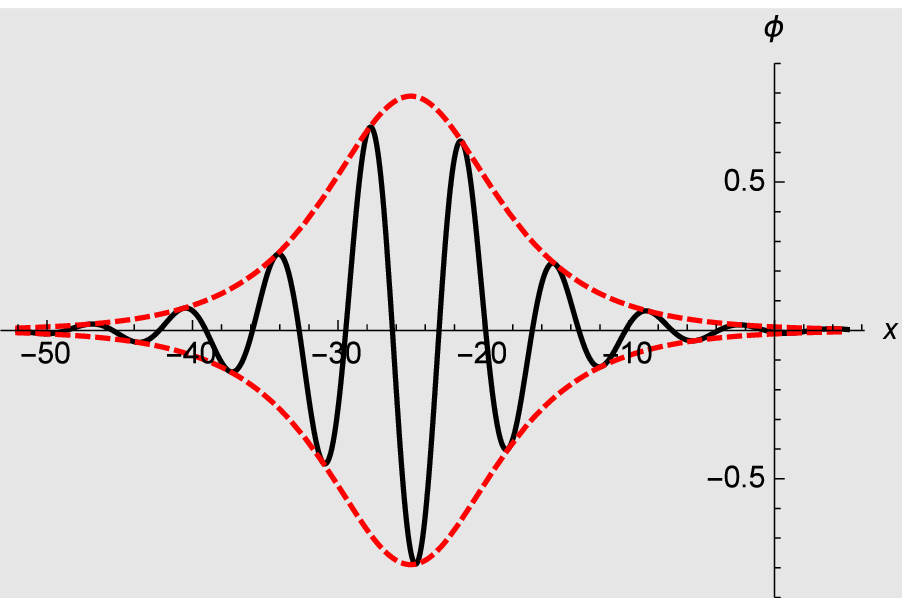} 
	\hspace{0.8cm}
	\includegraphics[width=0.4\linewidth]{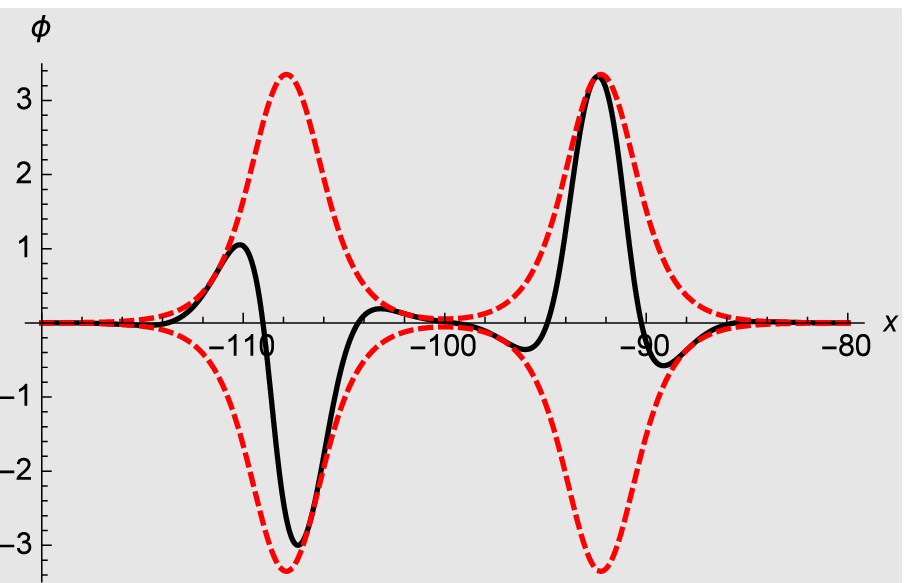}\\
	
	\caption{One-breather solution surrounded by enveloping function $\pm \phi _{\alpha,1/\alpha}^{\text{env}}$ at $t=25$ for $\theta=1/5$, left panel and degenerate two-breather solution surrounded by enveloping function $\pm \phi _{\alpha,\alpha,1/\alpha,1/\alpha}^{\text{env}}$ at $t=100$ for $\theta=4/3$, right panel.}
	\label{breatherenv}
\end{figure}

We compare this now with the breather solution $\phi_{\alpha,\alpha,1/\alpha,1/\alpha}$ constructed in Section 5.2.2. Taking for that solution $\sin\left[(t-x)/\sqrt{1+\theta ^{2}}\right]\rightarrow 0$ and $\cos\left[ (t-x)/\sqrt{1+\theta ^{2}}\right]\rightarrow 1$ we obtain the enveloping function
\begin{equation}
\phi_{\alpha,\alpha,1/\alpha,1/\alpha}^{\text{env}}=4\arctan \left[ \resizebox{.55\hsize}{!}{$\frac{4(t-x)\theta \sqrt{1+\theta ^{2}}\cosh\left( \frac{(t+x)\theta }{\sqrt{1+\theta ^{2}}}\right) }{2\theta^{2}(t-x)^{2}+2(t+x)^{2}+1+\theta ^{-2}+(1+\theta ^{-2})\cosh \left( \frac{2\theta (t+x)}{\sqrt{1+\theta ^{2}}}\right) }$}\right] .
\end{equation}

\noindent This function tends asymptotically to the maximal value of the one-breather enveloping function 
\begin{equation}
\lim\limits_{t\rightarrow \infty }\phi _{\alpha,\alpha,1/\alpha,1/\alpha}^{\text{env}}\left( -t\pm \Delta_{\alpha,\alpha,1/\alpha,1/\alpha},t\right) =4\arctan \left( \theta \right) ,
\end{equation}

\noindent when shifted appropriately with the time-dependent displacement
\begin{equation}
\Delta X_{\alpha,\alpha, 1/ \alpha, 1/ \alpha }(t)=\frac{1}{\theta }\sqrt{1+\theta ^{2}}\ln \left( \frac{4\theta^{2}t}{\sqrt{1+\theta ^{2}}}\right) .
\end{equation}

\noindent Similarly we may compute the displacements for the solutions $\phi_{\alpha, \alpha,\alpha, 1/ \alpha, 1/ \alpha, 1/ \alpha }$ etc.

\section{Conclusions}

We have seen the construction of various types of degenerate multi-soliton solutions for the KdV and SG equations based on BT, DCT and HDM. Many of them exhibit a compound behaviour on a small timescale, but their individual one-soliton constituents separate for large time. Exceptions are degenerate cnoidal kink solutions that we constructed
from shifted Lam\'{e} potentials for the SG equation. These type of solutions have constant speed and do not display any time-delay. 

Comparing the various methods, we see degeneracy has to be implemented in different ways for different equations and methods. For the KdV case, the key point to obtain regularized degenerate multi-soliton solutions is to add some complex shifts. The most straightforward and simplest way to obtain a KdV degenerate $N$-soliton solution is through DCT with Jordan states. Other methods such as HDM and BT involve guessing some appropriate constants which will allow non-trivial degenerate limiting.

For the SG case, we have a different story. Using the recurrence relations constructed from BT was found to be the most efficient way to obtain $N$-soliton solutions for large values of $N$. These equations are easily
implemented in computer calculations. By just requiring a simple solution to
the original NPDE they also have a relatively easy starting
point. However, the equations are less universal than those presented using Jordan states in DCT. The
disadvantage of DCT is that they also require the wave functions for the SG ZC representation. For large values of $N$ the computations become more involved than the recurrence relations for the BT.
The HDM for
the SG model follows similarly as for the KdV model. The degenerate limit for a $N$-soliton solution with different spectral parameters to one particular spectral parameter could only be taken with appropriate choice of some constants, for which a general systematic way has not been found up to now.

Looking at the asymptotic behaviour of the solutions, we present general analytical time-dependent expressions for displacements between the one-soliton solution and individual constituents of degenerate multi-soliton solutions. When expressed in terms of the soliton speed and spectral parameter, the expression found appears to be universal for the KdV and SG equations, although in general, the form of this is conjectural.

\chapter{Asymptotic and scattering behaviour for degenerate multi-soliton solutions in the Hirota equation}\label{ch_6}

In this chapter, we continue the study of degenerate multi-soliton solutions with the Hirota equation, a particular example of a higher order NLS equation. For the NLS equation, degenerate solutions have been studied in the context of the inverse scattering method \cite{olmedilla_multiple_1987,schiebold_asymptotics_2017} where they were referred to as multiple pole solutions, which make use of poles from kernels
of the Gel'fand-Levitan-Marchenko equations. In our previous analysis for the KdV and SG equations, we showed how to derive these type of solutions by employing HDM, BT, DCT or recursive equations derived from BT. Here we follow a similar approach for the Hirota equation in the construction of the degenerate multi-soliton solutions. Subsequently, we study their asymptotic and scattering behaviour at the origin \cite{cen_asymptotic_2019}.

\section{Degenerate multi-soliton solutions from Hirota's direct method}
Taking the Hirota bilinear form of the Hirota equation (\ref{bi1}-\ref{bi2}), exact multi-soliton solutions can be found in a recursive fashion by
terminating the formal power series expansions%
\begin{equation} 
f(x,t)=\dsum\limits_{k=0}^{\infty }\varepsilon ^{2k}f_{2k}(x,t) \quad \text{%
	and\quad }g(x,t)=\dsum\limits_{k=1}^{\infty }\varepsilon
^{2k-1}g_{2k-1}(x,t),  \label{psex}
\end{equation}%
at a particular order in $\varepsilon $, similarly as was done previously for other equations.

\subsection{One-soliton solution}

For $\varepsilon =1$ a one-soliton is obtained as%
\begin{equation}
\resizebox{.9\hsize}{!}{$q_{1}^{\mu }(x,t)=\frac{g_{1}^{\mu }(x,t)}{1+f_{2}^{\mu }(x,t)},\quad 
\text{with} \quad g_{1}^{\mu }(x,t)=c\tau _{\mu}\text{\quad and\quad }%
f_{2}^{\mu }(x,t)=\frac{\left\vert c\tau _{\mu }\right\vert ^{2}}{(\mu +\mu
	^{\ast })^{2}}$}.  \label{ones}
\end{equation}%
The building block is the function
\begin{equation}
\tau_{\mu
}(x,t):=e^{\mu x+\mu ^{2}(i\alpha -\beta \mu )t},
\end{equation}%
involving the complex constants $c$,$\mu \in \mathbb{C}$. More explicitly,
for $c=1$ we have%
\begin{equation}
\resizebox{.9\hsize}{!}{$q_{1}^{\mu }(x,t)=\frac{4\delta ^{2}e^{x(\delta +i\xi )+it(\delta +i\xi
		)^{2}(\alpha +i\beta \delta -\beta \xi )}}{4\delta ^{2}+e^{2\delta x-2\delta
		t\left[ 2\alpha \xi +\beta \left( \delta ^{2}-3\xi ^{2}\right) \right] }}, \quad \left\vert q_{1}^{\mu }(x,t)\right\vert =\frac{4\delta
	^{2}e^{\delta \left[ x-t\left( 2\alpha \xi +\beta \left( \delta ^{2}-3\xi
		^{2}\right) \right) \right] }}{4\delta ^{2}+e^{2\delta \left[ x-t\left(
		2\alpha \xi +\beta \left( \delta ^{2}-3\xi ^{2}\right) \right) \right] }}$}.
\label{q1mu}
\end{equation}%
with $\mu =\delta +i\xi $, $\delta $,$\xi \in \mathbb{R}$. Defining the real
quantities 
\begin{eqnarray}
A(x,t) &:=&x\xi +t\left[ \alpha (\delta ^{2}-\xi ^{2})+\beta \xi (\xi
^{2}-3\delta ^{2})\right] ,  \label{Ax} \\
x_{\pm }^{\delta ,\xi } &:=&t\left[ 2\alpha \xi +\beta (\delta ^{2}-3\xi
^{2})\right] \pm \frac{1}{\delta }\ln (2\delta ),
\end{eqnarray}%
we compute the maximum of the modulus for the one-soliton solution from 
\begin{equation}
q_{1}^{\mu }(x+x_{+}^{\delta ,\xi },t)=\delta \func{sech}(x\delta
)e^{iA(x+x_{+}^{\delta ,\xi },t)} \qquad \Rightarrow \qquad \left\vert q_{1}^{\mu
}(x_{+}^{\delta ,\xi },t)\right\vert_{\text{max}} =\left\vert \delta \right\vert .  \label{q1}
\end{equation}%
Thus while the real and imaginary parts of the one-soliton solution exhibit
a breather like behaviour, the modulus is a proper solitary wave with a
stable maximum value $\delta $. The solution $q_{1}^{\mu }$ becomes static in
the limit to the NLSE $\beta \rightarrow 0$ for real $\mu $, i.e. $\xi =0$,
and also in the limit to the mKdV equation $\alpha \rightarrow 0$ when $%
\delta ^{2}=3\xi ^{2}$.

\subsection{Nondegenerate and degenerate two-soliton solutions}

At the next order in $\varepsilon $ of the expansions (\ref{psex}) we
construct a general nondegenerate two-soliton solution as%
\begin{equation}
q_{2}^{\mu ,\nu }(x,t)=\frac{g_{1}^{\mu ,\nu }(x,t)+g_{3}^{\mu ,\nu }(x,t)}{%
	1+f_{2}^{\mu ,\nu }(x,t)+f_{4}^{\mu ,\nu }(x,t)},\quad \ \   \label{twos}
\end{equation}%
with functions%
\begin{eqnarray}
\hspace{-0.5cm}
g_{1}^{\mu ,\nu } \!\!\!&=&\!\!\!c\tau_{\mu }+\widetilde{c}\tau_{\nu},  \label{g1tau} \\
\hspace{-0.5cm}
g_{3}^{\mu ,\nu } \!\!\!&=&\!\!\!\frac{\left( \mu -\nu \right) ^{2}}{\left( \mu +\mu
	^{\ast }\right) ^{2}\left( \nu +\mu ^{\ast }\right) ^{2}}\widetilde{c}\tau_{\nu}\left\vert c\tau_{\mu}\right\vert ^{2}+\frac{\left( \mu -\nu \right)
	^{2}}{\left( \mu +\nu ^{\ast }\right) ^{2}\left( \nu +\nu ^{\ast }\right)
	^{2}}c\tau_{\mu}\left\vert \widetilde{c}\tau_{\nu}\right\vert ^{2}, \\
\hspace{-0.5cm}
f_{2}^{\mu ,\nu } \!\!\!&=&\!\!\!\frac{\left\vert c\tau_{\mu}\right\vert ^{2}}{\left(
	\mu +\mu ^{\ast }\right) ^{2}}+\frac{\widetilde{c}\tau _{\nu }c^{\ast}\tau _{\mu}^{\ast }}{\left( \nu +\mu ^{\ast }\right) ^{2}}+\frac{c\tau_{\mu}\widetilde{c}^{\ast}\tau_{\nu}^{\ast }}{\left( \mu +\nu ^{\ast }\right) ^{2}}+\frac{%
	\left\vert \widetilde{c}\tau_{\nu}\right\vert ^{2}}{\left( \nu +\nu ^{\ast
	}\right) ^{2}}, \\
\hspace{-0.5cm}
f_{4}^{\mu ,\nu } \!\!\!&=&\!\!\!\frac{\left( \mu -\nu \right) ^{2}\left( \mu ^{\ast
	}-\nu ^{\ast }\right) ^{2}}{\left( \mu +\mu ^{\ast }\right) ^{2}\left( \nu
	+\mu ^{\ast }\right) ^{2}\left( \mu +\nu ^{\ast }\right) ^{2}\left( \nu +\nu
	^{\ast }\right) ^{2}}\left\vert c\tau_{\mu}\right\vert ^{2}\left\vert
\widetilde{c}\tau_{\nu}\right\vert ^{2}.  \label{f4}
\end{eqnarray}%
We have set here also $\varepsilon =1$. As was noted previously in Chapter 5, the limit $\mu \rightarrow \nu $ to the
degenerate case cannot be carried out trivially for generic values of the
constants $c$, $\widetilde{c}$. However, we find that for the specific choice%
\begin{equation}
c=\frac{\left( \mu +\mu ^{\ast }\right) \left( \mu +\nu ^{\ast }\right) }{%
	\left( \mu -\nu \right) }, \quad \widetilde{c}=-\frac{\left( \nu +\nu ^{\ast
	}\right) \left( \nu +\mu ^{\ast }\right) }{\left( \mu -\nu \right) },
\label{cc}
\end{equation}%
the limit is nonvanishing for all functions in (\ref{g1tau})-(\ref{f4}). This
choice is not unique, but the form of the denominators is essential to
guarantee the limit to be nontrivial. With $c$ and $\widetilde{c}$ as in (\ref%
{cc}) the limit $\mu \rightarrow \nu $ leads to the new degenerate
two-soliton solution%
\begin{equation}
q_{2}^{\mu ,\mu }(x,t)=\frac{\left( \mu +\mu ^{\ast }\right) \tau%
	_{\mu }\left[ (2+\widehat{\tau}_{\mu })+(2-\widehat{\tau}_{\mu })\left\vert 
		\tau_{\mu }\right\vert ^{2}\right] }{1+(2+\left\vert \widehat{\tau}_{\mu
	}\right\vert ^{2})\left\vert \tau_{\mu }\right\vert ^{2}+\left\vert 
	\tau_{\mu }\right\vert ^{4}},  \label{dtwos}
\end{equation}%
where we introduced the function%
\begin{equation}
\widehat{\tau}_{\mu }(x,t):=x+\mu t(2i\alpha -3\beta \mu )\left( \mu +\mu ^{\ast
}\right) .
\end{equation}%
We observe the two different timescales in this solution entering through
the functions $\widehat{\tau}_{\mu}$ and $\tau_{\mu}$, in a linear and exponential
manner, respectively, which is a typical feature of degenerate solutions.

\section{Degenerate multi-soliton solutions from Darboux-Crum transformations}

In Section 2.5.3, we saw the construction of multi-soliton solutions to the Hirota equation using DCT. Degenerate solutions can be obtained in principle by taking the limit of all speed parameters to a particular speed parameter, which however, only leads to nontrivial solutions for some very specific choices of the constants as discussed in the previous section. The other method to achieve degeneracy
is to replace the standard solutions of the ZC representation with Jordan states, as explained in more detail in the previous chapter%
\begin{eqnarray}
\varphi _{2k-1} \!\!&\rightarrow &\!\!\partial _{\mu }^{k-1}\varphi \quad \text{, } \quad %
\varphi _{2k} \,\,\rightarrow\, - \partial _{\mu^{\ast} }^{k-1}\phi ^{\ast}\text{, } \\
\phi _{2k-1}\!\!&\rightarrow &\!\! \partial _{\mu }^{k-1}\phi  \quad \text{, } \quad %
\, \phi _{2k} \,\,\rightarrow\, \partial _{\mu^{\ast} }^{k-1}\varphi ^{\ast}\text{, }
\end{eqnarray}%

\noindent for $k=1,\ldots ,n$ where
\begin{eqnarray}
	\varphi&=&e^{\mu x+2\mu^{2}(i \alpha-2\beta \mu)t},\\
	\phi&=&e^{-\mu x-2\mu^{2}(i \alpha-2\beta \mu)t},\\
\end{eqnarray}
and the asterisk denotes conjugation. Explicitly, the first examples for the
matrices $\widetilde{D}_{n}$ and $\widetilde{W}_{n}$ related to the degenerate
solutions are%
\begin{equation}
\widetilde{D}_{1}=\left( 
\begin{array}{cc}
\varphi _{x} & \varphi \\ 
-\phi _{x}^{\ast } & -\phi^{\ast}%
\end{array}%
\right), \quad \widetilde{W}_{1}=\left( 
\begin{array}{cc}
\varphi & \phi \\ 
-\phi^{\ast} & \varphi^{\ast}%
\end{array}%
\right) ,
\end{equation}%

\begin{equation}
\hspace{-0.12cm}
\widetilde{D}_{2}\!=\!\!\left( \!\!
\begin{array}{cccc}
\phi & \varphi _{xx} & \varphi _{x} & \varphi \\ 
\varphi^{\ast} & -\phi _{xx}^{\ast } & -\phi _{x}^{\ast } & -\phi^{\ast} \\ 
\partial_{\mu}\phi & [\partial_{\mu}\varphi] _{xx} & [\partial_{\mu}\varphi]
_{x}& \partial_{\mu}\varphi \\ 
\partial_{\mu^{\ast}}\varphi^{\ast} & -[\partial_{\mu^{\ast}}\phi^{\ast}]_{xx} & -[\partial_{\mu^{\ast}}\phi^{\ast}]_{x} & -\partial_{\mu^{\ast}}\phi^{\ast}%
\end{array} \!\!
\right) ,
\end{equation}%
\begin{equation}
\hspace{-0.12cm}
 \widetilde{W}_{2}\!=\!\!\left( \!\!
	\begin{array}{cccc}
	\varphi _{x} & \varphi & \phi _{x} & \phi \\ 
	-\phi _{x}^{\ast } & -\phi^{\ast} & \varphi _{x}^{\ast } & \varphi ^{\ast} \\ 
	\left[\partial_{\mu}\varphi\right] _{x} & \partial_{\mu}\varphi & \left[\partial_{\mu}\phi\right] _{x} & \partial_{\mu}\phi \\ 
	-\left[\partial_{\mu^{\ast}}\phi^{\ast}\right]_{x} & -\partial_{\mu^{\ast}}\phi ^{\ast } & \left[\partial_{\mu^{\ast}}\varphi^{\ast}\right]_{x} & \partial_{\mu^{\ast}}\varphi^{\ast}%
	\end{array} \!\!\right) , \notag
\end{equation}%

The degenerate $n$-soliton solutions are then
computed as%
\begin{equation}
q_{n}^{n\mu }(x,t)=2\frac{\det \widetilde{D}_{n}}{\det \widetilde{W}_{n}},
\label{nsolution}
\end{equation}
where
\begin{eqnarray}
\hspace{-1cm}
(\widetilde{D}_{N})_{ij}\!\!\!\!&=&\!\!\!\! \left\{\!\!\!\begin{array}{c}
(i=2k-1)\,|\,\,\,[\partial _{\mu }^{k-1}\phi]^{(N-j-1)}, \,\,\, (j < N); \,\,\, [\partial _{\mu }^{k-1}\varphi]^{(2N-j)},\,\,\, (j \geq N)\\
(i=2k)\,|\,\,\, [\partial _{\mu^{\ast} }^{k-1}\varphi^{\ast}]^{(N-j-1)}, \,\,\, (j < N); \,\,\, -[\partial _{\mu^{\ast} }^{k-1}\phi^{\ast}]^{(2N-j)}, \,\,\, (j \geq N)
\end{array} \right.\\
\hspace{-1cm}
(\widetilde{W}_{N})_{ij}\!\!\!\!&=&\!\!\!\! \left\{\!\!\!\begin{array}{c}
(i=2k-1)\,|\,\,\,[\partial _{\mu }^{k-1}\varphi]^{(N-j)}, \,\,\, (j \leq N); \,\,\, [\partial _{\mu }^{k-1}\phi]^{(2N-j)},\,\,\, (j > N)\\
(i=2k)\,|\,\,\, [\partial _{\mu^{\ast} }^{k-1}\phi^{\ast}]^{(N-j)}, \,\,\, (j \leq N); \,\,\, [\partial _{\mu^{\ast} }^{k-1}\varphi^{\ast}]^{(2N-j)},\,\,\, (j > N)
\end{array} \right.
\end{eqnarray}
and for any function $f$, the derivatives with respect to $x$ are denoted as $[f]^{(m)}=\partial_{x}^{m}f$. The result is that only one spectral parameter, $\mu $, is left.

\section{Reality of charges for complex degenerate multi-solitons}

Recalling the ZC representation of the Hirota equation (\ref{hezc1}-\ref{hezc12}) from Section 2.5.3, the conserved quantities for this system are easily derived from an analogue of the Gardner transform for the KdV field \cite{miura_korteweg-vries_1968,miura_korteweg-vries_1976,kupershmidt_nature_1981,cen_time-delay_2017} and match the ones for the NLS hierarchy \cite{zakharov_on_1974}. Defining two new complex valued fields $T(x,t)$ and $\chi (x,t)$ in terms of the components of the auxiliary field $\Psi $ one trivially obtains a local conservation law 
\begin{equation}
T:=\frac{\varphi _{x}}{\varphi },\qquad \chi :=-\frac{\varphi _{t}}{\varphi }%
,\quad \Rightarrow T_{t}+\chi _{x}=0.  \label{TXI}
\end{equation}%
From the two first rows in the equations (\ref{hezc1}) we then derive%
\begin{equation}
T=q\frac{\phi }{\varphi }-i\lambda \text{,  } \quad \chi =-A-B\frac{\phi }{%
	\varphi },
\end{equation}%
so that the local conservation law in (\ref{TXI}) is expressed in terms of
the as yet unknown quantities $A$, $B$ and $T$%
\begin{equation}
T_{t}-\left( A+\frac{i\lambda B}{q}+\frac{B}{q}T\right) _{x}=0.  \label{LC}
\end{equation}%
The missing function $T$ is then determined by the Ricatti equation 
\begin{equation}
T_{x}=i\lambda \frac{q_{x}}{q}+rq-\lambda ^{2}+\frac{q_{x}}{q}T-T^{2},
\label{Ric}
\end{equation}%
which in turn is obtained by differentiating $T$ in (\ref{TXI}) with respect
to $x$. The Gardner transformation consists now on
expanding $T$ in terms of $\lambda $ and a new field $w$ as $T=-i\lambda
\lbrack 1-w/(2\lambda ^{2})]$. This choice is motivated by balancing the
first with the fourth and the third and the fifth term when $\lambda
\rightarrow \infty $. The factor on the field $w$ is just for convenience and
renders the following calculations in a simple form. Substituting this
expression for $T$ into the Ricatti equation (\ref{Ric}) with a further
choice $\lambda =i/(2\varepsilon )$, $\varepsilon \rightarrow 0$, made once more for convenience, yields%
\begin{equation}
w+\varepsilon \left( w_{x}-\frac{q_{x}}{q}w\right) +\varepsilon
^{2}w^{2}-rq=0.  \label{wnew}
\end{equation}%
Up to this point our discussion is entirely generic and the functions $%
r(x,t) $ and $q(x,t)$ can in principle be any function. Fixing their mutual
relation now to $r(x,t)=-q^{\ast }(x,t)$ and expanding the new auxiliary
density field as%
\begin{equation}
w(x,t)=\dsum\limits_{n=0}^{\infty }\varepsilon ^{n}w_{n}(x,t),
\end{equation}%
we can solve (\ref{wnew}) for the functions $w_{n}$ in a recursive manner
order by order in $\varepsilon $. Iterating these solutions yields%
\begin{equation}
w_{n}=\frac{q_{x}}{q}w_{n-1}-(w_{n-1})_{x}-\sum%
\limits_{k=0}^{n-2}w_{k}w_{n-k-2},\qquad \text{for }n\geq 1.  \label{wn}
\end{equation}%
We compute the first expressions to be%
\begin{eqnarray}
w_{0} &=&-\left\vert q\right\vert ^{2},  \label{w0} \\
w_{1} &=&\frac{1}{2}\left(
qq_{x}^{\ast }-q^{\ast }q_{x}\right)+\frac{1}{2}\left\vert q\right\vert _{x}^{2} , \\
w_{2} &=&\left\vert q_{x}\right\vert ^{2}-\left\vert q\right\vert ^{4}-\frac{%
	1}{2}\left( qq_{x}^{\ast }+q^{\ast }q_{x}\right) _{x}+\frac{1}{2}\left(
q^{\ast }q_{x}-qq_{x}^{\ast }\right)_{x} ,  \label{w2} \\
w_{3} &=&\frac{1}{2}\left( 3q\left\vert q\right\vert ^{2}q_{x}^{\ast }-3q^{\ast
}\left\vert q\right\vert ^{2}q_{x}+q_{x}^{\ast }q_{xx}-q_{x}q_{xx}^{\ast
}\right)
\label{w3} \\
&&+\left[ \frac{5}{4}\left\vert q\right\vert ^{4}+\frac{1}{2}\left(
qq_{xx}^{\ast }+q^{\ast }q_{xx}-\left\vert q_{x}\right\vert ^{2}\right) %
\right] _{x}+\frac{1}{2}\left( qq_{xx}^{\ast }-q^{\ast }q_{xx}\right) _{x} .  \notag
\end{eqnarray}%
When possible we have also extracted terms that can be written as
total derivatives, since they become surface terms in the expressions for the
conserved quantities. We note that with regard to the aforementioned $\mathcal{PT}$%
-symmetry we have $\mathcal{PT}(w_{n})=(-1)^{n}w_{n}$. Since $T$ is a
density of a local conservation law, also each function $w_{n}$ can be
viewed as a density.  We may then define a Hamiltonian density from the two
conserved quantities $w_{2}$ and $w_{3}$ as%
\begin{eqnarray}
\hspace{-1.2cm}
\mathcal{H}(q,q_{x},q_{xx}) \!\!&=&\!\!\mathcal{\alpha }w_{2}+i\beta w_{3}
\label{HH1} \\
\hspace{-1.2cm}
\!\!&=&\!\!\resizebox{.7\hsize}{!}{$\mathcal{\alpha }\left( \left\vert q_{x}\right\vert ^{2}-\left\vert
q\right\vert ^{4}\right) -i\frac{\beta }{2}\left( q_{x}q_{xx}^{\ast
}-q_{x}^{\ast }q_{xx}\right) -i\frac{3\beta }{4}\left[ \left( q^{\ast
}\right) ^{2}\left( q^{2}\right) _{x}-q^{2}\left( q^{\ast }\right) _{x}^{2}%
\right]$} ,  \label{HH2}
\end{eqnarray}%
with some real constants $\alpha $, $\beta $, where we have dropped all
surface terms in (\ref{HH2}). We also included an $i$ in front of the $w_{3}$-term to
ensure the overall $\mathcal{PT}$-symmetry of $\mathcal{H}$, which prompts
us to view the Hirota equation as a $\mathcal{PT}$-symmetric extension of
the NLSE. This form will ensure the reality of the total energy of the
system, defined by $E(q):=\int\nolimits_{-\infty }^{\infty }$ $\mathcal{H}%
(q,q_{x},q_{xx})dx$ for a particular solution. It is clear from our analysis
that the extension term needs to be of a rather special form as most terms,
even when they respect the $\mathcal{PT}$-symmetry, will destroy the
integrability of the model, see also \cite{fring_pt-symmetric_2013} for other models.

It is now easy to verify that functionally, the Hirota equation and its conjugate result
from varying the Hamiltonian $H=\int $ $\mathcal{H}dx$%
\begin{equation}
 iq_{t}\!=\frac{\delta H}{\delta q^{\ast }}=\sum\nolimits_{n=0}^{\infty
}(-1)^{n}\frac{d^{n}}{dx^{n}}\frac{\partial \mathcal{H}}{\partial
	q_{nx}^{\ast }},\quad iq_{t}^{\ast }\! =-\frac{\delta H}{\delta q}%
=\sum\nolimits_{n=0}^{\infty }(-1)^{n}\frac{d^{n}}{dx^{n}}\frac{\partial 
	\mathcal{H}}{\partial q_{nx}},
\end{equation}%
with Hamiltonian density (\ref{HH2}).%

At this point, noting that (\ref{2.1251}-\ref{2.1253}) for $r=-q^{\ast}$ are solutions to the ZC representation (\ref{hezc1}-\ref{hezc12}) if and only if the Hirota
equation holds. They serve to compute the function $\chi $ occurring
in the local conservation law (\ref{LC}).

\subsection{Real charges from complex solutions}

Let us now verify that all the charges resulting from the densities in (\ref%
{wn}) are real. Defining the charges as the integrals of the charge densities, that is%
\begin{equation}
Q_{n}=\dint\nolimits_{-\infty }^{\infty }w_{n}dx,  \label{Qn}
\end{equation}%
we expect from the $\mathcal{PT}$-symmetry behaviour $\mathcal{PT}%
(w_{n})=(-1)^{n}w_{n}$ that $Q_{2n}\in \mathbb{R}$ and $Q_{2n+1}\in i\mathbb{%
	R}$. Taking now $q_{1}$ to be in the form (\ref{q1}) and shifting $%
x\rightarrow x+x^{\delta, \xi}_{+}$ in (\ref{Qn}), we find from (\ref{wn}) that the only
contribution to the integral comes from the iteration of the first term,
that is%
\begin{equation}
Q_{n}=\dint\nolimits_{-\infty }^{\infty }\left( \frac{q_{x}}{q}\right)
^{n}w_{0}dx.  \label{Qnn}
\end{equation}%
It is clear that the second term in (\ref{wn}), $(w_{n-1})_{x}$, does not
contribute to the integral as it is a surface term. Less obvious is the
cancellation of the remaining terms, which can however be verified easily on a case-by-case basis.
For the one-soliton solution (\ref{q1}) the charges (\ref{Qnn}) become%
\begin{eqnarray}
Q_{n} &=&-\delta ^{2}\dint\nolimits_{-\infty }^{\infty }\left[ i\xi -\delta
\tanh (x\delta )\right] ^{n}\func{sech}^{2}(x\delta )dx \\
&=&-\left\vert \delta \right\vert \dint\nolimits_{-1}^{1}\left( i\xi -\delta
u\right) ^{n}du \\
&=&-\left\vert \delta \right\vert \dsum\nolimits_{k=0}^{n}\frac{n!}{%
	(k+1)!(n-k)!}\delta ^{k}(i\xi )^{n-k}\left[ 1+(-1)^{k}\right] .  \label{sum}
\end{eqnarray}%
Since only the terms with even $k$ contribute to the sum in (\ref{sum}), it
is evident from this expression that $Q_{2n}\in \mathbb{R}$ and $Q_{2n+1}\in
i\mathbb{R}$.

Of special interest is the energy of the system resulting from the
Hamiltonian (\ref{HH1}). For the one-soliton solution (\ref{q1}) we obtain 
\begin{equation}
E(q_{1}^{\mu })=\mathcal{\alpha }Q_{2}+i\beta Q_{3}=2\left\vert \delta
\right\vert \left[ \alpha \left( \xi ^{2}-\frac{\delta ^{2}}{3}\right)
+\beta \xi \left( \delta ^{2}-\xi ^{2}\right) \right] .
\end{equation}%
The energy is real and hence we can once again confirm the theory that $\mathcal{PT}$-symmetry guarantees reality despite being computed from a complex field.

The energy of the two-soliton solution (\ref{dtwos}) is computed to 
\begin{equation}
E(q_{2}^{\mu ,\mu })=2E(q_{1}^{\mu }).  \label{e1}
\end{equation}%
The doubling of the energy for the degenerate solution in (\ref{dtwos}) when
compared to the one-soliton solution is of course what we expect from the
fact that the model is integrable and the computation constitutes therefore
an indirect consistency check. We expect (\ref{e1}) to generalize to $%
E(q_{3}^{n\mu })=nE(q_{1}^{\mu })\,$, which we verified numerically for $n=3$
using the solution (\ref{nsolution}).

\section{Asymptotic properties of degenerate multi-soliton solutions}

Next we compute the asymptotic displacement in the scattering process in a
similar fashion as discussed in the previous chapter. The analysis relies on computing the
asymptotic limits of the multi-soliton solutions and comparing the results
with the tracked one-soliton solution. As a distinct point we track the
maxima of the one-soliton solution (\ref{q1mu}) within the two-soliton
solution. Similarly to the one-soliton, the real and imaginary parts of the
two-soliton solution depend on the function $A(x,t)$, as defined in (\ref{Ax}%
), occurring in the argument of the $\sin $ and $\cos $ functions. This
makes it impossible to track a distinct point with constant amplitude.
However, as different values for $A$ only produce an internal oscillation we
can fix $A$ to any constant value without affecting the overall speed.

We start with the calculation for the degenerate two-soliton solution and
illustrate the above behaviour in Figure \ref{Fig6.1} for a concrete choice of 
$A$.

\begin{figure}[h]
	\centering
	
	\includegraphics[width=0.48\linewidth,height=0.33\linewidth]{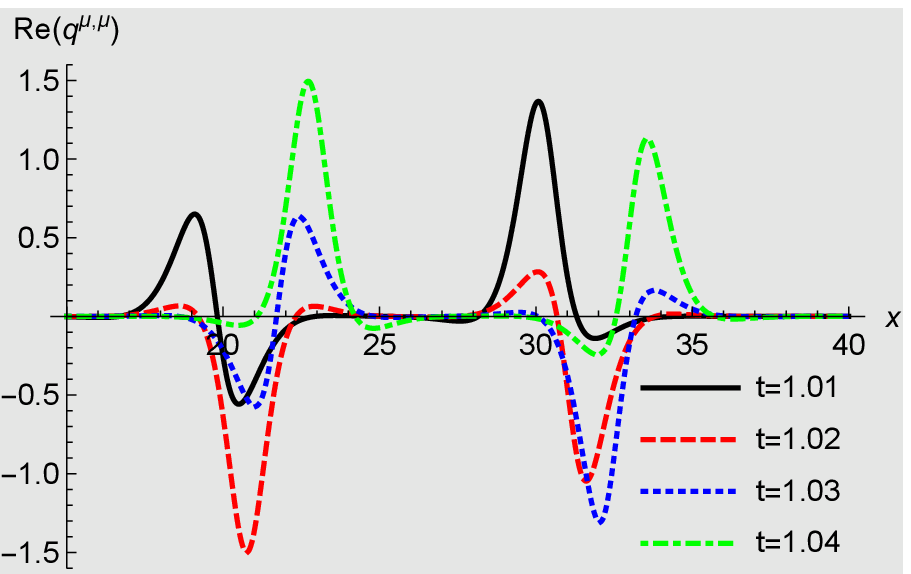} 
	\hspace{0cm}
	\includegraphics[width=0.48\linewidth]{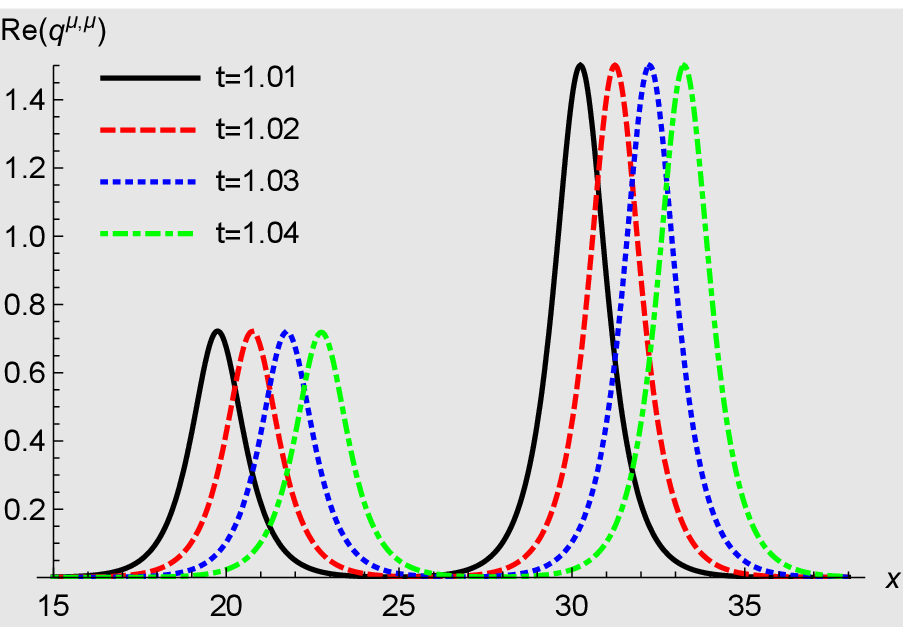}\\
	
	\caption{Real part of the degenerate two-soliton solution (\ref{dtwos}) for the Hirota equation at small values of times for $\alpha = 1$, $\beta=2$, 
		$\delta=3/2$, $\xi=1$ with generic $A(x,t)$ in the left panel and fixed $A=\pi/3$ in the right panel. For large values of time, the soliton constituents will reach the same heights.} \label{Fig6.1}
\end{figure}

The functions with constant values of $A$ can be seen as enveloping
functions similar to those employed for the computation of displacements in
breather functions, see e.g. \cite{cen_degenerate_2017}. Thus with $A(x,t)=A$ taken to
be constant we calculate the four limits 
\begin{eqnarray*}
	\hspace{-1cm}
\resizebox{.3\hsize}{!}{$	\lim_{t\rightarrow \pm \infty }\!q_{2}^{\mu ,\mu }(x_{+}^{\delta ,\xi
	}+\Delta X (t),t)$} \!\!&=&\!\!\resizebox{.6\hsize}{!}{$\pm \frac{\beta \delta ^{2}\cos A-\delta (\alpha -3\beta
		\xi )\sin A}{\sqrt{\beta ^{2}\delta ^{2}+(\alpha -3\beta \xi )^{2}}}\pm i%
	\frac{\delta (\alpha -3\beta \xi )\cos A+\beta \delta ^{2}\sin A}{\sqrt{%
			\beta ^{2}\delta ^{2}+(\alpha -3\beta \xi )^{2}}}$} \\
		\hspace{-1cm}
\resizebox{.3\hsize}{!}{$	\lim_{t\rightarrow \pm \infty }\!q_{2}^{\mu ,\mu }(x_{-}^{\delta ,\xi
	}-\Delta X (t),t)$} \!\!&=&\!\!\resizebox{.6\hsize}{!}{$\mp \frac{\beta \delta ^{2}\cos A+\delta (\alpha -3\beta
		\xi )\sin A}{\sqrt{\beta ^{2}\delta ^{2}+(\alpha -3\beta \xi )^{2}}}\pm i%
	\frac{\delta (\alpha -3\beta \xi )\cos A-\beta \delta ^{2}\sin A}{\sqrt{%
			\beta ^{2}\delta ^{2}+(\alpha -3\beta \xi )^{2}}}$}
\end{eqnarray*}%
with time-dependent displacement 
\begin{equation}
\Delta X (t)=\frac{1}{\delta }\ln \left[ 2\delta \left\vert t\right\vert \sqrt{%
	\beta ^{2}\delta ^{2}+(\alpha -3\beta \xi )^{2}}\right] .  \label{Dt}
\end{equation}%
Using the limits from above we obtain the same asymptotic value in all four
cases for the displaced modulus of the two-soliton solution 
\begin{equation}
\lim_{t\rightarrow \pm \infty }\left\vert q_{2}^{\mu ,\mu }(x_{\pm }^{\delta
	,\xi }\pm \Delta X (t),t)\right\vert =\delta .
\end{equation}%
In the limit to the NLSE, i.e. $\beta \rightarrow 0$, our expression for $%
\Delta X (t)$ agrees precisely with the result obtained in \cite{olmedilla_multiple_1987}.

We have here two options to interpret these calculations: As the compound
two-soliton structure is entirely identical in the two limits $t\rightarrow
\pm \infty $ and its individual one-soliton constituents are
indistinguishable we may conclude that there is no overall displacement for
the individual one-soliton constituents. Alternatively we may assume that
the two one-soliton constituents have exchanged their position and thus the
overall time-dependent displacement is $\pm 2\Delta X (t)$.

For comparison we compute next the displacement for the nondegenerate
two-soliton solution (\ref{twos}) with $c=\widetilde{c}=1$ and parametrisation $%
\mu =\delta +i\xi $, $\nu =\rho +i\sigma $ where $\delta $,$\xi $,$\rho $,$%
\sigma \in \mathbb{R}$. For definiteness we take $x_{+}^{\delta ,\xi
}>x_{+}^{\rho ,\sigma }$ and calculate the asymptotic limits%
\begin{eqnarray}
\lim_{t\rightarrow +\infty }\left\vert q_{2}^{\mu ,\nu }(x_{+}^{\delta ,\xi
}+\frac{1}{\delta }\widetilde{\Delta X },t)\right\vert &=&\lim_{t\rightarrow
	-\infty }\left\vert q_{2}^{\mu ,\nu }(x_{+}^{\delta ,\xi },t)\right\vert
=\delta , \\
\lim_{t\rightarrow +\infty }\left\vert q_{2}^{\mu ,\nu }(x_{+}^{\rho ,\sigma
},t)\right\vert &=&\lim_{t\rightarrow -\infty }\left\vert q_{2}^{\mu ,\nu
}(x_{+}^{\delta ,\xi }+\frac{1}{\rho }\widetilde{\Delta X },t)\right\vert =\rho ,
\end{eqnarray}%
with constant%
\begin{equation}
\widetilde{\Delta X }=\ln \left[ \frac{(\delta +\rho )^{2}+(\xi -\sigma )^{2}}{%
	(\delta -\rho )^{2}+(\xi -\sigma )^{2}}\right] .  \label{dtilde}
\end{equation}%
Thus, while the faster one-soliton constituent with amplitude $\delta $ is
advanced by the amount $\widetilde{\Delta X }/\delta $, the slower one-soliton
constituent with amplitude $\rho $ is regressed by the amount $\widetilde{\Delta X }%
/\rho $. We compare the two-soliton solution with the two one-soliton
solutions in Figure \ref{Fig6.2}.

\begin{figure}[h]
	\centering
	
	\includegraphics[width=0.48\linewidth]{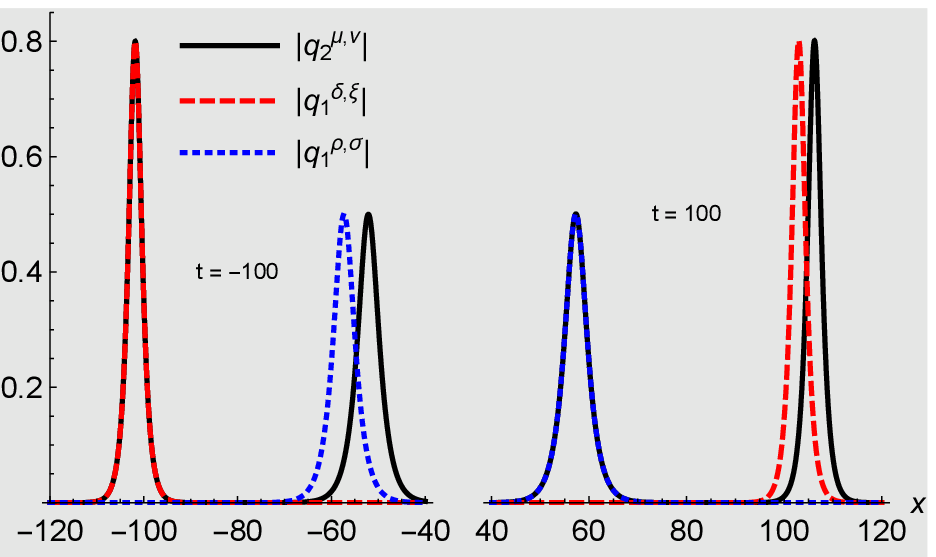} 
	\hspace{0cm}
	\includegraphics[width=0.48\linewidth]{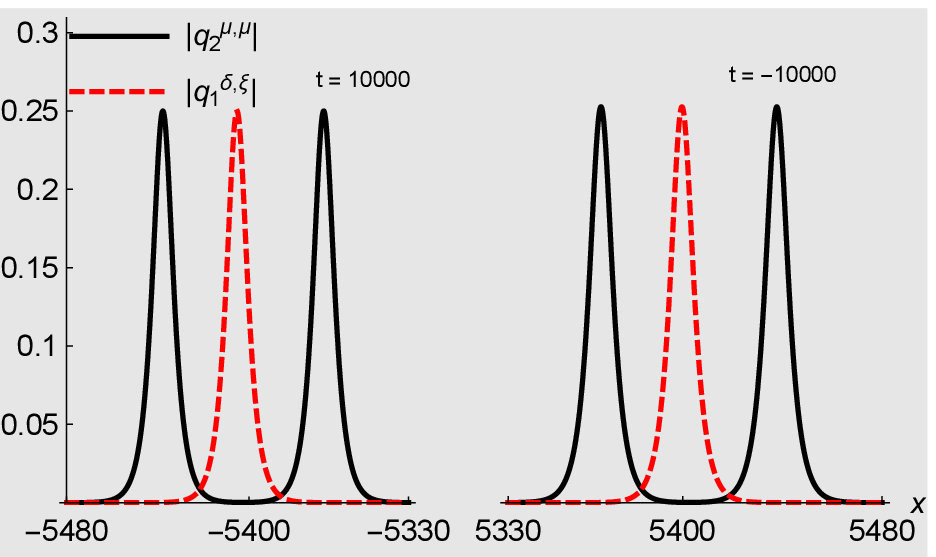}\\
	
	\caption{Nondegenerate two-soliton solution compared to two one-soliton solutions for large values of $|t|$ for $\alpha =1.1$, $\beta =0.9$, $\delta =0.8$, $\xi =0.4$, $\rho =0.5$, $\sigma =0.6$ in the left panel. Degenerate two-soliton solution compared to two one-soliton solutions for large values of $|t|$ for $\alpha =1.5$, $\beta =2.3$, 
		$\delta =0.25$, $\xi =0.6$ in the right panel.} \label{Fig6.2}
\end{figure}

We also observe that while the time-dependent displacement $\Delta X (t)$ in (%
\ref{Dt}) for the degenerate solution depends explicitly on the parameters $%
\alpha $ and $\beta $, the constant $\widetilde{\Delta X }$ in (\ref{dtilde}) is
the same for all values of $\alpha $ and $\beta $. In particular it is the
same in the Hirota equation, the NLSE and the mKdV equation. The values for $%
\alpha $ and $\beta $ only enter through $x_{+}^{\rho ,\sigma }$ in the
tracking process.

\section{Scattering properties of degenerate multi-soliton solutions}

Besides having a distinct asymptotic behaviour, the degenerate
multi-solitons also display very particular features during the actual
scattering event near $x=t=0$ when compared to the nondegenerate solutions.
For the nondegenerate two-soliton solution three distinct types of
scattering processes at the origin have been identified. Using the
terminology of \cite{anco_interaction_2011} they are {\large\bfseries \emph{merge-split}}
denoting the process of two solitons merging into one soliton and
subsequently separating while each one-soliton maintains the direction and
momentum of its trajectory, {\large\bfseries\emph{bounce-exchange}} referring to two-solitons
bouncing off each other while exchanging their momenta and {\large\bfseries\emph{absorb-emit}} characterizing the process of one soliton absorbing the other at its front
tail and emitting it at its back tail, see Figure \ref{ThreeProc}.

\begin{figure}
	\centering
	
	\includegraphics[width=0.48\linewidth]{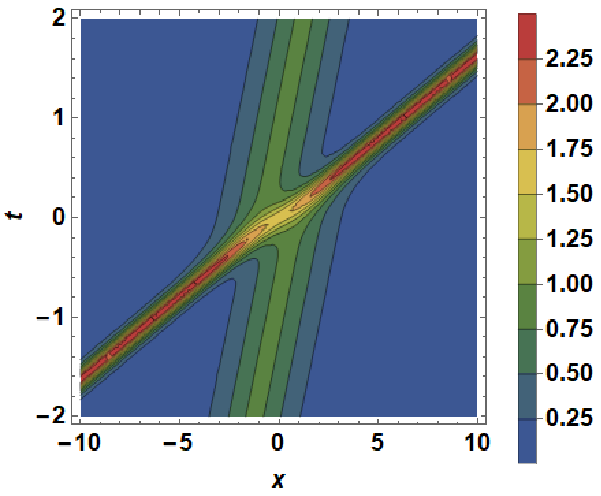}
\hspace{0cm}
	\includegraphics[width=0.48\linewidth]{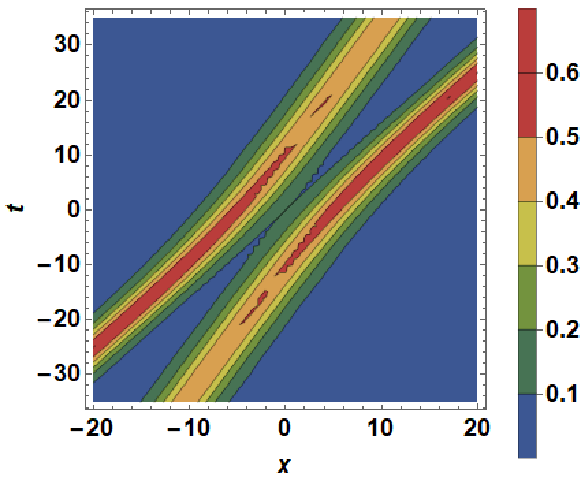}\\

	\includegraphics[width=0.48\linewidth]{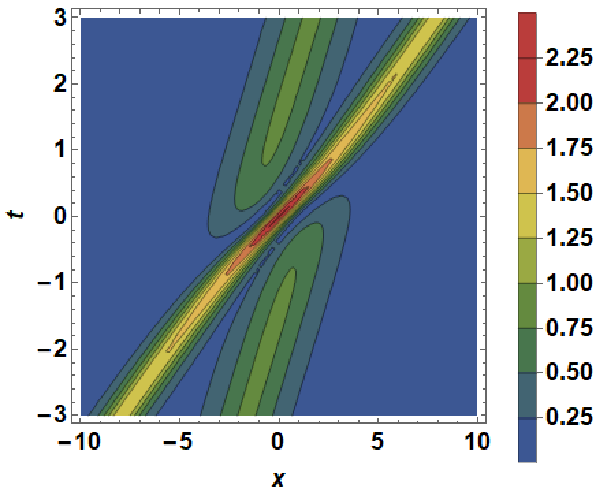}\\
	
	\caption{Different types of nondegenerate two-soliton scattering processes for the solution (\ref{twos}). Left
		panel: {\large\bfseries\emph{merge-split}} scattering with $\alpha =1.1$, $\beta =0.9$, $\rho =2.5$, 
		$\xi =0.4$, $\delta =-0.8$, $\sigma =0.6$. Middle panel: {\large\bfseries\emph{bounce-exchange}}
		scattering with $\alpha =1.1$, $\beta =0.9$, $\rho =-0.6$, $\xi =0.1$, $\delta =0.5$, $\sigma =0.2$. Right panel: {\large\bfseries\emph{absorb-emit}} scattering with $\alpha =1.1$, $\beta =0.9$, $\rho =-1.5$, $\xi =0.4$, $\delta =-0.8$, $\sigma =0.6$.} \label{ThreeProc}
\end{figure}

For the degenerate multi-soliton solutions the merge-split and
bounce-exchange scattering is not possible and only the {\large\bfseries\emph{absorb-emit}}
scattering process occurs as seen in Figure \ref{Degscat}.

\begin{figure}
	\centering
	
	\includegraphics[width=0.48\linewidth]{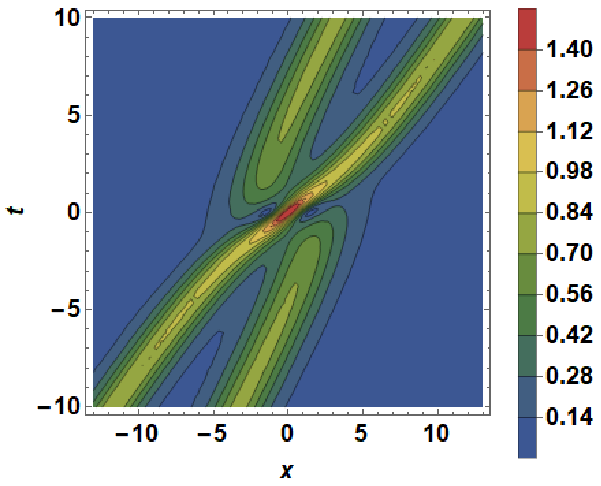} 
	\hspace{0cm}
	\includegraphics[width=0.48\linewidth]{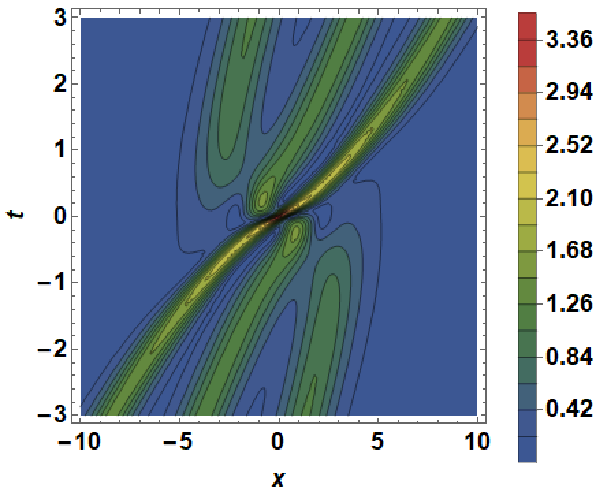}\\
	
	\caption{{\large\bfseries\emph{Absorb-emit}} scattering processes for degenerate two-solitons (\ref{dtwos}) with $\alpha =1.1$, $\beta =0.9$, $\delta =0.8$, $\xi =0.1$ (left panel) and three-solitons (\ref{nsolution}) 
		with $\alpha =1.1$, $\beta =0.9$, $\delta =0.6$, $\xi =0.4$ (right panel). } \label{Degscat}
\end{figure}

This feature is easy to understand when considering the behaviour of the
solution at $x=t=0$. As was argued in \cite{anco_interaction_2011} the
different behaviour can be classified as being either convex
downward or concave upward at $x=t=0$ together with the occurrence of additional
local maxima. For the degenerate two-soliton solution we find $\left.
\partial \left\vert q_{2}^{\mu ,\mu }(x,t)\right\vert /\partial x\right\vert
_{x=0,t=0}=0$ and $\left. \partial ^{2}\left\vert q_{2}^{\mu ,\mu
}(x,t)\right\vert /\partial x^{2}\right\vert _{x=0,t=0}=-10\left\vert \delta
\right\vert ^{3}$, which means this solution is always concave at $x=t=0$.
In addition, we find that $\func{Re}^{2}\left. q_{2}^{\mu ,\mu }(x,t)\right\vert
_{x=0,t=0}$ and $\func{Im}^{2}\left. q_{2}^{\mu ,\mu }(x,t)\right\vert _{x=0,t=0}
$ are always concave and convex at $x=t=0$, respectively. Hence, we always
have the emergence of additional local maxima, such that the behaviour must
be of the absorb-emit type. In Figure \ref{Degscat} we display this
scattering behaviour for the degenerate two and three-soliton solutions in
which the distinct features of the absorb-emit behaviour are clearly
identifiable.

We observe that the dependence on the parameters $\alpha $ and $\beta $ of
the degenerate and nondegenerate solution is now reversed when compared to
the asymptotic analysis. While the type of scattering in the nondegenerate
case is highly sensitive with regard to $\alpha $ and $\beta $, it is
entirely independent of these parameters in the degenerate case.

\section{Conclusions}
We constructed all charges resulting from the ZC representation (\ref{hezc1}) and (\ref{hezc12}) by means of a Gardner transformation, which matches the charges from the NLS equation. Furthermore, We computed a closed analytic expression
for all charges involving a particular one-soliton solution, verified for a high number of charges. Two of the charges were
used to define a Hamiltonian whose functional variation led to the Hirota
equation. The behaviour of these charges under $\mathcal{PT}$-symmetry
suggests to view the Hirota system as an integrable extended version of
NLSE. This point of view allows for confirmation of previous
arguments from Chapters 3 to 5 that guarantee the reality of the
energy to all higher order charges. 

Explicit multi-soliton solutions from HDM as well as the
DCT were derived and we showed how to construct
degenerate solutions in both schemes. As observed previously, the
application of HDM relies on choosing the arbitrary
constants in the solutions in a very particular way. When using DCT the degenerate solutions are obtained by replacing standard
solutions in the underlying auxiliary eigenvalue problem by Jordan states.

From the asymptotic behaviour of the degenerate two-soliton solution we
computed the new expression for the time-dependent displacement. As the
degenerate one-soliton constituents in the multi-soliton solutions are
asymptotically indistinguishable one cannot decide whether the two
one-solitons have actually exchanged their position and therefore the
time-dependent displacement can be interpreted as an advance or delay or
whether the two one-solitons have only approached each other and then
separated again. The analysis of the actual scattering event allows for both
views.

We showed that degenerate two-solitons may only scatter via an absorb-emit
process, that is by one soliton absorbing the other at its front tail and
subsequently emitting it at the back tail. Since the model is integrable all
multi-particle/soliton scattering processes may be understood as consecutive
two particle/soliton scattering events, so that the two-soliton scattering
behaviour (absorb-emit) extends to the multi-soliton scattering as we demonstrated.

\chapter{New integrable nonlocal Hirota equations}\label{ch_7}

When we compare the Hirota equation (\ref{HE}) with the NLS equation, (\ref{HE}) with $\beta=0$,
we notice that the additional term in the Hirota equation shares the same $\mathcal{PT}$-symmetry with the NLS equation, as it is invariant with respect to $\mathcal{PT}:x\rightarrow -x$, $t\rightarrow -t$, $i\rightarrow -i$, $q\rightarrow q$,
where $\mathcal{P}:x\rightarrow -x$ and $\mathcal{T}:$ $t\rightarrow -t$, $%
i\rightarrow -i$. Hence the Hirota equation may also be viewed as a $\mathcal{PT}$-symmetric extension of the NLS equation. Similarly as for many other $\mathcal{PT}$-symmetric nonlinear integrable systems \cite{fring_pt-symmetric_2013}, various other $\mathcal{PT}$-symmetric generalizations have been proposed and investigated
by adding terms to the original equation, e.g. \cite{abdullaev_solitons_2011,alexeeva_optical_2012,konotop_nonlinear_2016}. 

 A further option, that will be important here, was explored by Ablowitz and Musslimani \cite{ablowitz_integrable_2013,ablowitz_integrable_2016} who identified a new class of nonlinear integrable systems by exploiting various versions of $\mathcal{PT}$-symmetry present in the ZC condition/AKNS equations that relates fields in the theory to each other in a
nonlocal fashion. One particular type of these new systems that has attracted a lot of attention is the nonlocal NLS equation  \cite%
{khare_periodic_2015,li_dark_2015,valchev_mikhailovs_2015,fokas_integrable_2016,wen_dynamics_2016,caudrelier_interplay_2017,gerdjikov_complete_2017}.

Exploring this option below for the Hirota equation will lead us to new integrable systems with nonlocal properties \cite{cen_integrable_2019}.

\section{Zero-curvature and AKNS equations for the nonlocal Hirota equations}

As discussed in the introduction, the classical integrability of a model can be established by the explicit construction of its Lax pair \cite{lax_integrals_1968}, which is equivalent to the closely related ZC condition or AKNS equations \cite{ablowitz_nonlinear-evolution_1973} and constitutes a starting point for an explicit solution procedure. The reformulations of the equation of motion of the model in terms of the ZC condition allows for the construction of infinitely many conserved charges, which is roughly speaking synonymous with the model being classically integrable. We explore various symmetries in this reformulation that will lead us to new types of models exhibiting novel features.

In Section 2.5.3, we saw the ZC or AKNS equations for the Hirota equation. Next, one needs to make sure that these two equations are in fact compatible. Adapting now from \cite{ablowitz_integrable_2013,ablowitz_integrable_2016} the general idea that has
been applied to the NLSE to the current setting we explore various choices
and alter the $x,t$-dependence in the functions $r$ and $q$. For convenience
we suppress the explicit functional dependence and absorb it instead into
the function's name by introducing the abbreviations%
\begin{equation}
q:=q(x,t)\text{,\quad }\widetilde{q}:=q(-x,t)\text{,\quad }\widehat{q}:=q(x,-t)\text{%
	,\quad }\widecheck{q}:=q(-x,-t).\text{\quad }
\end{equation}%
All six choices for $r(x,t)$ being equal to $\widetilde{q}$, $\widehat{q}$, $\widecheck{q%
}$ or their complex conjugates $\widetilde{q}^{\ast }$, $\widehat{q}^{\ast }$, $%
\widecheck{q}^{\ast }$ together with some specific adjustments for the constants 
$\alpha $ and $\beta $ are consistent for the two AKNS equations, thus giving rise to six new types of integrable models
that have not been explored so far. We will first list them and then study
their properties, in particular their solutions, in the next chapters.

\vspace{0.5cm}

\noindent {\large \bfseries{{ The Hirota equation, a
	conjugate pair, $\mathbf{r(x,t)=\protect\kappa q^{\ast }(x,t)}$:}}}

\noindent The standard choice to achieve compatibility
between the two AKNS equations (\ref{2.126}-\ref{2.127}) is to take $r(x,t)=\kappa q^{\ast
}(x,t)$ with $\kappa =\pm 1$, such that
the equations acquire the forms%
\begin{eqnarray}
iq_{t} &=&\!-\alpha \left( q_{xx}-2\kappa \left\vert q\right\vert
^{2}q\right) \!-i\!\beta \!\left( q_{xxx}-6\kappa \left\vert q\right\vert
^{2}q_{x}\right) ,  \label{1} \\
-iq_{t}^{\ast } &=&-\alpha \left( q_{xx}^{\ast }-2\kappa \left\vert
q\right\vert ^{2}q^{\ast }\right) \!+i\!\beta \!\left( q_{xxx}^{\ast
}-6\kappa \left\vert q\right\vert ^{2}q_{x}^{\ast }\right) .\quad
\;\;\;\;\;\,  \label{2}
\end{eqnarray}%
Equation (\ref{1}) is the known Hirota 'local' equation. For $\alpha ,\beta \in \mathbb{R}$ equation (\ref{2}) is its
complex conjugate, respectively, i.e. (\ref{2})$^{\ast }=$(\ref{1}). When $%
\beta \rightarrow 0$ equation (\ref{1}) reduces to the NLS equation with conjugate (%
\ref{2}) and for $\alpha \rightarrow 0$ equation (\ref{1}) reduces to the
complex mKdV with conjugate (\ref{2}). The
aforementioned $\mathcal{PT}$-symmetry is preserved in these equations.

\vspace{0.5cm}

\noindent {\large \bfseries{{A parity transformed
	conjugate pair, $\mathbf{r(x,t)=\protect\kappa q^{\ast }(-x,t)}$:}}}

\noindent Taking now $r(x,t)=\kappa \widetilde{q}^{\ast }$ with $%
\kappa = \pm 1$ together with $\beta =i\delta $, $\alpha ,\delta \in 
\mathbb{R}$, the AKNS equations become 
\begin{eqnarray}
\!iq_{t}\! &=&\!-\alpha \left[ q_{xx}-2\kappa \widetilde{q}^{\ast }q^{2}\right]
\!+\delta \!\left[ q_{xxx}-6\kappa q\widetilde{q}^{\ast }q_{x}\right] ,
\label{new1} \\
-\!i\widetilde{q}_{t}^{\ast } &=&-\!\alpha \left[ \widetilde{q}_{xx}^{\ast }-2\kappa
q(\widetilde{q}^{\ast })^{2}\right] \!-\delta \!\left( \widetilde{q}_{xxx}^{\ast
}-6\kappa \widetilde{q}^{\ast }q\widetilde{q}_{x}^{\ast }\right) .\quad 
\label{new2}
\end{eqnarray}%
We observe that equation (\ref{new1}) is the parity transformed conjugate of
equation (\ref{new2}), i.e. $\mathcal{P}$(\ref{new1})$^{\ast }=$(\ref{new2}%
). We also notice that a consequence of the introduction of the nonlocality
is that the aforementioned $\mathcal{PT}$-symmetry has been broken.

\vspace{0.5cm}

\noindent {\large \bfseries{{ A time-reversed pair, $%
	\mathbf{r(x,t)=\protect\kappa q^{\ast }(x,-t)}$:}}}

\noindent Choosing $r(x,t)=\kappa \widehat{q}^{\ast }$ with $%
\kappa = \pm 1$ and $\alpha =i\widehat{\delta}$, $\beta =i\delta $, $\widehat{%
	\delta},\delta \in \mathbb{R}$ we obtain from AKNS equations the pair 
\begin{eqnarray}
\!iq_{t}\! &=&\!-i\widehat{\delta}\left[ q_{xx}-2\kappa \widehat{q}^{\ast }q^{2}%
\right] \!+\delta \!\left[ q_{xxx}-6\kappa q\widehat{q}^{\ast }q_{x}\right] ,
\label{m1} \\
\!i\widehat{q}_{t}^{\ast } &=&\!i\widehat{\delta}\left[ \widehat{q}_{xx}^{\ast }-2\kappa
q(\widehat{q}^{\ast })^{2}\right] \!+\delta \!\left( \widehat{q}_{xxx}^{\ast
}-6\kappa \widehat{q}^{\ast }q\widehat{q}_{x}^{\ast }\right) .\quad  \label{m27}
\end{eqnarray}%
Recalling here that the time-reversal map includes a conjugation, such that $%
\mathcal{T}:q\rightarrow \widehat{q}^{\ast },i\rightarrow -i$, we observe that (%
\ref{m1}) is the time-reversed of equations (\ref{m27}), i.e. $\mathcal{T}$(%
\ref{m27})$=$(\ref{m1}). The $\mathcal{PT}$-symmetry is also broken in this
case.

\vspace{0.5cm}

\noindent {\large \bfseries{{A $\mathcal{PT}$-symmetric
	pair, $\mathbf{r(x,t)=\protect\kappa q^{\ast }(-x,-t)}$: }}}

\noindent For the choice $r(x,t)=\kappa \widecheck{q}^{\ast }$
with $%
\kappa = \pm 1$ and $\alpha =i\widecheck{\delta}$, $\widecheck{\delta}%
,\beta \in \mathbb{R}$ the AKNS equations become 
\begin{eqnarray}
\!q_{t}\! &=&\!-\widecheck{\delta}\left[ q_{xx}-2\kappa \widecheck{q}^{\ast }q^{2}%
\right] \!-\beta \!\left[ q_{xxx}-6\kappa q\widecheck{q}^{\ast }q_{x}\right] ,
\label{pt1} \\
\!-\widecheck{q}_{t}^{\ast } &=&-\!\widecheck{\delta}\left[ \widecheck{q}_{xx}^{\ast
}-2\kappa q(\widecheck{q}^{\ast })^{2}\right] \!+\beta \!\left( \widecheck{q}%
_{xxx}^{\ast }-6\kappa \widecheck{q}^{\ast }q\widecheck{q}_{x}^{\ast }\right) .\quad
 \label{pt2}
\end{eqnarray}%
We observe that the overall constant $i$ has cancelled out and the two
equations are transformed into each other by means of a $\mathcal{PT}$%
-symmetry transformation $\mathcal{PT}$(\ref{pt2})$=$(\ref{pt1}). Thus,
while the $\mathcal{PT}$-symmetry for the equations (\ref{pt1}) is broken,
the two equations are transformed into each other by that symmetry.

\vspace{0.5cm}

\noindent {\large \bfseries{{ A real parity transformed
	conjugate pair, $\mathbf{r(x,t)=\protect\kappa q(-x,t)}$:}}}

\noindent We may also choose $q(x,t)$ to be real. For $%
r(x,t)=\kappa \widetilde{q}$ with $
\kappa = \pm 1$, $\widetilde{q}\in \mathbb{R}$ and $\beta
=i\delta $, \,\,$\alpha ,\delta \in \mathbb{R}$, the AKNS equations acquire the forms 
\begin{eqnarray}
\!iq_{t}\! &=&\!-\alpha \left[ q_{xx}-2\kappa \widetilde{q}q^{2}\right]
\!+\delta \!\left[ q_{xxx}-6\kappa q\widetilde{q}q_{x}\right] ,  \label{pr1} \\
\!-i\widetilde{q}_{t} &=&\!-\alpha \left[ \widetilde{q}_{xx}-2\kappa q\widetilde{q}^{2}%
\right] \!-\delta \!\left( \widetilde{q}_{xxx}-6\kappa \widetilde{q}q\widetilde{q}%
_{x}\right) .\quad   \label{pr2}
\end{eqnarray}%
The
equations (\ref{pr2}) and (\ref{pr1}) are related to each other by
conjugation and a parity transformation (\ref{new2}), i.e. $\mathcal{P}$(\ref%
{pr2})$^{\ast }=$(\ref{pr1}). However, the restriction to real values for $%
q(x,t)$ makes these equations less interesting as $q$ becomes static, which
simply follows from the fact that the left hand sides of (\ref{pr1}) and (%
\ref{pr2}) are complex valued, whereas the right hand sides are real valued.

\vspace{0.5cm}

\noindent {\large \bfseries{{A real time-reversed pair, $%
	\mathbf{r(x,t)=\protect\kappa q(x,-t)}$:}}}

\noindent For $r(x,t)=\kappa \widehat{q}$ with $%
\kappa = \pm 1$, $\widehat{q}%
\in \mathbb{R}$ and $\alpha =i\widehat{\delta}$, $\beta =i\delta $,\,\, $\widehat{\delta}%
,\delta \in \mathbb{R}$ we obtain from the AKNS equations 
\begin{eqnarray}
\!iq_{t}\! &=&\!-i\widehat{\delta}\left[ q_{xx}-2\kappa \widehat{q}^{\ast }q^{2}%
\right] \!+\delta \!\left[ q_{xxx}-6\kappa q\widehat{q}^{\ast }q_{x}\right] ,
\label{T1} \\
\!\!i\widehat{q}_{t}^{\ast } &=&\!i\widehat{\delta}\left[ \widehat{q}_{xx}^{\ast
}-2\kappa q(\widehat{q}^{\ast })^{2}\right] \!+\delta \!\left( \widehat{q}%
_{xxx}^{\ast }-6\kappa \widehat{q}^{\ast }q\widehat{q}_{x}^{\ast }\right) .\quad 
\label{T2}
\end{eqnarray}%
Again we observe the same behaviour as in the complex variant, namely that
the two equations (\ref{T1}) and (\ref{T2}) become their time-reversed
counterparts, i.e. $\mathcal{T}$(\ref{T2})$=$(\ref{T1}) and vice versa.

\vspace{0.5cm}

\noindent {\large \bfseries{{ A conjugate $\mathcal{PT}$%
	-symmetric pair, $\mathbf{r(x,t)=\protect\kappa q(-x,-t)}$:}}}

\noindent For our final choice $r(x,t)=\kappa \widecheck{q}$ with $%
\kappa = \pm 1$, we
have $\alpha ,\beta \in 
\mathbb{C}$, i.e. no restriction on the constants and the AKNS equations become 
\begin{eqnarray}
\!q_{t}\! &=&\!i\alpha \left[ q_{xx}-2\kappa \widecheck{q}q^{2}\right] \!-\beta
\!\left[ q_{xxx}-6\kappa q\widecheck{q}q_{x}\right] ,  \label{F1} \\
\!-\widecheck{q}_{t} &=&\!i\alpha \left[ \widecheck{q}_{xx}-2\kappa q\widecheck{q}^{2}%
\right] \!+\beta \!\left( \widecheck{q}_{xxx}-6\kappa \widecheck{q}q\widecheck{q}%
_{x}\right) .\quad  \label{F2}
\end{eqnarray}%
These two equations are transformed into each other by means of a $\mathcal{%
	PT}$-symmetry transformation and a conjugation $\mathcal{PT}$(\ref{F2})$%
^{\ast }=$(\ref{F1}). A comment is in order here to avoid confusion. Since a
conjugation is included into the $\mathcal{T}$-operator, the additional
conjugation of (\ref{F1}) when transformed into (\ref{F2}) means that we
simply carry out $x\rightarrow -x$ and $t\rightarrow -t$. \ 

The paired up equations (\ref{m1})-(\ref{F2}) are all new integrable nonlocal systems and we summarise the cases with the table below.

\vspace{0.5cm}

\begin{tabular} { | c | c | c | }
				
				\specialrule{1.5pt}{0pt}{0pt}
				\begin{minipage}{1.6in} \vspace{0.1in} 	
					\begin{center}
						$\mathcal{P}$arity transformed conjugate pair
					\end{center} \vspace{0.01in}
				\end{minipage}
				
				&\begin{minipage}{1.6in} \vspace{0.1in}	
					\begin{center}
						$\mathcal{T}$ime-reversed pair
					\end{center} \vspace{0.01in}
				\end{minipage}
				
				&\begin{minipage}{1.6in} \vspace{0.1in}
					\begin{center}
						$\mathcal{PT}$-symmetric pair
					\end{center} \vspace{0.01in}
				 \end{minipage}\\ 
				
				\hline
				$r=\pm q^{*}(-x,t)$ & 
				$r= \pm q^{*}(x,-t)$ & 
				$r=\pm q^{*}(-x,-t)$ \\ 
				
				$\footnotesize{\alpha \in \mathbb{R}, \beta \in i \mathbb{R}}$ & 
				$\footnotesize{\alpha \in i \mathbb{R}, \beta \in i \mathbb{R}}$ & 
				$\footnotesize{\alpha \in i \mathbb{R}, \beta \in \mathbb{R}}$\\
				
				\specialrule{1.5pt}{0pt}{0pt}
				
				\begin{minipage}{1.6in} \vspace{0.1in}
						\begin{center} $\mathcal{P}$arity transformed   conjugate real pair\end{center} \vspace{0.01in} \end{minipage} & 
				
				\begin{minipage}{1.6in} \vspace{0.1in} 
					 	\begin{center} $\mathcal{T}$ime-reversed   real pair\end{center} \vspace{0.01in} \end{minipage} & 
				
				\begin{minipage}{1.6in} \vspace{0.1in} 	\begin{center}$\mathcal{PT}$-symmetric  conjugate pair\end{center} \vspace{0.01in} \end{minipage}\\
				
				\hline
				$r=\pm q(-x,t)$ &
				$r=\pm q(x,-t)$ & 
				$r=\pm q(-x,-t)$\\[5pt]
				
				$\footnotesize{\alpha \in \mathbb{R}, \beta \in i \mathbb{R}}$ &
				$\footnotesize{\alpha \in i \mathbb{R}, \beta \in i \mathbb{R}}$ & $\footnotesize{\alpha \in \mathbb{C}, \beta \in \mathbb{C}}$\\
				\specialrule{1.5pt}{0pt}{0pt}	
				
		\end{tabular}
	
	\vspace{0.5cm}

Let us now discuss solutions and properties of these equations. Since the
two equations in each pair are related to each other by a well identified
symmetry transformation involving combinations of conjugation, reflections
in space and reversal in time, it suffices to focus on just one of the
equations.

\section{The nonlocal complex parity transformed Hirota equation}

In this case the compatibility between the AKNS equations is achieved by the choice $r(x,t)=\kappa q^{\ast }(-x,t)$. As $x$
is now directly related to $-x$, we expect some nonlocality in space to
emerge in this model.

\subsection{Soliton solutions from Hirota's direct method}

Let us now consider the new nonlocal integrable equation (\ref{new1}) for $%
\kappa =-1$. We factorize again $q(x,t)=g(x,t)/f(x,t)$, but unlike in the
local case we no longer assume $f(x,t)$ to be real but allow $%
g(x,t),f(x,t)\in \mathbb{C}$. We then find the identity 
\begin{eqnarray}
\hspace{-1cm}
&&f^{3}\widetilde{f}^{\ast }\left[ iq_{t}\!+\alpha q_{xx}+2\alpha \widetilde{q}%
^{\ast }q^{2}-\delta \!\left( q_{xxx}+6q\widetilde{q}^{\ast }q_{x}\right) \right]
= \\
\hspace{-1cm}
&&\resizebox{.8\hsize}{!}{$f\widetilde{f}^{\ast }\left[ iD_{t}g\cdot f+\alpha D_{x}^{2}g\cdot f-\delta
D_{x}^{3}g\cdot f\right] +\left( \widetilde{f}^{\ast }D_{x}^{2}f\cdot f-2fg%
\widetilde{g}^{\ast }\right) \left( \frac{3\delta }{f}D_{x}g\cdot f-\alpha
g\right)$} .  \notag
\end{eqnarray}%
When comparing with the corresponding identity in the local case (\ref{BINLSE}), we notice that this equation is of higher degree in the functions
involved, in this case $g$,$\widetilde{g}^{\ast }$,$f$,$\widetilde{f}^{\ast }$,
having increased from three to four. The left hand side vanishes when the
local Hirota equation (\ref{new1}) holds and the right hand side vanishes
when demanding%
\begin{equation}
iD_{t}g\cdot f+\alpha D_{x}^{2}g\cdot f-\delta D_{x}^{3}g\cdot f=0,
\label{HI1}
\end{equation}%
together with 
\begin{equation}
\widetilde{f}^{\ast }D_{x}^{2}f\cdot f=2fg\widetilde{g}^{\ast }.  \label{43}
\end{equation}%
We notice that equation (\ref{43}) is still trilinear. However, it may be
bilinearised by introducing the auxiliary function $h(x,t)$ and requiring
the two equations%
\begin{equation}
D_{x}^{2}f\cdot f=hg,\qquad \text{and\qquad }2f\widetilde{g}^{\ast }=h\widetilde{f}%
^{\ast },  \label{HI2}
\end{equation}%
to be satisfied separately. In this way we have obtained a set of three
bilinear equations (\ref{HI1}) and (\ref{HI2}) instead of two. These
equations may be solved systematically by using an additional formal power series expansion 
\begin{equation}
h(x,t)=\dsum\nolimits_{k}\varepsilon ^{k}h_{k}(x,t).
\end{equation}%
For vanishing deformation parameter $\delta \rightarrow 0$ the equations (%
\ref{HI1}) and (\ref{HI2}) constitute the bilinearisation for the nonlocal
NLSE. As our equations differ from the ones recently proposed for that model
in \cite{stalin_nonstandard_2017} we will comment below on some solutions related to that
specific case. The local equations are
obtained for $\widetilde{f}^{\ast }\rightarrow f$, $\widetilde{g}\rightarrow g$, $%
h\rightarrow g^{\ast }$ as in this case the two equations in (\ref{HI2})
combine into the one equation (\ref{bi2}).

\vspace{0.5cm}

\noindent {\large \bfseries{{{Two types of one-soliton solutions}}}}

Let us now solve the bilinear equations (\ref{HI1}) and (\ref{HI2}). First
we construct the one-soliton solutions. Unlike the local case we have
here several options, obtaining different types. Using the truncated
expansions%
\begin{equation}
f=1+\varepsilon ^{2}f_{2},\qquad g=\varepsilon g_{1},\qquad h=\varepsilon
h_{1},  \label{ans}
\end{equation}%
we derive from the three bilinear forms in (\ref{HI1}) and (\ref{HI2}) the
constraining equations 
\begin{eqnarray}
0 &=&\resizebox{.75\hsize}{!}{$\varepsilon \left[ i\left( g_{1}\right) _{t}+\alpha \left( g_{1}\right)
_{xx}-\delta (g_{1})_{xxx}\right] +\varepsilon ^{3}\left[ 2\left(
f_{2}\right) _{x}\left( g_{1}\right) _{x}-g_{1}\left[ \left( f_{2}\right)
_{xx}+i\left( f_{2}\right) _{t}\right] \right. $}   \label{c1} \\
&&\left. +if_{2}\left[ \left( g_{1}\right) _{t}+i\left( g_{1}\right) _{xx}%
\right] \right] ,  \notag \\
0 &=&\varepsilon ^{2}\left[ 2(f_{2})_{xx}-g_{1}h_{1}\right] +\varepsilon ^{4}%
\left[ 2f_{2}(f_{2})_{xx}-2(f_{2})_{x}^{2}\right] ,  \label{c2} \\
0 &=&\varepsilon \left[ 2\widetilde{g}_{1}^{\ast }-h_{1}\right] +\varepsilon ^{3}%
\left[ 2f_{2}\widetilde{g}_{1}^{\ast }-\widetilde{f}_{2}^{\ast }h_{1}\right] .
\label{c3}
\end{eqnarray}%
At this point we pursue two different options. At first we follow the
standard Hirota procedure and assume that each coefficient for the powers in 
$\varepsilon $ in (\ref{c1})-(\ref{c3}) vanishes separately. We then easily
solve the resulting six equations by%
\begin{equation}
g_{1}=\lambda \tau _{\mu ,\gamma },\qquad f_{2}=\frac{\left\vert \lambda
	\right\vert ^{2}}{(\mu -\mu ^{\ast })^{2}}\tau _{\mu ,\gamma }\widetilde{\tau}%
_{\mu ,\gamma }^{\ast },\qquad h_{1}=2\lambda ^{\ast }\widetilde{\tau}_{\mu
	,\gamma }^{\ast },
\end{equation}%
with constants $\gamma $, $\lambda $, $\mu \in \mathbb{C}$. Setting then $%
\varepsilon =1$ we obtain the exact one-soliton solution 
\begin{equation}
q_{\text{st}}^{(1)}=\frac{\lambda (\mu -\mu ^{\ast })^{2}\tau _{\mu ,\gamma }%
}{(\mu -\mu ^{\ast })^{2}+\left\vert \lambda \right\vert ^{2}\tau _{\mu
		,\gamma }\widetilde{\tau}_{\mu ,\gamma }^{\ast }}.\quad  \label{sol2}
\end{equation}%
where
\begin{equation}
	\tau_{\mu\gamma}=e^{\mu x+i\mu^{2}(\alpha-\delta \mu)t+\gamma}.
\end{equation}
Next we only demand that the coefficients in (\ref{c1})-(\ref{c2}) vanish
separately, but deviate from the standard approach by requiring (\ref{c3})
only to hold for $\varepsilon =1$. This is of course a new option that was
not at our disposal for the standard local Hirota equation, since in that
case the third equation did not exist. In this setting we obtain the
solution 
\begin{equation}
g_{1}=(\mu +\nu )\tau _{\mu ,i\gamma },\qquad f_{2}=\tau _{\mu ,i\gamma }%
\widetilde{\tau}_{-\nu ,-i\theta }^{\ast },\qquad h_{1}=2(\mu +\nu )\widetilde{\tau}%
_{-\nu ,-i\theta }^{\ast },
\end{equation}%
so that this one-soliton solution becomes%
\begin{equation}
q_{\text{nonst}}^{(1)}=\frac{(\mu +\nu )\tau _{\mu ,i\gamma }}{1+\tau _{\mu
		,i\gamma }\widetilde{\tau}_{-\nu ,-i\theta }^{\ast }}.  \label{sol1}
\end{equation}

The standard solution (\ref{sol2}) and the nonstandard solution (\ref{sol1})
exhibit qualitatively different behaviour. Whereas $q_{\text{st}}^{(1)}$
depends on one complex spectral and one complex shift parameter, $q_{\text{%
		nonst}}^{(1)}$ depends on two real spectral parameters and two real shift
parameters. Hence the solutions cannot be converted into each other. Taking
in (\ref{sol2}) for simplicity $\lambda =\mu -\mu ^{\ast }$ the modulus
squared of this solution becomes 
\begin{equation}
\resizebox{.8\hsize}{!}{$\left\vert q_{\text{st}}^{(1)}\right\vert ^{2}=\frac{(\mu -\text{$\mu $}%
	^{\ast })^{2}e^{x(\mu +\text{$\mu $}^{\ast })}}{2\cosh \left[ (\mu -\text{$%
		\mu $}^{\ast })x\right] -2\cosh \left[\gamma +\gamma ^{\ast }+i\alpha
	 \left[ \mu ^{2}-\left( \mu ^{\ast }\right) ^{2}\right]t -i\delta \left[
	\mu ^{3}-\left( \mu ^{\ast }\right) ^{3}\right]t \right] }.$}
\label{qst}
\end{equation}%
This solution is therefore nonsingular for $\func{Re}\gamma \neq 0$ and
asymptotically nondivergent for $\func{Re}\mu =0$. We depict a regular
solution in the left panel of Figure \ref{regdiv} and observe the expected
nonlocal structure in form of periodically distributed static breathers.

In contrast, the nonstandard solution (\ref{sol1}) is unavoidably singular.
We compute%
\begin{equation}
\left\vert q_{\text{nonst}}^{(1)}\right\vert ^{2}=\frac{(\mu +\nu
	)^{2}e^{x(\mu -\nu )}}{2\cosh \left[ (\mu +\nu )x\right] +2\cos \left[
	\gamma +\theta +\alpha (\mu ^{2}-\nu ^{2})t-\delta \left( \mu
	^{3}+\nu ^{3}\right)t  \right] }.  \label{qnonst}
\end{equation}%
which for $x=0$ becomes singular for any choice of the parameters
involved at 
\begin{equation}
t_{\text{s}}=\frac{\gamma +\theta +(2n-1)\pi }{\alpha \left( \nu ^{2}-\mu
	^{2}\right) +\delta \left( \mu ^{3}+\nu ^{3}\right) }, \qquad n\in 
\mathbb{Z}\text{.}  \label{rogue}
\end{equation}%
We depict a singular solution in the right panel of Figure \ref{regdiv} with
a singularity developing at $t_{\text{s}}\approx -0.689751$. We only zoomed
into one of the singularities, but it is clear from equation (\ref{rogue})
that this structure is periodically repeated so that we can speak of a 
nonlocal rogue wave \cite{kharif_physical_2003,chabchoub_rogue_2011}.

\begin{figure}[h]
	\centering
	
	\includegraphics[width=0.48\linewidth,height=0.335\linewidth]{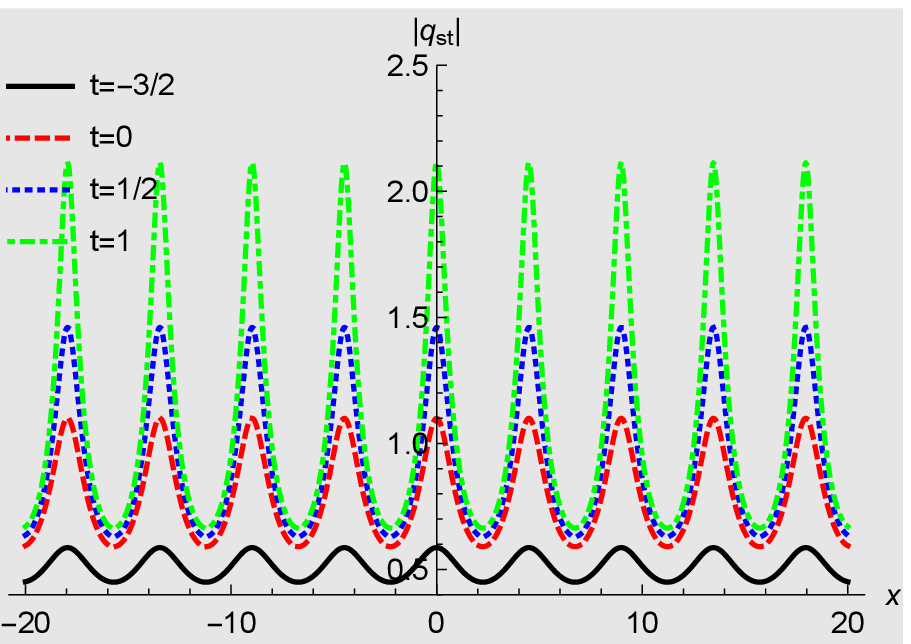} 
	\hspace{0cm}
	\includegraphics[width=0.48\linewidth]{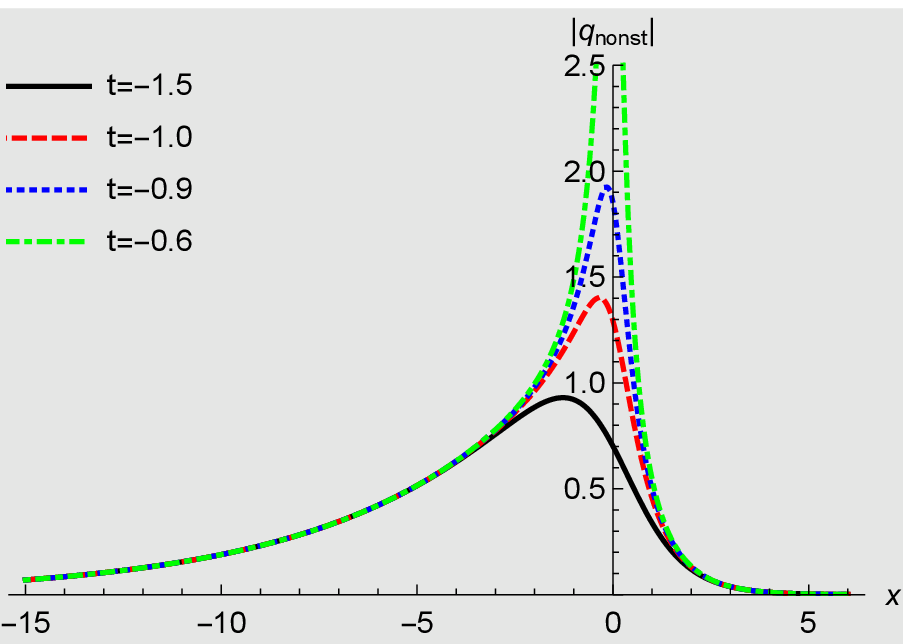}\\
	
	\caption{Nonlocal regular one-soliton solution (\ref{qst}) for the nonlocal Hirota equations obtained from the standard HDM at different times for $\alpha = 0.4$, $\delta=0.8$, 
		$\gamma=0.6+i 1.3$ and $\mu=i0.7$, $\lambda = i 1.7$ (left panel). Nonlocal rogue wave one-soliton solution (\ref{qnonst}) for the nonlocal Hirota equations obtained from the nonstandard HDM at different times for 
		$\alpha = 0.4$, $\delta=1.8$, $\gamma=0.5$, $\theta=0.1$, $\mu=0.2$ and $\nu=1.2$ (right panel).} \label{regdiv}
\end{figure}

Notice that for $\alpha \rightarrow -1$ and $\delta \rightarrow 0$ the
system (\ref{new1}) reduces to the nonlocal NLSE studied in \cite%
{ablowitz_integrable_2013}. For this case the solution (\ref{sol1}) acquires exactly the
form of equation (22) in \cite{ablowitz_integrable_2013} when we set $\nu \rightarrow
-2\eta _{1}$, $\mu \rightarrow -2\eta _{2}$, $\gamma \rightarrow \theta _{2}$
and $\theta \rightarrow \theta _{1}$. There is no equivalent solution to the
regular solution (\ref{sol2}) reported in \cite{ablowitz_integrable_2013}, so that $q_{%
	\text{st}}^{(1)}$ for $\delta \rightarrow 0$ is a also new solution for the
nonlocal NLSE.

\vspace{0.5cm}

\noindent {\large \bfseries{{The standard (two-parameter) two-soliton solution}}}

As in the local case we expand our auxiliary functions two orders further in
order to construct the two-soliton solution. Using the truncated expansions 
\begin{equation}
f=1+\varepsilon ^{2}f_{2}+\varepsilon ^{4}f_{4},\qquad g=\varepsilon
g_{1}+\varepsilon ^{3}g_{3},\qquad h=\varepsilon h_{1}+\varepsilon ^{3}h_{3},
\end{equation}%
to solve the bilinear equations (\ref{HI1}) and (\ref{HI2}), we find%
\begin{eqnarray}
g_{1} &=&\tau _{\mu ,\gamma }+\tau _{\nu ,\zeta },  \label{g1} \\
g_{3} &=&\resizebox{.7\hsize}{!}{$\frac{\left( \mu -\nu \right) ^{2}}{\left( \mu -\mu ^{\ast }\right)
	^{2}\left( \nu -\mu ^{\ast }\right) ^{2}}\tau _{\mu ,\gamma }\tau _{\nu
	,\zeta }\widetilde{\tau}_{\mu ,\gamma }^{\ast }+\frac{\left( \mu -\nu \right)
	^{2}}{\left( \mu -\nu ^{\ast }\right) ^{2}\left( \nu -\nu ^{\ast }\right)
	^{2}}\tau _{\mu ,\gamma }\tau _{\nu ,\zeta }\widetilde{\tau}_{\nu ,\zeta
}^{\ast },$} \\
f_{2} &=&\frac{\tau _{\mu ,\gamma }\widetilde{\tau}_{\mu ,\gamma }^{\ast }}{%
	\left( \mu -\mu ^{\ast }\right) ^{2}}+\frac{\tau _{\nu ,\zeta }\widetilde{\tau}%
	_{\mu ,\gamma }^{\ast }}{\left( \nu -\mu ^{\ast }\right) ^{2}}+\frac{\tau
	_{\mu ,\gamma }\widetilde{\tau}_{\nu ,\zeta }^{\ast }}{\left( \mu -\nu ^{\ast
	}\right) ^{2}}+\frac{\tau _{\nu ,\zeta }\widetilde{\tau}_{\nu ,\zeta }^{\ast }%
}{\left( \nu -\nu ^{\ast }\right) ^{2}}, \\
f_{4} &=&\frac{\left( \mu -\nu \right) ^{2}\left( \mu ^{\ast }-\nu ^{\ast
	}\right) ^{2}}{\left( \mu -\mu ^{\ast }\right) ^{2}\left( \nu -\mu ^{\ast
	}\right) ^{2}\left( \mu -\nu ^{\ast }\right) ^{2}\left( \nu -\nu ^{\ast
	}\right) ^{2}}\tau _{\mu ,\gamma }\widetilde{\tau}_{\mu ,\gamma }^{\ast }\tau
_{\nu ,\zeta }\widetilde{\tau}_{\nu ,\zeta }^{\ast }, \\
h_{1} &=&2\widetilde{\tau}_{\mu ,\gamma }^{\ast }+2\widetilde{\tau}_{\nu ,\zeta
}^{\ast }, \\
h_{3} &=&\resizebox{.7\hsize}{!}{$\frac{2\left( \mu ^{\ast }-\nu ^{\ast }\right) ^{2}}{\left( \mu
	-\mu ^{\ast }\right) ^{2}\left( \nu ^{\ast }-\mu \right) ^{2}}\widetilde{\tau}%
_{\mu ,\gamma }^{\ast }\widetilde{\tau}_{\nu ,\zeta }^{\ast }\tau _{\mu ,\gamma
}+\frac{2\left( \mu ^{\ast }-\nu ^{\ast }\right) ^{2}}{\left( \mu ^{\ast
	}-\nu \right) ^{2}\left( \nu -\nu ^{\ast }\right) ^{2}}\widetilde{\tau}_{\mu
	,\gamma }^{\ast }\widetilde{\tau}_{\nu ,\zeta }^{\ast }\tau _{\nu ,\zeta }.$}
\label{h3}
\end{eqnarray}%
So that for $\varepsilon =1$ we obtain from (\ref{g1})-(\ref{h3}) the
two-soliton solution%
\begin{equation}
q_{\text{nl}}^{(2)}(x,t)=\frac{g_{1}(x,t)+g_{3}(x,t)}{1+f_{2}(x,t)+f_{4}(x,t)%
}  \label{2nl}
\end{equation}%
As for the one-soliton solution (\ref{sol2}) we recover the solutions to the
local equation by taking $\widetilde{\tau}\rightarrow \tau $ and $\mu ^{\ast
}\rightarrow -\mu ^{\ast }$, $\nu ^{\ast }\rightarrow -\nu ^{\ast }$ in the
pre-factors. In Figure \ref{nonlocH} we depict the solution (\ref{2nl}) at
different times.

\begin{figure}[h]
	\centering
	
	\includegraphics[width=0.48\linewidth]{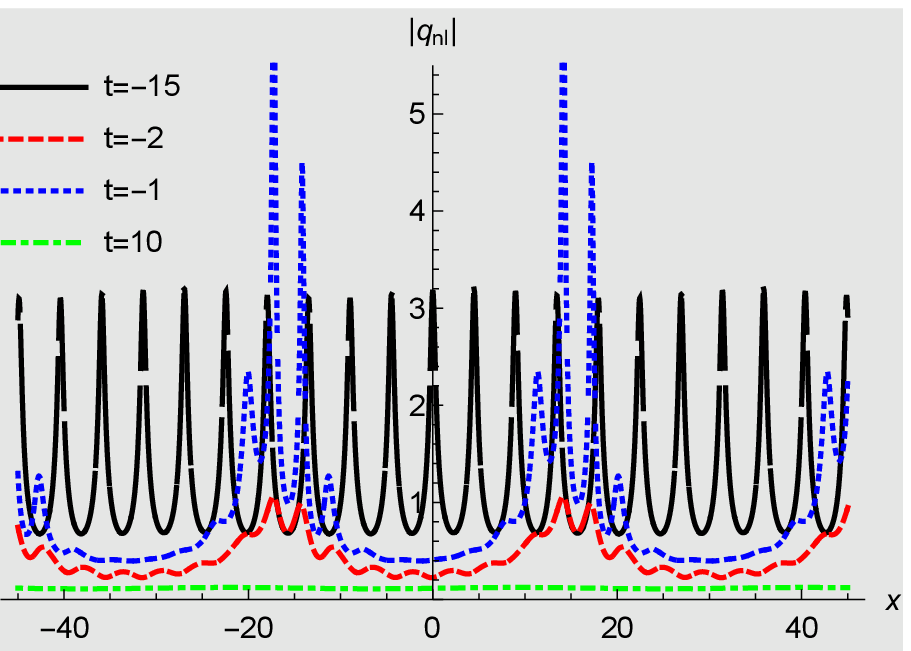} 
	\hspace{0cm}
	\includegraphics[width=0.48\linewidth]{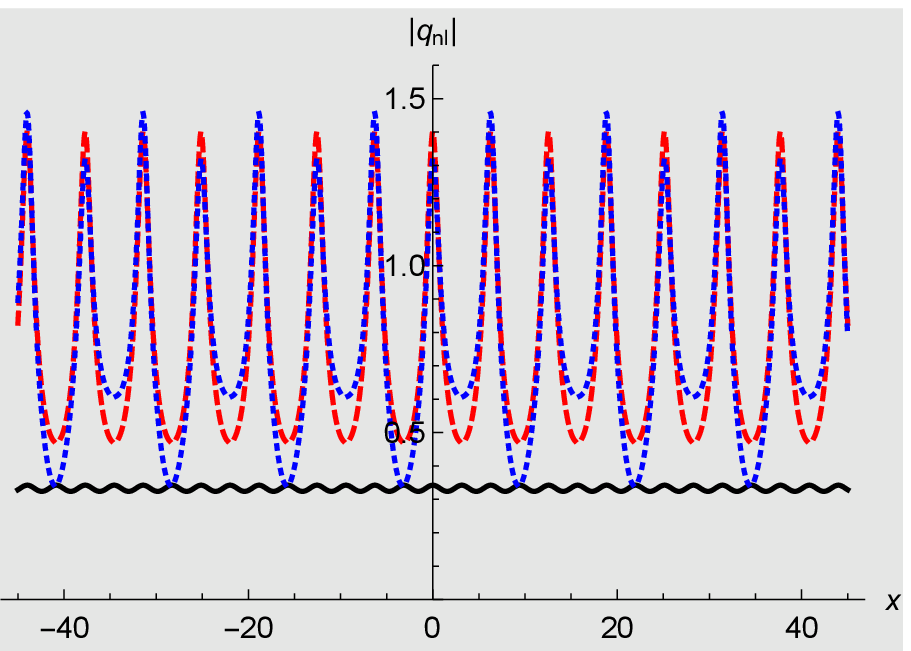}\\
	
	\caption{Nonlocal regular two-soliton solution (\ref{2nl}) for the nonlocal Hirota equations obtained from the standard HDM at different times for $\alpha = 0.4$, $\delta=0.8$, 
		$\gamma_1=0.6+i 1.3$, $\mu_1=i0.7$, $\gamma_2=0.9+i 0.7$, $\mu_2=i0.9$ (left panel). Nonlocal regular two one-soliton solution (\ref{sol2}) for the nonlocal Hirota equations $\gamma_1=0.6+i 1.3$, $\mu_1=i0.7$ (red) and $\gamma_2=0.9+i 0.7$, 
		$\mu_2=i0.9$ (black) versus the blue nonlocal regular two-soliton solution (\ref{2nl}) at the same values at time $t=2.5$  (right panel).} \label{nonlocH}
\end{figure}

In the left panel, we observe the evolution of the two-soliton solution
producing a complicated nonlocal pattern. In the right panel we can see that the two-soliton solution appears to be a result from the interference
between two nonlocal one-solitons.

As in the construction of the one-soliton solutions we can also pursue the
option to solve equation (\ref{c3}) only for $\varepsilon =1$ leading to a
second type of two-soliton solutions. We will not report them here, but
instead discuss how they emerge when using DCT.

\subsection{Soliton solutions from Darboux transformation}

Taking the DT prescription from Section 2.5.3, we start again by choosing the vanishing seed functions $q=r=0$ and solve
the linear equations from the ZC representation with $\lambda \rightarrow i\lambda $, with the additional constraint $\beta =i\delta $ by 
\begin{equation}
\widetilde{\Psi}_{1}(x,t;\lambda )=\left( 
\begin{array}{c}
\varphi _{1}(x,t;\lambda ) \\ 
\phi _{1}(x,t;\lambda )%
\end{array}%
\right) =\left( 
\begin{array}{c}
e^{\lambda x+2i\lambda ^{2}(\alpha -2\delta \lambda )t+\gamma _{1}} \\ 
e^{-\lambda x-2i\lambda ^{2}(\alpha -2\delta \lambda )t+\gamma _{2}}%
\end{array}%
\right) .  \label{P1}
\end{equation}%
In the construction of $\Psi _{2}$ we implement now the constraint $%
r(x,t)=\kappa q^{\ast }(-x,t)$, with $\kappa =\pm 1$, that gives rise to the
nonlocal equations (\ref{new1}) and (\ref{new2}). As suggested from the
previous section we expect to obtain two different types of solutions.
Indeed, unlike in the local case we have now two options at our disposal
to enforce the constraint. The standard choice consists of taking $\varphi
_{2}=\kappa \widetilde{\phi}_{1}^{\ast }$, $\phi _{2}=\widetilde{\varphi}_{1}^{\ast }$
for complex parameters which is very similar to the approach in the local case.
Alternatively we can choose here $\phi _{1}=\widetilde{\varphi}_{1}^{\ast }$, $%
\phi _{2}=\kappa \widetilde{\varphi}_{2}^{\ast }$. Evidently the first equation in
the latter constraint holds when $\gamma _{2}^{\ast }=\gamma _{1}$ in (\ref%
{P1}). It is also clear that the second option is not available in the local
case.

For the first choice we obtain therefore 
\begin{equation}
\!
\resizebox{.9\hsize}{!}{$
\widetilde{\Psi}_{2}(x,t;\lambda )=\left( 
\begin{array}{c}
\varphi _{2}(x,t;\lambda ) \\ 
\phi _{2}(x,t;\lambda )%
\end{array}%
\right) =\left( 
\begin{array}{c}
 \phi _{1}^{\ast }(-x,t;\lambda ) \\ 
-\kappa\varphi _{1}^{\ast }(-x,t;\lambda )%
\end{array}%
\right) =\left( 
\begin{array}{c}
 e^{\lambda ^{\ast }x+2i(\lambda ^{\ast })^{2}(\alpha -2\delta \lambda
	)t+\gamma _{2}^{\ast }} \\ 
-\kappa e^{-\lambda ^{\ast }x-2i(\lambda ^{\ast })^{2}(\alpha -2\delta \lambda
	^{\ast })t+\gamma _{1}^{\ast }}%
\end{array}%
\right)$} ,  \label{P2}
\end{equation}%
with $\lambda $,$\gamma _{1},\gamma _{2}\in \mathbb{C}$ and from (\ref{hirotaq}) we have 
\begin{equation}
q_{\text{st}}^{(1)}(x,t)=\frac{2(\text{$\lambda $}^{\ast }-\lambda )e^{2%
		\text{$\lambda $}x+4i\text{$\lambda $}^{2}(\alpha -2\delta 
		\text{$\lambda $})t+\text{$\gamma $}_{1}-\text{$\gamma $}%
		_{2}}}{1+\kappa e^{2(\text{$\lambda $}-\lambda^{\ast } )x+4i \lambda^{2} (\alpha-2\delta \lambda)t-4i(\lambda^{\ast})^{2} (\alpha-2\delta\lambda^{\ast})t+\gamma _{1}-\gamma
		_{2}+\text{$\gamma $}_{1}^{\ast }-\text{$\gamma $}_{2}^{\ast }}}.
\end{equation}%
For the second choice we take $\Psi _{1}(x,t;\mu )$ with $\mu \in \mathbb{R}$
and $\gamma _{2}=\gamma _{1}^{\ast }$ in (\ref{P1}). In this choice the
second wavefunction decouples entirely from the first and we may therefore
also choose different parameters. For $\kappa=1$ we take 
\begin{equation}
\widetilde{\Psi}_{2}(x,t;\nu )=\left( 
\begin{array}{c}
\varphi _{2}(x,t;\nu ) \\ 
\phi _{2}(x,t;\nu )%
\end{array}%
\right) =\left( 
\begin{array}{c}
e^{\nu x+2i\nu ^{2}(\alpha -2\delta \nu )t+\gamma _{3}} \\ 
-e^{-\nu x-2i\nu ^{2}(\alpha -2\delta \nu )t+\gamma _{3}^{\ast }}%
\end{array}%
\right) 
\end{equation}%
where $\nu \in \mathbb{R}$ and hence (\ref{hirotaq}) yields 
\begin{equation}
q_{\text{nonst}}^{(1)}(x,t)=\frac{2(\nu -\mu )e^{\gamma _{1}-\gamma
		_{1}^{\ast }+2\mu x+4i\mu ^{2}(\alpha -2\delta \mu )t}}{1+e^{2(\mu -\nu
		)x+4i\alpha( \mu ^{2}-\nu ^{2})t-2i\delta( \mu ^{3}- \nu
		^{3})t+\gamma _{1}-\gamma _{1}^{\ast }-\gamma _{3}+\gamma _{3}^{\ast }}}.
\end{equation}%
The $N$-soliton solutions are obtained considering the set 
\begin{equation}
\resizebox{.9\hsize}{!}{$\widetilde{S}_{2N}^{\text{st}}=\left\{ \widetilde{\Psi}_{1}(x,t;\lambda _{1}),\widetilde{%
	\Psi}_{2}(x,t;\lambda _{1}),\widetilde{\Psi}_{1}(x,t;\lambda _{2}),\widetilde{\Psi}%
_{2}(x,t;\lambda _{2}),...,\widetilde{\Psi}_{1}(x,t;\lambda _{n}),\widetilde{\Psi}%
_{2}(x,t;\lambda _{n})\right\} $}
\end{equation}%
for the standard case or%
\begin{equation}
\resizebox{.9\hsize}{!}{$\widetilde{S}_{2N}^{\text{nonst}}=\left\{ \widetilde{\Psi}_{1}(x,t;\mu _{1}),\widetilde{%
	\Psi}_{2}(x,t;\nu _{1}),\widetilde{\Psi}_{1}(x,t;\mu _{2}),\widetilde{\Psi}%
_{2}(x,t;\nu _{2}),...,\widetilde{\Psi}_{1}(x,t;\mu _{n}),\widetilde{\Psi}%
_{2}(x,t;\nu _{n})\right\} $}
\end{equation}%
for the non-standard case, with (\ref{P1}) and (\ref{P2}) and the formulae (\ref{hirotaqn}).

Investigating the nonlocal complex $\mathcal{PT}$-transformed Hirota equation, we observe nonlocality in time rather than space, displaying a time crystal like structure \cite{wilczek_quantum_2012,shapere_classical_2012}. The other cases have similar nonlocal properties respectively.

\section{Conclusions}
We exploited various possibilities involving different combinations of
parity, time-reversal and complex conjugation to achieve compatibility
between the two AKNS equations resulting from the
ZC condition for the Hirota equation, which is closely related to the lax representation. In this sense, along with the sense that we have provided various methods on how to obtain soliton solutions, each possibility
corresponds to a new type of integrable system. Solving these new nonlocal
equations by means of HDM, we encountered various new
features. Instead of having to solve two bilinear equations, these new
systems correspond to three bilinear equations involving an auxiliary
function. We solved these equations in the standard fashion by using a
formal expansion parameter that in the end can be set to any value when the
expansions are truncated at specific orders. In addition, the new auxiliary
equation allows for a new option for this equation to be solved for a
specific value of the expansion parameter, thus leading to a new type of
solution different from the one obtained in the standard fashion. We also
identified the mechanism leading to this second type of solution within the
approach of using DCT. In that context the nonlocal
relations between $q$ and $r$ allow for different options in (\ref{hirotaq}).

\chapter{Nonlocal gauge equivalence: Hirota versus ECH and ELL equations}\label{ch_8}

Having explored nonlocality for the Hirota equation in the previous chapter, we explore here how nonlocality can be
implemented into extended versions of the continuous limit of the Heisenberg
(ECH) equation \cite{nakamura_solitons_1974,lakshmanan_dynamics_1976,tjon_solitons_1977,takhtajan_integration_1977,demontis_effective_2018}
and extended Landau-Lifschitz (ELL) equations  \cite{landau_zur_1935,baryakhtar_landau-lifshitz_2015}. We exploit the gauge equivalence of these
systems and investigate how the nonlocality property of one system is
inherited by the other \cite{cen_gauge}. For the NLS equation, the gauge equivalence to the ECH equation in the local case has been known for some time \cite{zakharov_equivalence_1979} and nonlocal case recently explored in \cite{gadzhimuradov_towards_2016}. In there, gauge equivalence was explored for a particular case of nonlocal NLS soliton solution. In this chapter we will not only extend this investigation for the Hirota equation, but also show how the nonlocality is implemented in the ECH equation through DT, finding new types of solutions in the nonlocal setting, which have no counterpart in the local case.

The local version of the original Landau-Lifschitz equation famously describes the precession of the magnetization in a solid when subjected to a torque resulting from an effective external magnetic field. Various extended versions have been proposed, such as for instance the Landau-Lifshitz-Gilbert equation \cite{gilbert_a_1955} to take damping into account. The nonlocal versions of this equation studied here provide further extensions with complex components. 

\section{Gauge equivalence}

In Section 2.5.4, we introduced the gauge correspondence between two ZC representations and showed how a special choice of the gauge operator leads to DCT (an auto-gauge correspondence), the iteration to produce infinitely many solutions of the ECH equation with an initial seed solution. In this chapter, we will show another usage of gauge correspondence. This is to take our investigations of the nonlocal Hirota equation from the previous chapter to the analogous nonlocal ECH equation through utilising the gauge equivalence between the two gauge equivalent ZC conditions.

Here, we first present the gauge equivalence of the nonlocal Hirota and ECH equations. 

\subsection{The nonlocal Hirota system}
Let us first rewrite the ZC representation for the Hirota equation from Section 2.5.3 expanded in terms of spectral parameter $\lambda$, which is convenient later in finding the gauge equivalence 
\begin{equation}
	[\Psi _{H}]_{t}=V_{H}\Psi _{H} \text{,  } \qquad [\Psi
	_{H}]_{x}=U_{H}\Psi _{H},
\end{equation}
\noindent with $U_{H}$, $V_{H}$ to be of the form
\begin{equation}
U_{H}=A_{0}+\lambda A_{1},\qquad V_{H}=B_{0}+\lambda B_{1}+\lambda
^{2}B_{2}+\lambda ^{3}B_{3},\qquad  \label{U1}
\end{equation}%
where%
\begin{eqnarray}
\hspace{-1cm}
A_{0} &=&\left( 
\begin{array}{cc}
0 & q(x,t) \\ 
r(x,t) & 0%
\end{array}%
\right) ,A_{1}=\left( 
\begin{array}{cc}
-i & 0 \\ 
0 & i%
\end{array}%
\right) =-i\sigma _{3}, \\
\hspace{-1cm}
B_{0} &=&i\alpha \left[ \sigma _{3}\left( A_{0}\right) _{x}-\sigma
_{3}A_{0}^{2}\right] +\beta \left[ 2A_{0}^{3}+\left( A_{0}\right)
_{x}A_{0}-A_{0}\left( A_{0}\right) _{x}-\left( A_{0}\right) _{xx}\right] , \\
\hspace{-1cm}
B_{1} &=&2\alpha A_{0}+2i\beta \sigma _{3}\left[ \left( A_{0}\right)
_{x}-A_{0}^{2}\right] , \\
\hspace{-1cm}
B_{2} &=&4\beta A_{0}-2i\alpha \sigma _{3}, \\
\hspace{-1cm}
B_{3} &=&-4i\beta \sigma _{3},  \label{5}
\end{eqnarray}%
with $\sigma _{i}$, $i=1,2,3$ denoting the Pauli spin matrices. One can check with the
explicit expressions (\ref{U1})-(\ref{5}), that the ZC condition becomes equivalent to the following AKNS system for the fields $q(x,t)$ and $r(x,t)$:
\begin{eqnarray}
q_{t}-i\alpha q_{xx}+2i\alpha q^{2}r+\beta \left[ q_{xxx}-6qrq_{x}\right]
&=&0,  \label{akns1} \\
r_{t}+i\alpha r_{xx}-2i\alpha qr^{2}+\beta \left( r_{xxx}-6qrr_{x}\right)
&=&0.  \label{akns2}
\end{eqnarray}
As seen in the previous chapter, we can reduce the AKNS system to one equation through various choices of $r(x,t)$, in particular, exploiting $\mathcal{PT}$-symmetry, various nonlocal cases come out such as $r(x,t)=\kappa q^{\ast
}(-x,t) $, $r(x,t)=\kappa q^{\ast }(x,-t)$, $r(x,t)=\kappa q^{\ast }(-x,-t)$%
, $r(x,t)=\kappa q(-x,t)$, $r(x,t)=\kappa q(x,-t)$ or $r(x,t)=\kappa
q(-x,-t) $ with $\kappa =\pm 1$ and a suitable condition on the
parameters $\alpha $ and $\beta $. For the rest of this chapter, we focus our investigations for the nonlocal case  $r(x,t)=\kappa q^{\ast }(-x,t)$ with $\beta=i \delta$ as other cases will follow similarly.

\subsection{The nonlocal ECH equation}
Now, let us take another system $U_{E}$, $V_{E}$ as the gauge equivalent ZC representation to the Hirota system,
\begin{equation}
[\Psi _{E}]_{t}=V_{E}\Psi _{E} \text{,  } \qquad [\Psi
_{E}]_{x}=U_{E}\Psi _{E}.
\end{equation}
Taking the two systems with operators $U_{H}$, $V_{H}$ and $U_{E}$, $V_{E}$ and considering their corresponding solutions $\Psi _{H}$ to $\Psi _{E}$ are related by a gauge operator $G$ as $\Psi_{H}=G\Psi_{E}$, then $U_{H}$, $V_{H}$ and $U_{E}$, $V_{E}$ are related as%
\begin{equation}
U_{H}=GU_{E}G^{-1}+G_{x}G^{-1},\qquad \text{and\qquad }%
V_{H}=GV_{E}G^{-1}+G_{t}G^{-1},  \label{U12}
\end{equation}%
\noindent or
\begin{equation}
U_{E}=G^{-1}U_{H}G-G^{-1}G_{x},\qquad \text{and\qquad }%
V_{E}=G^{-1}V_{H}G-G^{-1}G_{t}.  \label{U13}
\end{equation}
Employing the
expansion (\ref{U1}), we obtain from (\ref{U13}) the expressions 
\begin{equation}
U_{E}=-i\lambda G^{-1}\sigma _{3}G,\qquad V_{E}=\lambda G^{-1}B_{1}G+\lambda
^{2}G^{-1}B_{2}G+\lambda ^{3}G^{-1}B_{3}G,  \label{U2}
\end{equation}%
together with%
\begin{equation}
G_{x}=A_{0}G,\qquad \text{and\qquad }G_{t}=B_{0}G.  \label{10}
\end{equation}%
Given $A_{0}$ and $B_{0}$, it is the solution for these two equations
in (\ref{10}) that determines the precise form of $G$ for a particular set
of models.

An interesting and universally applied equation emerges when we use the
gauge field $G$ to define a new field operator 
\begin{equation}
S:=G^{-1}\sigma _{3}G\text{.}  \label{SGhei}
\end{equation}%
The following properties follow directly from above: 
\begin{eqnarray}
S^{2}&=&1,\quad S_{x}=2G^{-1}\sigma _{3}A_{0}G,\quad\\
SS_{x}&=&-S_{x}S=2G^{-1}A_{0}G,\quad \left[ S,S_{xx}\right] =2\left(
SS_{xx}+S_{x}^{2}\right) .  \label{S}
\end{eqnarray}%
Next we notice that instead of expressing the operators $U_{E}$ and $V_{E}$
in terms of the gauge field $G$, one can express them entirely in terms
of the operator $S$ as%
\begin{eqnarray}
\hspace{-1cm}
U_{E}&=&-i\lambda S,\qquad\\
\hspace{-1cm}
 V_{E}&=&\alpha \left( \lambda SS_{x}-\lambda
^{2}2iS\right) +\beta \left[ \lambda \left( i\frac{3}{2}SS_{x}^{2}+iS_{xx}%
\right) +\lambda ^{2}2SS_{x}-\lambda ^{3}4iS\right] .  \label{UV2}
\end{eqnarray}%
Using this variant we evaluate the ZC condition to obtain the equation of motion for
the $S$-operator%
\begin{eqnarray}
S_{t} &=&i\alpha \left( S_{x}^{2}+SS_{xx}\right) -\beta \left[ \frac{3}{2}%
\left( SS_{x}^{2}\right) _{x}+S_{xxx}\right]  \label{St} \\
&=&\frac{i}{2}\alpha \left[ S,S_{xx}\right] -\frac{\beta }{2}\left(
3S_{x}^{3}+S\left[ S,S_{xxx}\right] \right) .
\end{eqnarray}%
For $\beta =0$ this equation reduces to the well-known continuous limit of
the Heisenberg spin chain \cite{nakamura_solitons_1974,lakshmanan_dynamics_1976,tjon_solitons_1977,takhtajan_integration_1977,demontis_effective_2018}
and for $\beta \neq 0$ to the first member of the corresponding hierarchy 
\cite{wang_darboux_2005}. We refer to this equation as the ECH equation. Taking $r(x,t)=\kappa q^{\ast}(-x,t)$ with $\beta=i \delta$, we obtain the nonlocal ECH equation
\begin{equation}
	S_{t}=\frac{i}{2}\alpha \left[ S,S_{xx}\right] -i\frac{\delta }{2}\left(
	3S_{x}^{3}+S\left[ S,S_{xxx}\right] \right) , \label{nst}
\end{equation}
where nonlocality will appear in the entries of the $S$ matrix, as we shall see in the next section. The equation (\ref{St}) is rather universal as it also
emerges for other types of integrable higher order equations of NLS type, such as the mKdV equation \cite{kundu_landaulifshitz_1984,ma_soliton_2017} or the Sasa-Satsuma equation \cite{kundu_landaulifshitz_1984,ma_integrability_2018}. The
distinction between specific models of this general type is obtained by
specifying $G$.

Given the above gauge correspondence one may now obtain solutions to the
nonlinear equations of a member of the NLS hierarchy
from the equations of motion of the corresponding member the continuous
Heisenberg hierarchy, or vice versa. For instance, given a solution $q(x,t)\,$%
\ and $r(x,t)$ to the Hirota equations (\ref{akns1}), (\ref{akns2}) one may
use equation (\ref{10}) to construct the gauge field operator $G$ and
subsequently simply compute $S$, that solves (\ref{St}) by construction, by
means of the relation (\ref{SGhei}). Conversely, from a solution $S$ to (\ref%
{St}) we may construct $G$ by (\ref{SGhei}) and subsequently $q(x,t)\,$\ and $%
r(x,t)$ from (\ref{10}). We elaborate below on the details of this
correspondence.

\section{Nonlocal multi-soliton solutions for the ECH equation from Darboux-Crum transformation}

In Section 2.5.4, we introduced the method of DCT for the ECH equation. In this section, we explain particularly how nonlocality is naturally introduced into these systems through seed solutions to the nonlocal ZC representation equations.

\subsection{Nonlocal one-soliton solution}

We start with a simple constant solution to the ECH equation (\ref{St}) as in Section 2.5.4 describing the free case%
\begin{equation}
S=\left( 
\begin{array}{rr}
-w & u \\ 
v & w%
\end{array}%
\right) ,\quad \text{with} \quad w=1\text{, }u=v=0.  \label{S0}
\end{equation}%
In order to define the matrix operator $H$ as in (\ref{H}), we need to
construct the seed solution $\psi (\lambda )$ to the spectral problem (\ref%
{ECHzc}) and evaluate it for two different nonzero spectral parameters $%
\psi (\lambda _{1})=(\varphi _{1},\phi _{1})$ and $\psi (\lambda
_{2})=(\varphi _{2},\phi _{2})$.

For nonlocality, we impose the symmetry condition 
\begin{equation}
	S(x,t)=\kappa S^{\dagger}(-x,t) \quad \text{with} \quad \beta=i \delta
\end{equation}
 and choose $v=\kappa u^{\ast
}(-x,t)$, which leads to the constraints $w=\kappa w^{\ast
}(-x,t)$,%
\begin{equation}
\varphi _{2}(x,t)= \phi _{1}^{\ast }(-x,t),\quad \phi
_{2}(x,t)=\kappa \varphi _{1}^{\ast }(-x,t),\quad \text{with } \quad \lambda _{2}=\lambda
_{1}^{\ast } \text{.}  \label{loco}
\end{equation}%
We can now solve the spectral problem (\ref{ECHzc}) with $S$ for $\psi
(\lambda _{1}=\lambda )$ in the form%
\begin{equation}
\psi _{1}(\lambda )=\left( 
\begin{array}{l}
e^{Z(x,t) +\gamma _{1}} \\ 
e^{-Z(x,t) +\gamma _{2}}%
\end{array}%
\right) ,  \label{psis1}
\end{equation}%
where we introduced the function%
\begin{equation}
Z(x,t) =i\lambda x+2\lambda ^{2}(i\alpha -2\delta \lambda )t
\end{equation}%
and the additional constants $\gamma _{1},\gamma _{2}\in \mathbb{C}$ to
account for boundary conditions. The second solution is then simply obtained
from the constraint (\ref{loco}) to be%
\begin{equation}
\psi _{2}(\lambda ^{\ast })=\left( 
\begin{array}{l}
\phi^{\ast}(-x,t) \\ 
-\kappa \varphi^{\ast}(-x,t)%
\end{array}%
\right)=\left( 
\begin{array}{l}
 e^{-Z^{\ast }(-x,t)+\gamma _{2}^{\ast }} \\ 
-\kappa e^{Z^{\ast }(-x,t)+\gamma _{1}^{\ast }}%
\end{array}%
\right) .  \label{psis2}
\end{equation}%
Notice that $\psi _{2}(\lambda ^{\ast })$ is the solution to the parity
transformed and conjugated spectral problem (\ref{ECHzc}). Given these
solutions we can now compute the functions in the iterated $S^{(1)}$ matrix for a nonlocal ECH one-soliton solution.

\subsection{Nonlocal $N$-soliton solution}

We proceed further in the same way for the nonlocal multi-soliton solutions. In general, for a nonlocal $N$-soliton solution we take $2N$ non-zero spectral parameters with constraints
\begin{equation}
\lambda _{2k}=\lambda _{2k-1}^{\ast } \quad k=1,2,\ldots ,N,
\end{equation}%
and the seed functions computed at these values as%
\begin{eqnarray}
\hspace{-1cm}
\psi _{2k-1}(\lambda _{2k-1}) \!\!\!&=&\!\!\!\left( 
\begin{array}{l}
\varphi _{2k-1} \\ 
\phi _{2k-1}%
\end{array}%
\right) \!\!=\!\!\left( 
\begin{array}{l}
e^{Z _{{2k-1}}(x,t)+\gamma _{2k-1}} \\ 
e^{-Z _{{2k-1}}(x,t)+\gamma _{2k}}%
\end{array}%
\right) ,  \label{seed1} \\
\hspace{-1cm}
\psi _{2k}(\lambda _{2k})\!\!\! &=&\!\!\!\left( 
\begin{array}{l}
\varphi _{2k} \\ 
\phi _{2k}%
\end{array}%
\right)\!\! =\!\!\left( 
\begin{array}{l}
\phi^{\ast} _{2k-1}(-x,t) \\ 
\kappa \varphi^{\ast} _{2k-1}(-x,t)%
\end{array}%
\right)\!\!=\!\!\left( 
\begin{array}{l}
 e^{-Z_{2k-1}^{\ast }(-x,t)+\gamma _{2k}^{\ast }} \\ 
\kappa e^{Z _{2k-1}^{\ast }(-x,t)+\gamma _{2k-1}^{\ast }}%
\end{array}%
\right)  \label{seed2}
\end{eqnarray}%
\noindent where
\begin{equation}
Z_{j}(x,t) =i\lambda_{j} x+2\lambda_{j} ^{2}(i\alpha -2\delta \lambda_{j} )t.
\end{equation}%
We may then use (\ref{sng}) to
evaluate $u_{N}$, $v_{N}$, and $w _{N}$ for $S^{(N)}$. We find the nonlocality
property $v_{N}(x,t)=\kappa u_{N}^{\ast }(-x,t)$ for all solutions.

\section{Nonlocal solutions of the Hirota equation from the ECH equation}

Let us now demonstrate how to obtain nonlocal solutions for the Hirota
equation from those of the ECH equation. For this purpose, with $S$
being parametrised as in (\ref{suvwn}), we solve equation (\ref{SGhei}) for $G$ to find
\begin{equation}
G=\left( 
\begin{array}{rr}
a(x,t) & a(x,t)\frac{w_{N} +1}{v_{N}} \\ 
c(x,t) & c(x,t)\frac{w_{N} -1}{v_{N}}%
\end{array}%
\right) ,
\end{equation}%
where the functions $a(x,t)$ and $c(x,t)$ remain unknown at this point. They
can be determined when substituting $G$ into the equations (\ref{10}).
Solving the first equation for $q_{N}(x,t)\,$and $r_{N}(x,t)$ we find
\begin{eqnarray}
\hspace{-1cm}
q_{N}(x,t) \!\!\!&=&\!\!\!\resizebox{.75\hsize}{!}{$\frac{1}{2}\left( \frac{[v_{N}]_{x}}{v_{N}}+\frac{w_{N} [v_{N}]_{x}-[w_{N}]
	_{x}v_{N}}{v_{N}}\right) \exp \left[ \int\frac{w_{N} [v_{N}]_{x}-[w_{N}]
	_{x}v_{N}}{v_{N}}dx\right]$} ,  \label{qint} \\
\hspace{-1cm}
r_{N}(x,t) \!\!\!&=&\!\!\!\resizebox{.75\hsize}{!}{$\frac{1}{2}\left( \frac{[v_{N}]_{x}}{v_{N}}-\frac{w_{N} [v_{N}]_{x}-[w_{N}]
		_{x}v_{N}}{v_{N}}\right) \exp \left[ - \int\frac{w_{N} [v_{N}]_{x}-[w_{N}]
		_{x}v_{N}}{v_{N}}dx\right]$} .  \label{rint}
\end{eqnarray}%
Notice that
the integral representations (\ref{qint}) and (\ref{rint}) are valid for 
any solution to the ECHE (\ref{St}). Next we demonstrate how to
solve these integrals. Using the expression in (\ref{uvw1})-(\ref{uvw3}),
with suppressed subscripts $N$ and $S$ chosen as in (\ref{S0}), we can
re-express the terms in (\ref{qint}) and (\ref{rint}) via the components of
the intertwining operator $\widetilde{\mathcal{Q}}^{(N)}$ as 
\begin{eqnarray}
\frac{w_{N} [v_{N}]_{x}-[w_{N}] _{x}v_{N}}{v_{N}} &=& \partial
_{x}\ln \left( \frac{C_{N}}{D_{N}}\right) , \\
\frac{[v_{N}]_{x}}{v_{N}} &=&\partial
_{x}\ln \left( C_{N}D_{N}\right),
\end{eqnarray}%
where we used the property (\ref{AB}). With these relations the integral
representations (\ref{qint}), (\ref{rint}) simplify to 
\begin{eqnarray}
q_{N}(x,t) &=&\frac{\left( C_{N}\right) _{x}}{D_{N}}=
	 \frac{W_{2}\left[\det \Omega^{(N)}, \det \widetilde{\mathcal{Q}}_{21}^{(N)}\right]}{\det \Omega^{(N)} \det \widetilde{\mathcal{Q}}_{22}^{(N)}} ,  \label{qn} \\
r_{N}(x,t) &=&\frac{\left( D_{N}\right) _{x}}{C_{N}}=
 \frac{W_{2}\left[\det \Omega^{(N)}, \det \widetilde{\mathcal{Q}}_{22}^{(N)}\right]}{\det \Omega^{(N)} \det \widetilde{\mathcal{Q}}_{21}^{(N)}}. \label{rn}
\end{eqnarray}%
Thus, we have now obtained a simple relation between the spectral problem of
the ECH equation and the solutions to the Hirota
equation. It appears that this is a novel relation even for the local
scenario. The nonlocality property of the solutions to the ECHE is then
naturally inherited by the solutions to the Hirota equation. Using the
nonlocal choices for the seed functions as specified in (\ref{seed1}) and (%
\ref{seed2}) we may compute directly the right hand sides in (\ref{qn}) and (%
\ref{rn}). Crucially these solutions satisfy the nonlocality property%
\begin{equation}
r_{N}(x,t)=\kappa q_{N}^{\ast }(-x,t).
\end{equation}%

\section{Nonlocal solutions of the ECH equation from the Hirota equation}

For the nonlocal choice $r(x,t)=\kappa q^{\ast }(-x,t)$ the first equation
in (\ref{10}) implies that
\begin{equation}
	G=\left( 
	\begin{array}{rr}
	a & \kappa \widetilde{b}^{\ast} \\ 
	b & \widetilde{a}^{\ast}
	\end{array}%
	\right).
\end{equation} 
\noindent We adopt here the notation from \cite{cen_integrable_2019} and suppress the explicit dependence on $(x,t)$, indicating
the functional dependence on $(-x,t)$ by a tilde, i.e. $\widetilde{q}:=q(-x,t)$. The first equation in (%
\ref{10}) then reduces to the two equations%
\begin{equation}
a_{x}=b q,\qquad b_{x}=\kappa a \widetilde{q}^{\ast}.
\label{nl1}
\end{equation}%
\noindent If we take $b=\kappa c \widetilde{q}^{\ast}$, then $a=c\partial_{x}\ln \widetilde{q}^{\ast}$. Having specified the gauge transformation $G$, we can compute the corresponding nonlocal solution to the ECH equation
\begin{equation}
S=G^{-1}\sigma_{3}G=\left( 
\begin{array}{rr}
-w & u \\ 
v & w
\end{array}%
\right) \quad \text{with} \quad
G=c\left( 
\begin{array}{rr}
\partial_{x}\ln \widetilde{q}^{\ast} & q \\ 
\kappa \widetilde{q^{\ast}} & \partial_{x}\ln q
\end{array}%
\right),
\end{equation}
\noindent where
\begin{equation}
\hspace{-0.05cm}
\resizebox{.9\hsize}{!}{$u=-\frac{2 (q\widetilde{q}^{\ast})q_{x}}{\kappa(q\widetilde{q}^{\ast})^{2}+(q\widetilde{q}^{\ast})_{x}}, \quad 
v=-\frac{2\kappa(q\widetilde{q}^{\ast}) \widetilde{q}^{\ast}_{x}}{\kappa(q\widetilde{q}^{\ast})^{2}+(q\widetilde{q}^{\ast})_{x}}, \quad 
w=-\frac{\kappa (q\widetilde{q}^{\ast})^{2}-(q\widetilde{q}^{\ast})_{x}}{\kappa(q\widetilde{q}^{\ast})^{2}+(q\widetilde{q}^{\ast})_{x}}.$}
\end{equation} 
We can check the solution satisfies $v=\kappa u^{\ast}(-x,t)$, $w=\kappa w^{\ast}(-x,t)$ and $S(x,t)=\kappa S^{\dagger}(-x,t)$, as expected from gauge equivalence.
\section{Nonlocal soliton solutions to the ELL equation}
\subsection{Local ELL equation}
Given the solutions to the ECH equation (\ref{St}), it is now also
straightforward to construct solutions to the ELL equation (\ref{ELLE}) from them simply by using the representation $S_{N}=\textbf{\overrightharp{s}$_{N}$}\cdot \textbf{\overrightharp{$\mathbf{\sigma}$}}$ with $S_{N}$ taken to be in the parametrisation (\ref{suvwn}).
Suppressing the index $N$, a direct expansion then yields%
\begin{equation}
s_{1}=\frac{1}{2}(u+v),\qquad s_{2}=\frac{i}{2}(u-v),\qquad s_{3}=-w
.
\end{equation}%
For the local choice $v(x,t)=u^{\ast }(x,t)$ these function are evidently
real%
\begin{equation}
s_{1}(x,t)=\func{Re}u,\qquad s_{2}=-\func{Im}u,\qquad s_{3}=\pm \sqrt{%
	1-\left\vert u\right\vert ^{2}}.
\end{equation}%
Thus, since $\textbf{\overrightharp{s}}$ is a real unit vector function and $\mathbf{\textbf{\overrightharp{s}}\cdot \textbf{\overrightharp{s}}%
}=1$.

We briefly discuss some of the key characteristic behaviours of $\mathbf{s}$
for various choices of the parameters. When $\beta =0$, the solutions
correspond to the one-soliton solutions of the NLS
equation. For pure imaginary parameter $\lambda $, we obtain the well known periodic
solutions to the ELLE as seen in the left panel of Figure \ref{Fig8.1}.
However, when the parameter $\lambda $ is taken to be complex we obtain
decaying solutions tending towards a fixed point as in the right panel.

\begin{figure}
	\centering
	\includegraphics[width=0.49\linewidth]{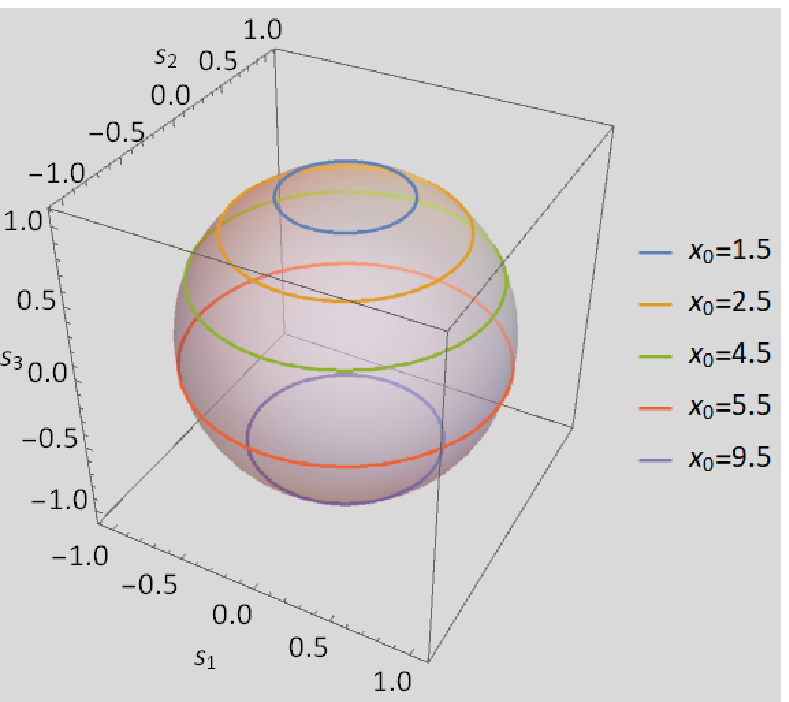}
	\hspace{0cm}
	\includegraphics[width=0.49\linewidth]{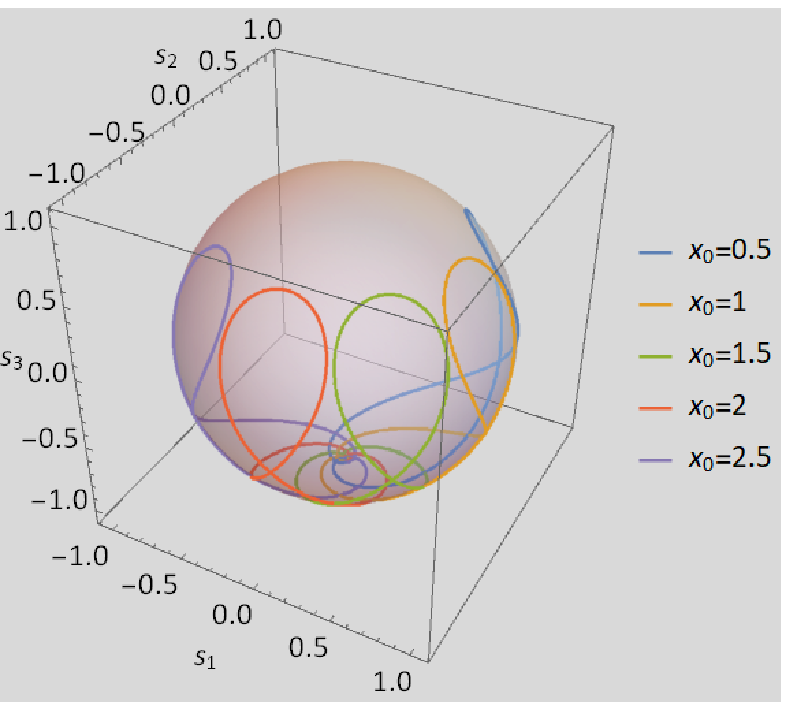}\\
	
	\caption{Local solutions to the ELLE (\ref{ELLE}) from a gauge
		equivalent one-soliton solution (\ref{ones}) of the NLS
		equation for different initial values $x_{0}$, complex shifts $\gamma_{1} =0.1+0.6i$, $\gamma_{2} =0.3i$ and $\alpha =5$, $\beta =0$. In the left panel the spectral parameter is pure imaginary, $\lambda =0.1i$, and
		in the right panel it is complex, $\lambda =0.2+0.5i$.}
	\label{Fig8.1}
\end{figure}

When taking $\beta \neq 0$, that is the solutions to the Hirota equation,
even for pure imaginary values $\lambda$, the behaviour of the trajectories is drastically, as they become more knotty and
convoluted as seen in the left panel of Figure \ref{Fig8.2}. Complex values of 
$\lambda$ are once more decaying solutions tending towards a fixed point.

\begin{figure}
	\centering
	\includegraphics[width=0.49\linewidth]{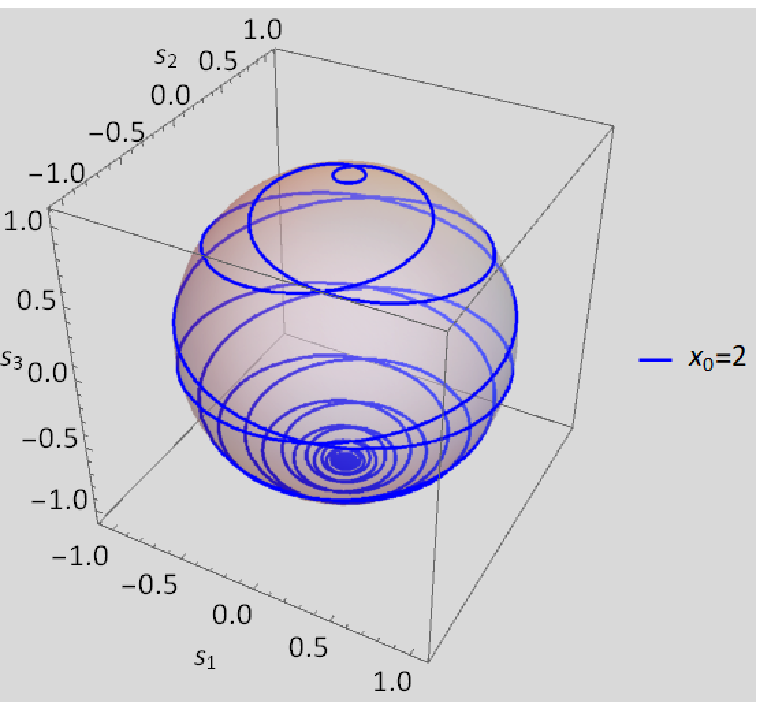}
	\hspace{0cm}
	\includegraphics[width=0.49\linewidth]{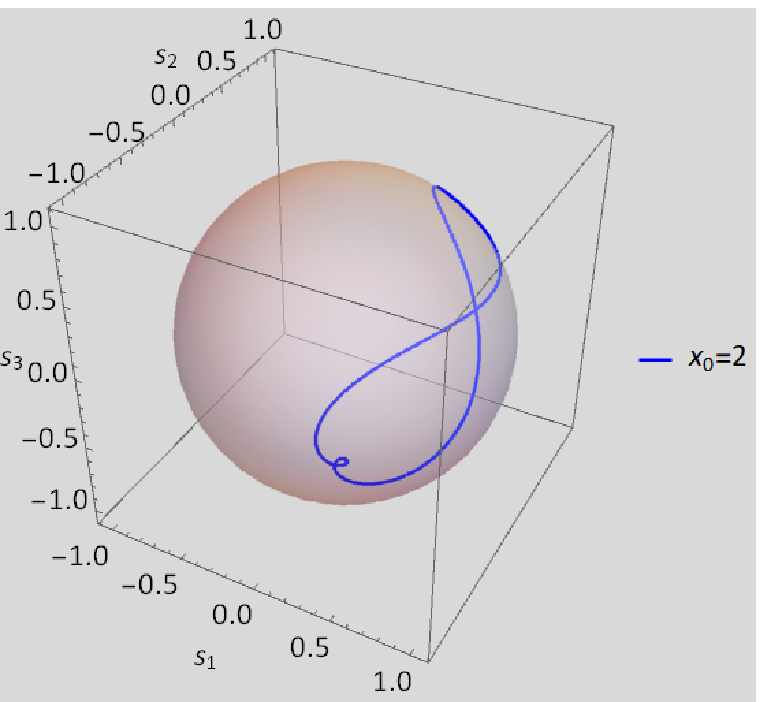}\\
	
	\caption{Local solutions to the ELLE (\ref{ELLE}) from a gauge equivalent
		one-soliton solution (\ref{ones}) of the Hirota equation for a fixed value of $x_{0}$, complex shifts $\gamma_{1} =0.1+0.6i$, $\gamma_{2} =0.3i$ and $\alpha =5$, $\beta =2$. In
		the left panel the spectral parameter is pure imaginary, $\lambda =0.1i$, and in the right
		panel it is complex, $\lambda =0.02+0.05i$.}
	\label{Fig8.2}
\end{figure}
\subsection{Nonlocal ELL equation}
For the nonlocal choice $v(x,t)=\kappa u^{\ast }(-x,t)$ with $\beta=i \delta$, which also results to $w^{\ast}(-x,t)=w$ and $S^{\dagger}(-x,t)=S$, the vector function $%
\mathbf{s}$ is no longer real so that we may decompose it into $\textbf{\overrightharp{s}}=%
\textbf{\overrightharp{m}}+i\,\textbf{\overrightharp{l}}$, where now $\textbf{\overrightharp{m}}$ and $\textbf{\overrightharp{l}}$ are real
valued vector functions. From the relation $\mathbf{\textbf{\overrightharp{s}}\cdot \textbf{\overrightharp{s}}}=1$ it follows
directly that $\textbf{\overrightharp{m}}^{2}-$ $\textbf{\overrightharp{l}}\,^{2}=1$ and that these vector
functions are orthogonal to each other $\textbf{\overrightharp{m}}\cdot \textbf{\overrightharp{l}}=0$. The ELL equation (\ref{ELLE}) then becomes a set
of coupled equations for the real valued vector functions $\textbf{\overrightharp{m}}$ and $%
\textbf{\overrightharp{l}}$ 
\begin{eqnarray}
\hspace{-1cm}
\textbf{\overrightharp{m}}_{t}\!\!\! &=&\!\!\!\resizebox{.8\hsize}{!}{$\alpha \left( \mathbf{\textbf{\overrightharp{l}}\times \textbf{\overrightharp{l}}}_{xx}-\mathbf{\textbf{\overrightharp{m}}\times \textbf{\overrightharp{m}}}%
	_{xx}\right) +\frac{3}{2}\delta \left[ \left( \textbf{\overrightharp{m}}_{x}\cdot \textbf{\overrightharp{m}}%
	_{x}\right) \textbf{\overrightharp{m}}_{x}+2\left( \textbf{\overrightharp{l}}_{x}\cdot \textbf{\overrightharp{m}}%
	_{x}\right) \textbf{\overrightharp{m}}_{x}-\left( \textbf{\overrightharp{l}}_{x}\cdot \textbf{\overrightharp{l}}_{x}\right) 
	\textbf{\overrightharp{l}}_{x}\right]$}  \label{NELL1} \\
\hspace{-1cm}
\!\!\!&&\!\!\!\resizebox{.8\hsize}{!}{$+\delta \left[ \mathbf{\textbf{\overrightharp{l}}\times }\left( \mathbf{\textbf{\overrightharp{l}}\times \textbf{\overrightharp{l}}}_{xxx}\right) -%
	\mathbf{\textbf{\overrightharp{m}}\times }\left( \mathbf{\textbf{\overrightharp{l}}\times \textbf{\overrightharp{m}}}_{xxx}\right) -\mathbf{\textbf{\overrightharp{m}}\times }%
	\left( \mathbf{\textbf{\overrightharp{m}}\times \textbf{\overrightharp{l}}}_{xxx}\right) -\mathbf{\textbf{\overrightharp{l}}\times }\left( \mathbf{%
		\textbf{\overrightharp{m}}\times \textbf{\overrightharp{m}}}_{xxx}\right) \right]$} ,  \notag \\
	\hspace{-1cm}
\textbf{\overrightharp{l}}_{t} \!\!\!&=&\!\!\!\resizebox{.8\hsize}{!}{$-\alpha \left( \mathbf{\textbf{\overrightharp{l}}\times \textbf{\overrightharp{m}}}_{xx}+\mathbf{\textbf{\overrightharp{m}}\times \textbf{\overrightharp{l}}}%
	_{xx}\right) +\frac{3}{2}\delta \left[ \left( \textbf{\overrightharp{l}}_{x}\cdot \textbf{\overrightharp{l}}%
	_{x}\right) \textbf{\overrightharp{m}}_{x}+2\left( \textbf{\overrightharp{l}}_{x}\cdot \textbf{\overrightharp{m}}_{x}\right) 
	\textbf{\overrightharp{l}}_{x}-\left( \textbf{\overrightharp{m}}_{x}\cdot \textbf{\overrightharp{m}}%
	_{x}\right) \textbf{\overrightharp{m}}_{x}\right]$}   \label{NELL2} \\
\hspace{-1cm}
\!\!\!&&\!\!\!\resizebox{.8\hsize}{!}{$+\delta \left[ \mathbf{\textbf{\overrightharp{m}}\times }\left( \mathbf{\textbf{\overrightharp{m}}\times \textbf{\overrightharp{m}}}_{xxx}\right) -%
	\mathbf{\textbf{\overrightharp{l}}\times }\left( \mathbf{\textbf{\overrightharp{m}}\times \textbf{\overrightharp{l}}}_{xxx}\right) -\mathbf{\textbf{\overrightharp{l}}\times }%
	\left( \mathbf{\textbf{\overrightharp{l}}\times \textbf{\overrightharp{m}}}_{xxx}\right) -\mathbf{\textbf{\overrightharp{m}}\times }\left( \mathbf{%
		\textbf{\overrightharp{l}}\times \textbf{\overrightharp{l}}}_{xxx}\right) \right]$} .  \notag
\end{eqnarray}%

\noindent Clearly despite the fact that $\mathbf{\textbf{\overrightharp{s}}\cdot \textbf{\overrightharp{s}}}=1$, the real and imaginary
components no longer trace out a curve on the unit sphere.

Let us analyse how $\textbf{\overrightharp{m}}$ and $\textbf{\overrightharp{l}}$ behave in this case. As
expected, the trajectories will not stay on the unit sphere. However, for
certain choices of the parameters it is possible to obtain well localised
closed three dimensional trajectories that trace out curves with fixed
points at $t=\pm \infty $ as seen for an example in Figure \ref{Fig8.3}. Thus the nonlocal nature of the solutions to the Hirota equation has apparently
disappeared in the setting of the ELL equation.
However, not all solutions are of this type as some of them are now
unbounded.

\begin{figure}[h]
	\centering
	\includegraphics[width=0.495\linewidth]{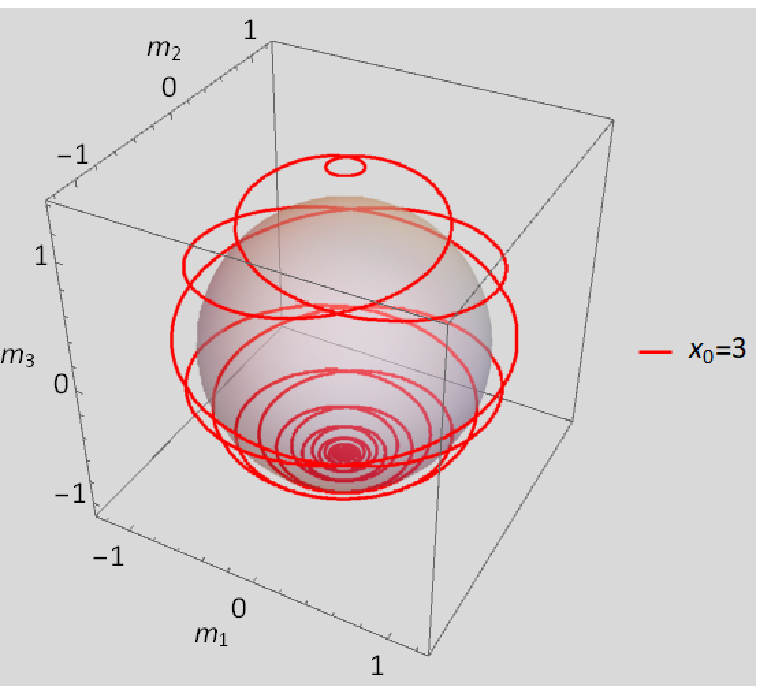}
	\hspace{0cm}
	\includegraphics[width=0.485\linewidth]{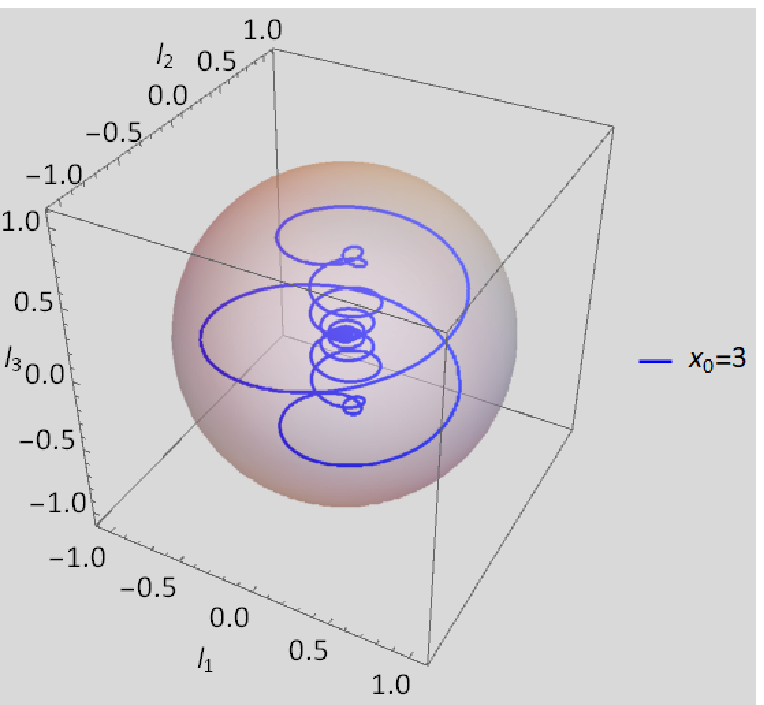}\\
	
	\caption{Nonlocal solutions to the ELLE (\ref{NELL1}) and (\ref{NELL2}) from a gauge equivalent
		nonlocal parity transformed
		conjugate one-soliton solution of the Hirota equation for a fixed value of
		$x_{0}$ with $\gamma_{1} =2.1$, $\gamma_{2} =0$, $\lambda = 0.2$, $\alpha =1$ and $\delta =0.2$.}
	\label{Fig8.3}
\end{figure}

\section{Conclusions}
In this chapter, we took our nonlocal Hirota integrable system, in particular the parity transformed
conjugate pair, to find the gauge equivalent ECH and ELL systems and the corresponding nonlocal soliton solutions. Furthermore, we developed a direct scheme using DCT to find nonlocal multi-soliton solutions of the nonlocal ECH equation making use of nonlocality of the seed solutions, similar in concept as for nonlocal Hirota case. Likewise, taking our new nonlocal ECH soliton solutions, we carried out gauge transformations and found the solution matches the corresponding solution for the Hirota case. Making use of the vector variant of the ECH equation, namely the ELL equation, we are able to observe diagrammatically differences between local and nonlocal solutions.

\chapter{Time-dependent Darboux transformations for non-Hermitian quantum systems}\label{ch_9}

In previous chapters, we have seen that DTs are very efficient tools to construct soliton
solutions of NPDEs, such as for instance the
KdV equation, the SG equation or the Hirota equation. The classic example we have seen is a second order differential
equation of Sturm-Liouville type or time-independent Schr\"{o}dinger
equation. In this context the DT relates
two operators that can be identified as isospectral Hamiltonians. This
scenario has been interpreted as the quantum mechanical analogue of
supersymmetry \cite{witten_dynamical_1981,cooper_supersymmetry_1995,bagrov_darboux_1997}. Many potentials with
direct physical applications may be generated with this technique, such as
for instance complex crystals with invisible defects \cite%
{longhi_invisible_2013,correa_pt-symmetric_2015}.
 
Initially DTs were developed for stationary equations,
so that the treatment of the full time-dependent (TD) Schr\"{o}dinger equation was not possible. Evidently the
latter is a much more intricate problem to solve, especially for
non-autonomous Hamiltonians. Explicitly, DTs for TD Schr\"{o}dinger equation with TD potential was introduced briefly by Matveev and Salle \cite{matveev_darboux_1991} and
subsequently, Bagrov and Samsonov explored the reality condition for the iteration of the potentials \cite{bagrov_supersymmetry_1996}. Generalization to other types of TD systems have also been explored since, \cite{finkel_on_1999,song_generalization_2003,schulze-halberg_darboux_2006,suzko_darboux_2009,tian_analytic_2012}. 

The limitations of the generalization from the
time-independent to the TD Schr\"{o}dinger equation were that
the solutions considered in \cite{bagrov_supersymmetry_1996} force the Hamiltonians involved
to be Hermitian. One of the central purposes of this chapter is to
demonstrate how we can overcome this shortcoming and propose fully TD DTs that deal directly with the TD Schr\"{o}dinger equation involving non-Hermitian
Hamiltonians \cite{cen_time-dependent_2019}, with or without potentials. As an alternative scheme we also discuss the
intertwining relations for Lewis-Riesenfeld invariants for Hermitian as well
as non-Hermitian Hamiltonians. These quantities are constructed as auxiliary
objects to convert the fully TD Schr\"{o}dinger equation into an eigenvalue equation that is easier
to solve and subsequently allows to tackle the TD Schr\"{o}dinger equation. The class of
non-Hermitian Hamiltonians we consider here is the one of $\mathcal{PT}$%
-symmetric/quasi-Hermitian ones \cite{scholtz_quasi-hermitian_1992,bender_making_2007,mostafazadeh_pseudo-hermitian_2010} i.e. they
remain invariant under the antilinear transformation $\mathcal{PT}:$ $%
x\rightarrow -x$, $p\rightarrow p$, $i\rightarrow -i$, that are
related to a Hermitian counterpart by means of the TD Dyson
equation \cite%
{figueira_de_morisson_faria_time_2006,mostafazadeh_time-dependent_2007,znojil_time-dependent_2008,gong_time-dependent_2013,fring_unitary_2016,fring_exact_2017,fring_metric_2018,fring_mending_2017,fring_solvable_2018,mostafazadeh_energy_2018,fring_quasi-exactly_2018}%
.

Given the interrelations of the various quantities in the proposed scheme
one may freely choose different initial starting points. A quadruple of
Hamiltonians, two Hermitian and two non-Hermitian ones, is related by two
TD Dyson equations and two intertwining relations in form of a commutative diagram. This
allows to compute all four Hamiltonians by solving either two intertwining
relations and one TD Dyson equation or one intertwining relations and two TD Dyson equations, with the
remaining relation being satisfied by the closure of the commutative
diagram. We discuss the working of our proposal by taking two concrete
non-Hermitian systems as our starting points, the Gordon-Volkov Hamiltonian
with a complex electric field and a reduced version of the Swanson model.

\section{Time-dependent Darboux and Darboux-Crum transformations}

\subsection{Time-dependent Darboux transformation for Hermitian systems}

Before introducing the TD DTs for
non-Hermitian systems we briefly recall the construction for the Hermitian
setting. This revision will not only establish our notation, but it also
serves to highlight why previous suggestions are limited to the treatment of
Hermitian systems. 

The TD Hermitian standard intertwining relation for potential Hamiltonians introduced in \cite{bagrov_supersymmetry_1996} reads 
\begin{equation}
\ell^{(1)} \left( i\partial _{t}-h_{0}\right) =\left( i\partial _{t}-h_{1}\right)
\ell^{(1)}  ,  \label{HI}
\end{equation}%
where the Hermitian Hamiltonians $h_{0}$ and $h_{1}$ involve explicitly
TD potentials $v_{j}\left( x,t\right) $ 
\begin{equation}
h_{j}\left( x,t\right) =p^{2}+v_{j}\left( x,t\right) ,\qquad j=0,1.
\label{HamiltonianForm}
\end{equation}%
The intertwining operator $\ell^{(1)}  $ is taken to be a first order differential
operator 
\begin{equation}
\ell^{(1)}  \left( x,t\right) =\ell _{0}\left( x,t\right) +\ell _{1}\left(
x,t\right) \partial _{x}.  \label{ll}
\end{equation}%
In general we denote by $\phi _{j}$, the solutions to the two
partner TD Schr\"{o}dinger equations 
\begin{equation}
 i\partial _{t}\phi _{j}=h_{j}\phi _{j}, \qquad j=0,1.
\end{equation}
Throughout this chapter we use the convention $\hbar =1$ and $p=-i \partial_{x}$. Taking $u$ as a particular solution to $[i\partial _{t}-h_{0}]u=0$, the constraints imposed
by the intertwining relation (\ref{HI}) can be solved by%
\begin{equation}
\hspace{-0.05cm}
\resizebox{.9\hsize}{!}{$\ell _{1}\left( x,t\right) =\ell _{1}\left( t\right) ,\quad \ell _{0}\left(
x,t\right) =-\ell _{1}\frac{u_{x}}{u},\quad v_{1}=v_{0}+i\frac{\left( \ell
	_{1}\right) _{t}}{\ell _{1}}+2\left[\left( \frac{u_{x}}{u}\right) ^{2}-\frac{%
	u_{xx}}{u}\right],$}  \label{v1}
\end{equation}%
where, as indicated, $\ell _{1}$ must be an arbitrary function of $t$ only.
At this point the new potential $v_{1}$ might still be complex, however, when one imposes as in \cite{bagrov_supersymmetry_1996}
\begin{equation}
\ell _{1}(t)=\exp \left[ -2\int^{t}\func{Im}\left[\left( \frac{u_{x}}{%
	u}\right) ^{2}-\frac{u_{xx}}{u}\right] ds\right] ,  \label{h1}
\end{equation}%
this forces the new potentials $v_{1}$ to be real
\begin{equation}
v_{1}=v_{0}+2\func{Re}\left[\left(\frac{ u_{x}}{u}\right) ^{2}-\frac{u_{xx}}{u}\right] .
\end{equation}
 Notice that one
might not be able to satisfy (\ref{h1}), as the right-hand side must be
independent of $x$. If the latter is not the case, the partner Hamiltonian $h_{1}$ does not exist. In the case where Hamiltonian $h_{1}$ does exist, the resulting form is
\begin{equation}
	h_{1}=h_{0}+2\func{Re}\left[\left( \frac{ u_{x}}{u}\right) ^{2}-\frac{u_{xx}}{u}\right].
\end{equation} 
However, besides mapping the coefficient functions, the main practical purpose of the DT is that one also obtains exact solutions $\phi _{1}$
for the partner TD Schr\"{o}dinger equation $i\partial_{t}\phi_{1}=h_{1}\phi _{1}$ by employing the intertwining operator. The new solution is computed as
\begin{equation}
\phi_{1}=\ell^{(1)}  \phi_{0}	\quad \text{where} \quad \ell^{(1)} =\ell_{1}(t)(\partial_{x}-\frac{u_{x}}{u}). \label{second}
\end{equation}
When $u$ is linearly dependent on $\phi_{0}$, the solution to the second TD Schr\"{o}dinger equation $i\partial _{t}\phi _{1}=h_{1}\phi _{1}$ becomes trivial, $\phi_{1}=0$. To obtain a non-trivial solution we have seen various ways as presented in earlier chapters, for instance by taking a different spectral parameter, taking other linearly independent solutions or by using Jordan states in the case of the same parameter. The key is to find a solution linearly independent to $\phi_{0}$ that satisfies (\ref{h1}) for reality. In \cite{bagrov_supersymmetry_1996}, some nontrivial solutions satisfying the reality condition were proposed as
\begin{equation}
\widehat{\phi}_{1}=\frac{1}{\ell _{1}\phi_{0}^{\ast }},\qquad \text{and\qquad }\widecheck{%
	\phi}_{1}=\widehat{\phi}_{1}\int^{x}\left\vert \phi_{0}\right\vert ^{2}dy.
\label{ntsol}
\end{equation}%

\subsection{Time-dependent Darboux-Crum transformation for Hermitian systems}

The iteration procedure of the DTs i.e. DCT, will lead also in the TD
case to an entire hierarchy of exactly solvable TD Hamiltonians $%
h_{0}$, $h_{1}$, $h_{2}$, \ldots\ for the TD Schr\"{o}dinger equations $i\partial _{t}\phi
^{(k)}=h_{k}\phi ^{(k)}$ related to each other by intertwining operators $%
\ell ^{(k)}$\ 
\begin{equation}
\ell ^{(k)}\left( i\partial _{t}-h_{k-1}\right) =\left( i\partial
_{t}-h_{k}\right) \ell ^{(k)},\qquad k=1,2,\ldots  \label{iter}
\end{equation}%
Taking $\phi_{0}=\phi_{0}(\gamma _{0})$, a solution of the TD Schr\"{o}dinger equation for $h_{0}$ and the linearly independent solutions $u_{k}=u(\gamma _{k})$ by a choice of different parameter values $\gamma_{k} $ with $k=1,2,\dots,N$, we employ here the Wronskian $W_{N}[u_{1},u_{2},\ldots ,u_{N}]=\det
\omega $ with matrix entries $\omega _{ij}=\partial ^{i-1}_{x}u_{j}$ for $i,j=1,\ldots ,N$, which allows us to write the expressions of the intertwining operator and Hamiltonians in the hierarchy in a very compact form. Iterating these equations we obtain the compact closed form for the intertwining operators 
\begin{equation}
\ell ^{(k)}=\ell _{k}\left( t\right)\left(\partial_{x}-\frac{u^{(k-1)}_{x}}{u^{(k-1)}}\right) \quad \text{where} \quad u^{(k-1)}=\frac{W_{k}[u_{1},u_{2},\ldots ,u_{k}]%
}{W_{k-1}[u_{1},u_{2},\ldots ,u_{k-1}]}  \label{DCa}
\end{equation}%
for $ k=1,2,\ldots,N$. We can in addition, in a compact way, write also the intertwining relation between Hamiltonians $h_{0}$ and $h_{N}$ and their solutions utilising $\mathcal{L} ^{(N)}=\ell^{(N)}\cdots\ell^{(1)}$ as
\begin{equation}
\mathcal{L} ^{(N)} (i \partial_{t}-h_{0})=(i \partial_{t}-h_{N})\mathcal{L} ^{(N)} \quad \text{with} \quad \phi^{(N)}=\mathcal{L} ^{(N)}\phi_{0}.
\end{equation}
The TD Hamiltonians we derive are%
\begin{equation}
h_{N}=h_{0}-2\left[ \ln W_{N}\left( u_{1},u_{2},\ldots ,u_{N}\right) %
\right] _{xx}+i\partial_{t}\left[ \ln \left(\prod_{k=1}^{N}\ell _{k}\right)\right].  \label{hnn}
\end{equation}%
Solutions to the related TD Schr\"{o}dinger $%
i\partial _{t}\phi ^{(N)}=h_{N}\phi ^{(N)}$ are then obtained as%
\begin{eqnarray}
\phi ^{(N)}&=&\mathcal{L} ^{(N)}(\phi_{0}),\\
&=&\left(\ell^{(N)}\cdots\ell^{(1)}\right)\phi_{0},\\
&=& \left(\prod_{k=1}^{N}\ell_{k}\right)\frac{W_{N+1}[u_{1},\ldots ,u_{N},\phi_{0}]%
}{W_{N}[u_{1},\ldots ,u_{N}]}\text{,}  \label{nsol9}
\end{eqnarray}
 The reality condition (\ref{h1}) becomes
\begin{equation}
	\prod_{k=1}^{N}\ell_{k}(t)=e^{2\int \text{Im} \, [\partial_{x}^{2} \ln W_{N}\left(u_{1}\dots u_{N}\right)]dt}.
\end{equation}
For $N=1$, we can match the DT scheme presented in the previous Section 9.1.1 by identifying $u_{1}=u$ and $\phi^{(1)}=\phi_{1}$, which we will use interchangeably in this chapter. 

Again, instead of using the same solution $u_{k}$ of the TD Schr\"{o}dinger equation for $h_{0}$ at
different parameter values in the closed expression, it is also possible to replace some of the solutions $u_{k}$ by other linear independent
solutions at the same parameter values, leading to degeneracy. Closed form expressions for DCT built from the solutions (\ref{ntsol}) can be found in \cite{bagrov_supersymmetry_1996}.
 
\subsection{Darboux scheme with Dyson maps for time-dependent non-Hermitian systems}

Before we extend our Darboux scheme, let us first fix some notation through looking at TD DCT for TD Schr\"{o}dinger equations 
\begin{equation}
i\partial _{t}\psi^{(k)}=H_{k}\psi^{(k)},
\end{equation}
with TD non-Hermitian Hamiltonians $H_{k}$ for $k=0,1,\dots$.

\vspace{0.3cm}

\noindent {\large{\bfseries{Time-dependent Darboux-Crum transformations for non-Hermitian systems}}}

The iteration procedure for the non-Hermitian system goes along the same
lines as for the Hermitian case, albeit with different intertwining
operators $L^{(N)}$. The iterated systems are

\begin{equation}
	L^{(N)}\left(i \partial_{t}-H_{N-1}\right)=\left(i \partial_{t}-H_{N}\right)L^{(N)},\qquad N=1,2,\ldots  \label{H1}
\end{equation}
The intertwining operators read in this case 
\begin{equation}
L^{(N)}=L_{N}\left(\partial_{x}-\frac{U^{(N-1)}_{x}}{U^{(N-1)}}\right) \quad \text{with} \quad U^{(N-1)}=\frac{W_{N}[U_{1},U_{2},\ldots ,U_{N}]%
}{W_{N-1}[U_{1},U_{2},\ldots ,U_{N-1}]}
\end{equation}
denoting $\psi_{0}=\psi_{0}(\gamma _{0})$ and the linearly independent solutions $U_{k}=U(\gamma _{k})$, of the TD Schr\"{o}dinger equation for $H_{0}$ by different parameters $k=1,2,\dots,N$, the TD Hamiltonians are%
\begin{equation}
H_{N}=H_{0}-2\left[ \ln W_{N}\left( U_{1},U_{2},\ldots ,U_{N}\right) %
\right] _{xx}+i\partial_{t}\left[ \ln \left(\prod_{k=1}^{N}L _{k}\right)\right].
\end{equation}
Nontrivial solutions to the related TD Schr\"{o}dinger equation are then obtained as
\begin{eqnarray}
	\psi^{(N)}&=&\mathcal{L}^{(N)}_{H}\psi_{0}, \label{H4}\\
	&=&\left(L^{(N)}\cdots L^{(1)}\right)\psi_{0}, \notag\\
	&=& \left(\prod_{k=1}^{N}L_{k}\right)\frac{W_{N+1}[U_{1},\ldots ,U_{N},\psi_{0}]%
	}{W_{N}[U_{1},\ldots ,U_{N}]}\text{.}\notag
\end{eqnarray}
Note the key difference from the scheme with TD Hermitian systems is that no restrictions are required, as our potentials of interest are no longer restricted to the real case.

\vspace{0.5cm}

Now we extend our analysis and develop here a new Darboux scheme for TD non-Hermitian Hamiltonians, and especially ones that are $\mathcal{PT}$-symmetric/quasi-Hermitian \cite{scholtz_quasi-hermitian_1992,bender_making_2007,mostafazadeh_pseudo-hermitian_2010}, through making use of the TD Dyson equation \cite%
{figueira_de_morisson_faria_time_2006,mostafazadeh_time-dependent_2007,znojil_time-dependent_2008,gong_time-dependent_2013,fring_unitary_2016,fring_exact_2017,fring_metric_2018,fring_mending_2017,fring_solvable_2018,mostafazadeh_energy_2018,fring_quasi-exactly_2018}. This scheme provides a powerful network for the hierarchy of TD Hermitian and TD non-Hermitian systems. 

To illustrate, we focus first on the pairs of TD Hermitian Hamiltonians $h_{0}(t)$, $h_{1}(t)$ and TD non-Hermitian Hamiltonians $H_{0}(t)$, $H_{1}(t)$ 
\begin{equation}
h_{j}=\eta _{j}H_{j}\eta _{j}^{-1}+i\left(\eta _{j}\right)_{t}\eta
_{j}^{-1},\qquad j=0,1.  \label{Dysoneq}
\end{equation}%
The TD Dyson maps $\eta _{j}(t)$ relate the solutions of the
TD Schr\"{o}dinger equation $i\partial _{t}\psi _{j}=H_{j}\psi_{j}$ to the previous ones for $\phi
_{j}$ as%
\begin{equation}
\phi _{j}=\eta _{j}\psi_{j},\qquad j=0,1.
\end{equation}%
Using (\ref{Dysoneq}) in the intertwining relation (\ref{HI}) yields 
\begin{equation}
\ell^{(1)} \left( i\partial _{t}-\eta _{0}H_{0}\eta _{0}^{-1}-i \left(\eta
_{0}\right)_{t}\eta _{0}^{-1}\right) =\left( i\partial _{t}-\eta _{1}H_{1}\eta
_{1}^{-1}-i\left(\eta _{1}\right)\partial _{t}\eta _{1}^{-1}\right) \ell^{(1)} .  \label{aux9}
\end{equation}%
Multiplying (\ref{aux9}) from the left by $\eta _{1}^{-1}$ and acting to the
right with $\eta _{0}$ on both sides of the equation,
\begin{equation}
\hspace{-0.05cm}
\resizebox{.9\hsize}{!}{$\eta _{1}^{-1}\ell^{(1)} \left[ i\partial _{t}-\eta _{0}H_{0}\eta _{0}^{-1}-\left( \eta _{0}\right)_{t}\eta _{0}^{-1}%
\right] \eta_{0}= 
 \eta _{1}^{-1}\left[ i\partial _{t}-\eta _{1}H_{1}\eta
_{1}^{-1}-i\left(\eta _{1}\right) _{t}\eta _{1}^{-1}\right] \ell^{(1)} \eta_{0}$}.
\end{equation}%
and rearranging the time derivative terms and removing the test function, we
derive the new intertwining relation for non-Hermitian Hamiltonians%
\begin{equation}
L^{(1)}\left( i\partial _{t}-H_{0}\right) =\left( i\partial _{t}-H_{1}\right) L^{(1)},
\label{IH}
\end{equation}%
where we introduced the new intertwining operator 
\begin{equation}
L^{(1)}=\eta _{1}^{-1}\ell^{(1)} \eta _{0}.  \label{DarbouxGeneral}
\end{equation}%
We note that $H_{j}-p^{2}$ is in general not only no longer real and might
also include a dependence on the momenta, i.e. $H_{j}$ does not have to be a
natural potential Hamiltonian. In summary, our quadruple of Hamiltonians is
related as depicted in the commutative diagram%
\begin{equation}
\begin{array}{ccccc}
& \fcolorbox{blue}{white}{${\color{blue}{H_{0}}}$} & \underrightarrow{\eta _{0}} & h_{0} &  \\ 
{\color{blue}{L^{(1)}=\eta _{1}^{-1}\ell^{(1)} \eta _{0}}} & {\color{blue}{\vert}} &  & 
\downarrow &\ell^{(1)} \\ 
& \fcolorbox{blue}{white}{${\color{blue}{H_{1}}}$} & \underleftarrow{\eta _{1}^{-1}} & h_{1} & 
\end{array}
\label{DH}
\end{equation}
from a TD non-Hermitian system $H_{0}$ to another, $H_{1}$. An interesting result of this new scheme is that without an explicit solution to $H_{0}$, we can still carry out DT to find another TD non-Hermitian  $H_{1}$. For instance, taking $H_{0}$, we can find a Dyson map $\eta_{0}$ to a Hermitian system $h_{0}$, then carry out DT as in Section 9.1.1 to a new Hermitian system $h_{1}$ and take the second Dyson map $\eta_{1}$ to a non-Hermitian system, $H_{1}$.

One may of course also try to solve the intertwining relation (\ref{IH}) directly as shown with DCT for non-Hermitian Hamiltonians above and build the intertwining operator $L^{(1)}$ from a known solution for the TD Schr\"{o}dinger equation for $H_{0}$ to find $H_{1}$. To make sense of these Hamiltonians one still needs to construct the Dyson maps $\eta _{0}$ and $\eta _{1}$ to find the corresponding Hermitian counterparts $h_{0}$ and $h_{1}$. In the case in which the TD Dyson equation has been solved for $\eta _{0}$, $H_{0}$, $h_{0}$ and $H_{1}$, $h_{1}$ have been constructed with intertwining operators build from
the solutions of the respective TD Schr\"{o}dinger equation, we address the question of whether it is possible to close our diagram for our quadruple of Hamiltonians, that is making it commutative. For this to be possible we require $\eta_{1}=\eta_{0}$. The diagram becomes 
\begin{equation}
\begin{array}{ccccc}
& \fcolorbox{blue}{white}{${\color{blue}{H_{0}}}$} & \underrightarrow{\eta _{0}} & h_{0} &  \\ 
L_{U}^{(1)}=L_{1}\left(\partial_{x}-\frac{U_{1x}}{U_{1}}\right) & \downarrow &  & \downarrow & \ell^{(1)}=\ell_{1}\left(\partial_{x}-\frac{u_{1x}}{u_{1}}\right) \\ 
& \fcolorbox{blue}{white}{${\color{blue}{H_{1}}}$} & {\color{blue}{\underline{{\color{blue}{\eta_{1}=\eta_{0}}}}}} & h_{1} & 
\end{array}
\label{dia}
\end{equation}
It is easy to verify that $L_{U}^{(1)} =\eta _{1}^{-1}\ell^{(1)}\eta
 _{0}$ holds if and only if $\eta _{1}\mathcal{=}\eta _{0}$.

\section{Intertwining relations for Lewis-Riesenfeld invariants}

As previously argued \cite{pedrosa_exact_1997,maamache_pseudo-invariants_2017,fring_solvable_2018,fring_quasi-exactly_2018}, the most
efficient way to solve the TD Dyson equation (\ref{Dysoneq}), as well as the TD Schr\"{o}dinger equation, is to
employ the Lewis-Riesenfeld invariants \cite{lewis_exact_1969}. They are operators $I(t)$ satisfying 
\begin{equation}
	\frac{d I(t)}{d t}\equiv\frac{\partial I(t)}{\partial t}+\frac{1}{i \hbar}\left[I,H\right]=0.
\end{equation}

 The steps in this
approach consists of first solving the evolution equation for the invariants
of the Hermitian and non-Hermitian system separately and subsequently
constructing a similarity transformation between the two invariants. By
construction the map facilitating this transformation is the Dyson map
satisfying the TD Dyson equation.

Here we need to find four TD Lewis-Riesenfeld invariants $I_{j}^{h}(t)$ and $%
I_{j}^{H}(t)$, $j=0,1$, that solve the equations%
\begin{equation}
 \left(I_{j}^{H}\right)_{t}=i\left[ I_{j}^{H},H_{j}\right] ,\quad \text{%
	and\quad }\left(I_{j}^{h}\right)_{t}=i\left[ I_{j}^{h},h_{j}\right] 
\text{.}  \label{LRin}
\end{equation}%
The solutions $\phi _{j}$, $\psi _{j}$ to the respective TD Schr\"{o}dinger equations are
related by a phase factor $ \phi _{j} =e^{i\alpha
	_{j}} \,\,\phi^{I}_{j} $, \,\, $ \psi
_{j} =e^{i\alpha _{j}}\,\, \psi^{I}%
_{j} $ to the eigenstates of the invariants 
\begin{equation}
I_{j}^{h}\,\, \phi^{I}_{j} =\Lambda
_{j}\,\, \phi^{I}_{j}
,\,\,\,\,I_{j}^{H}\,\, \psi^{I}_{j} =\Lambda
_{j}\,\, \psi^{I}_{j} ,\text{\quad with \quad }\dot{%
	\Lambda}_{j}=0.  \label{LR1}
\end{equation}%
Subsequently, the phase factors can be computed from 
\begin{equation}
\dot{\alpha}_{j}=\left\langle \phi^{I}_{j}\right\vert i\partial
_{t}-h_{j}\left\vert \phi^{I}_{j}\right\rangle =\left\langle 
\psi^{I}_{j}\right\vert \eta _{j}^{\dagger }\eta _{j}\left[
i\partial _{t}-H_{j}\right] \left\vert \psi^{I}_{j}\right\rangle .
\end{equation}%
As has been shown \cite{maamache_pseudo-invariants_2017,fring_solvable_2018,fring_quasi-exactly_2018}, the two
invariants for the Hermitian and non-Hermitian system obeying the TD Dyson equation are
related to each other by a similarity transformation 
\begin{equation}
I_{j}^{h}=\eta _{j}I_{j}^{H}\eta _{j}^{-1}\text{.}  \label{simhH}
\end{equation}%
Here we show that the invariants $I_{0}^{H}$, $I_{1}^{H}$ and $I_{0}^{h}$, $%
I_{1}^{h}$ are related by the intertwining operators $L^{(1)}$ in (\ref%
{DarbouxGeneral}) and $\ell^{(1)} $ in (\ref{ll}), respectively. We have 
\begin{equation}
L^{(1)}I_{0}^{H}=I_{1}^{H}L^{(1)},\qquad \text{and}\qquad \ell^{(1)} I_{0}^{h}=I_{1}^{h}\ell^{(1)} .
\label{II}
\end{equation}%
This is seen from computing%
\begin{equation}
i\partial _{t}\left( L^{(1)}I_{0}^{H}-I_{1}^{H}L^{(1)}\right) =H_{1}\left(
L^{(1)}I_{0}^{H}-I_{1}^{H}L^{(1)}\right) -\left( L^{(1)}I_{0}^{H}-I_{1}^{H}L^{(1)}\right) H_{0},
\label{LHH}
\end{equation}%
where we used (\ref{IH}) and (\ref{LRin}) to replace time-derivatives of $L^{(1)}$
and $I_{0}^{H}$, respectively. Comparing (\ref{LHH}) with (\ref{IH}) in the
form $i\partial _{t}L^{(1)}=H_{1}L^{(1)}-L^{(1)}H_{0}$, we conclude that $%
L^{(1)}=L^{(1)}I_{0}^{H}-I_{1}^{H}L^{(1)}$ or $L^{(1)}I_{0}^{H}=I_{1}^{H}L^{(1)}$. The second relation in (%
\ref{II}) follows from the first when using (\ref{DarbouxGeneral}) and (\ref%
{simhH}). Thus schematically the invariants are related in the same manner
as depicted for the Hamiltonians in (\ref{DH}) with the difference that the
TD Dyson equation is replaced by the simpler adjoint action of the Dyson map. Given the
above relations we have no obvious consecutive orderings of how to compute
the quantities involved. For convenience we provide a summary of the above
in the following diagram to illustrate schematically how different
quantities are related to each other

\begin{figure} [h]
	\centering
	\thispagestyle{empty} \setlength{\unitlength}{1.0cm} 
	\resizebox{.8\hsize}{!}{$ \begin{picture}(14.48,9.0)(-2.2,6.5)
	\thicklines
	\put(-0.6,12.0){\LARGE{$H_0  \quad \longleftrightarrow \quad h_0 \quad \longleftrightarrow \quad h_1 \quad \longleftrightarrow \quad H_1$}}
	\put(-0.6,10.0){\LARGE{$I_0^H  \quad \longleftrightarrow \quad I_0^h \quad \longleftrightarrow \quad I_1^h \quad \longleftrightarrow \quad I_1^H$}}
	
	\put(-0.6,8.0){\LARGE{$\psi^{I}_0 \quad \longleftrightarrow \quad \phi^{I}_0 \quad \longleftrightarrow \quad \phi^{I}_1 \quad \longleftrightarrow \,\,\, \psi^{I}_1$}}
	\put(-0.6,14.0){\LARGE{${\psi}_0 \quad \longleftrightarrow \quad {\phi}_0 \quad \longleftrightarrow \quad {\phi}_1 \quad \longleftrightarrow \,\, {\psi}_1$}}
	
	\put(-0.3,11.8){\vector(0,-1){1.2}}	
	\put(3.,11.8){\vector(0,-1){1.2}}
	\put(6.1,11.8){\vector(0,-1){1.2}}
	\put(9.3,11.8){\vector(0,-1){1.2}}
	
	\put(-0.3,9.7){\vector(0,-1){1.1}}	
	\put(3.,9.7){\vector(0,-1){1.1}}
	\put(6.1,9.7){\vector(0,-1){1.1}}
	\put(9.3,9.7){\vector(0,-1){1.1}}
	
	\put(-0.3,12.5){\vector(0,1){1.1}}	
	\put(3.,12.5){\vector(0,1){1.1}}
	\put(6.1,12.5){\vector(0,1){1.1}}
	\put(9.3,12.5){\vector(0,1){1.1}}
	
	\put(1.2,14.6){\LARGE{$\eta_0 $}}
	\put(1.2,12.6){\LARGE{$\eta_0 $}}
	\put(1.2,9.6){\LARGE{$\eta_0 $}}
	\put(1.2,7.6){\LARGE{$\eta_0 $}}
	
	\put(4.5,14.4){\LARGE{$\ell^{(1)} $}}
	\put(4.5,12.4){\LARGE{$\ell^{(1)} $}}
	\put(4.5,9.4){\LARGE{$\ell^{(1)} $}}
	\put(4.5,7.4){\LARGE{$\ell^{(1)} $}}

	\put(7.6,14.6){\LARGE{$\eta_1 $}}
	\put(7.6,12.6){\LARGE{$\eta_1 $}}
	\put(7.6,9.6){\LARGE{$\eta_1 $}}
	\put(7.6,7.6){\LARGE{$\eta_1 $}}
	
	\thicklines
	\put(0.14,11.67){\vector(-3,2){0.2}}
	\put(0.14,10.72){\vector(-3,-2){0.2}}
	\put(9.0,11.68){\vector(3,2){0.2}}
	\put(9.0,10.71){\vector(3,-2){0.2}}
	
	\put(-0.9,13.8){\vector(2,3){0.2}}
	
	\put(9.91,13.8){\vector(-2,3){0.2}}

	\qbezier(0.1, 11.7)(4.8, 9.9)(9.0,11.7)
	\qbezier(0.1, 10.7)(4.8, 12.5)(9.0,10.7)
	\put(4.3,10.9){\LARGE{$L^{(1)}$}}
	
	\qbezier(9.8, 14.0)(11.1, 11.0)(9.8,8.0)
	\put(10.5,11.0){\LARGE{$\alpha_1 $}}
	
	\qbezier(-0.8, 14.0)(-2.1, 11.0)(-0.8,8.0)
	\put(-2.2,11.0){\LARGE{$\alpha_0 $}}
	
	\end{picture} $}
	
	\caption{ Schematic representation
		of Dyson maps $\eta _{0}$,$\eta _{1}$ and intertwining operators $\ell^{(1)} $,$L^{(1)}$
		relating quadruples of Hamiltonians $h_{0}$,$h_{1}$,$H_{0}$,$H_{1}$ and
		invariants $I_{0}^{h}$,$I_{1}^{h}$,$I_{0}^{H}$,$I_{1}^{H}$ together with
		their respective eigenstates $\phi _{0}$,$\phi _{1}$,$\psi _{0}$,$\psi _{1}$
		and $\phi^{I}_{0}$,$\phi^{I}_{1}$,$\psi^{I}_{0}$,$\psi^{I}_{1}$ that are related by phases $\alpha _{0}$,$\alpha _{1}$.}
	\label{fig9.0}
\end{figure}
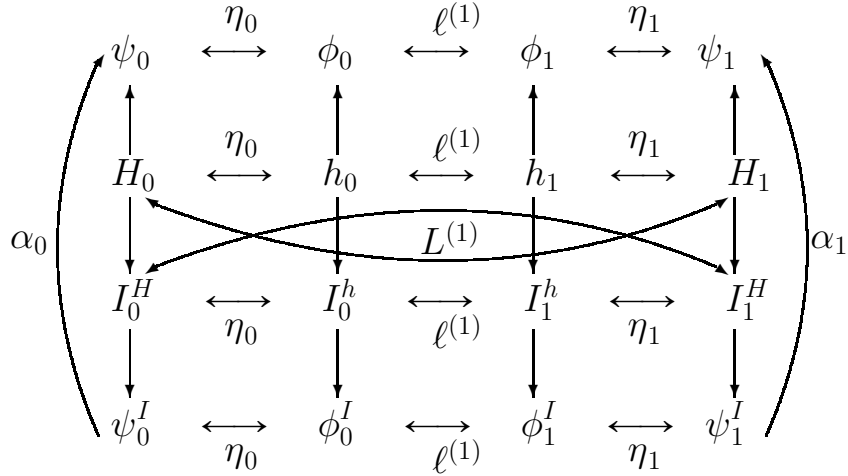

\section{Solvable time-dependent trigonometric potentials from the complex Gordon-Volkov Hamiltonian}

We will now discuss how the various elements in Figure \ref{fig9.0} can be computed.
Evidently the scheme allows to start from different quantities and compute
the remaining ones by following different indicated paths, that is we may
solve intertwining relations and TD Dyson equation in different orders for different
quantities. As we are addressing here mainly the question of how to make
sense of non-Hermitian systems, we always take a non-Hermitian Hamiltonian $%
H_{0}$ as our initial starting point and given quantity. Subsequently we
solve the TD Dyson equation (\ref{Dysoneq}) for $h_{0}$,$\eta _{0}$ and thereafter close
the commutative diagrams in different ways.

We consider a complex version of the Gordon-Volkov Hamiltonian \cite{gordon_comptoneffekt_1926,volkov_class_1935}
\begin{equation}
H_{0}=H_{GV}=p^{2}+iE\left( t\right) x,
\end{equation}%
in which $iE\left( t\right) \in i\mathbb{R}$ may be viewed as a complex
electric field. In the real setting $H_{GV}$ is a Stark Hamiltonian with
vanishing potential term around which a perturbation theory can be build in
the strong field regime, see e.g. \cite{figueira_de_morisson_faria_analytical_1999}. Such type of potentials are
also of physical interest in the study of plasmonic Airy beams in linear
optical potentials \cite{liu_plasmonic_2011}. Even though the Hamiltonian $H_{GV}$ is
non-Hermitian, it belongs to the interesting class of $\mathcal{PT}$%
-symmetric Hamiltonians, i.e. it remains invariant under the antilinear
transformation $\mathcal{PT}:$ $x\rightarrow -x$, $p\rightarrow p$, $%
i\rightarrow -i$.

In order to solve the TD Dyson equation (\ref{Dysoneq}) involving $H_{0}$ we make the ansatz 
\begin{equation}
\eta _{0}=e^{\alpha \left( t\right) x}e^{\beta \left( t\right) p},
\label{e0}
\end{equation}%
with $\alpha \left( t\right) $, $\beta \left( t\right) $ being some
TD real functions. The adjoint action of $\eta _{0}$ on $x$, $p$
and the TD term of Dyson equation form are easily computed to be
\begin{eqnarray}
\eta _{0}x\eta _{0}^{-1}&=&x-i\beta ,\qquad\\
 \eta _{0}p^{2}\eta _{0}^{-1}&=&p^{2}+2i\alpha p-\alpha^{2}
,\quad\\
 i\dot{\eta}_{0}\eta _{0}^{-1}&=&i\dot{\alpha}x+i\dot{\beta}\left(
p+i\alpha \right) .
\end{eqnarray}%
We use now frequently overdots as an abbreviation for partial derivatives
with respect to time. Therefore the right-hand side of the TD Dyson equation (\ref%
{Dysoneq}) yields 
\begin{equation}
h_{0}=h_{GV}=p^{2}+ip\left( 2\alpha +\dot{\beta}\right) -\alpha
^{2}+ix\left( E+\dot{\alpha}\right) +E\beta -\dot{\beta}\alpha .
\end{equation}%
Thus, for $h_{0}$ to be Hermitian we have to impose the reality constraints 
\begin{equation}
\dot{\alpha}=-E,\quad \dot{\beta}=-2\alpha ,
\end{equation}%
so that $h_{0}$ becomes a free particle Hamiltonian with an added real
TD field 
\begin{equation}
h_{0}=h_{GV}=p^{2}+\alpha ^{2}+E\beta =p^{2}+\left[ \int^{t} E%
\left( s\right) ds\right] ^{2}+2E\left( t\right)
\int^{t}\int^{s} E\left( \tau\right) d\tau ds.
\label{freeparticle}
\end{equation}%
There are numerous solutions to the TD Schr\"{o}dinger equation $i\partial _{t}\phi_{0}=h_{GV}\phi
_{0}$, with each of them producing different types of partner potentials $%
v_{1}$ and hierarchies. We will discuss below an example using a trigonometric type solution.

We start by considering the scenario as depicted in the commutative diagram (\ref{dia}). Thus we start with a solution to the TD Dyson equation in form of $h_{0}$, $%
H_{0}$, $\eta _{0}$ as given above and carry out the intertwining relations
separately using the intertwining operators $\ell^{(1)} $
and $L_{U}^{(1)} $ for the construction of $h_{1}$ and $H_{1}$%
, respectively. As indicated in the diagram (\ref{dia}), in this scenario, the expression for the
second Dyson map is dictated by the closure of the diagram to be $\eta
_{1}=\eta _{0}$. 

We construct our intertwining operator from the simplest
solutions to the TD Schr\"{o}dinger equation for $h_{0}=h_{GV}$%
\begin{equation}
\phi _{0}\left( m\right) =\cos (mx)e^{-im^{2}t-i\int^{t}(\alpha
	^{2}+E\beta) ds}  \label{fm}
\end{equation}%
with continuous parameter $m$. A second linearly independent solution $\phi_{0}\left( m\right)$ could be obtained by replacing the $\cos $ in (\ref{fm}) by $\sin $. However, for our iteration, we take a second linearly independent solution by replacing the continuous parameter $m$ with a different one, $m_{1}$ to obtain $u_{1}=\phi _{0}\left( m_{1}\right)$, then we compute
\begin{eqnarray}
	\hspace{-1cm}
\ell^{(1)} &=&\partial_{x}-\frac{(u_{1})_{x}}{u_{1}} =\ell _{1}\left( t\right) \left[
\partial _{x}+m_{1}\tan (m_{1}x)\right] ,\\
\hspace{-1cm}
h_{1} &=&p^{2}+2m_{1}^{2}\func{sec}^{2}(m_{1}x)+\alpha ^{2}+E\beta +i\frac{\left(
	\ell _{1}\right) _{t}}{\ell _{1}} ,\\
\hspace{-1cm}
\phi_{1} &=&\ell^{(1)}\phi _{0},\\
\hspace{-1cm}
&=&\resizebox{.7\hsize}{!}{$\ell _{1}(t) \left[ m_{1}\cos (mx)\tan (m_{1}x)-m\sin
(mx)\right] e^{-i\left[m^{2}t+\int^{t}(\alpha ^{2}+E\beta) ds\right]}$}. \notag
\end{eqnarray}%
Evidently $\ell _{1}(t)$ must be constant for $h_{1}$ to be Hermitian, so
for convenience we set $\ell _{1}(t)=1$. We can also directly solve
the intertwining relation (\ref{IH}) for $H_{0}$ and $H_{1}$ using an intertwining operator built from a solution for the TD Schr\"{o}dinger equation of $H_{0}$, i.e. $L_{U}^{(1)} =\partial_{x}-\frac{(U_{1})_{x}}{U_{1}}$, where $U_{1}=\eta^{-1}_{0}u_{1}$ to obtain%
\begin{eqnarray}
H_{1} &=&p^{2}+iE\left(
t\right) x+2m_{1}^{2}\func{sec}^{2}\left[ m_{1}(x+i\beta )\right] , \\
\psi _{1} &=&\eta_{0}^{-1}\phi_{1}=e^{-\alpha (x+i\beta )}\phi _{1}(x+i\beta,t ).
\end{eqnarray}%
We verify that the TD Dyson equation for $h_{1}$ and $H_{1}$ is solved by $\eta _{1}=\eta
_{0}\,$, which is enforced by the closure of the diagram (\ref{dia}).

We can extend our analysis to the DCT and compute
the two hierarchies of solvable TD trigonometric Hamiltonians $%
H_{0} $,$H_{1}$,$H_{2}$,$\ldots $ and $h_{0}$,$h_{1}$,$h_{2}$,$\ldots $
directly from the expressions (\ref{DCa})-(\ref{H4}). For instance, we
calculate%
\begin{equation}
H_{2}=p^{2}+iE\left( t\right) x+\frac{(m_{1}^{2}-m_{2}^{2})\left[\left(m_{1}^{2}-m_{2}^{2}\right) +m_{1}^{2}\cos (2m_{2}\xi%
	)-m_{2}^{2}\cos (2m_{1}\xi)\right] }{\left[ m_{1}\cos (m_{2}\xi%
	)\sin (m_{1}\xi)-m_{2}\cos (m_{1}\xi)\sin (m_{2}\xi)\right]
	^{2}}
\end{equation}%
with $\xi=x+i\beta $. The solutions to the corresponding TD Schr\"{o}dinger equation are
directly computable from the general formula (\ref{H4}).

\section{Reduced Swanson model hierarchy}

Next we consider a model that is built from a slightly more involved TD Dyson map. We proceed as outlined in the commutative diagram (\ref{DH}). This is a good example to show the power of our new Darboux iteration scheme for a TD non-Hermitian system, where an explicit solution to the system is not needed to perform the iteration. Our simple starting point is a non-Hermitian, but $\mathcal{PT}$-symmetric, Hamiltonian that may be viewed as reduced version of the well-studied Swanson model \cite{swanson_transition_2004} 
\begin{equation}
H_{0}=H_{RS}=ig\left( t\right) xp.
\end{equation}%
We follow the same procedure as before and solve at first the TD Dyson equation for $\eta
_{0}$ and $h_{0}$ with given $H_{0}$. In this case the arguments in the
exponentials of the TD Dyson map can no longer be linear and we make the ansatz 
\begin{equation}
\eta _{0}=e^{\lambda \left( t\right) xp}e^{\frac{\zeta \left( t\right)}{2} p^{2}}.
\label{eta0}
\end{equation}%
The right-hand side of the TD Dyson equation (\ref{Dysoneq}) is then computed to be
\begin{equation}
h_{0}=h_{RS}=\left[ \left( g\zeta +i\frac{\dot{\zeta}}{2}\right) \cos
(2\lambda )+\left( ig\zeta -\frac{\dot{\zeta}}{2}\right) \sin (2\lambda )%
\right] p^{2}+i(g+\dot{\lambda})xp.
\end{equation}%
Thus for $h_{0}$ to be Hermitian we have to impose 
\begin{equation}
\dot{\lambda}=-g,\quad \dot{\zeta}=-2g\zeta \tan 2\lambda .  \label{const}
\end{equation}%
These reality
constraints (\ref{const}) can be solved by 
\begin{equation}
\lambda (t)=-\int^{t} g\left( s\right) ds,\quad \text{and \quad }\zeta
(t)=\sec \left( 2\int^{t} g\left( s\right) ds\right) ,
\end{equation}%

\noindent so that we obtain a free particle Hamiltonian with a TD mass $%
m(t)$%
\begin{equation}
h_{0}=h_{RS}=\frac{1}{2m(t)}p^{2},\qquad \text{with \quad }m(t)=\frac{\cos^{2}(2\lambda )}{2g}.  \label{freep}
\end{equation}%
TD masses have been proposed as a possible mechanism to explain
anomalous nuclear reactions which cannot be explained by existing
conventional theories in nuclear physics, see e.g. \cite{davidson_variable_2015}.  

An exact solution
to the TD Schr\"{o}dinger equation for $h_{RS}$ can be found for instance in \cite{pedrosa_exact_1997}
when setting in there the TD frequency to zero%
\begin{eqnarray}
\hspace{-1cm}
\phi_{0}\left(n\right) \!\!\!&=&\!\!\!\frac{e^{i\alpha _{0}(n,t)}}{%
	\sqrt{\varrho }}\exp \left[ m\left( i\frac{\dot{\varrho}}{%
	\varrho }-\frac{1}{m\varrho ^{2}}\right) \frac{x^{2}}{2}\right]
\mathcal{H}_{n}\left[ \frac{x}{\varrho }\right] ,  \label{Ped} \\
\alpha _{0}(n,t) \!\!\!&=&\!\!\! -\dint\nolimits\frac{\left( n+\frac{1}{2}\right) }{%
	m\varrho ^{2}}dt,  \label{al}
\end{eqnarray}%
where $\mathcal{H}_{n}\left[x\right]$, denotes the Hermite polynomials of $x$. For (\ref{Ped}) to be a solution, the auxiliary function $\varrho =\varrho(t)$ needs
to obey the dissipative Ermakov-Pinney equation with vanishing linear term, that is 
\begin{equation}
\ddot{\varrho}+\frac{\dot{m}}{m}\dot{\varrho}=\frac{1}{m^{2}\varrho ^{3}}.
\label{EP2}
\end{equation}%
We derive an explicit solution for this equation in Appendix C. Evaluating
the formulae in (\ref{v1}), with $h_{0}$ and $h_{1}$ divided by $2m(t)$, we
obtain the intertwining operators and the partner Hamiltonians 
\begin{eqnarray}
\hspace{-0.5cm}
\ell ^{(1)}(s) \!\!\!&=&\!\!\!\ell _{1} \left[\partial _{x}+ \frac{x}{\varrho ^{2}}-%
\frac{2s\mathcal{H}_{s-1}\left[ \frac{x}{\varrho} \right] }{\varrho \mathcal{H}_{s}\left[ \frac{x}{\varrho}  %
	\right] }-im\frac{\dot{\varrho}}{\varrho }x\right] ,
\label{ln} \\
\hspace{-0.5cm}
h_{1}(s) \!\!\!&=&\!\!\!\resizebox{.8\hsize}{!}{$h_{0}+\frac{4s}{m\varrho ^{2}}\left[ \frac{s\mathcal{H}_{s-1}^{2}\left[
	\frac{x}{\varrho} \right] -(s-1)\mathcal{H}_{s-2}\left[ \frac{x}{\varrho} \right] \mathcal{H}_{s}\left[
	\frac{x}{\varrho} \right] }{\mathcal{H}_{s}^{2}\left[ \frac{x}{\varrho} \right] }\right] +\frac{1}{%
	m\varrho ^{2}}+i\partial_{t}\left[\ln \frac{\ell_{1}}{\varrho}\right] ,$}  \notag  \label{hn}
\end{eqnarray}%
respectively. As in the previous section, the imaginary part of the
Hamiltonian only depends on time and can be made to vanish with the
choice $\ell _{1}=\varrho $. For concrete values of $n$ we obtain for
instance the TD Hermitian Hamiltonians 
\begin{eqnarray}
\hspace{-1cm}
h_{1}(0) &=&\frac{p^{2}}{2m}+\frac{1}{m\varrho ^{2}},\quad\! h_{1}(1)=h_{1}(0)+%
\frac{1}{mx^{2}},\quad\! \resizebox{.3\hsize}{!}{$ h_{1}(2)=h_{1}(0)+\frac{4(\varrho ^{2}+2x^{2})}{%
	m(\varrho ^{2}-2x^{2})^{2}},$} \\
\hspace{-1cm}
h_{1}(3) &=&h_{1}(0)+\frac{3(3\varrho ^{4}+4x^{4})}{m(2x^{3}-3x\varrho
	^{2})^{2}},\quad\! \resizebox{.45\hsize}{!}{$h_{1}(4)=h_{1}(0)+\frac{8\left( 9\varrho
	^{6}-12x^{4}\varrho ^{2}+18x^{2}\varrho ^{4}+8x^{6}\right) }{m\left(
	3\varrho ^{4}-12x^{2}\varrho ^{2}+4x^{4}\right) ^{2}}$}.
\end{eqnarray}%
Notice that all these Hamiltonians are singular at certain values of $x$ and 
$t$ as $\varrho $ is real. Solutions to the TD Schr\"{o}dinger equation for the Hamiltonian $%
h_{1}(s) $ can be computed for $s\neq n$, taking
\begin{equation}
u_{1}=\phi_{0}(s) \quad \text{and} \quad	\ell^{(1)}(s)=\ell_{1}\left(\partial_{x}-\frac{[\phi_{0}(s)]_{x}}{\phi_{0}(s)}\right),
\end{equation}
then according to (\ref{second})%
\begin{eqnarray}
\phi _{1}\left(s,n\right) &=&\ell^{(1)}(s)\left[\phi_{0}(n)\right]\\
&=&\frac{1}{\sqrt{n-s}}\left[ \frac{n\mathcal{H}_{n-1}\left[ \frac{x}{\varrho} \right] }{\mathcal{H}_{n}%
	\left[ \frac{x}{\varrho} \right] }-\frac{s\mathcal{H}_{s-1}\left[ \frac{x}{\varrho} \right] }{\mathcal{H}_{s}%
	\left[ \frac{x}{\varrho} \right] }\right] \phi_{0}(n). \notag
\end{eqnarray}%
Both $\phi_{0}(n)$ and $\phi_{1}(s,n)$ are square integrable
functions with $L^{2}(\mathbb{R})$-norm equal to $1$. In Figure \ref{fig9.4}
we present the computation for some typical probability densities obtained
from these functions. Notice that demanding $m(t)>0$ we need to impose some
restrictions for certain choices of $g(t)$.

\begin{figure}[h]
	\centering
	
	\includegraphics[width=0.51\linewidth]{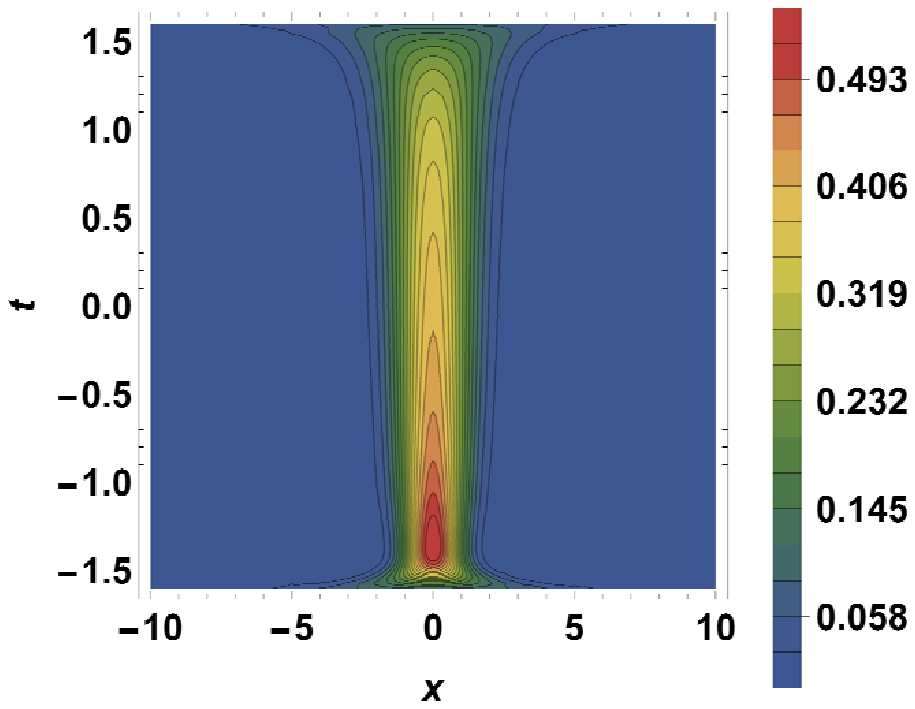} 
	\includegraphics[width=0.48\linewidth]{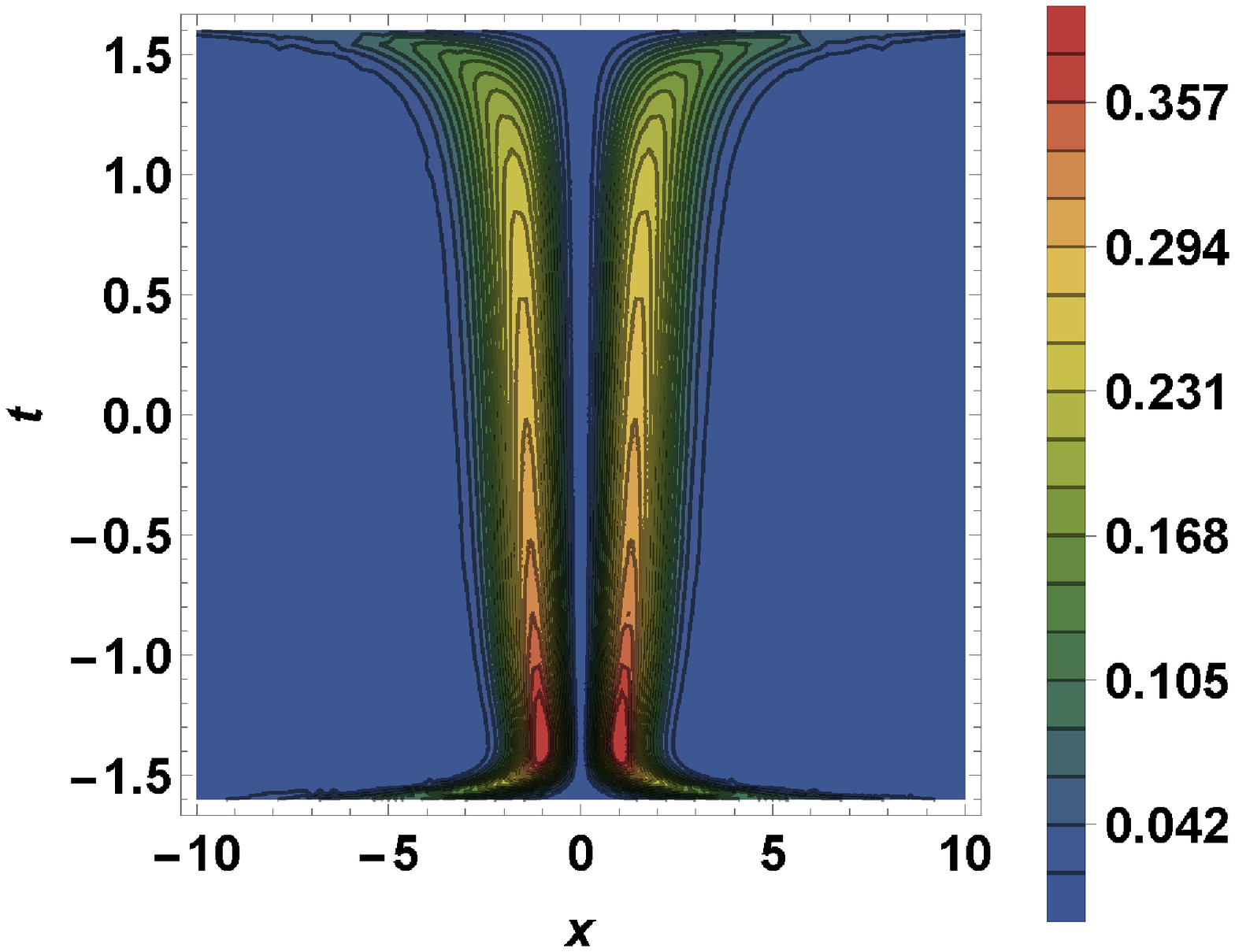}
	\includegraphics[width=0.48\linewidth]{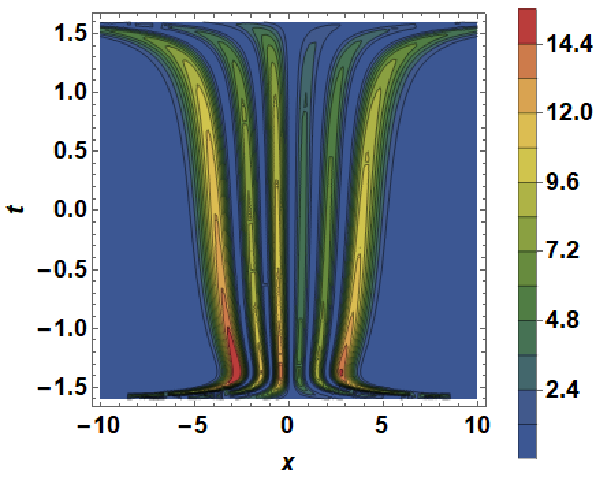}\\
	
	\caption{Probability densities $\left\vert \phi _{0}(0)\right\vert ^{2}$, $%
		\left\vert \phi _{0}(1)\right\vert ^{2}$, $\left\vert \phi_{1}(5,1)\right\vert ^{2}$ from left to right for $g(t)=(1+t^{2})/4$, $%
		m(t)=\left[ 1+\cos (t+t^{3}/3)\right] /(1+t^{2})$, $\varrho (t)=\sqrt{%
			1+[C+B\tan (t/2+t^{3}/6)]^{2}}$ with $B=1/2$ and $C=1$. }
	\label{fig9.4}
\end{figure}

Next we compute the non-Hermitian counterpart $H_{1\text{ }}$with a concrete
choice for the second Dyson map. Taking 
\begin{equation}
\eta _{1}=e^{\gamma(t) x}e^{\delta(t) p} \label{eta1}
\end{equation}
the non-Hermitian Hamiltonian becomes in our case
\begin{equation}
H_{1}(1)=\frac{p^{2}}{2m}+\frac{1}{m(x+i\delta )^{2}}-i\dot{\gamma}x+\frac{1}{%
	m\varrho ^{2}}-\frac{\gamma ^{2}}{2m}+\dot{\gamma}\delta ,
\end{equation}%
where we have also imposed the constraint $\dot{\delta}=-\gamma /m$ to
eliminate a linear term in $p$, hence making the Hamiltonian a natural potential
one. The corresponding solution for the TD Schr\"{o}dinger equation is%
\begin{equation}
\psi_{1}(s,n)=\eta _{1}^{-1}\phi_{1}(s,n).
\end{equation}%

\subsection{Lewis-Riesenfeld invariants}

Having solved the TD Dyson equation for $\eta _{0}$ and $\eta _{1}$ we can now also
verify the various intertwining relations for the Lewis-Riesenfeld
invariants as derived in Section 9.2. We proceed here as depicted in the
following commutative diagram 
\begin{equation}
\begin{array}{ccccc}
& \fcolorbox{blue}{white}{${\color{blue}{I_{0}^{H}}}$} & \underleftarrow{\eta _{0}^{-1}} & I_{0}^{h} &  \\ 
{\color{blue}{L^{(1)}=\eta_{1}^{-1}\ell^{(1)}\eta_{0}}} & {\color{blue}{\vert}} & & \downarrow & \ell^{(1)} \\ 
& \fcolorbox{blue}{white}{${\color{blue}{I_{1}^{H}}}$} & \underleftarrow{\eta _{1}^{-1}} & I_{1}^{h} & 
\end{array}
\label{di2}
\end{equation}%
See also the more general schematic representation in Figure \ref{fig9.0}. We start with the Hermitian invariant $I_{0}^{h}$ from which we compute the
non-Hermitian invariant $I_{0}^{H}$ using the Dyson map $\eta _{0}$ as
specified in (\ref{eta0}). Subsequently we use the intertwining operator $\ell^{(1)}$ in (\ref{ln}) to compute the Hermitian invariants $%
I_{1}^{h}$ for the Hamiltonians $h_{1}$. The invariant $I_{1}^{H}$ is
then computed from the adjoint action of $\eta _{1}^{-1}$ as specified in (\ref{eta1}). Finally, the intertwining relation between the non-Hermitian
invariants $I_{0}^{H}$ and $I_{1}^{H}$ is just given by the closure of the
diagram (\ref{di2}), $L^{(1)}$.

The invariant for the Hermitian Hamiltonian $h_{0}$ has been computed
previously in \cite{pedrosa_exact_1997}\footnote{%
	We corrected a small typo in there and changed the power $1/2$ on the $%
	x/\rho $-term into $2$.} as%
\begin{equation}
I_{0}^{h}=A_{h}(t)p^{2}+B_{h}(t)x^{2}+C_{h}(t)\{x,p\},
\end{equation}%
where the TD coefficients are%
\begin{equation}
A_{h}=\frac{\varrho ^{2}}{2},\quad B_{h}=\frac{1}{2}\left( \frac{1}{\varrho
	^{2}}+m^{2}\dot{\varrho}^{2}\right) ,\quad C_{h}=-\frac{1}{2}m\varrho \dot{%
	\varrho}.
\end{equation}%
It then follows from%
\begin{equation}
\resizebox{.9\hsize}{!}{$\left[ I_{0}^{h},h_{0}\right] =\frac{2i}{m}\left( C_{h}p^{2}+\frac{1}{2}%
	B_{h}\{x,p\}\right) ,\quad \dot{A}_{h}=-\frac{2}{m}C_{h},\quad \dot{B}%
	_{h}=0,\quad \dot{C}_{h}=-\frac{1}{m}B_{h},$}
\end{equation}%
that the defining relation (\ref{LRin}) for the invariant is satisfied by $%
I_{0}^{h}$. According to the relation (\ref{simhH}), the non-Hermitian
invariant $I_{0}^{H}$ for the non-Hermitian Hamiltonian $H_{0}$ is simply
computed by the adjoint action of $\eta _{0}^{-1}$ on $I_{0}^{h}$. Using the
expression (\ref{eta0}) we obtain%
\begin{equation}
I_{0}^{H}=\eta _{0}^{-1}I_{0}^{h}\eta
_{0}=A_{H}(t)p^{2}+B_{H}(t)x^{2}+C_{H}(t)\{x,p\},
\end{equation}%
with%
\begin{equation}
\resizebox{.9\hsize}{!}{$A_{H}=\frac{1}{2}e^{-2i\lambda }\rho ^{2}-\zeta ^{2}B_{H}-i\zeta m\rho \dot{%
		\rho},\quad B_{H}=\frac{e^{2i\lambda }\left( 1+m^{2}\rho ^{2}\dot{\rho}%
		^{2}\right) }{2\rho ^{2}},\quad C_{H}=i\zeta B_{H}-\frac{1}{2}m\rho \dot{\rho%
	}.$}
\end{equation}%
We verify that $I_{0}^{H}$ is indeed an invariant for $H_{0}$ according to
the defining relation (\ref{LRin}), by computing%
\begin{equation}
\resizebox{.9\hsize}{!}{$\left[ I_{0}^{H},H_{0}\right] =2g\left( A_{H}p^{2}-B_{H}x^{2}\right) ,\quad 
	\dot{A}_{H}=2igA_{H},\quad \dot{B}_{h}=-2igB_{H},\quad \dot{C}_{H}=0,$}
\end{equation}%
using the constraints (\ref{const}) and (\ref{EP2}).

Given the intertwining operators $\ell^{(1)}$ in (\ref{ln}) and the
invariant $I_{0}^{h}$, we can use the intertwining relation (\ref{II}) to
compute the invariants $I_{1}^{h}$ for the Hamiltonians $h_{1}$ in (\ref%
{hn}). Solving (\ref{II}) we find%
\begin{equation}
I_{1}^{h}(s)=I_{0}^{h}+1+4s^{2}\frac{\mathcal{H}_{s-1}^{2}\left[ \frac{x}{\varrho }\right] ^{2}%
}{\mathcal{H}_{s}^{2}\left[ \frac{x}{\varrho } \right] ^{2}}-4s(s-1)\frac{\mathcal{H}_{s-2}\left[
	\frac{x}{\varrho } \right] }{\mathcal{H}_{s}^{2}\left[ \frac{x}{\varrho } \right] }.
\end{equation}%
We verify that this expression solves (\ref{LRin}). The last invariant in
our quadruple is%
\begin{equation}
I_{1}^{H}(s)=\eta _{1}^{-1}I_{1}^{h}(s,x,p)\eta
_{1}=I_{1}^{h}(s,x+i\delta ,p-i\gamma ).
\end{equation}%
Finally we may also verify the eigenvalue equations for the four invariants.
Usually this is of course the first consideration as the whole purpose of
employing Lewis-Riesenfeld invariants is to reduce the TD Schr\"{o}dinger equation to the much
easier to solve eigenvalue equations. Here this computation is simply a
consistency check. With%
\begin{eqnarray}
\phi_{0}^{I}\left(n\right) &=&e^{-i\alpha _{0}\left(n\right)}\phi _{0}\left(n\right),\quad 
	\phi_{1}^{I}\left(s,n\right)=e^{-i\alpha _{0}\left(s\right)}\phi _{1}\left(s,n\right),\\
\psi_{0}^{I}\left(n\right)&=&e^{-i\alpha _{0}\left(n\right)}\psi _{0}\left(n\right),\quad 
	\psi_{1}^{I}\left(s,n\right)=e^{-i\alpha _{0}\left(s\right)}\psi_{1}\left(s,n\right),
\end{eqnarray}%
and $\alpha _{0}\left(n\right)$ as specified in equation (\ref{al}) we compute 
\begin{eqnarray}
\hspace{-1cm}
I_{0}^{h}\phi_{0}^{I}\left(n\right) \!\!\!&=&\!\!\!\left( n+\frac{1}{2}\right) \phi%
_{0}\left(n\right)\text{,}  \quad I_{1}^{h}\left(s\right)\phi_{1}^{I}\left(s,n\right)=\left( s+\frac{1}{2}\right) 
\phi_{1}\left(s,n\right) \text{,} \\
\hspace{-1cm}
I_{0}^{H}\psi_{0}^{I}\left(n\right) \!\!\!&=&\!\!\!\left( n+\frac{1}{2}\right) \psi%
_{0}\left(n\right)\text{,}  \quad I_{1}^{H}\left(s\right)\psi_{1}^{I}\left(s,n\right)=\left( s+\frac{1}{2}\right) 
\psi_{1}\left(s,n\right) \text{.}
\end{eqnarray}%
All eigenvalues are time-independent as shown in \cite{pedrosa_exact_1997}.

\section{Conclusions}
We have generalized the scheme of TD DTs to
allow for the treatment of non-Hermitian Hamiltonians that are $\mathcal{PT}$%
-symmetric/quasi-Hermitian. It was essential to employ intertwining
operators different from those used in the Hermitian scheme previously
proposed. We have demonstrated that the quadruple of Hamiltonians, two
Hermitian and two non-Hermitian ones, can be constructed in alternative
ways, either by solving two TD Dyson equations and one intertwining relation or by
solving one TD Dyson equation and two intertwining relations.
We extended the scheme to the construction of the
entire TD Darboux-Crum hierarchies. We also showed that the
scheme is consistently adaptable to construct Lewis-Riesenfeld invariants by
means of intertwining relations. Here we verified this for a concrete system
by having already solved the TD Schr\"{o}%
dinger equation, however, evidently it should also be
possible to solve the eigenvalue equations for the invariants first and
subsequently construct the solutions to the TD Schr\"{o}dinger equation.

\chapter{Conclusions and outlook}\label{ch_10}

Given that we have summarised in each chapter the key findings, here we will give a more general overview of the results and contributions from this thesis both in nonlinear classical and non-Hermitian quantum systems, whilst discussing some of many interesting open problems.

In mathematics, it has been known for a long time that extending real numbers to complex numbers gives us a deeper insight of the real domain. In quantum mechanics, difficult problems in the Hermitian regime sometimes become easier to solve in the non-Hermitian regime. In addition, with the success of extensions from Hermitian to non-Hermitian with $\mathcal{PT}$-symmetries in quantum mechanics, we are motivated to explore the analogy with classical nonlinear integrable systems. 

In particular, we extended real integrable systems to the complex and multicomplex regimes through extending the solution field. As a result, we also obtain equivalent systems of multi-coupled real equations. Up to now, we have three ways to solve these systems, by taking a complex shift, using combined imaginary unit or idempotent bases. This has solved the origin of $\mathcal{PT}$-symmetric complex solutions from \cite{khare_novel_2015}, however it remains an open problem to investigate the origin of new $\mathcal{PT}$-symmetric complex solutions from \cite{khare_novel_2016}. Besides the many interesting properties, a valuable application is that the newly constructed complex solitons helped regularise singularities that form when taking degeneracy in the real regime. Moreover, in the multicomplex regime, a new type of degeneracy appears as we discover some $2N$-parameter $N$-soliton solutions. 

In addition to generalising some well-established methods to construct complex and multicomplex soliton solutions, another fascinating discovery is that the newly found complex solitons solutions, although complex, they admit real conserved charges. Through detailed scattering and asymptotic analysis for lateral displacements or time-delays, we find the reasoning for guaranteed reality is due to  $\mathcal{PT}$-symmetry with integrability. Hence for all multi-solitons, $\mathcal{PT}$-symmetric or not, as long as they are composed of $\mathcal{PT}$-symmetric one-soliton solutions in the asymptotes, then reality is guaranteed. For multi-complex extensions, we have increased variety of $\mathcal{PT}$-symmetries to play with, but arguments are similar.

A more challenging extension would be to complexify and multi-complexify also
the variables $x$ and $t$, which then also impacts on the definition of the derivatives with
respect to these variables.

Extending scattering and asymptotic analysis for degenerate multi-soliton solutions, we find further interesting physically different properties from non-degenerate cases, such as time-delays not being constant but TD expressions and a universal general form is found for the KdV and SG cases. For the Hirota case, we compare non-degenerate and degenerate scatterings behaviours to find the degenerate case only admits one of three types of scattering behaviour from the non-degenerate case. In the investigations of ways to implement degeneracy in various methods, we derived for the SG case the simplest and most convenient method to construct degenerate multi-soliton solutions by means of a 'recursive' formula. Besides degenerate solitons possessing similar properties to the famous tidal bore phenomenon, another physical application which is left open to investigate is the statistical behaviour of a degenerate soliton gas along the lines of, for instance \cite{gupta_investigation_1976,mertens_soliton-gas_1981,sasaki_soliton-breather_1986,shurgalina_nonlinear_2016}, which should certainly exhibit different characteristics as the underlying statistical distributions would be based on indistinguishable rather than distinguishable particles.

With the growing recent interest to investigate classical integrable nonlocal systems, with nonlocality of space and or time in the fields of the system, we followed \cite{ablowitz_integrable_2013} to investigate another type of $\mathcal{PT}$-symmetric deformations. These are $\mathcal{PT}$-symmetric reductions of the AKNS equations. Furthermore, we developed new methods to construct 'nonlocal' solutions. In the process, discovering a new type of solution, which is a $2N$ parameter nonlocal $N$-soliton solution. From the different space and or time nonlocalities we see various nonlocal periodic distributed breathers or rogue waves, of particular interest are time crystal like structures from being nonlocal in time. To see this type of nonlocality implemented in the quantum regime would be interesting.

In our construction of various types of complex, multicomplex, degenerate and nonlocal systems, we define many of them to be integrable in the sense that we have various methods to construct soliton solutions for them. In some cases, we have also found the system possessing infinitely many conserved quantities or ZC representations. More thorough investigation on integrability can be carried out in many cases, for example to find infinite many conserved quantities or local commuting symmetries for all our systems. 

Another natural direction to investigate are combinations of various types of extensions for systems, such as implementing nonlocality in multicomplex systems, degeneracy in nonlocal systems etc. Furthermore, as our approaches are entirely model independent to extend investigations for other models. With the many newly discovered systems and solutions, the most interesting challenge is to investigate whether these solutions can be realised experimentally.

In a step closer to the quantum regime, we look at gauge equivalence of the nonlocal Hirota system to an ECH spin model, then to an ELL model. When independently developing nonlocal solution method for ECH equation, an unanticipated finding is that the spin matrix has an internal pair of nonlocal $\mathcal{P}$-symmetry and itself nonlocal $\mathcal{P}$-symmetric. Utilising gauge equivalence between the models allowed us to find the connection between the corresponding solutions of the systems.

Knowing how DTs are useful tools in constructing new solvable systems from previous chapters on soliton constructions, we extended the application of DTs with Dyson maps to develop a powerful new scheme in which we have a fully connected network of infinitely many solvable TD non-Hermitian Hamiltonians with corresponding infinitely many solvable Hermitian Hamiltonians and Lewis-Riesenfeld invariants. This network is powerful especially for TD non-Hermitian Hamiltonians, in which exact solutions are usually difficult to find. 

As in \cite{bagrov_supersymmetry_1996} for the Hermitian case,
our scheme allows to treat TD systems directly instead of having
to solve the time-independent system first and then introducing time by
other means. The latter is not possible in the context of the Schr\"{o}%
dinger equation, unlike as in the context of NPDEs that admit soliton solutions, where a time-dependence can be
introduced by separate arguments, such as for instance using Galilean
invariance. Naturally it will be very interesting to apply our scheme to the construction of
multi-soliton solutions.


\begin{appendices}
\chapter{}
Here, we present a derivation of the identity (\ref{Id1}). We start by
considering the limit for $N=2$, which is true as 
\begin{eqnarray}
& &	\lim_{\beta \rightarrow \alpha }\frac{\beta+\alpha  }{\beta-\alpha  }\tan
	\left( \frac{\phi _{\beta }-\phi _{\alpha }}{4}\right)  \label{a1}\\
	&=&\lim_{\beta\rightarrow \alpha}\left[\tan \left( \frac{\phi _{\beta}-\phi _{\alpha }}{4}\right)+\frac{\alpha+\beta}{4}\sec^{2}\left(\frac{\phi_{\beta}-\phi_{\alpha}}{4}\right)\frac{d}{d\beta}\phi_{\beta}\right],\\
	&=&\frac{\alpha }{2}\frac{d}{d\alpha }\phi _{\alpha }.  
\end{eqnarray}

Using this expression
the non-degenerate and degenerate two-soliton solutions may be written as%
\begin{equation}
\hspace{-0cm}
	\phi _{\alpha \beta }\!=\!-4\arctan \left[ \frac{ \beta+\alpha }{ \beta-\alpha }%
	\tan \left( \frac{\phi _{\beta }-\phi _{\alpha }}{4}\right) \right] \,\,\,\,
	\text{and \,\,\,\,}\phi _{\alpha \alpha }\!=\!-4\arctan \left[ \frac{\alpha }{2}%
	\frac{d}{d\alpha }\phi _{\alpha }\right] ,  \label{ndegdeg}
\end{equation}%
respectively.

Before carrying the next step of our derivation for higher order $N$, we first introduce an intermediate identity
\begin{equation}
	\lim_{\beta \rightarrow \alpha }\frac{d }{d\beta }\phi _{\alpha^{m} \beta}=\frac{1}{2m} \frac{d }{d\alpha }\phi _{\alpha^{m+1}}. \label{A4}
\end{equation}%
This identity can be proved by induction. To begin, let us first show the identity is true for $m=1$ by carrying out the derivative $\beta$ then taking the limit 
\begin{eqnarray}
	\lim_{\beta \rightarrow \alpha }\frac{d}{d\beta}\phi _{\alpha \beta }&=&\lim_{\beta \rightarrow \alpha }\frac{d}{d\beta}\left[-4 \arctan \left(\frac{\beta+\alpha}{\beta-\alpha}\tan\frac{\phi_{\beta}-\phi_{\alpha}}{4}\right)\right],\\
	&=&\frac{-2\frac{d}{d\alpha}\phi_{\alpha}-2\alpha\frac{d^{2}}{d\alpha^{2}}\phi_{\alpha}}{2+\frac{\alpha^{2}}{2}\left(\frac{d}{d\alpha}\phi_{\alpha}\right)^{2}},\\
	&=&\frac{1}{2}\frac{d}{d\alpha}\phi_{\alpha\alpha}.
\end{eqnarray}%
Now assuming $m=k-1$ is true, $m=k$ will be
\begin{eqnarray}
\lim_{\beta \rightarrow \alpha }\frac{d}{d\beta}\phi _{\alpha^{k} \beta }
&=&\lim_{\beta \rightarrow \alpha }\left[\frac{-2\frac{d}{d\beta}\phi_{\alpha^{k-1}\beta}-2\alpha\frac{d^{2}}{d\beta^{2}}\phi_{\alpha^{k-1}\beta}}{2+\frac{\alpha^{2}}{2}\left(\frac{d}{d\beta}\phi_{\alpha^{k-1}\beta}\right)^{2}}\right],\\
&=&\frac{-2\frac{d}{d\alpha}\left[\frac{\phi_{\alpha^{k}}}{2(k-1)}\right]-2\alpha\frac{d^{2}}{d\alpha^{2}}\left[\frac{\phi_{\alpha^{k}}}{2(k-1)}\right]}{2+\frac{\alpha^{2}}{2}\left(\frac{d}{d\alpha}\left[\frac{\phi_{\alpha^{k}}}{2(k-1)}\right]\right)^{2}},\\
&=&\frac{1}{2k}\frac{d}{d\alpha}\phi_{\alpha^{k+1}},
\end{eqnarray}
where we used the identity for $m=k-1$ from the first to second equality. Hence identity (\ref{A4}) proved. 

Now we can continue our derivation of (\ref{Id1}) for $N \geq 3$, starting with the left hand side of (\ref{Id1})
\begin{eqnarray}
\hspace{-1cm}
	&&\lim_{\beta\rightarrow \alpha}\frac{\beta +\alpha}{\beta-\alpha}%
	\tan \left[ \frac{\phi _{\alpha^{N-2} \beta}-\phi _{\alpha^{N-1} }}{4}\right],\\
\hspace{-1cm}
	\!\!\!&=&\!\!\!\frac{\alpha}{2}\lim_{\beta \rightarrow \alpha } \frac{d}{d\beta}\phi_{\alpha^{N-2}\beta},\\
\hspace{-1cm}
	\!\!\!&=&\!\!\!\frac{\alpha }{2}\left[\frac{1}{2(N-2)}\frac{d}{d\alpha }\phi _{\alpha
		^{N-1}}\right] , \\
\hspace{-1cm}
	\!\!\!&=&\!\!\!\frac{\alpha }{2(N-1)}\frac{d}{d\alpha }\phi _{\alpha^{N-1} }  ,
\end{eqnarray}%
using the identity
(\ref{A4}) from the first to second equality, which consequently simplifies to the right hand side of (\ref{Id1}).
\chapter{}

Identity (\ref{axt}) is a simple variable transformation based on the
assumption that $\phi _{\alpha^{N} }(x,t)$ can always be expressed as $\phi
_{\alpha^{N} }(\xi _{+},\xi _{-})$. So we compute%
\begin{eqnarray}
	\frac{\partial \phi _{\alpha^{N} }}{\partial x} &=&\alpha \frac{\partial \phi
		_{\alpha^{N} }}{\partial \xi _{+}}-\alpha \frac{\partial \phi _{\alpha^{N} }}{%
		\partial \xi _{-}},  \label{i1} \\
	\frac{\partial \phi _{\alpha^{N} }}{\partial t} &=&\frac{1}{\alpha }\frac{%
		\partial \phi _{\alpha^{N} }}{\partial \xi _{+}}+\frac{1}{\alpha }\frac{%
		\partial \phi _{\alpha^{N} }}{\partial \xi _{-}},  \label{i2} \\
	\frac{\partial \phi _{\alpha^{N} }}{\partial \alpha } &=&(x-\frac{t}{\alpha ^{2}%
	})\frac{\partial \phi _{\alpha^{N} }}{\partial \xi _{+}}-(x+\frac{t}{\alpha ^{2}%
	})\frac{\partial \phi _{\alpha^{N} }}{\partial \xi _{-}}.  \label{i3}
\end{eqnarray}%
Comparing (\ref{i1}), (\ref{i2}) and (\ref{i3}) we can eliminate the
derivatives with respect to $\xi _{+}$, $\xi _{-}$ and obtain (\ref{axt}).

\chapter{}

We briefly explain how to solve the Ermakov-Pinney equation with dissipative
term (\ref{EP2})%
\begin{equation}
	\ddot{\varrho}+\frac{\dot{m}}{m}\dot{\varrho}=\frac{1}{m^{2}\varrho ^{3}}.
	\label{C1}
\end{equation}%
The solutions to the standard version of the equation \cite{ermakov_transformation_1880,pinney_nonlinear_1950} 
\begin{equation}
	\ddot{ \sigma }+\lambda (t)\sigma =\frac{1}{\sigma ^{3}}  \label{EPtime}
\end{equation}%
are well known to be of the form \cite{pinney_nonlinear_1950}%
\begin{equation}
	\sigma (t)=\left( Au^{2}+Bv^{2}+2Cuv\right) ^{1/2},  \label{EPsol}
\end{equation}%
with $u(t)$ and $v(t)$ denoting the two fundamental solutions to the
equation $\ddot{\sigma}+\lambda (t)\sigma =0$ and $A$, $B$, $C$ are
constants constrained as $C^{2}=AB-W^{-2}$ with Wronskian $W=u\dot{v}-v\dot{u%
}$. The solutions to the equation (\ref{EPtime}) with an added dissipative term
proportional to $\dot{\sigma}$ are not known in general. However, the
equation of interest here, (\ref{C1}), which has no linear term in $\varrho$, may
be solved exactly. For this purpose we assume $\varrho (t)$ to be of the form%
\begin{equation}
	\varrho (t)=f[q(t)],\qquad \ \ \ \text{with }q(t)=\dint\nolimits^{t}\frac{1}{%
		m(s)}ds.  \label{fq}
\end{equation}%
Using this form, equation (\ref{C1}) transforms into%
\begin{equation}
	\frac{d^{2}f}{dq^{2}}=\frac{1}{f^{3}},
\end{equation}%
which corresponds to (\ref{EPtime}) with $\lambda (t)=0$. Taking the linear
independent solutions to that equation to be $u(t)=1$ and $v(t)=q$, we obtain%
\begin{equation}
	f(q)=\frac{\pm 1}{\sqrt{B}}\sqrt{1+(Bq+C)^{2}}
\end{equation}%
and hence with (\ref{fq}) a solution to (\ref{C1}).

\end{appendices}


\end{document}